**UNIVERSITÀ DEGLI STUDI DI MILANO**

**DIPARTIMENTO DI INFORMATICA**
**“Giovanni Degli Antoni”**

## Corso di Dottorato in Informatica

**Settore Scientifico-Disciplinare: INF/01**

**XXXIV Ciclo**

# Music Interpretation Analysis

***A Multimodal Approach to Score-Informed Resynthesis of Piano Recordings***

*Federico Simonetta*

**Tutor**: *Prof. Stavros Ntalampiras*

**Co-Tutor**: *Prof. Federico Avanzini*

**Coordinatore del corso**: *Prof. Paolo Boldi*

A.A. 2020/2021

Laboratorio di Informatica Musicale

# Abstract

This Thesis discusses the development of technologies for the automatic resynthesis of music recordings using digital synthesizers. First, the main issue is identified in the understanding of how Music Information Processing (MIP) methods can take into consideration the influence of the acoustic context on the music performance. For this, a novel conceptual and mathematical framework named "Music Interpretation Analysis" (MIA) is presented. In the proposed framework, a distinction is made between the "performance" – the physical action of playing – and the "interpretation" – the action that the performer wishes to achieve. Second, the Thesis describes further works aiming at the democratization of music production tools via automatic resynthesis: 1) it elaborates software and file formats for musicological archives and multimodal machine-learning datasets; 2) it explores and extends MIP technologies; 3) it presents the mathematical foundations of the MIA framework and shows preliminary evaluations to demonstrate the effectiveness of the proposed approach.

*There is no scientific work*
*which only one man can write*
G. Galilei, “Life of Galileo”
B. Brecht

# Acknowledgments[1,2]

Not many thanks should be given for this work.

One special thank for making my Ph.D. possible should be given to Trenord delays, which were a remarkable presence; hopefully, delay cancellation methods are still not enough accurate for train-related applications.

Another relevant contribution to this Thesis came from the scientific publishers. The biggest opportunity for any research project is the free access to platforms such as Sci-H... ahem Scopus. Thank you Elsevier, thank you Clarivate!

I should not forget the funding received for this Ph.D. It is mandatory to say thank to the manifold ministers that committed themselves to the flowering of the public scientific research. They were undoubtedly fundamental for the abundant funds received for this work.

Not least, this and all the research activities in Europe should say thank to the CPR, expulsions, and special methods reserved by the EU police forces for the other human races. Thanks to you, I can write these lines today.

[1] I really thank anyone that I have met during the last 3 years and that has encouraged me towards this unforgivable experience. Of course, a huge, enormous thank must go to my two supervisors, Federico and Stavros, whose help, patience, and constancy have been fundamental for every single sentence in this Thesis. If I had been in them, I would not have bet anything on me working on this project, but they did, and – I can say now – they were right. However, they were not the only ones that professionally encouraged me. Many thanks to all the LIM people; the LIM have been my professional home for 3 years and has continuously pushed me towards experimenting and discovering new music-computing stuffs. A special thank to the three reviewers that worked on this Thesis as well. You provided so many important advice, helping me make this work much better.
A few careful words for people that everyday put up with me and still stay there. The friends who stay there everyday or that show up once per year, but still do. My family, of course, included those 4 little "birbanti", and the others that will come in the next years. Giulia, who is a small dot on this new white paper.
You all have contributed to this little thing. I don't know what it is, but I feel it's something. It's that you are there, between the lines, and it makes me feel good.

f

[2] A special thank must also be spent for Modartt and NVidia who provided me a Pianoteq license and GPUs free of charges; these tools were fundamental for the main achievements of my Ph.D.

# Preface

This work originated from my 3-years Ph.D. at the LIM - Laboratorio di Informatica Musicale of the University of Milan under the supervision of proff. Stavros Ntalampiras and Federico Avanzini. The original project with which the Ph.D. was started was not completely outlined, and only towards the end of the first year I realized the main long-term application that underlies the Thesis. It was only during the third year that the concept of interpretation became totally clear.

The Ph.D. was started in October 2018 and should have ended in December 2021, but due to the COVID-19 pandemic crisis, the end of the h.D. was postponed to April 2022.

A few scientific contributions were born during the doctoral studies.

**(P1)** Federico Simonetta, Stavros Ntalampiras, and Federico Avanzini. "Multimodal Music Information Processing and Retrieval: Survey and Future Challenges". In: *Proceedings of 2019 International Workshop on Multilayer Music Representation and Processing*. Int. Work. on Multilayer Music Representation and Processing. Milan, Italy: IEEE Conference Publishing Services, 2019, pp. 10–18

**(P2)** Federico Simonetta, Carlos Cancino-Chacón, Stavros Ntalampiras, and Gerhard Widmer. "A Convolutional Approach to Melody Line Identification in Symbolic Scores". In: *20th Int. Conf. on Music Information Retrieval Conference (ISMIR)*. 2019

**(P3)** Luca Andrea Ludovico, Adriano Baratè, Federico Simonetta, and Davide Andrea Mauro. "On the Adoption of Standard Encoding Formats to Ensure Interoperability of Music Digital Archives: The IEEE 1599 Format." In: *6th International Conference on Digital Libraries for Musicology*. ACM, Nov. 2019, pp. 20–24

**(P4)** Federico Simonetta, Stavros Ntalampiras, and Federico Avanzini. "ASMD: An Automatic Framework for Compiling Multimodal Datasets with Audio and Scores". In: *Proceedings of the 17th Sound and Music Computing Conference*. Torino, 2020

**(P5)** Federico Simonetta, Stavros Ntalampiras, and Federico Avanzini. "Audio-to-Score Alignment Using Deep Automatic Music Transcription". In: *Proceeddings of the IEEE MMSP 2021*. 2021

**(P6)** Federico Simonetta, Federico Avanzini, and Stavros Ntalampiras. "A Perceptual Measure for Evaluating the Resynthesis of Automatic Music Transcriptions". In: *Multimed. Tools Appl.* (2022)

**(P7)** Federico Simonetta. "Towards Faithful Automatic Music Resynthesis". In: *Comput. Music J.* (2022)

Part I is composed by **(P1)**, that has been entirely used in Chapter 1, and **(P7)**, that is the main source for Chapter 2.

Part II has been created by reusing materials from publications **(P3)**, **(P4)**, and **(P5)**. Specifically, Chapter 4 has been built around publication **(P4)** and updated with content from **(P5)**, while publication **(P3)** has inspired Chapter 3.

Part III is composed by Chapter 6, that is mainly modeled around publication **(P5)**, and Chapter 5, that has been written by using publication **(P2)**.

Finally, Part IV is written by using publication **(P6)** for Chapter 7. Chapter 8 is an original contribute that is planned to be submitted for publication during 2022.

# Contents

# Introduction

## Problem formulation and main contributions

This Thesis discusses the development of score-informed technologies for the automatic faithful re-synthesis of music recordings using modern computer-controlled instruments or digital synthesizers based on physical models, concatenative synthesis, or neural networks. While working on such problem, I realized that it required the resolution of a second connected task, namely the understanding of how Music Information Processing (MIP) methods can take into consideration the influence of the acoustic context on the music performance. These two problems have a particular interesting application in a third field: the restoration of degraded music recordings, especially with the objective of democratizing music production tools. Modern mobile devices, indeed, allow people to record music with inexpensive transducers that produce low quality data in respect to the expensive professional technology. The latter task worked as main motivation for the work behind this Thesis, while the first two problems were considered as the main challenges to be faced.

Manifold issues arise when considering the proposed problems: on one side, there is the need for archiving, preserving and retrieving music documents; then, there exist several technological gaps in both the analysis and synthesis of music, which are still prone to significant errors when a fully automated process is considered; finally, a major problem is the ethical issue concerning the possible interference of the restoration on the original artistic intention of the performer.

Accordingly, this work proceeds in three directions. First, it elaborates and proposes software and file formats for historical music archiving and multimodal machine-learning datasets. Second, it explores the technological limits of Music Performance Analysis and digital synthesis by analyzing and extending Multimodal MIP technologies with the aim of improving traditional MIP tasks by exploiting multiple information sources. Third, it proposes a methodological perspective to deal with the artistic content of a music recording. Altogether, the Thesis proposes a novel framework that I refer to with “Music Interpretation Analysis” (MIA). MIA is both a conceptual and mathematical framework able to describe the transformation of the sound signal during the analysis of a music performance and its faithful resynthesis. In the proposed framework, accordingly with existing research, a distinction is made between the “performance”, that corresponds to the physical acts that take place while a musician plays, and the “interpretation”, that instead corresponds to the acts that the performer wishes to achieve. I think that the originality of the latter direction is worth of giving the main title to the Thesis.

## Structure of the Thesis

Given the novelty of the proposed application to the music computing field, a large overview of every connected problem is needed in order to obtain a complete understanding of all problem's dimensions. This is the aim of Part I, which surveys the existing literature and highlights the principal issues to be addressed. First, I focus on Multimodal MIP, with a particular attention on multimodal approaches with the intention of understanding how multiple modalities – e.g. audio and score – can restrict the number of errors occurring in monomodal methods. Chapter 1 provides a broad overview and proposes the very first definition of the field. Then, Chapter 2 presents an overview of the main basic concepts for audio resynthesis and restoration. Various approaches are discussed in both their ethical and technical implications, and a first introduction to the MIA framework is proposed. Possible solutions to the context-aware automatic resynthesis problem are analyzed. Since the most promising approach seems to be the resynthesis of the output of Automatic Music Transcription (AMT) models, the rest of the Thesis mainly focuses on AMT-based resynthesis.

Part II focuses on the problem of archiving music documents. We believe that an effort in the standardization of music representation deriving from multiple and potentially heterogeneous sources is of paramount importance for many categories of users ranging from music enthusiasts to musicians and musicologists, while including the Music Information Retrieval (MIR) community. In this sense, the IEEE 1599 format, as explained in this Thesis, represents the perfect match, as it provides the ability to collect and represent in a synchronized way various kinds of information related to a single music piece within a multi-layer environment. Chapter 3 focuses on how IEEE 1599 can be exploited in a distributed infrastructure to connect different sources and information modalities for the sake of merging the scientific efforts of different teams without breaking copyright rules. Chapter 4, then, discusses the archiving of machine-learning datasets and proposes a novel framework to ease the development of multimodal MIP technologies. The framework, named *Audio and Score Meta Dataset [4] (ASMD)*, is written in Python and based on a JSON data format. It allows compiling, distributing and using multimodal music datasets including audio and score information.

Part III deals with the attempt of extending existing MIP technologies. It proposes two real-world applications that allow to explore multimodal MIP from two different perspectives. The first work with the aim of enhancing MIP technologies on which I committed during the doctoral studies is presented in Chapter 5. Here, the source-separation problem is tackled in the symbolic domain and a method for melody separation in music scores is presented. The method can be used to assist multimodal melody separation in the audio domain, serving as guidance to any related application; as explained in the Chapter, AMT can also be seen as a source-separation task and thus the proposed method can help AMT models as well. The second work in this Part is in Chapter 6, where I propose an audio-to-score alignment method that improves the state-of-art for piano music by leveraging feature extraction performed via AMT. The proposed method allows to infer the onset and offset of all the notes annotated in a music score, while disregarding possible pitch errors of AMT models. In other words, the proposed audio-to-score method can be used by a musicologist to control the process of music transcription by conditioning the AMT model on the music pitches really played in the recording and disregarding the pitch errors produced in the transcription process.

Finally, in Part IV, the MIA framework is studied from a narrower perspective. Chapter 7 presents an attempt to perceptually evaluate the limit of the present technologies both in terms of transcription and synthesis. In this Chapter, the concept of "interpretation" as opposed to the one of "performance" is perceptually evaluated from the listener perspective with a precise methodology that allows to optimally cover the entire space of possible note combinations. The outcomes of the test are that 1) the usual format for music performance representations (*Standard MIDI Format*) is not able to grasp the artistic content of music performance, and that 2) existing AMT systems are not perceptually effective when the recording and re-synthesis acoustic contexts are different. It is in Chapter 8 that the concept is formalized within a mathematical groundwork that allows to define rigorous approaches for future studies. There, the MIA theoretical framework is presented and it is shown that AMT models can be improved if the acoustic context is considered during training. The framework will work as basis for further experiments and application developments.

# Part I.

# Filling the literature gaps

# 1
# Multimodal MIP

Multimodal Music Information Processing (MIP) is an emerging field that concerns the exploitation of multiple source of information for music processing tasks. In this chapter, a general review about Multimodal MIP is presented.

The chapter is organized as follows: in Section 1.1 we present the motivations that lead us to study Multimodal MIP; in Section 1.2, we give some basic definition and discuss previous reviews on similar topics to explain categorization and the taxonomy we used. Sections 1.3 to 1.7 describe the different tasks faced with multimodal approaches, the various features extracted the preprocessing steps and the fusion approaches adopted in literature; in Section 1.8 we express our idea about how the multimodal paradigm can be enhanced.

## 1.1. Introduction

Beginning with the oldest evidence of music notation, music has been described in several form [8]. Such descriptions have been used by computational systems for facilitating music information computing tasks. Interestingly, when observing the history of music, at least in the Western culture, one can see how the various descriptive forms have gradually emerged with a strict dependence both on technology advancements and changes in music practices.

Initially, no written description systems for music existed besides text. Between the 6th-7th cen., Isidore of Seville, Archbishop and theologian, wrote that no melody could be written. Indeed, the first systems to memorize music were based solely on lyrics and only later some signs over the words appeared. Such notation, called *neumatic*, evolved in more complex forms, which differed from region to region. Due to the need of more powerful tools to express music features, new notation systems, called *pitch specific*, took place, such as the *alphabetic* and the *staff*-based notations. In particular, the system introduced by Guido d'Arezzo (10th-11th cen.) was particularly successful and similar conventions spread all over Europe. Music notation was now able to represent text, pitches and durations at the same time. During the following centuries, other types of symbols were introduced addressing directly the performer towards peculiar colors or sentiments. At the crossing of the 16th and 17th cen., Opera was born in Italy, after a long tradition of plays, including Greek drama, medieval entertainers and renaissance popular plays (both liturgic and profane) [9]. The tremendous success of the Opera in Italy and then in the rest of Europe, determined a fundamental way to connect music and visual arts for the future centuries. A turning point in the history of music description systems was the invention of the *phonograph cylinder* by Thomas Edison in 1877 and the *disc phonograph* diffused by Emile Berliner ten years later [10]. In the same years, Edison and the Lumière brothers invented the first devices to record video [11]. Since then, a number of technologies were born paving the way for new music description systems. With the invention of computers and the beginning of the digital era, the elaboration of sound signals highlighted the need for more abstract information characterizing audio recordings. Thus, researchers started proposing *mid-level* representation [12], with reference to *symbolic* and *physical* level [13]. Nowadays, the availability of vast, easily accessible quantities of data, along with appropriate modern computational technologies, encourages the collection of various types of *meta-data*, which can be either *cultural* or *editorial* [14].

From a cognitive point of view, the connecting, almost evolutionary, element between the above-mentioned representations is that each one relates to a different abstraction level. Psychology, indeed, is almost unanimous in identifying an abstraction process in our music cognition [15]: we can recognize music played on different instruments, with different timings, intensity changes, various metronome markings, tonalities, tunings, background noises and so on. The different descriptions of music developed by the western culture in different era or contexts can be seen as an answer to the necessity of representing new modalities – such as the visual one – or new unrevealed abstraction levels – such as the audio recordings and the mid-symbolic levels, or the pitch specific notation compared to the neumatic one.

Aside from these historical and cognitive considerations, it is a fact that in the last two decades researchers have obtained better results through multimodal approaches in respect to single-modalities approache [16, 17]. As Minsky said [18]:

> *To solve really hard problems, we'll have to use several different representations.*

It can be argued that music processing tasks can benefit profoundly from multimodal approaches, and that a greater focus is needed by the research community in creating such a synergistic framework. A fundamental step would be the study and design of suitable algorithms through which different modalities can collaborate. Then, a particular effort should be devoted in developing the needed technologies. In fact, given the course of history summarized above, we could expect that in the future, new disparate music representations will be born.

## 1.2. Definitions, taxonomy and previous reviews

We have found no univocal definition of modality. In the music computing literature, authors use the word *multimodal* in two main contexts:

- in computational psychology, where *modality* refers to a human sensory channel;
- in music information retrieval, where *modality* usually refers to a source of music information;

Since we are focusing on music information retrieval methods, to the purpose of the present paper, with *modality* we mean a specific way to digitize music information. Different modalities are obtained through different transducers, in different places or times, and/or belong to different media. Examples of modalities that may be associated to a single piece of music include audio, lyrics, symbolic scores, album covers, and so on.

Having defined what we mean by modality, we define *multimodal music information processing* as an MIR [19] approach which takes as input multiple modalities of the same piece of music. All the papers which we are going to discuss show methods which take as input various music representations. Conversely, we are not considering those approaches which exploit features derived through different methods from the same modality: an example is pitch, rhythmic and timbral features, when they are all derived from the audio [20]. Similarly we are not considering approaches which process multiple representations of the same modality: an example is spectrograms (treated as 2D images) and traditional time-domain acoustic features [21], which are both derived from the audio. The reason for this choice is to analyze how music representations born for different use-cases than MIP can be effectively exploited in computer-based technologies. Such an effort is coherent with the cognitive perspective illustrated in Section 1.1 and aims at representing music in a way that is able to grasp not only the sound but also other aspects that effectively constitute the music phenomenon. The underlying idea is that, if music *often* deals with sound, it *always* includes other modalities as well.

Moreover, we do not focus on general multimodal sound processing: the idea which moves our effort is that music is characterized by the *organization* of sounds in time; thus, we are interested in exploiting this organization, which is not available in general sound processing.

One previous review on multimodal music processing was written in 2012 [22]. However, that work was more focused on a few case studies rather than on an extensive survey. The authors recognized a distinction between "the effort of characterizing the *relationships* between the different modalities", which they name *cross-modal processing*, and "the problem of efficiently combining the information conveyed by the different modalities", named *multimodal fusion*. To our analysis, this distinction is useful if with *cross-modal processing* we mean the end-user systems that offer an augmented listening experience by providing the user with additional information. If this is the case, we are primarily interested in *multimodal fusion*; nevertheless, some synchronization algorithms that are classified as *cross-modal processing* by the previous authors [22] are used as preprocessing steps in other works. Because of this ambiguous distinction, we base our classification on the performed task rather than on the processing stage – see Section 1.3.

Almost all authors dealing with multimodal information fusion talk about two approaches: *early fusion* and *late fusion*. Figure 1.1 shows the main difference between the two approaches: in *early fusion*, data is used "as is" in one single processing algorithm which fuse the data representation, while in *late fusion* data from each modality is first processed with specific algorithms and then all the outputs are merged, so that it is the output to be fused and not the data. Because of this, *early fusion* is also called *feature-level fusion*, and *late fusion* is also called *decision-level fusion*, even if *hybrid fusion* for multimedia analysis, but we have found no example in the music domain.

Finally, we have found useful to introduce a new diagram to represent the data flow in retrieval systems – see Figure 1.2. Indeed, in most of these systems, one modality is used to query a database for retrieving another modality; in such cases, no fusion exists, but just a data conversion and a similarity computation.

An exhaustive table, which summarizes all the works reviewed in this chapter, is available on-

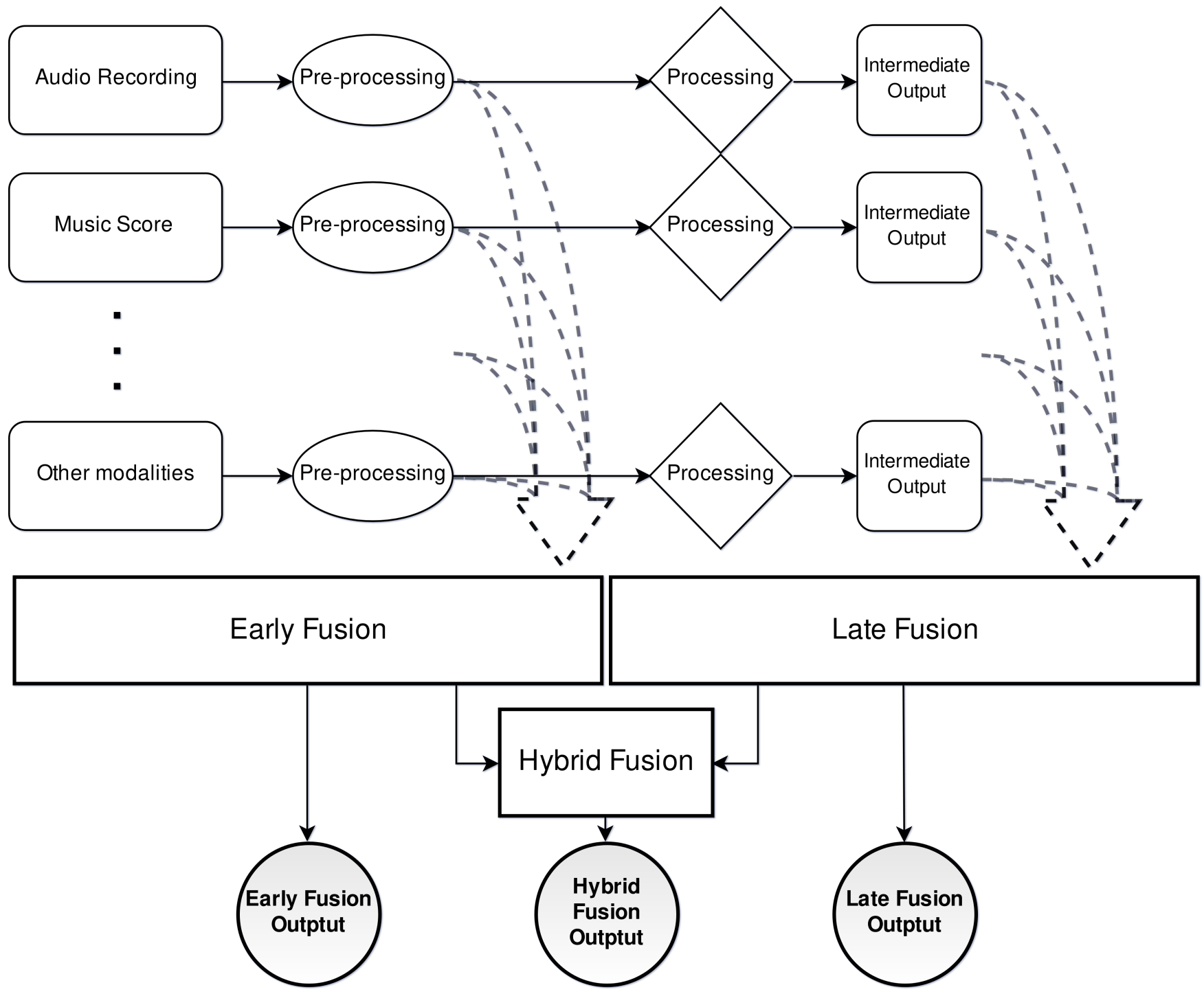

**Figure 1.1.:** *Diagram showing the flow of information in* early*-fusion and* late*-fusion. Early fusion process takes as input the output of the pre-processing of the various modalities, while the late fusion takes as input the output of specific processing for each modality.* Hybrid fusion*, instead, uses the output of both* early *and* late *fusion.*

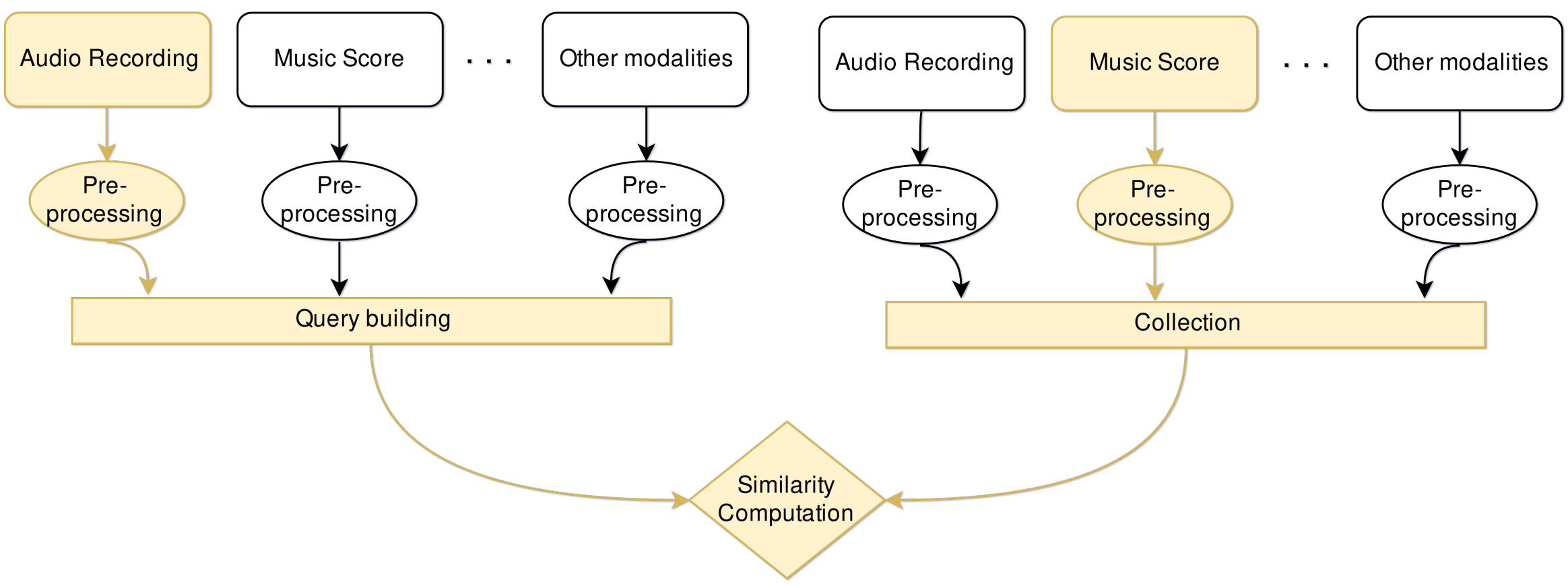

**Figure 1.2.:** *Multimodal retrieval: usually, the query and the collection contain different modalities, so that the diagram should be collapsed to the highlighted elements; however a more general case is possible [23], in which both the query and the collection contain multiple modalities.*

line.[1]

## 1.3. Multimodal music processing tasks

To date, several tasks have been experimented in multimodal approaches. We found two possible categorizations for the application level:

- **less** vs **more** studied tasks: some tasks have been extensively studied with a multimodal approach, such as *audio-to-score alignment*, *score-informed source separation*, *music segmentation*, *emotion* or *mood* recognition; other tasks, instead, have been little explored and are worth of more attention.
- **macro-task** based categorization: we identified 4 different macro-tasks, that are a partial re-elaboration of a previous effort [19]: *classification* of music, *synchronization* of different representations, *similarity* computation between two or more modalities, and *time-dependent representation*.

Figure 1.3 outlines all the tasks that we found in the literature. Here, instead, we are going to briefly describe each task and how it has been fulfilled by exploiting a multimodal approach.

### 1.3.1. Synchronization

Synchronization algorithms aim at aligning in time or space different modalities of music, i.e. creating associations between points in different modalities. They can be performed both in real-time and offline. In the real-time case, the challenge is to predict if a new event discovered in a real-time modality – e.g. an onset in the audio – corresponds to an already known event in another off-line modality – e.g. a new note in the score. Off-line synchronization, instead, is usually referred to as *alignment* and involves the fusion of multiple modalities by definition. Well-studied alignment algorithms include *audio-to-score* alignment [24], *audio-to-audio* alignment [24] and *lyrics-to-audio* alignment [25]. An interesting task is to align the audio recording to the images, without using any symbolic data [26]. Often, alignment algorithms are a fundamental pre-processing step for other algorithms – see Section 1.4.

### 1.3.2. Similarity

With *similarity*, we mean the task of computing the amount of similarity between the information content of different modalities. Often, this task has the purpose of retrieving documents from a collection through a query, which can be explicitly expressed by the user or implicitly deduced by the system. The multimodal approach, here, can exist either in the different modalities between the query and the retrieved documents or in the query itself. A common example of explicit queries for retrieving another modality is *query-by-humming* or *query-by example*, in which the query is represented by an audio recording and the system retrieves the correct song; this task is usually performed with two main approaches: by using a collection of recordings in a single-modality fashion, or by exploiting multimodality with a collection of symbolic data [27, 28]. An example of *implicit* query systems, instead, are reccommender systems and playlist generators, where the user is usually not aware of which specific parameters are used for the recommendations; most of the recent research in this field tries to exploit multimodal approaches – also called *hybrid* – involving *metadata*, user *context*, *audio* features [29, 30].

[1] See Appendix A.

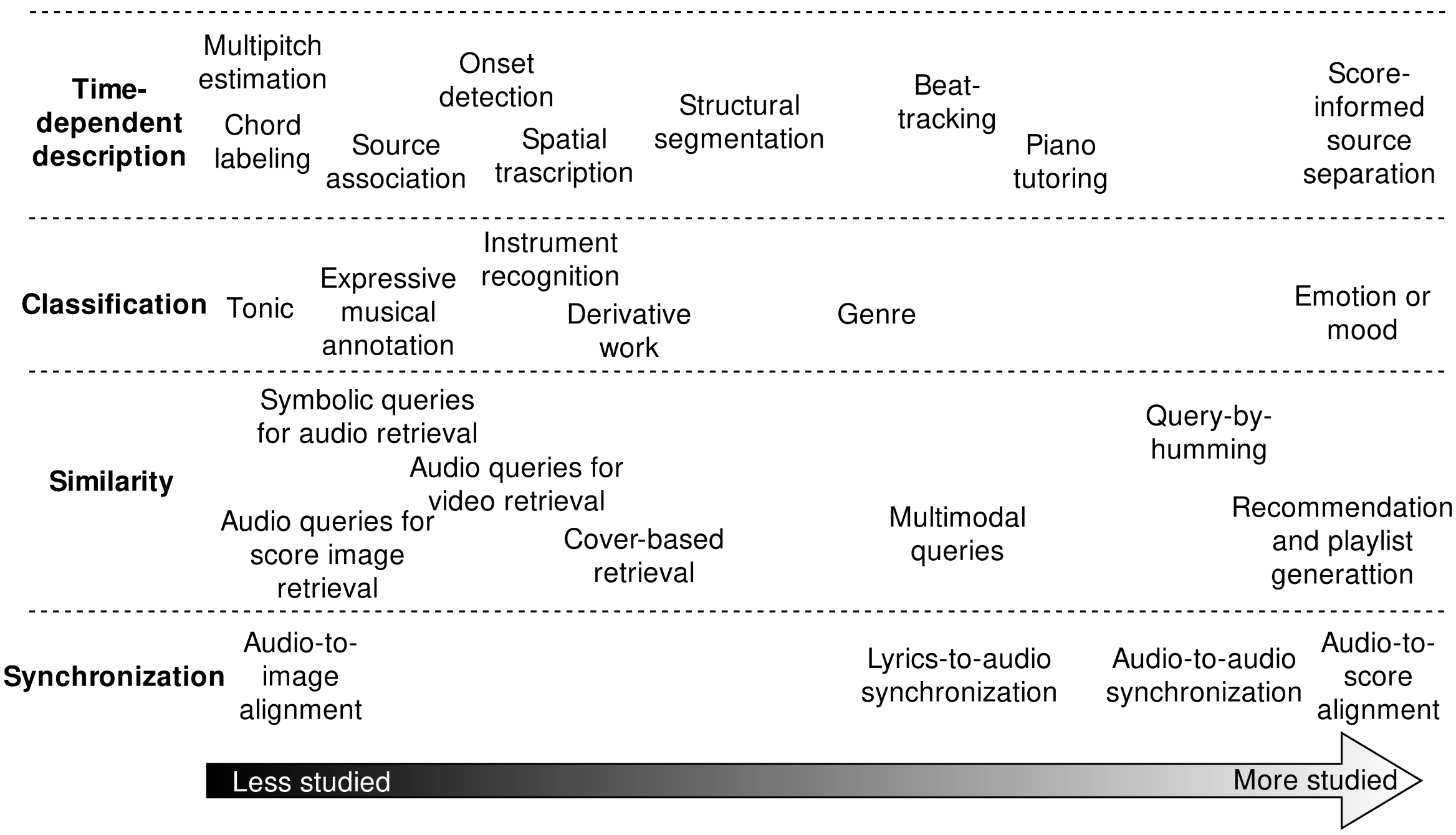

**Figure 1.3.:** *The tasks identified in literature, divided in 4 macro-tasks and plotted along a* less - more *studied axis. Tasks for which only one paper has been found appear at the left-side (*less studied*); at the rightmost side are tasks for which extensive surveys are already available; the other tasks are placed in the remaining space proportionally to the number of corresponding papers found in literature. All references to these tasks can be found in the discussion and in the online spreadsheet – see footnote 1. Note that labels refer to the multimodal approach at hand and not to generic MIR tasks – e.g.* genre *classification task is intended to be performed with a multimodal approach and thus it has been less studied than* emotion or mood *classification in the context of multimodal approaches.*

A method consisting of two axes was proposed to categorize query-by-example systems. *Specificity* axis refers to the similarity level of the documents retrieved by the systems; for instance, a song identification system has high specificity since the retrieved song should be exactly the one from which the query was extracted, while a recommender system has low specificity because the retrieved documents will contain music not identical to the query. *Granularity* axis, instead, refers to the temporal matching of the query, which can represent a little fragment of the searched documents – fragment-level – or it can refer to characteristics of the whole looked documents – document-level.

An emerging field in the retrieval context is the so-called *multimodal queries*, where the user can explicitly create a query by using different parameters for different modalities [23, 31]. Following this line of thought, some researchers devised and studied novel tasks in the context of multimodal music retrieval. Some example are: a system for retrieving music score images through audio queries [26]; an algorithm to retrieve the cover of a given song [32]; systems to retrieve audio recordings through symbolic queries [33, 34]; an approach to query a music video database with audio queries [35].

### 1.3.3. Classification

The *classification* process consists in taking as input a music document and returning one or more labels. A popular multimodal classification task is the *mood* or *emotion* recognition [36], while an emerging one is *genre* classification [37, 38, 39, 40, 41, 42, 43, 44]. Both these two tasks can take advantage of audio recordings, lyrics, cover arts and meta-tags. Additionally, emotion recognition can exploit EEG data, while for genre classification one can use music video and generic text such as critic reviews. Usually, just one modality is considered in addition to audio recordings, but an interesting work [44] tries to exploit more than two modalities. Other multimodal classification tasks found in the literature are:

- *artist* identification, through lyrics and audio fusion [45];
- *derivative works* classification of youtube video through audio, video, titles and author [46];
- *instrument* classification by exploiting audio recordings and performance video [47, 48];
- *tonic* identification, that is: given an audio recording and the note level, find the tonic [49];
- *expressive musical description*, which consists in associating a musical annotation to an audio recording by extracting features with the help of symbolic level [50].

### 1.3.4. Time-dependent representation

With *time-dependent representation*, we mean the creation of a time-dependent description of the music data, created by merging and processing multiple modalities. Possibly the most studied task within this family is *score-informed source separation* [24], in which symbolic music data and audio recordings of a musical ensemble are used to create different audio recordings for each different instrument. A number of researchers have also tried to use audio and video recordings of a music performance or of a dancer to extract *beat tracking* information [51, 52, 53, 54, 55]. An emerging task is *piano tutoring*, which consists in the tracking of errors in a piano performance: to this end, the audio recording, the instrument timbre and the symbolic score can be exploited [56, 57, 58, 59, 60, 61, 62]. Less studied tasks are:

- *music segmentation*, in which audio and video, lyrics or note level can be exploited to identify the music piece structure [63, 64, 65];
- *spatial transcription*, that is the inference, starting from audio and video, of the note level of songs for fretted instruments, so that the resulting score includes the annotation of fingering [66, 67];
- *onset* detection through audio and performer video [68] or rhythmic structure knowledge;
- *chords* labeling, by comparing multiple audio recordings of the same work [69];
- *source association*, that is the detection of which player is active time by time by exploiting audio, video and music score [70, 71];
- *multi-pitch* estimation, that is the transcription of parts being played simultaneously, with the help of performance video to detect play-nonplay activity of the various instruments [72].

## 1.4. Data pre-processing

*Data pre-processing* is the elaboration of data to the end of transforming their representation to a more suitable format for the subsequent steps. We have identified a number of possible non-exclusive types of pre-processing :

- *Synchronization*: the synchronization process described in Section 1.3.1 is sometime used as pre-processing step to align multiple modalities; thus, the pre-processing itself can be multimodal. For example, in *piano tutoring* and *score-informed source separation*, an *audio-to-score* alignment is performed; *audio-to-audio* synchronization is a fundamental pre-processing step in tasks requiring comparison of multiple recordings of the same piece [69]; *audio-to-score* alignment is also used in several previously cited works [33, 34, 50, 65];
- *Feature extraction*: usually, music representations are not used as they are, but a number of features are extracted – see Section 1.5.
- Other pre-processing steps include:
  - *conversion* from one modality to the other, such as in *query-by-humming*, which includes a conversion from audio to the symbolic level, or in *audio-to-score* alignment, where symbolic scores can be converted to audio through a synthesis process or audio can be converted to score via transcription.
  - *feature selection* through *Linear Discriminant Analysis* (LDA) [35] or Relief [50]
  - *normalization* of the extracted feature [55]

- *source-separation* in lyrics-to-audio alignment and source association [70, 71]
- chord labeling on audio only [69]
- multi-pitch estimation on audio onlyj [72]
- video-based hand tracking [66]
- *tf-idf*-based statistics – see Section 1.5.3 – adapted for audio [45]

Finally, we think that a step worthy of a particular attention is the *conversion to a common space* of the extracted features to make them comparable. We will talk about this step in Section 1.6. The accompanying online table (see footnote 1) contains a short description of the pre-processing pipeline adopted in each cited paper.

## 1.5. Feature extraction in multimodal approaches

Various types of features can be extracted from each modality. In this section, we provide a general description for audio, video, textual and symbolic score features.

### 1.5.1. Audio features

Various categorizations for audio features have been proposed until today, including [73]:

- abstraction level: how many abstraction layers are needed to conceive a certain feature with the physical sound wave being the lowest possible level;
- temporal scope: the length of the time segment a feature refers to can be used as a discriminating aspect – e.g. global features refers to the whole audio under study, while local features refers to a narrower audio segment;
- music elements: features can be connected to theoretical music elements such as harmony, melody, rhythm, tonality, etc.

Regarding these three different classifications, we note that most of the features can be used for both global and local descriptions of audio and that music-theory-based descriptions are strongly biased towards academic western music culture. Classifications based on the abstraction level, instead, while being largely applicable to various features, are connected to their implementation and make the reason why they are used obscure. In this work, we will refer to a more recent categorization [74] which subdivides audio features in *physical* and *perceptual*. We found such a classification more practical than the previous ones because it highlights the reason why a feature should be used. For instance, if the aim is studying sound in its physical aspects, physical features will be used; if, instead, sound is studied from a perceptual or cognitive perspective, perceptual features should be preferred.

#### 1.5.1.1. Physical features

Physical features can be computed in various domains, such as time, frequency or wavelet. Time-domain features can be computed directly on the digitally recorded audio signal and include *zero-crossing rate*, *amplitude*, *rhythm* and *power-based* features, such as the *volume*, the *MPEG-7 tem-* *poral centroid* or the *beat histogram*. Frequency-domain features are the richest category; they are usually computed through a Short-Time Fourier Transform (STFT) or an autoregression analysis and can be subdivided in: *autoregression-based*, *STFT-based* and *brightness*, *tonality*, *chroma* or *spectrum shape* related. Features in the Wavelet-domain are computed after a Wavelet transform, which has the advantage of being able to represent discontinuous, finite, non-periodic or non-stationary functions. Image-domain features are computed through a graphic elaboration of the spectrogram, that is a matrix that can be represented as a one-channel image computed with the STFT; often, spectrogram is used as input for a convolutional neural network (CNN) that is trained to compute *ad-hoc* features, which lack straightforward interpretation.

#### 1.5.1.2. Perceptual features

Perceptual features try to integrate human sound cognition in the feature extraction stage or in the elaboration of physical audio features. Most of them aim at mapping certain measurements to a perceptual-based scale and/or metrics. For example, *Mel Frequency Cepstral Coefficients* (MFCC) are derived by mapping the Fourier transform to a Mel-scale, thus improving the coherence with human perception. *Perceptual wavelet packets* [75] employ a perceptually motivated critical-band based analysis to characterize each component of the spectrum using wavelet packets. *Loudness* is computed from the Fourier transform with the aim of providing a psychophysically motivated measure of the intensity of a sound.

In the light of the discussion in Section 1.1, music-theory-based features may be classified as a subclass of perceptual features.

### 1.5.2. Video and image features

This section is mainly written with reference to a previous work [55]. Video features used in the music domain are similar to visual features used in general purpose video analysis. Image features can be based on the *color space* (RGB or HSV), on *edges* detection, on the *texture* – such as the LBP –, or on the *moment* of a region. In video, motion detection is also possible and can be performed with *background detection* and *subtraction*, *frame difference* and *optical flow*. *Object tracking* has been also used to detect hand movements, for example in *piano-tutoring* applications. *Object tracking* can happen by exploiting the difference between frames of the detected object contours, by using deviations frame-to-frame of whole regions or generic features. In video, one can also detect *shots*, for example by analyzing the variation of the color histograms in the video frames, using the Kullbach-Leibler distance [76] or other metrics.

In genre and mood related analysis, other features can also be exploited [77]. The use of *tempo* is essential to express emotions in video clips and can be analyzed through features related to motion and length of video shots. Another relevant factor is lighting, which can be measured through brightness-based features. Colors have an affective meaning too, and color features are consequently useful for genre or emotion recognition.

Finally, images can also be used as they are as input of CNNs.

### 1.5.3. Text features

This section is written with reference to a previous review [78]. The most common text representations are based on *tf-idf*. In this context, $\mathrm{tf}(d, t)$ is the *term frequency* and is computed as the number of occurrences of a term $t$ in a document $d$. Conversely, $\mathrm{idf}(d, t)$ is a short for *inverse document frequency* and is needed to integrate the discrimination power of the term $t$ for the document $d$, considering the whole collection; it is related to the inverse ratio between the number of documents containing $t$ at least once and the total number of documents in the considered collection:

$$\mathrm{idf} = \frac{\text{docs in collection}}{\text{docs containing t}} \tag{1.1}$$

Usually, *tf-idf* takes the following form:

$$\text{tf-idf}(d, t) = \mathrm{tf}(d, t) \times \log[\mathrm{idf}(d, t)] \tag{1.2}$$

Features based on *tf-idf* are often used in Bag-of-Words (BoW) models, where each document is represented as a list of words, without taking care of the cardinality and order of words. In order to make BoW and *tf-idf* models effective, a few preliminary steps are usually performed, such as *stemming* and removal of *punctuation* and *stop-words*. More sophisticated methods are also available, allowing topic- or semantics-based analysis, such as *Latent Dirichlet Allocation* (LDA), *Latent Semantic Analysis* (LSA), *Explicit Semantic Analysis* (ESA) [79] and CNN feature extraction.

For lyrics analysis, other types of features can be extracted, like rhymes or positional features. Finally, when the available text is limited, one can extend it with a semantic approach consisting of *knowledge boosting* [44].

### 1.5.4. Symbolic score features

Symbolic music scores have been rarely used in feature extraction approaches. Most of the papers that deal with symbolic scores use MIDI-derived representations, such as the pianoroll [24] or inter-onset intervals (IOI )[65]. To the end of audio-symbolic comparison, one can compute chromograms, that are also derivable from the audio modality alone. However a number of representation exist and have been tested in Music Information Retrieval applications, such as *pitch histograms*, *Generalized Pitch Interval Representation (GPIR)*, *Spiral Array*, *Rizo-Iñesta trees*, *Pinto graphs*, *Orio-Rodà graphs* and others. A brief review of the music symbolic level representations is provided in a previous work [80].

## 1.6. Conversion to common space

The conversion of the extracted features to a *common space* is often a mandatory step in *early fusion* approaches. Nevertheless, almost no authors emphasize this aspect. Thus, we think that greater attention should be posed on this step of the pre-processing pipeline.

The conversion to a common space consists in the mapping of the features coming from different modalities to a new space where they are comparable. This can be needed in single-modality approaches too, when the features refer to different characteristics of the signal. Indeed, many papers describe techniques that include a mapping of the features to a common space, both in the pre-processing and in the processing stages, but no particular attention is put on the conversion itself. Common methods include:

- *normalization*, that is the most basic approach;
- *conversion from one modality to another*, so that features can be computed in the same units;
- *machine learning algorithms* such as CNNs or SVMs: SVMs compute the best parameters for a kernel function that is used to transform the data into a space where they are more easily separable; CNNs, instead, can be trained to represent each input modality in a space so that the last network layers can use as input the concatenation of these representations;
- *dimensionality reduction* algorithms, which usually search for a new space where data samples are representable with a fewer number of dimensions without losing the ability to separate them; examples are *Principal Component Analysis* (PCA) and *Linear Discriminant Analysis* (LDA).

It must be said that some types of features are suitable for multimodal fusion without any conversion step. For example, *chroma features* can be computed from both the audio recordings and the symbolic scores and thus can be compared with no additional processing.

A possible categorization of the conversion to common space methods is the identification of two classes, *coordinated* and *joint*. In the former type, the mapping function takes as input a unimodal representation, while in the latter type it takes as input a multimodal representation [17]. In other words, *coordinated* conversion learns to map each modality to a new space trying to minimize the distance between the various descriptions of the same object, while *joint* conversion learns the best mapping function that uses all the modalities and optimizes the subsequent steps – e.g. SVM.

## 1.7. Information fusion approaches

Two major information fusion approaches exist: *early fusion* and *late fusion* – see Figure 1.1. Some authors also report a *hybrid* approach [16], which consists in fusing information both in a *early* and *late* fashion and in adding a further step to fuse the output of the two approaches. Nevertheless, we did not find any existing application to the music domain. Before discussing in detail the two approaches, we recall that no fusion is usually needed in *similarity* tasks, but just a comparison of the various modalities and, thus, a conversion to a common space. The accompanying online table (see footnote 1) contains a short description of the fusion approach used in all the cited papers. To our understanding the main difference between *early* and *late* fusion is about their efficiency and ease of development; however authors disagree about which one is the more effective.

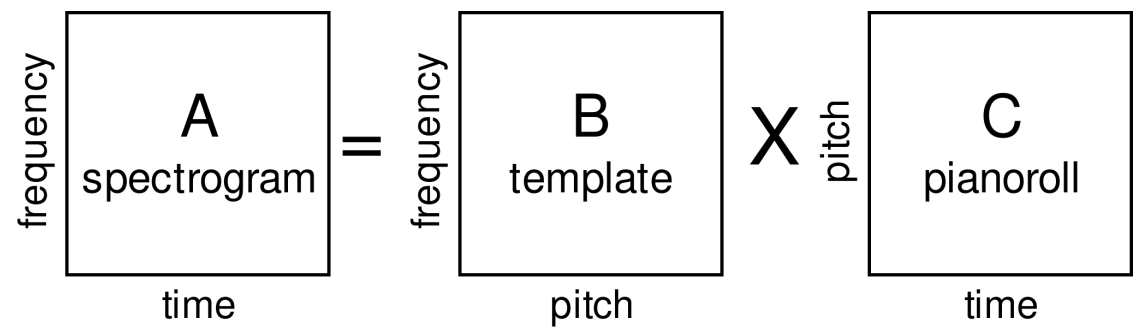

**Figure 1.4.:** *Exemplification of Non-negative Matrix Factorization for music transcription.*

## 1.7.1. Early fusion

*Early fusion* consists in the fusion of the features of all the modalities, using them as input in one single processing algorithm. Although the development of such techniques is more straightforward, they need a more careful treatment because the features extracted from various modalities are not always directly comparable.

One of the most used methods that allow feature fusion at an early stage are synchronization algorithms, that are usually needed when features are local descriptors of different modalities – see Section 1.5. Note that synchronization can be a multimodal MIP task by itself 1.3.1. The most used approach exploits Dynamic Time Warping [81] (DTW) for synchronizing sequences of features extracted from different modalities. DTW is a well-known technique based on a similarity matrix between two sorted sets of points, for example two time-sequences. By using a dynamic programming algorithm, one can exploit the similarity matrix to find the best path which connects the first point in one modality to the last point in the same modality and which satisfies certain conditions. This path will indicate the corresponding points between the two modalities. Other common methods for synchronization purposes are HMM [25, 65] where hidden states represent points in one modality and observations represent points in a second modality; this is particularly effective for real-time alignment or generic sequence fusion such as in *time-dependent descriptions*.

Aside HMMs, various additional machine learnins [82] approaches are used to perform early fusion: SVM, GMM, CNN, and Particle Filters are the most used techniques.

Another interesting method is Non-negative Matrix Factorization (NMF), through which audio and symbolic scores can be exploited to the end of precise performance transcription, as in *score-informed source separation* and *piano tutoring* application [24]. In NMF, a matrix $A$ is decomposed in two components $C$ and $B$, so that $A = B \times C$. If $A$ is a spectrogram and $B$ is a *template matrix* dependent on the instrumentation, then we can think to $C$ as a pianoroll matrix – see Figure 1.4. Consequently, one can use an optimization algorithm to minimize a loss function between $A$ and $B \times C$, by initializing $C$ with a symbolic score; at the end of the optimization, $C$ will be a precise transcription of the performance contained in $A$.

Finally, feature fusion can also happen at the feature selection stage [45, 50].

## 1.7.2. Late fusion

Unlike *early fusion*, *late fusion* is the fusion of the output of various *ad-hoc* algorithms, one for each modality. Some author refers to late fusion as *decision-level* fusion, even if a decision process is not mandatory. The main advantage of *late fusion* is that it allows for a more adjustable processing of each modality. However, it is usually more demanding in terms of development costs.

In *classification* and *time-dependent description* tasks, the most used types of late fusion are *rule-based*. Rules can include voting procedure [69, 71], linear combination [48, 83], maximum and minimum operations [48, 83]. Many authors have developed sophisticated algorithms to execute this step, such as in *beat tracking*, *piano tutoring* and *structural segmentation* [64], multi-pitch estimation [72] and tonic identification [49].

In *synchronization* tasks, instead, no *late-fusion* approach is possible, since the task consists in creating associations between points in different modalities and, thus, the process must take as input all the modalities, eventually in some common representation.

## 1.8. Future directions

In this Chapter, the literature on multimodal music information processing and retrieval has been analyzed. Based on such study, the following concluding remarks are proposed.

First of all, the unavailability of datasets of suitable size is a main problem. This issue is usually addressed with various methods such as *co-learning* approaches [17], that has the side-effect of impoverishing the goodness of the developed algorithms. Although a few datasets have been recently created [44, 84, 85, 86], a great effort should still be carried out in this direction. Indeed, existing multimodal music datasets are usually characterized by limited size and only rarely include a wide range of modalities. However an exhaustive list of the available datasets is out of the scope of this Chapter. It can be argued that this limit is due to two main reasons: first, the precise alignment of various modalities is a hard computational task and should be controlled by human supervision; second, no largely adopted standard exists for multimodal music representation. About the first point, more effort should be devoted to the development of algorithms for the alignment of various sequences. The representation of the intrinsic music multimodality, instead, is faced by the *IEEE 1599*[2] standard and the *Music Encoding Initiative*[3]; moreover, the *W3C* group is currently working on a new standard with the purpose of enriching *MusicXML* with multimodal information[4]. The course of history described in Section 1.1 and the rapid technology advancements of our times suggest that new representation modalities could be needed in the future and that multimodal representation standards should also focus on this challenge.

Another challenge that multimodal music researchers should face in the next years is the exploration of various techniques already used in multimodal processing of multimedia data, that have not been tested in the musical domain. According to previous surveys [16, 17], multimodal methods never applied to the music domain include: the hybrid approach, the Dempster-Shafer theory, Kalman filters, the maximum entropy model, Multiple Kernel Learning and Graphical Models. Moreover, we have found only one paper in which the information fusion happens during the feature extraction itself [65] and not afterwards. This approach should be explored more deeply.

Finally, we suggest that the conversion to a common space where modalities are comparable – see Section 1.6 – should be more rigorously addressed. For this sake, transfer learning technologies could be explored towards forming a synergistic feature space able to meaningfully represent multiple modalities [87, 88]. Such a direction may include the use of an existing feature space characterizing a specific modality, or the creation of a new one where multiple modalities are represented. Such a space could satisfy several desired properties, such as sparseness, reduced dimensionality, and so on.

[2]IEEE 1599 website: http://ieee1599.lim.di.unimi.it/
[3]MEI website: https://music-encoding.org/
[4]W3C music notation group website: https://www.w3.org/community/music-notation/

# 2
# Towards Automatic Music Resynthesis

After having analyzed the state-of-art in multimodal Music Information Processing (MIP), in this Chapter the topic of music resynthesis is analyzed closer. First, approaches to Music Restoration and Synthesis are reviewed, with a particular focus to technologies for music production. The traditional Digital Signal Processing (DSP) techniques are discussed while considering both technical aspects and ethical implications. Then, the limitations of traditional methods for audio restoration and resynthesis are reviewed and a novel automated approach, which is able to address the previous ethical issues, is described. The proposed method is based on the resynthesis of extracted features and possible solutions are discussed. Empirical evaluations of such an approach are provided throughout the Thesis, especially in Chapters 7 and 8.

## 2.1. Introduction

A definition of Automatic Music Resynthesis (AMR) that is relevant to Computer Science studies is provided next.

The first aspect that should be considered is that AMR is a task dealing with *music*, which can be handled using multimodal representations and technologies [1]. Of course, the most used modality for music representation is the acoustic one, and this is the main modality under study in this Chapter. However, it must be stressed out that music representations include other typologies of data, such as music score, video, images, and texts.

While focusing on music recordings, another important characteristic that should be taken into account is the difference between music and generic sound signals. From our information processing perspective, the aspect that allows computational systems to better exploit the information contained in music signals is the organization of the sounds in the time and, often but not necessarily, in the frequency domain.

The second relevant word in AMR is *resynthesis*. It refers to the act of making accessible the original information with a renewed sound via synthesis process. Even if resynthesis is usually applied with artistic intents to make some art piece accessible in a renewed way, it can also be useful to improve the quality of successive information processing methods, for instance acoustic scene analysis or sound classification. In the latter context, resynthesis can be considered as a restoration approach, a task that, in conjunction with the Digital Humanities field [89], has gained popularity in recent years, especially for the fruition of historical arts. Music itself can be considered an art and, especially in the Western tradition, it is connected to an expressive intent of composers and performers.

Finally, the word “automatic” places the attention on automated resynthesis processes. Automating complex tasks have recently become popular thank to the advancements in the machine learning field, especially in Neural Network approaches. Automated methods can both help the processing on large archives of data, making human work much less intensive, and help non-experts in achieving good results. Automated approaches for the resynthesis of artistic products is particularly useful in the context of the World Wide Web and mobile devices. On one side, the Web is a large resource of low quality videos, images, and audios; on the other side, modern mobile devices allow people to record music and take photos with inexpensive transducers that produce low quality data in respect to the expensive professional technology. For instance, an AMR system could be used to record digital sounds with a smartphone and transform it in a studio-quality audio recording.

For these reasons, if restoration processes usually target old and deteriorated operas, potentially recorded on some archaic support and often accompanied by multimodal data, restoration of contemporary art is a relevant objective as well. Designing automated algorithms that make deteriorated information more accessible, thus, has not only the aim of enjoying our historical heritage; it can indeed pave the way for a democratization of art production via economical devices. We think that both these two objectives can be partially achieved with proper automatic resynthesis technologies.

Motivated by the previous observations, with AMR we refer to automated methods for the resynthesis of music with two main democratization objectives: making music production cheaper and allowing the rapid restoration of large music archives. Moreover, some AMR methods are designed for enhancing Music Information Retrieval (MIR) and Processing (MIP) tasks.

The contributions of this Chapter are:

- an overview of the ethical discussion around music recording restoration;
- an overview of existing Digital Signal Processing methods for audio restoration, with a particular focus on music production tools;
- a new approach for music restoration based on sound resynthesis that increases the chances of automating the restoration process;
- a discussion of possible technologies useful for the proposed approach, consisting in an overview about vocoder-like music synthesis methods for faithful reconstruction of sounds.

## 2.2. Music Restoration

### 2.2.1. Philosophy and ethical discussions

Traditionally, two main directions proposed by William Storm [90] guided the ethical discussion about the restoration of historical music recordings. According to Storm, a first approach to audio restoration (Type I) consists in the recreation of the sound as it was heard by the people of the specific era. The reasons that would motivate such an approach are mainly focused on the historical meaning of the sound source. For instance, it is known that the invention of phonograph influenced the way in which musicians performed and composed music [91]. The second approach (Type II), instead, would consist in the restoration of "the sound of the artist", that is, the sound heard in the room where the recording happened.

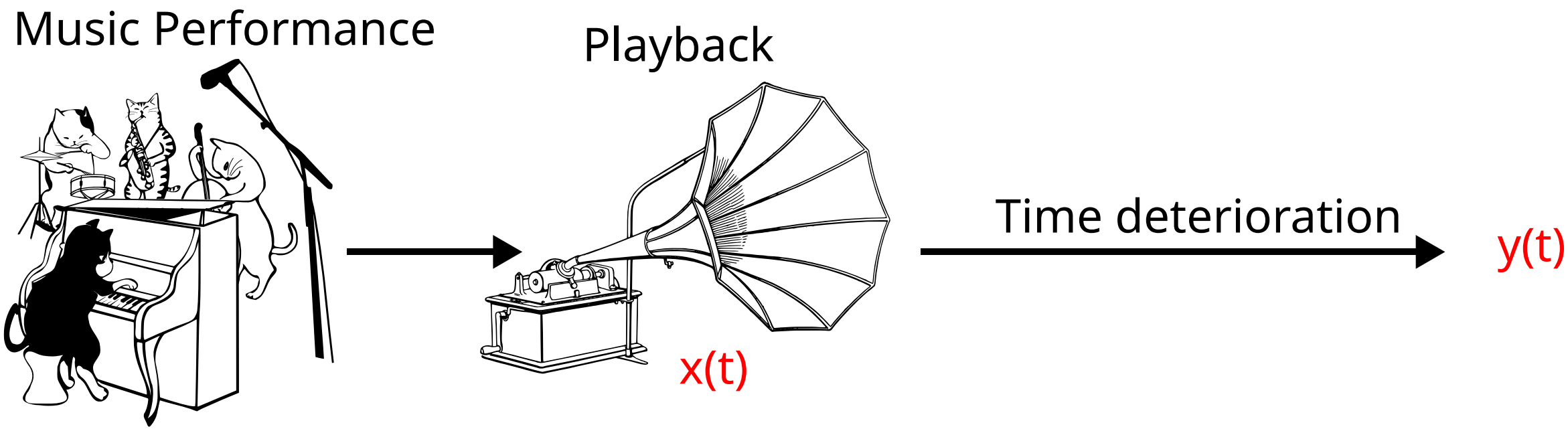

**Figure 2.1.:** *Schematic representation of Type I approach. $y(t)$ is the deteriorated signal, while $x(t)$ is the true signal that should be restored.*

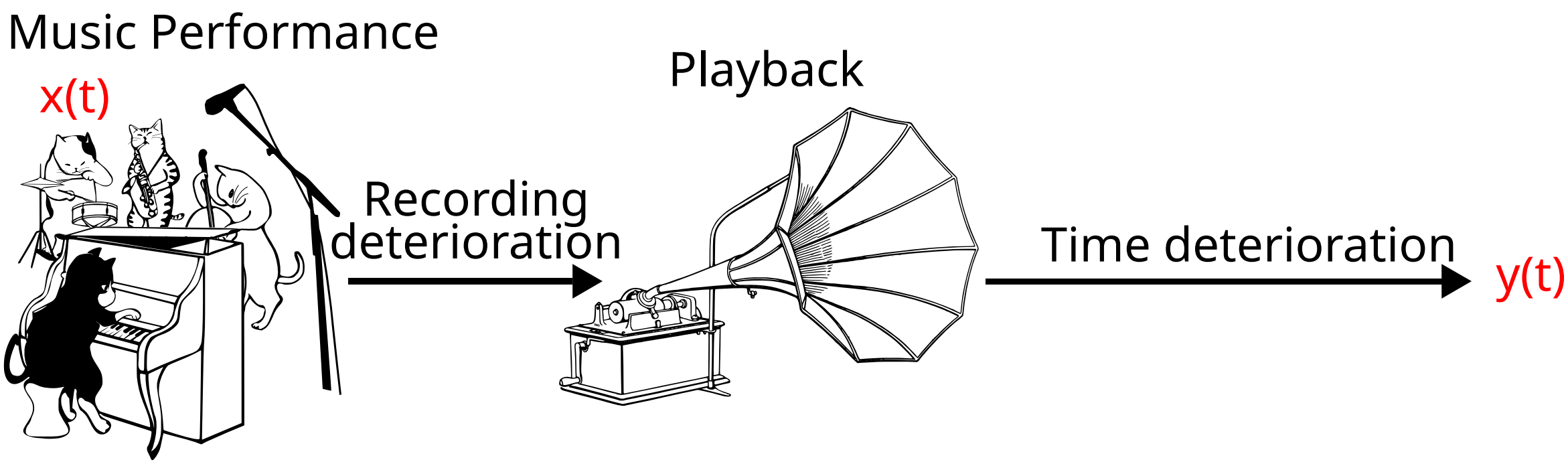

**Figure 2.2.:** *Schematic representation of Type II approach. $y(t)$ is the deteriorated signal, while $x(t)$ is the true signal that should be restored.*

As shown in Figures 2.1 and 2.2, the element that marks the difference between the two approaches is the recording procedure: restoration of Type I takes into account only the reproduction equipment for which the sound source was designed, while restoration of Type II analyzes the recording environment as well. Even though Type II restoration looks more attractive, it is tremendously difficult and ambitiously targeted. The issues that hinder approaches of Types II are manifold: microphones compression apply nonlinear distortions, noises added by the circuitry are rather unpredictable, as well as noises coming from the recording environment; generic degradation due to poor conservation is hard to completely reconstruct too. All these issues make almost impossible to reconstruct the sound heard in the recording room. As a direct consequence, Type II restoration suffers from subjective effects that can bias and alter the original source towards the restorer aesthetic sensibility.

Other methodologies suggest hybrid approaches for Type I restoration, allowing some alteration of the recorded source and the use of modernly manufactured equipment while still relying on the same type of technology for which the sound source was designed [92]. Recently, Or-calli [93] identified five different approaches to audio restoration summarized in Table 2.1. Considering the previous discussion, Orcalli's classification of restoration approaches allows to better distinguish the effects of aesthetic trends from historical aims by dividing the Type II approach in two separated classes. Looking at our democratizing aims, Orcalli's classification also apply to music production procedures based on accessible devices; this use-case scenario fits reasonably well into the *Aesthetic* category.

| **Orcalli Nomenclature** | **Storm Nomenclature** | **Question answered** | **Operational description** |
|---|---|---|---|
| *Preservative* | – | "what is the document like?" | use the new digital medium (e.g., photos, audio, video) to "represent the information and material characteristics of the original document as it came to us" |
| *Documentary* | – | "where is the sound fabric of the document?" | analyze and document "the source editions, the production equipment and techniques, the compositional practice and the authorial bases" |
| *Sociological* | *Type I* | "how was the published document perceived?" | focus on the "listening habits" of the time by studying "the characteristics of historical systems of storage and distribution of sound" |
| *Reconstructive* | *Type II* | how can the intention of the author be re-generated in the sound-fabric?[1] | merging of source editions using "multi-track (master) recordings" and studying "traditions, data provided by the music score, and authoritative mixing notes" |
| *Aesthetic* | *Type II* | "how can the document be transformed?" | transforms the sound fabric by considering "the potential of the work" in relation both to its "commercial" use and its "performance" in the contemporaneity |

**Table 2.1.:** *Table that summarize the approaches described in [93]. Quotes indicate the expressions used by Orcalli.*

[1]This question was formulated by the author of the present work

The remainder of this section will give a general overview of the methods used for restoration, briefly schematized in Figure 2.3. Since Type I restoration is hardly achievable with automated approaches, we will focus on Type II approaches usable for the restoration of various degradation factors, with special attention to disturbs associated to modern cheap devices.

Rather than providing an accurate survey of existing methods, our aim is to understand the main strengths and weaknesses of the methodologies available for Type II restoration. The main reference for the following two Subsections is [94].

### 2.2.2. Local Degradation

Local degradation affects only portions of the source audio track. Typical local degradation are *clicks*, whose name is onomatopoeic and consist of impulsive disturbances of variable amplitude and duration. Another common type of local degradation are *low-frequency transients*, that influence low-frequency band with a "thump"-like sound produced by discontinuities in the medium support [94]. These types of degradation are typical of old support formats and rarely occur in modern devices because they are not usually connected with the recording procedure but with the old reproduction equipment. Since we are more interested to cheap modern devices, we will only briefly review the DSP methodologies useful for local degradation.

Local degradation are usually modeled using an additive model:

$$y(t) = x(t) + i(t) \times n(t), \tag{2.1}$$

where $x(t)$ is the true sound signal, $n(t)$ is the distortion component and $i(t)$ is a rectangular window that turns on and off $n(t)$. Consequently, the problem is split in two parts: first, the automated method should predict $i(t)$, that is the absence or presence of distortions; then, it should estimate the distortion $n(t)$ and add the inverse of the estimation. A slightly different approach consists in estimating the distortion indirectly, by reconstructing the full original signal from scratch.

Click removal is by far the most studied task and is approached with techniques similar to declipping – see Section 2.2.3. Regarding the thump-like sounds in the low-frequency bands, Godsill [94] discusses two approaches based on Autoregressive models and on template reconstruction in the wave domain [95], stating that, at the time of his writing (1998), only those two methods existed. More recently, new approaches were proposed, exploiting source-separation methods such as Non-negative Matrix Factorization [96] and Huang's Empirical Mode [97], or machine learning models such as Bayesian Gaussian fitting [98], and Neural Networks [99].

It is also possible, albeit less common, to correct local impulsive degradation in a single step by merging the detection and estimation phase [100].

### 2.2.3. Global Degradation

Global degradation refer to disturbances that influence the audio recording as a whole [94].

The most typical and known degradation is the "hiss", an onomatopoeic alias for broad-band noise. The classical solution is to exploit an additive signal model of the form of Eq. (2.1) when

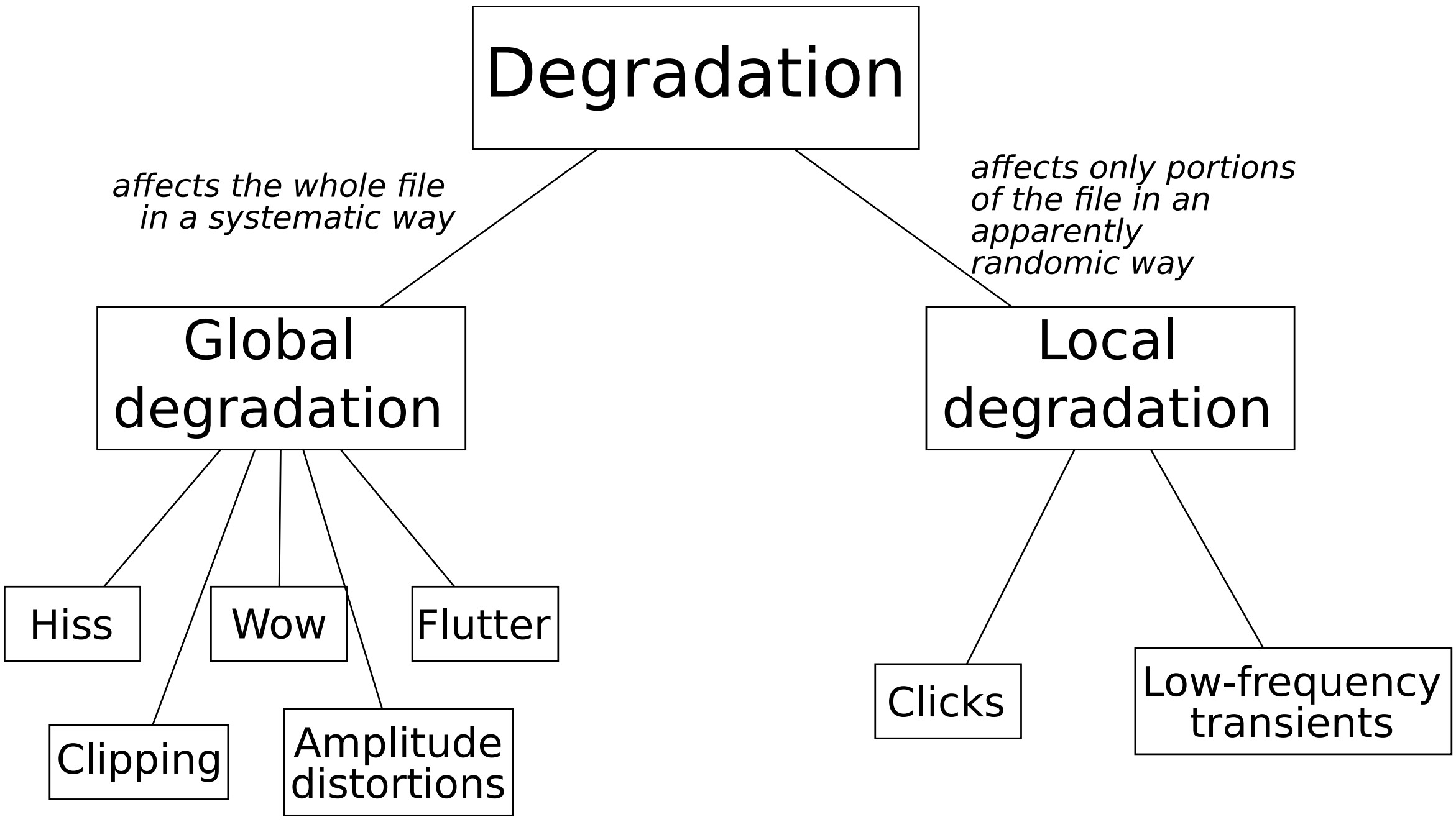

**Figure 2.3.:** *A scheme of degradation types.*

$i(t) \equiv 1$, hence:

$$y(t) = x(t) + n(t). \tag{2.2}$$

Specifically, the global characteristic of the deterioration makes $n(t)$ estimable from any silent portion of the track. The estimated $\hat{n}(t)$ can then be subtracted from $y(t)$ to obtain the estimated $\hat{x}(t)$. Such an estimation usually happens in the spectral domain: the deteriorated sound signal is first converted in the DFT domain, then a statistical model of the noise spectrum is computed and subtracted from the original spectrum. Inverse DFT can then be used to resynthesize the cleaned signal by using the original phase component – even if there is no proof that the noise component is phase invariant, the human ear is usually thought as phase insensitive. The classical correction aims at minimizing the Mean Squared Error in the time domain and uses the so-called "Wiener" formulation [94], that is:

$$\hat{X}(m) = \begin{cases} \frac{\mathcal{S}_Y(m) - \mathcal{S}_{\hat{N}}(m)}{\mathcal{S}_Y(m)} Y(m), & \mathcal{S}_Y(m) > \mathcal{S}_{\hat{N}}(m) \\ 0, & \text{otherwise} \end{cases} \tag{2.3}$$

where $m$ is the index of the frequency bin, $\hat{X}$ is the estimated spectrum of the clean signal, $\hat{N}$ is the estimated spectrum of the noise component, $Y$ is the spectrum of the degraded audio at our disposal, and $\mathcal{S}_\alpha$ is the power spectrum of $\alpha$. Other variants to the Wiener formulation exist, such as the Spectral and Power Subtraction methods [94].

Recently, however, the trends in hiss-reduction research have seen a boost of attention due the increasing importance of speech and acoustics scene analysis [101]. Source-separation approaches have been used for instance in various works [102, 103, 104, 105], while Neural Networks represent the most promising approach for generic noise reduction as for many other audio processing

tasks [103, 105, 106, 107, 108, 109, 110, 111, 112, 113]. The number of works tackling problems related to hiss-reduction is constantly increasing and a dedicated survey would be needed to properly review the entirety of new methods.

Another well-studied global disturbance is "clipping", consisting of the saturation of the signal produced during the recording or the reproduction of the audio. A recent work has reviewed, categorized and evaluated the main approaches and we refer to that work for further information [114]. To the purpose of the present discussion, we only note that the declipping task bears some similarities with the click removal task. The differences between the two disturbs are that: (1) the clipped region contains saturated values that substitute the true *higher* values, while the click *masks* the true value with an *additional* sound source; (2) the clipping phenomenon comes from inherently ill recording/reproduction methodologies that potentially affect the whole audio recording in a systematic way (hence the word *global*), while clicks come from degradation that only occur locally and unpredictably.

Interpolation of the missing signal can happen via traditional DSP methods such as median filtering [115], Least Squares regression, Autoregressive models [116] and Maximum Likelihood or Maximum A Posteriori estimation. More advanced methods include Bayesian approaches for the detection of distortion $\hat{\imath}(t)$ and its estimation $\hat{n}(t)$, such as Markov Chains Monte Carlo methods [94]. More recently, advances have been obtained by source-separation approaches that work in the time-frequency domain, using Non-negative Matrix Factorization and Binary Masks [117, 118], but these methods only apply to click removal. The most original contribution in the latest decade is probably coming from image processing and is the *audio inpainting*; generically, this expression refers to the task of reconstructing a missing part of a signal, so that interpolation can be considered a particular sub-task. Actually, *inpainting* procedures have their roots in image processing [119] and were first introduced in the audio processing world only ten years later [120]. With the increasing effectiveness of neural networks for processing two-dimensional data, audio inpainting is nowadays moving to time-frequency representations and deep-learning approaches [121, 122, 123, 124, 125].

Generic nonlinear amplitude distortion has been faced in only a few approaches [126] and never tackled from a restoration perspective. Hence, it still remains one of the most difficult tasks to solve.

Other global disturbs are "wow" and "flutter", that consist of a broad-band pitch modulation (FM) due to problems in the reproducing system. Such problems has been little explored as of today, with the notable exception of a few approaches presented in an existing review [127]. More recent approaches by the same authors were proposed to exploit the information carried by the photos of the tape to detect shrinked regions [128]. In general, however, this type of degradation has attracted little interest by the research community.

## 2.3. Music Resynthesis

### 2.3.1. Traditional Resynthesis

Traditional resynthesis methods in literature try to minimize some cost function or to leverage synthesis-by-analysis workflows [129]. Minimizing a cost function between a target and a resynthesized signal is what "corpus-based concatenative synthesis" and "cross-synthesis" techniques try to achieve. Synthesis-by-analysis, instead, is the main underlying idea of vocoding techniques.

"Corpus-based concatenative synthesis" and "cross-synthesis" are approaches to sound synthesis aimed at the transformation of an input sound signal so that it acquires characteristics of one or more different signals. More specifically, "corpus-based concatenative synthesis" consists in the analysis of the input signal and in the resynthesis of it using concatenative synthesis based on samples coming from a predefined corpus [130]. Corpus-based synthesis deals with the analysis, the indexing, and the distance estimation between sound-signals. "Cross-synthesis", instead, is a wider definition and includes both dictionary-based methods [131] and other more creative approaches [132, 133, 134, 135]. However, such methods have been mainly explored in artistic settings and never applied to faithful resynthesis of music performances. Consequently, it is unlikely that the existing methodologies can properly produce signals resynthesized according to Equation (2.4) without introducing any artifact.

The second main traditional methodology relevant for resynthesis are vocoders, a classical approach for analysis, processing and synthesis of audio signals first introduced for speech in 1939 [136]. Vocoders are based on two steps, namely feature extraction – $\phi(\cdot)$ – and synthesis – $\sigma(\cdot)$. They were extensively used for both speech and music processing with a multitude of synthesis models, such as signal plus noise [137], source-filter [138], and neural networks – see Section 2.3.2. An experimental comparison of the perceptual effectiveness of various types of vocoders for speech synthesis is available in [139]. Notably, phase vocoders [140] have been successful in applications on music signal processing, with time-stretching and pitch-shifting being the most typical use-cases. Sound reconstruction with vocoders has been first studied in the context of speech synthesis as a method for compressing sound signals in telecommunication applications [141, 142, 143]. Looking at more recent approaches, an implementation of Equation (2.4) may be found in Statistical Parametric Speech Synthesis (SPSS) [144], where the analysis is performed using a trained statistical model such as Gussian Mixture Models or Hidden Markov Models and the synthesis is performed in the target context; in any case, speech reconstruction is not the most typical application of SPSS – see for instance [145, 146] – and most of SPSS studies focus on text-to-speech synthesis. It is therefore hard to guess if such an approach could be successful for music signals.

The reasons for which vocoders are not applied to music for realistic resynthesis are mainly two. First, phase vocoders usually work on harmonic signals that are insensitive to phase change but they have limitations with impulsive components. Second, existing SPS vocoders only perform well with monophonic signals and need to source-separate input polyphonic sounds to successfully analyze them. However, the source-separation step introduces artifacts that are then reflected in the resynthesized signal. As of today, some promising directions are given by statistical models obtained via Neural Networks, that will be discussed in the next section 2.3.2.

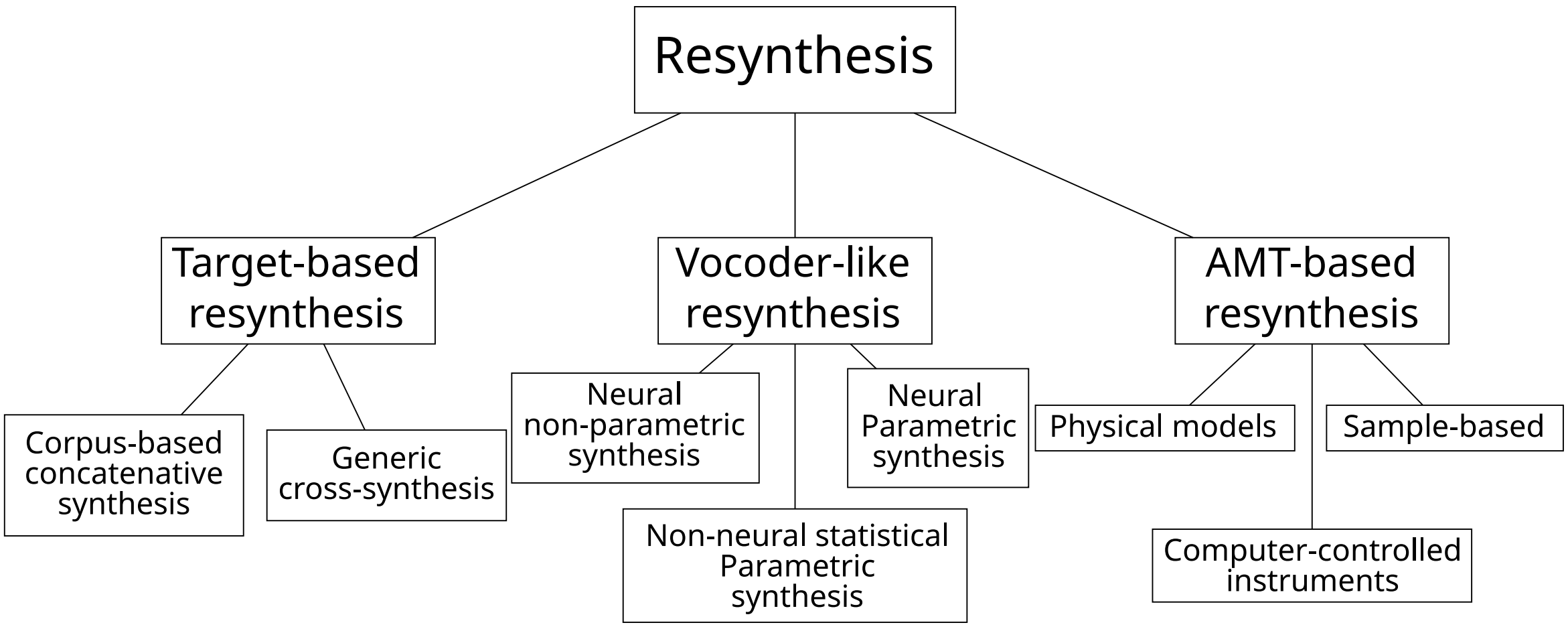

**Figure 2.4.:** *A scheme of the considered synthesis types.*

### 2.3.2. Neural Resynthesis

Starting with the *WaveNet* model [147] in 2016, a novel way of approaching audio synthesis emerged thanks to the effort of improving automated music generation. Specifically, deep neural networks were successfully applied to audio synthesis. In the following, a short review of neural synthesis is presented, mainly based on another work [148].

*WaveNet* is based on a dilated convolutional neural network and can produce short audio excerpts using an autoregressive model. In other words, the output is conditioned on the previously inferred audio samples. The direct successor of *WaveNet* was based on an autoencoder architecture [149] learning a latent representation. Unfortunately, *WaveNet*-based models are slow to train, have limited time-coherence, and the reconstructed audio is noisy due to a variety of artifacts. Due to these limitations, other models aimed at solving these issues were proposed [150, 151, 152], but their effectiveness is still far from what a user would expect in a restoration application.

Other Autoencoder-based methods were proposed, such as Vector Quantised-Variational Autoencoder (VQ-VAE) [153], and applied to various instruments, e.g. piano [154] and drum [155]. Recurrent Neural Networks (RNN) have also proven to be useful for modeling time signals and have been used for synthesizing audio. SampleRNN [156] uses RNN to analyze and synthesize audio in a hierarchic manner with multiple frequency resolutions. With such an architecture, SampleRNN achieves improved time-coherence, but, despite the attempts to make the processing more efficient [157], training and inference are associated with a prohibiting computational cost.

Another interesting method borrowed from image generation are Generative Adversarial Networks (GAN). An initial time-domain model, *WaveGAN*, was proposed in 2019 [158], predicting all the samples at once by exploiting the high GPU parallelism. Since then, other models were proposed, mainly based on time-frequency representations: *GANSynth* [159] uses log-spectrogram, while *MelNet* [160] uses mel-spectrogram. Finally, drum sound has been also approached via GAN [161]. Models based on GAN are usually faster and more effective than *WaveNet*-based models, but the output is still in low sample-rate and bit-rate while the duration of the predicted samples is limited.

Given that the above-mentioned works produce low-quality sound in respect to the input, recently more attention was placed on the reinterpretation of traditional DSP methods in the context of neural networks. In the context of sing synthesis, parametric neural synthesis was used to synthesize good quality audio, but still in low sample and bit rates [162]. The method presented in [163] uses Autoencoders and Conditional Variational Autoencoders for reconstructing, rather than the full audio waveform or spectrogram, a spectral envelope representation obtained via non-neural processing. Consequently, resynthesis is also free of neural networks. Such an approach allows to train models with fewer samples, higher generalization ability, and improved sound quality. The same arguments also apply to the Differentiable Digital Signal Processing [164] (DDSP) framework that is built on the differentiability of several DSP operations. In particular, the traditional sinusoidal plus noise model only uses differentiable operations for synthesizing sound starting from a limited number of parameters – the sine amplitudes, frequencies and phases. By exploiting this fact, a DDSP vocoder can be trained by using a neural network for inferring the vocoder parameters. As of the time of writing, DDSP vocoder [164] and *VaPar Synth* [163] probably produce the highest quality resynthesis results ever achieved by vocoder-like technologies, but systematic listening tests would be needed for a more rigorous evaluation.

### 2.3.3. Faithful music synthesis

As of today, realistic synthesis of music happens via simulation software such as virtual instruments or computer-controlled acoustic instruments of which Disklavier[2], Spirio[3] and the old CEUS system[4] are the most famous examples [165].

Virtual instruments used in the music production industry usually consist of three types: (1) instruments based on large sample libraries and concatenative synthesis, (2) physical models targeted for the simulation of specific instruments, and (3) signal synthesizers for sound design [166]. The last type of instruments is not relevant to this discussion unless the synthesizer parameters can be automatically tuned to emulate a target instrumental sound, e.g. through machine learning procedures as in DDSP. The first two categories are instead of more interest. Physically-informed instrument models have been applied to music restoration [167] by estimating the sound characteristics of the underlying sound source for a better modeling of the disturbs. Unfortunately, music instruments are complex systems that are hard to model in an understandable way as physical models try to do. Consequently, physical models are not usually comparable to real recorded sounds [166] with few notable exceptions such as piano models [168], and their evaluation from a perceptual perspective is still challenging [169]. This is why, as of today, the most used methodology in music production are sample-based virtual instruments that concatenate individual samples based on specific set of features given as input. Usually, the input features of both model-based and sample-based instruments have a direct correspondence with the MIDI protocol.

[2]https://web.archive.org/web/20200612115251/https://www.disklavier.com/history-of-the-disklavier

[3]https://web.archive.org/web/20211029183044/https://www.steinway.com/spirio

[4]https://web.archive.org/web/20161013082642/http://www.boesendorfer.com/en/ceus-reproducing-system

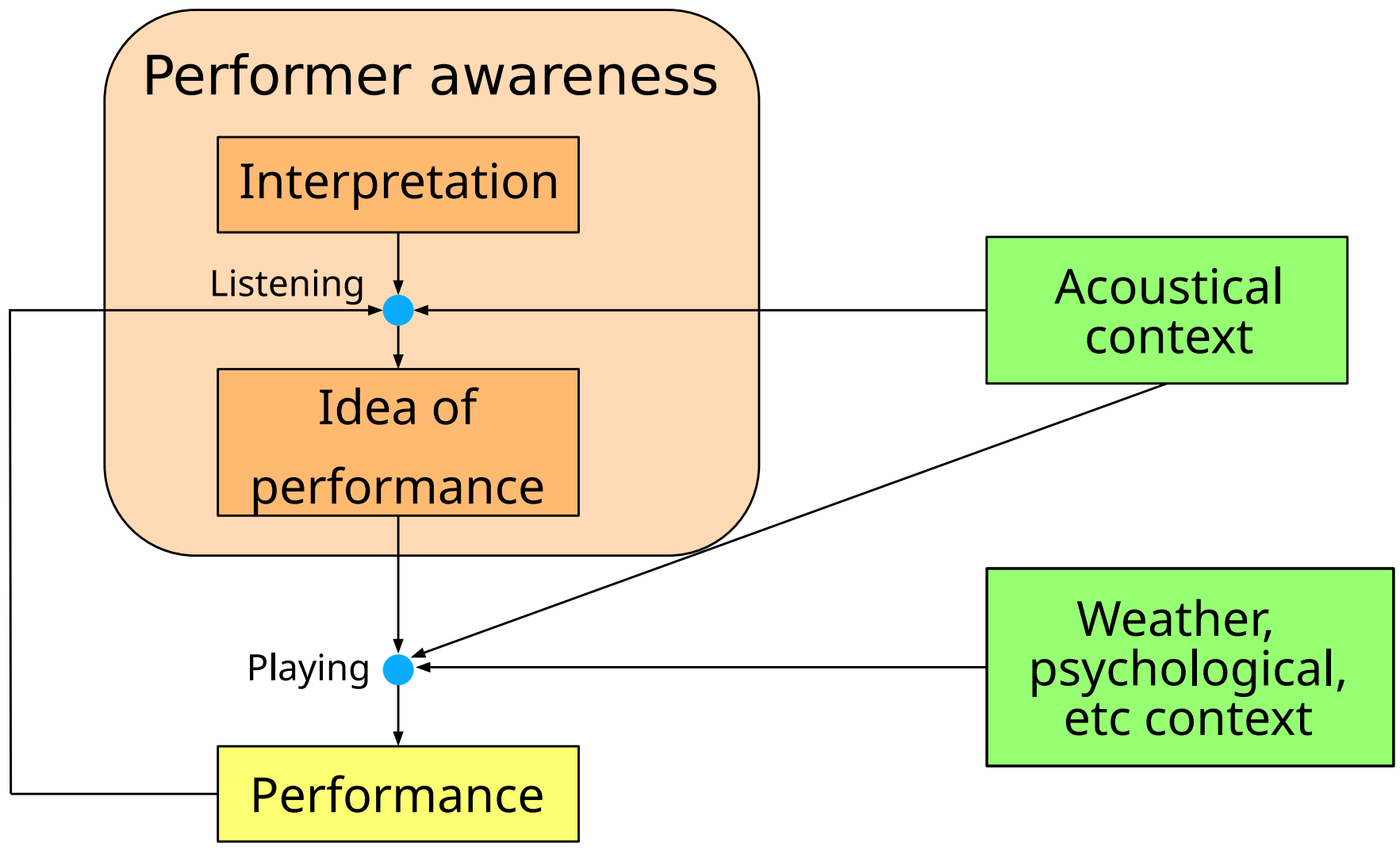

**Figure 2.5.:** *Diagram of our proposed model for context-based music performance analysis. Blue circles represent adaptation functions: a first adaptation happens in conscious way considering the acoustics of the environment and the feedback coming from the performance that is being played; a second adaptation happens unconsciously due to factors about which the performer is not aware.*

It is important to note that the MIDI protocol was modeled around digital keyboards and has thus many similarities with model-based parameters; consequently, input and output data of sample-based piano synthesizers and physically-informed piano models are essentially identical and they can be considered as the same black-box. The same should also apply to non-piano instruments, with the difference that the MIDI standard does not provide rigorously defined descriptions for non-keyboard instruments and no unified representation has been able to establish itself as de-facto standard until today.

The input parameters of synthesizers can be studied from the two perspectives of generating and analyzing music. Whereas the former is studied in the context of virtual instrument design, the latter is studied in the context of Music Performance Analysis [170] and, more interestingly, of Automatic Music Transcription (AMT) [171]. Specifically, in this discussion, we will refer to AMT as a signal processing task that converts an input audio to the corresponding synthesizer parameters. The lack of a unified representation for physical model parameters of non-piano instruments and the difficulty to inspect non-piano instruments with non-invasive techniques lead to problems in the analysis of non-piano music performances. The consequence is that, at the time of writing, AMT models are not as effective for non-piano music – see Section 6.6 [5].

## 2.4. A novel approach for music resynthesis

The approaches described in Section 2.2 suffer from several limitations. First, no specific method exists for a *fully-automated* restoration. This is probably due to the fact that all the previously described tasks are intended to be supervised by an expert user. Moreover, many of the discussed methods assume only one type of deterioration in the source track, while, in the real world, mul-tiple types of deterioration occur on the same audio. A fully-automated algorithm should run type-specific methods in an iterative way and each specific method should be re-designed with this purpose in mind. The second limitation is that they cannot take advantage of modern and sophisticated recording equipment, but propose an elaboration of existing low-fidelity recordings.

In this work, a new approach leveraging music resynthesis methods is proposed, so that the list in Table 2.1 can be extended and the two previous issues can be solved.

### 2.4.1. Face the facts: context-based resynthesis

Classical approaches consider music performance as a deterministic event, where the expressive intention of the performer is completely conveyed. However, recent research shows that music performers adapt their execution to the surrounding environment based on a number of factors.

The interest in the influence of the room acoustics on the performance dates back to 1968 [172], but it has not received significant attention by the research community in the successive decades with the exception of a limited amount of works [173, 174, 175]. All these studies showed that musicians (orchestra, choir, piano, and percussion players) adapt their music performance to the acoustic environment in which they perform. Several music psychologists hypothesized the existence of an interior representation of the sound that the musician wants to convey [176]. Such a perspective was further elaborated by various authors in an attempt to understand how musicians adapt their performance to various acoustic environments. First, subjective tests on musicians playing in different virtualized acoustic settings were explored and a circular feedback model between the performer and acoustic environment was proposed [177, 178]. Then, in the last decade, a few studies attempted to tackle the problem with objective evaluations. In 2010 and 2015, the same authors proposed two new studies in which physical features extracted from audio recordings were compared with the subjective self-evaluation of musicians and of listeners [179, 180]. From the comparison of objective and subjective evaluations, they argued that the feedback process was conscious. Other researchers attempted to understand which factors of the room acoustics influence the performance and how [181], arguing that the way in which musicians change their execution is performer-specific. In recent years, the research in the room influence on the performance has continued with the analysis of singers [182, 183] and trumpet players [184].

The overall contribution of the previous studies is that the adaptations applied by musicians influence the timbre, the amplitude dynamics, and the timing. An overview of existing works and methodologies has been recently published [185]. However, all the existing studies, are directed towards the understanding of the factors characterizing room acoustics. At the same time, they rarely consider the listener perception and never take into account indirect factors that can effectively change the acoustics of the instrument, such as the temperature and the humidity.

A different perspective to the same problem was presented in our previous work [6] – see also Chapter 7 – where we tackled the acoustic context issue regarding piano music from the listener viewpoint and analyzed how Automatic Music Transcription (AMT) models are able to deal with the variation of the acoustic context. In that work, we showed that listeners can recognize an acoustic context change and that the perceptual evaluation of AMT models is affected by that.

Another work attempted to transfer music performances across different contexts by studying the sound effect associated to MIDI parameters in each environment and then looking for MIDI parameters that minimize the difference between the produced sound [186].

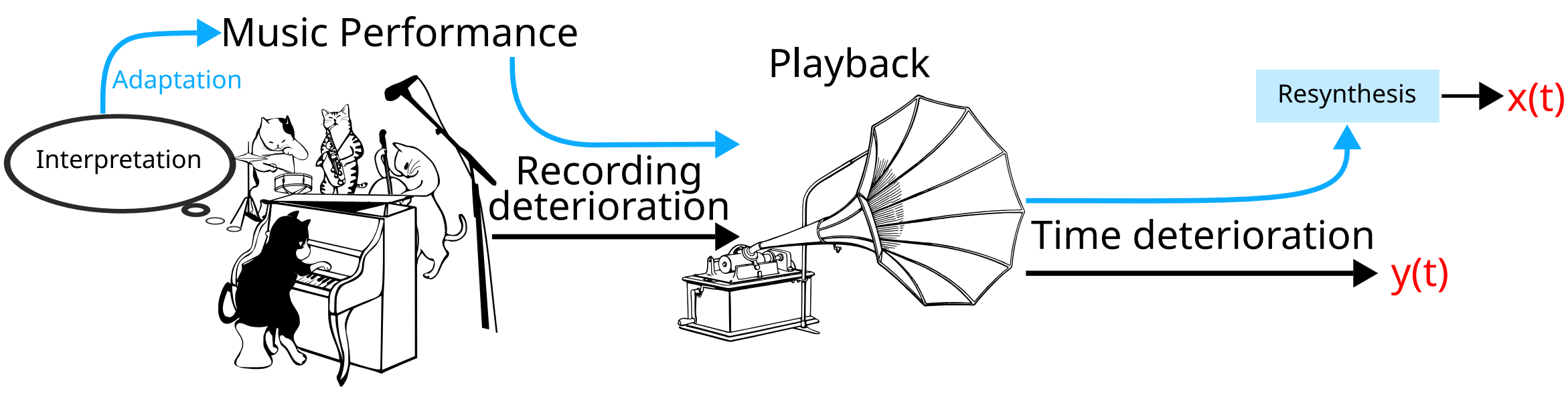

**Figure 2.6.:** *Schematic representation of Type III approach. $y(t)$ is the deteriorated signal, while $x(t)$ is the true signal that should be restored. Blue arrows indicate the flow of the interpretation information, that is subject to the audio deterioration. The resynthesis process allows to reconstruct a sound signal based on the survived interpretation information.*

Two theoretical concepts have been developed in the related literature, as outlined above:

(i) the existence a *circular feedback* between the performer and the surrounding environment [179, Fig. 1]; in Figure 2.5, an extension of such concept is proposed so that acoustic factors that are not known to the performer can be included;

(ii) the existence of an *interior representation* of the music performance that has to be realized; this idea was proposed by psychological studies [176], and developed in the above-mentioned literature so that it is referred to with the word "interpretation" [6, 187].

In this work, we will still adopt this terminology, remarking that our use does not depend on the musicological debate about how and when the concept of interpretation was born, but it has the sole scope of distinguishing the interior and ideal representation of the performance, from the real performance that was actually realized. In Figure 2.5, the difference between "interpretation" and "performance" is clearly outlined: if the "performance" indicates the set of physical events that constitute the act of playing, the "interpretation" refers to the ideal performance that the musicians want to convey.

More in general, in order to better define the phenomenon of interpretation adaptation, we also postulate the existence of unknown factors that can induce an unconscious adaptation of the interpretation. Such factors could include, for instance, the psycho-physiological conditions of the performer or weather conditions such as humidity and temperature that can affect the acoustic properties of the instrument. It is important to observe that such unknown factors are included in the proposed framework to achieve a more complete description of the phenomenon, but no experimental proof of their actual relevance is known to the authors as of the time of writing. The adaptation phenomenon is illustrated in Figure 2.5, which is based on [188, Fig. 1].

The proposed framework is in line with the state-of-art research that considers the music performance as an adaptation of another ideal execution. Following this line of thought, the proposed approach, that we will call Type III for convenience, aims at restoring the intended expressive intention of the performer. However, as depicted in Figure 2.6, to avoid the pitfalls of Type II, the new approach seeks to reconstruct the original interpretation as it survived and is perceivable in the available sound source. In other words, Type III is not concerned with inferring the part of the expressive intention that was lost and thus not restorable with complete certainty. The restoration in Type III approach takes place by modeling the adaptation that performers apply to their interpretation, reverting it, and finally applying a new adaptation in a newer different synthesis context. The newer context can be a computer-controlled instrument [165] or a virtual instrument emulating a real one.

Type III approach comes with various benefits. First, it is not prone to subjective bias in the way Type II approach is. Indeed, if the aim of Type II is to restore the original artistic intention by reconstructing "the true sound", Type III instead tries to restore the *survived* "interpretation", thus avoiding the challenging reversal of all the distortion factors happened during deterioration process.

A second strength of approach III is that, in contrast to Type I and Type II, it allows to exploit modern recording and playback equipment by resynthesizing the inferred survived interpretation. This aspect is advantageous with respect to Orcalli's "aestethic" approach – Table 2.1 – while still being useful for the historicist "reconstructive" one. The latter type, indeed, can be achieved by using historically meaningful synthesizers – e.g. virtual instruments modeled around original ones – and/or recording equipment. In this sense, Type III approach is similar to the historicist execution of early music that aims at historically-informed performance with historically-informed instruments built using contemporary technologies [189, 190].

The third asset of Type III approach is the robustness to various kinds of distortions. Indeed, the methods discussed in Section 2.2 are only applicable for selected types of disturbances, and a joint correction of multiple combined distortions is of significant difficulty. Type III approach, instead, concentrates efforts in the extraction of interpretation information which is still available despite the disturbances. Wow, flutter, hiss, clipping, clicks, and thump sounds only marginally degrade the interpretation information. The main reason is that interpretation is an abstract idealization of a music performance and is consequently generally describable via abstract music representations such as Standard MIDI Files or similar alternatives – for a systematic review see [80].

In this work, we propose Type III approach as an alternative to Type II. Namely, the restoration process is split in two steps:

- $f = \phi(y(t))$: extraction of features $f$ from the raw deteriorated audio $y(t)$, through the function $\phi$;
- $x(t) = \sigma(f)$: resynthesis $x(t)$ in a new context from the extracted features, through the function $\sigma$.

As such, the restoration process consists in the following minimization problem:

$$\min_{\Theta_\phi} \delta\Big(x(t), y(t)\Big),\ x(t) = \sigma(\phi(y(t); \Theta_\phi)) \tag{2.4}$$

In Eq. (2.4), $x(t)$ and $y(t)$ are the clean and the deteriorated signal, $\Theta_\phi$ are the parameters of the extraction function, and $\delta(\cdot)$ is a loss function between sound signals. Note that $\sigma(\cdot)$ has no parameters: under this formalization, the only component that must be searched – possibly learned

– is $\phi(\cdot)$. In other words, the extracted features $f$ must be optimal for the synthesis function $\sigma(\cdot)$. This also means that $\phi(\cdot)$ must also include the modeling of acoustic context adaptation.

In the following the feasibility of Type III approach is analyzed in light of possible resynthesis techniques that allow to minimize Equation (2.4). A diagram of the considered synthesis types is shown in Figure 2.4.

### 2.4.2. Using performance analysis for vocoder-like resynthesis

Coming back to the purpose of Automatic Music Resynthesis, we can identify a vocoder-like pipeline, where the sound signal is first encoded in a compressed information consisting of the transcribed synthesizer parameters, and then it is resynthesized using the target synthesizer [191]. Using the notation in Equation (2.4), $\phi(\cdot)$ would be the music transcription function while $\sigma(\cdot)$ would be the synthesizer itself. In the case of vocoders and autoencoders the training needs to be split in two parts to condition the extraction function $\phi(\cdot)$ on the synthesizer $\sigma(\cdot)$. In the case considered here, instead, the synthesizer is already designed and fixed. Moreover, $\sigma(\cdot)$ is not easily differentiable and as consequence it is also hardly trainable. There are three options to solve such issue:

(i) make use of an ad-hoc derivative-free optimization method such as Bayeasian Optimization [192] or generic Surrogate Modeling [193], but these are techniques that usually solve the optimization problem with less effectiveness and efficiency than differentiable methods;

(ii) rewrite the synthesizer functions so that they are differentiable and use neural-based synthesis;

(iii) find a way to condition $\phi(\cdot)$ on the final synthesizer without using $\sigma(\cdot)$ during training.

Regarding option (ii), neural vocoder and autoencoder approaches are useful for the restoration case under study, because they are easy to adapt to Equation (2.4) by considering $\phi(\cdot)$ to represent the encoder part and $\sigma(\cdot)$ the decoder network. In a real-world application, the training should be split in two phases: first, an encoder and a decoder should be trained on high quality audio; then, the decoder should be frozen while the encoder learns a latent space able to optimize Equation (2.4). However, these neural parametric synthesis methods only work on monophonic audio and have limited time-coherence. The computational resources required for a real-case scenario could substantially decrease the effectiveness of such methods. It is reasonable to expect further improvements in neural synthesis that may make the restoration via neural networks a viable option. For now, however, it still seems a distant goal.

A possible way to tackle option (iii) is to include sounds resynthesized with $\sigma(\cdot)$ in the training set for $\phi(\cdot)$. One could indeed condition the transcription function $\phi(\cdot)$ by training it on the resynthesized version of the ground-truth. In other words, whereas in the usual AMT pipeline the input are recorded sounds $y(t)$ and the output are synthesizer parameters $f(t)$, we propose to use $\sigma(f(t))$ as input and $y(t)$ as target. However, such a procedure would lead to a transcription function that only works on synthesized samples.

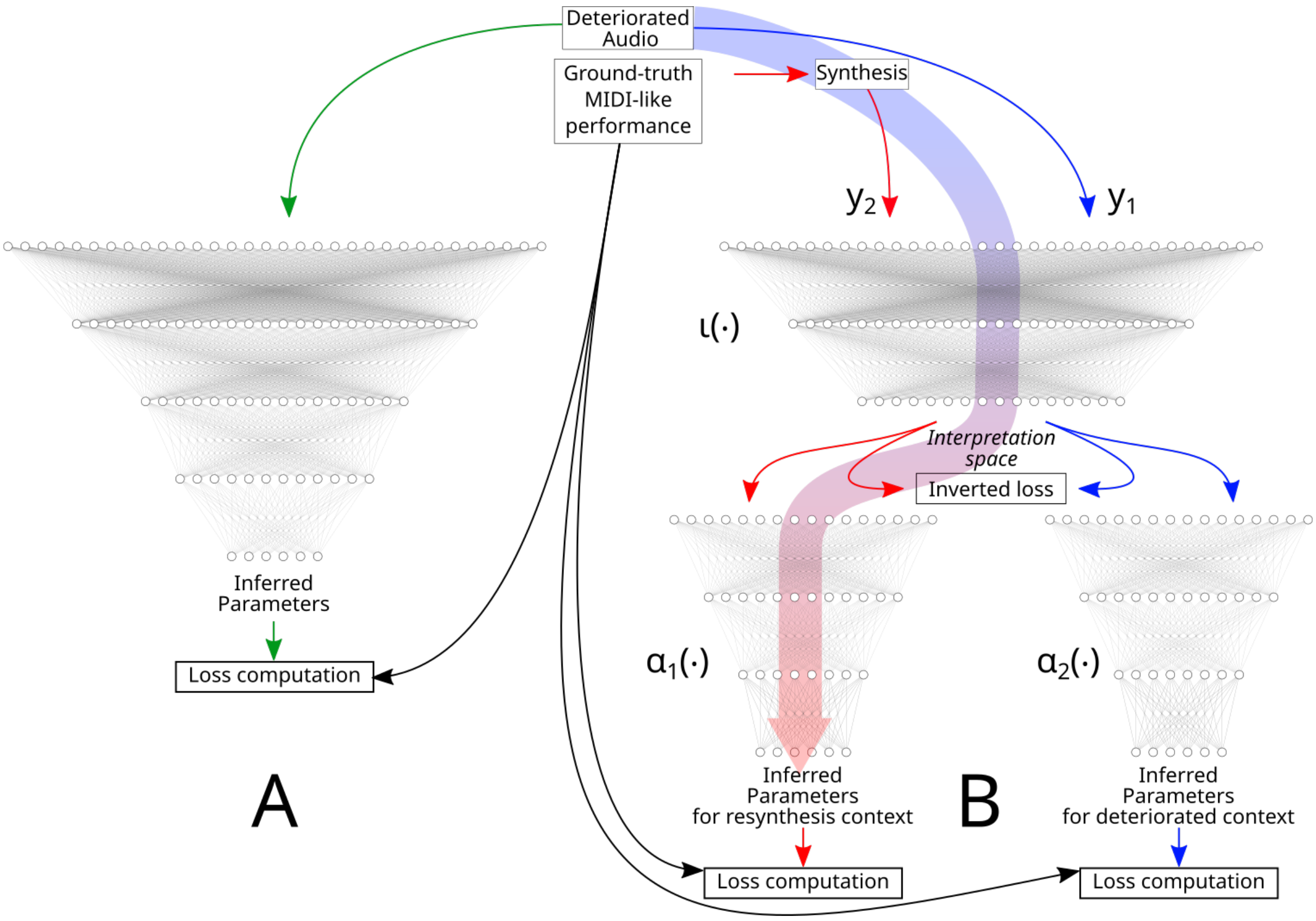

**Figure 2.7.:** *An example of the proposed workflow. On the left (A) the usual AMT workflow. On the right (B) the proposed workflow that conditions on the target synthesizer. Green, red and blue thin arrows represent the flow of the information data during training. The big shaded arrow represents the flow of information during the restoration process: the degraded audio is transcribed using the adaptation function specific to the target synthesizer. The "inverted loss function" allows to maximize the distance between same MIDI-like data synthesized in different contexts in the interpretation space.* $y_1$ *and* $y_2$ *are respectively the degraded sound and the resynthesized ground-truth;* $\iota(\cdot)$ *is the interpretation function;* $\alpha_1(\cdot)$ *and* $\alpha_2(\cdot)$ *are the adaptation functions.*

### 2.4.3. Restoring the Interpretation

To solve this problem, let us consider the definition of "interpretation" given in Section 2.4.1: the interpretation is the "ideal" performance that the musician wants to realize and, as such, it does not vary across different contexts. The performance, instead, varies across different environments as it is an adaptation of the interpretation to a specific context. In order to explicitly account for the concepts of interpretation and performance in our framework, we assume the transcription $\phi(\cdot)$ to be the composition of two functions:

$$\phi(\cdot) = \alpha_c(\iota(\cdot)), \tag{2.5}$$

where $\iota(\cdot)$ is an *interpretation* function and $\alpha_c(\cdot)$ is an adaptation function. Consistently with our definitions, $\iota(\cdot)$ can be forced to be constant across the deteriorated and the target environment, while $\alpha_c(\cdot)$ can be forced to be specific to the context $c$ – e.g. the deteriorated context or the resynthesized one.

An example of the proposed workflow is shown in Figure 2.7. There, one model learns the underlying performance of the deteriorated recording $y_1$, while another model learns the same performance data from the resynthesized audio $y_2$. The inference can happen by using the output of $\iota(y_1)$ as input to $\alpha_c(\cdot)$, where $c$ is the resynthesized context. Note that the output of $\iota(\cdot)$ obtained from the same performance data synthesized in different contexts with no adaptation should be different, because the underlying interpretation is likely different. Instead, the output of the adaptation functions $\alpha_c(\cdot)$ should be the same because the underlying performance – the ground-truth – is the same.

A preliminary empirical evaluation of such workflow is presented in Chapter 8.

### 2.4.4. Drawbacks

Given the high quality of the synthesized audio, virtual instruments and computer-controlled pianos are probably the most promising tool for the proposed approach to music resynthesis. However, to this day, the transcription process performs well only for piano-based music. Moreover, while AMT models are rather good at identifying note onsets, they can introduce pitch and offset transcription errors that alter the restored interpretation [194]. In a real-world application, an expert user could mitigate such phenomenon by using Audio-to-Score alignment [4] to constrain the transcription process.

Another downside of using Music Performance Analysis for resynthesis is that there is no complete agreement about the proper way of representing performances. For instance, the usual AMT output is based on MIDI-like representation, but MIDI is not directly applicable to music involving instruments other than keyboards. Moreover, even for piano music, some perceptual tests suggest the importance of considering non-MIDI parameters for explaining timbre variations in piano music [195, 196]; hopefully, recent investigations found that it is possible to describe such timbre variations by combining usual MIDI descriptors with the addition of few indicators that sensorized piano usually provide [197, 198, 199]. Finally, extraction of all relevant features [197] is still a challenge even for piano music, but recent advancements make such objective conceiv-

able [191, 194].

## 2.5. Conclusions

In this work, a generic overview of Music Restoration was presented, with particular focus on automatized methods and the aim of democratizing the access to historical music recordings and to music production tools.

Traditional DSP methods suffer from two main problems: first, they are hard to fully automatize as all of them solve only one specific degradation type at a time; second, they all use the original sound track as main reference for enhancing the audio, forbidding the improvement of the sound fabric.

A new approach (Type III) was proposed, consisting in the resynthesis of the "interpretation" information survived until today. The proposed approach allows to exploit modern recording and playback equipment, is potentially robust to various noise, is easily automatized, and may serve both historical and production purposes.

Various technologies useful for Type III approach were reviewed and discussed. The "Resynthesis via Performance Analysis" was found to be the most promising approach but only for where Automatic Music Transcription (AMT) models are effective. Neural parametric synthesis such as DDSP and *VarPar Synth* are likely to be an option worth of investigation for more generic tools.

The remaining of this Thesis will focus on the development of AMT-based methods for AMR. Empirical evaluations of the approach presented in this Chapter are provided in Chapters 7 and 8.

# Part II.

# Archiving multimodal music documents

# 3 Multimodal music archives in the age of web cloud

With the purpose of creating multimodal datasets and archives easily accessible, this Chapter discusses how archiving and accessing music documents could be improved in the future. The ability to access music archives is of paramount importance for the democratization objectives that motivates this Thesis – see Section 2.1 and Preface. Indeed, in the era of data-driven approaches, there can be no music democratization process without the free availability of large music collections, because it would be impossible to develop efficient methods. Moreover, in a multimodal scenario as the one described in earlier chapters, the end-users will need to be able to access multimodal music information over the Web to increase the accuracy of the Automatic Music Resynthesis (AMR) pipeline.

Among the various international file formats for music representation, the IEEE 1599 standard is analyzed as it provides the ability to collect and represent in a synchronized way various kinds of information related to a single music piece, resulting in a perfect match for the intent of this Thesis. This Chapter discusses how IEEE 1599 can be used in cloud and distributed environments that can be easily leveraged for research, music production, and recreational purposes.

The Chapter is structured as follows: Section 3.2.2 describes the key features of the IEEE 1599 standard, both in its current form and in its expected evolution; Section 3.3 discusses the applicability of this format to different categories of musical assets' stakeholders; Section 3.4 proposes an architecture to federate distributed information sources; finally, Section 3.5 discusses the most relevant advantages deriving by such an effort, draws the conclusions and sheds some light on future developments.

## 3.1. Introduction

Catalogue metadata, scores, audio tracks, computer-based formats are only a few examples of the heterogeneity to be managed in order to describe a single music piece in all of its aspects. On one side, such a complexity – if properly managed – can pave the way to a number of innovative and advanced applications, ranging from a more complete and satisfactory music experience to the possibility to conduct multi-layer content analysis. On the other side, even if digitization campaigns have originated a great number of music-related digital objects, the adoption of non-interoperable encoding formats and the geographical distribution across multiple institutions (libraries, archives, repositories, etc.) have often limited the possibility to enjoy heterogeneous information as a whole.

In this sense, the IEEE 1599 format, as explained in this chapter, represents the perfect match, as it provides the ability to collect and represent in a synchronized way various kinds of information related to a single music piece within a multi-layer environment.

## 3.2. The IEEE 1599

### 3.2.1. The 2008 Standard

IEEE 1599 encodes all music-related information in XML, which is a hierarchical, extensible, portable, and machine and human-readable language. This choice makes it similar to other standards originally conceived to describe and interchange musical notation, such as MusicXML and the Music Encoding Initiative. The main goal of IEEE 1599, however, is different: supporting and synchronizing multi-layered music information, as defined in [200], [201] and [202].

The milestones that brought to the standardization of IEEE 1599 are the establishment of the *IEEE Computer Society Task Force on Computer Generated Music* (1992), the constitution of the *IEEE Technical Committee on Computer Generated Music* (1994), the approval by IEEE Standard Association of the *Recommended practice for the Definition of a Commonly Acceptable Musical Application Using the XML Language* by the IEEE Standard Association, and, consequently, the creation of the *IEEE Standards Association Working Group on Music Application of XML* (2001). The balloting phase ended in 2008, thus making IEEE 1599 an internationally recognized standard.

During the following 10 years, a number of applications have been developed, both to make the production of materials quicker and easier, and to show the potential of the format when applied to multimedia fruition, formal music education, gamification and edutainment, computational musicology, promotion of intangible cultural heritage, etc. Besides, about 50 scientific papers and a book [203] dealing with IEEE 1599 and its applications have been published. On the other hand, some critical issues have emerged as well, as remarked in [204].

In conclusion, if on one side the interest of the scientific community towards the multi-layer approach has been and is still high, the penetration of the format in industry has been unsatisfactory so far.

IEEE 1599 uses XML to organize music information in 6 different layers. The defined layers are:

(i) General – Music-related metadata, i.e. catalog information about the piece;

(ii) Logic – A score description in terms of music symbols;

(iii) Structural – Identification of music objects and their mutual relationships;

(iv) Notational – Graphical representations of the score;

(v) Performance – Computer-based descriptions and music performances through languages such as MIDI or MPEG4;

(vi) Audio – Digital or digitized recordings, including video clips and movies.

Compared to other multi-layer music formats under development such as MNX [205], IEEE 1599 offers the structural layer, which includes musicological information – such as motives, themes and form-related data – and features usable in Music Information Retrieval (MIR) applications. This reason by itself could justify the choice of the standard for the purposes of the Thesis.

Generally, IEEE 1599 is more adaptable to different needs than MNX and it can take references to different representations of the same paradigm – i.e. different score renderings or different recordings, including original manuscripts or performances. On the other side MNX, by adopting SVG as graphic format, allows real-time rendering – e.g. by changing in real-time the correspondence rules of the mapping between symbolic and SVG elements. Moreover the performance data used by MNX follows a completely new schema, while the de facto standard in this field is SMF (Standard MIDI File).

Another relevant asset of IEEE 1599 is the concept of spine, a container for a sorted list of elements, each corresponding to a musical event – like a note-on or a rest – and each one attached with identifier, timing information and vertical alignment. Then, each element in other layers can be associated to a spine element through its identifier. The spine structure is contained in the logic layer, next to the Logical Organized Symbols (music symbols such as agogic and embellishments) and layout information. In MNX, instead, each graphical element in an SVG file can be mapped onto an element of a MNX-Common file – containing "semantic" data; then, performance audio and data can be mapped to graphical elements in the SVG file. MNX lacks a clear mapping structure and uses graphical information for two different purposes – graphic and map – sometimes with incoming links (to map audio to graphical symbols) and sometimes with outgoing links (to map graphic elements to semantic data), breaking the Single Responsibility Principle.

On the other hand, MNX achieves a better separation between model and view through CSS, but it is biased towards CMWN at the moment of writing. IEEE 1599 instead is more general and usable with any notation idiom. Finally, MNX provides the ability to store performance rules, that are rules to be followed by a machine performer like Sibelius or Finale automatic playback.

### 3.2.2. Key features of the standard

IEEE 1599 is a language for a comprehensive description of music, standardized by IEEE Standards Association in 2008. IEEE 1599 provides a meta-representation of music information within a multilayered environment, which achieves integration among the general, structural, notational, computer-driven performance, and audio layers [206].

Music information encoding often adopts several distinct reference formats for audio (e.g., CD-DA, DVD-A, FLAC, MP3, AAC), for computer-driven performance (e.g., MIDI, MPEG, SASL/SAOL), for music scores (score editors' proprietary formats, MEI, MusicXML). Some of these are formal standards, others represent *de-facto* practices. Each of these deals with musical information only in a restricted sector, and not addressing all its aspects simultaneously. Conversely, in our opinion there is a strong need for the integration of the various layers musical information is made of (audio, performance, music notation, musical forms, metadata), in order to provide access to all these layers interactively and as an integrated whole. This would enable, for instance, the navigation of score notation while listening to the corresponding audio (score following), the real-time comparison of different graphical representation as well as audio performances, and the interaction with musical contents within a multimodal environment. IEEE 1599 integrates music representation with already defined and commonly accepted standards and formats.

Another key feature of the format is the possibility to support multiple digital objects for each layer, and to achieve synchronization among instances, both belonging to the same layer and across layers, based on time and space dimensions.

The format has been conceived not only to act as an aggregator for heterogeneous music-related contents, but also as an interchange format among different applications. In this sense, the standard addresses any kind of software dealing with music information, e.g. digital score editors, optical music recognition (OMR) systems, web and mobile apps, musical databases and archives, and performance, composition and musicology-oriented applications. IEEE 1599 paves the way for novel applications for music enjoyment, publishing and research, such as innovative multimedia products [207], music-oriented educational platforms [208], and software tools for the maintenance and exploitation of cultural heritage [209].

### 3.2.3. Expected Evolutions

In March 2018, 10 years after the standardization of the first version, IEEE started a new recommended practice to update the standard. The reasons for this new initiative are multiple: first, technology has evolved and other similar initiatives have appeared; moreover, during the last years, both requests for improvement in the definition of the standard and new needs emerged; finally, the research group who defined the format was planning to extend the representation possibilities offered by the original standard [204].

The new recommended practice strives to generalize the organization of the document into layers, letting the user define custom layers too. In this way, IEEE 1599 could provide support to representation domains that are currently unpredictable, overcoming the original rigid 6-layer structure. Other change requests are more focused on specific aspects, nevertheless their impact could be crucial to the affirmation of the standard in an interoperability context. For instance, a future generalization of the Logic layer should natively foster the integration of currently-supported notational formats with new ones. Besides, external media materials linked from the XML document could be organized in a better way, maybe enclosed in a suitable file format to deliver both the XML and the attached binary objects. Another important advancement is the expected integration of Digital Rights Management representation within the IEEE 1599 multilayer asset. It is worth remembering that the format does not aim at substituting the various format or standards already available but rather at providing a meaningful integration.

These possibilities have been investigated and proposed to the scientific community during the 1st International Workshop on Multilayer Music Representation and Processing (MMRP19), held in Milano, Italy on 24-25 January 2019 [210]. On that occasion, the steering committee has organized 3 panels dealing with IEEE 1599, focusing on:

(i) history, technical notes, and demos, (ii) the goals of the new PAR WG1599, and (iii) the structure of the Working Group and a draft of workplan.

The discussion, which involved a worldwide audience of experts and scholars, resulted in the constitution of 5 sub-working groups, dealing with

(i) Descriptional framework extensions, (ii) Automatic recognition, (iii) Intellectual property and Digital Right Management, (iv) Platform improvements, and (v) Demo deliverables, respectively.

## 3.3. Applicability to Music Digital Libraries, Repositories and Datasets

The aim of this Chapter is to investigate the potential offered by IEEE 1599 to the stakeholders of musical heritage, and, specifically: (1) *music digital libraries*, (2) *archives*, and (3) *dataset repositories*.

Analyzing music and music-related objects hosted by these categories of institutions, a great heterogeneity emerges. In the following, we will treat such a multifaceted subject by applying the multilayer approach mentioned above. The goal is to show how the extensive adoption of IEEE 1599 both as an encoding format and as an information interchange standard could federate multiple geographically-distributed sources in order to provide users with a comprehensive experience of music.

Examples of *music libraries* of our interest include:

(i) Traditional libraries with a digital catalogue – This is the case of the *Biblioteca Marciana* in Venice,[1] that holds manuscript scores by Antonio Vivaldi, but publishes online catalogue information only;[2]

(ii) Libraries sharing (also) digital objects – An example is the "Music Classics" section of the *BEIC – Biblioteca Europea di Informazione e Cultura*, where, in addition to catalogue information, excerpts of musical masterpieces are publicly available;[3]

(iii) Thematic on-line repositories – A relevant example focusing on free public-domain sheet music is *IMSLP – International Music Score Library Project*, also known as *Petrucci Music Library*.[4]

According to the 6-layer layout proposed in IEEE 1599, all the mentioned cases provide general information, i.e. metadata, while Case 2 involves also the audio layer, and Case 3 the notational one.

Digitized *music archives* are another relevant (and often heterogeneous) source of information. For instance, the *Teatro alla Scala* of Milan has created an integrated management system for its heritage: La Scala DAM (Digital Asset Management), that represents the digital archive of all the available material, from the second decade of the '900 through to the present, in various archives, warehouses and safes. The theatre's activity is documented in 5000 tapes of La Scala recordings of operas, ballets and symphonic concerts from 1950 onwards (circa 10,000 hours of music), in 17,000 posters and subsequent detailed chronologies, and in more than one million photographs taken from the stage, rehearsals and back-stage. Besides, *La Scala*'s archives comprises about 24,000 sketches and costume designs, 45,000 costumes completed by 60,000 accessories (jewels, clothing, footwear, wigs and hats), 80,000 props. A rich selection of these materials is freely available at ArchivioLaScala, the web page of the DAM Digital Archive.[5] In this case, a single multimedia database can provide information for multiple layers (general, notational, audio), and also enrich the description of a music piece with additional digital objects (e.g., on-stage photos, playbills, etc.).

[1] https://marciana.venezia.sbn.it/

[2] https://polovea.sebina.it/SebinaOpac/query/antonio%20vivaldi

[3] https://www.beic.it/it/articoli/classici-della-musica

[4] https://imslp.org/

Concerning the third typical case of music stakeholders, many *datasets* exist for MIR tasks, but they often lack the ability of inter-operation. For instance, [211] contains source-separated and mixed audio and video tracks, MIDI scores and frame-level transcriptions; in [212] a dataset containing audio recordings, music scores and sheet music images is used; another interesting multimodal dataset containing time-aligned notes, audio and lyrics is presented in [84]; audio recordings, notes and expressive markings were recently collected in [213]. To date, each of these datasets used its own format for representing the synchronization of music along the various modalities. The researcher who wants to use more than one dataset needs to implement multiple reading utilities. Moreover, these *ad-hoc* music formats are not able to deal with the future developments of music technologies, which continuously demands for the representation of new kinds of music information. For instance, the same music piece is often used in multiple libraries, but it is hard to merge the annotations coming from the various archives because they were digitized in an agnostic manner – that is, digitization/labeling efforts were conducted without reference to standard annotations for synchronization.

## 3.4. The Proposed Architecture

An in-depth explanation of the mechanisms adopted by IEEE 1599 to support multiple descriptions of music events and keep them mutually synchronized goes beyond the scope of this Thesis. It is just worth mentioning that each IEEE 1599 document, in a sub-part of the Logic layer, has a list of identifiers for music events, called the *spine*. While the logical descriptions of these events are expressed in XML format and contained inside the document itself, multimedia descriptions (e.g., images and audio tracks) are demanded to external digital objects encoded in usual formats (e.g., TIFF and MP3 files). In other words, the IEEE 1599 document does not embed multimedia, but contains in the suitable layers the information required to retrieve the occurrences of music events inside external files. For instance, the Audio layer links a number of audio performances of the piece, and the position of music events within each of these instances is expressed within the XML document as a time offset from the beginning of the track. Similarly, the Notational layer links external graphical files containing notation, and the music symbols corresponding to spine events are identified through bounding boxes expressed in pixels in the XML code.

This structure where synchronization information is internal to the document and multimedia information is external presents a number of advantages due to the adoption of commonly ac-cepted formats for digital objects: supporting a huge number of already-available media types and formats, relying on their characteristics to encode information efficiently and effectively, avoiding the need to re-encode information into the verbose XML language, etc.

[5] http://www.archiviolascala.org

In the context of our research, however, the main advantage is the possibility to link geographically distributed files with no change to the standard. This approach opens the way to a number of innovative applications based on information sharing and integration. The possibilities emerging from the establishment of a network of relationships both internal to an IEEE 1599 document and external, i.e. involving other information sources, has been already addressed in [214] in the context of the Semantic Web.

In the present work we propose a different vision, involving a number of musical assets' stakeholders sharing information in IEEE 1599 format. A possible architecture of the environment is shown in Figure 3.1. According to this star model, there is a *Central Node* that asks/receives information to/from both *Peripheral Nodes* and *File Repositories*. The former category is made by stakeholders that hold XML and multimedia information, whereas the latter by nodes that share only digital objects. The role of the *Central Node* is to coordinate such a multiplicity of actors, thus presenting music information to the *Web Server* as if it were contained in a single IEEE 1599 document. Similarly, the *Application Server* allows to add and remove multimedia objects, edit metadata, perform synchronizations, etc. operating on distributed objects in a transparent way. Concerning the physical location of IEEE 1599 documents, they could be hosted by the *Central Node* (centralized solution) or scattered across *Peripheral Nodes* as well, provided that the *Central Node* knows the location of the master document for each music piece.

Please note that, for the sake of simplicity, this diagram deliberately ignores problems such as redundancy, network topologies with the presence of firewalls and DMZs, etc.

## 3.5. Discussion and Final Remarks

One more example that can benefit from the adoption of the IEEE 1599 standard is the emerging field of multimodal analysis, that in recent years has raised more and more attention by the music research community. According to [215], which extensively reviewed the Multimodal Music Information Retrieval (MIR), the concept of *modality* can be defined as a music representation digitized in a particular place and time; multiple modalities can originate by digitizing music information in different places or times. This renewed interest in multimodal analysis was due to the occurrence of multiple factors, such as an increased computational power allowing for more complex approaches in everyday computers and the spread of machine-learning methods which are able to deal with heterogeneous kinds of data. In general, multimodal methods are useful in Computer Science where one modality is not enough for describing the whole object under study; music, indeed, is a complex process in which multiple abstraction levels cooperate. Despite the good results obtained through Multimodal MIR, some issue still remains for the field. The main bottleneck for these methods is the lack of federated datasets, which represents a limit for machine learning approaches. IEEE 1599, thanks to the possibility to provide multiple descriptions (also coming from different sources) and synchronize them, can be the answer.

Now, let us analyze the main advantages of the proposed approach from the stakeholder per-spective, thus omitting the already-mentioned benefits for different categories of final users. First, stakeholders would keep the control over their assets, since metadata and digital objects are physically hosted on *Peripheral Nodes* and *File Repositories*.

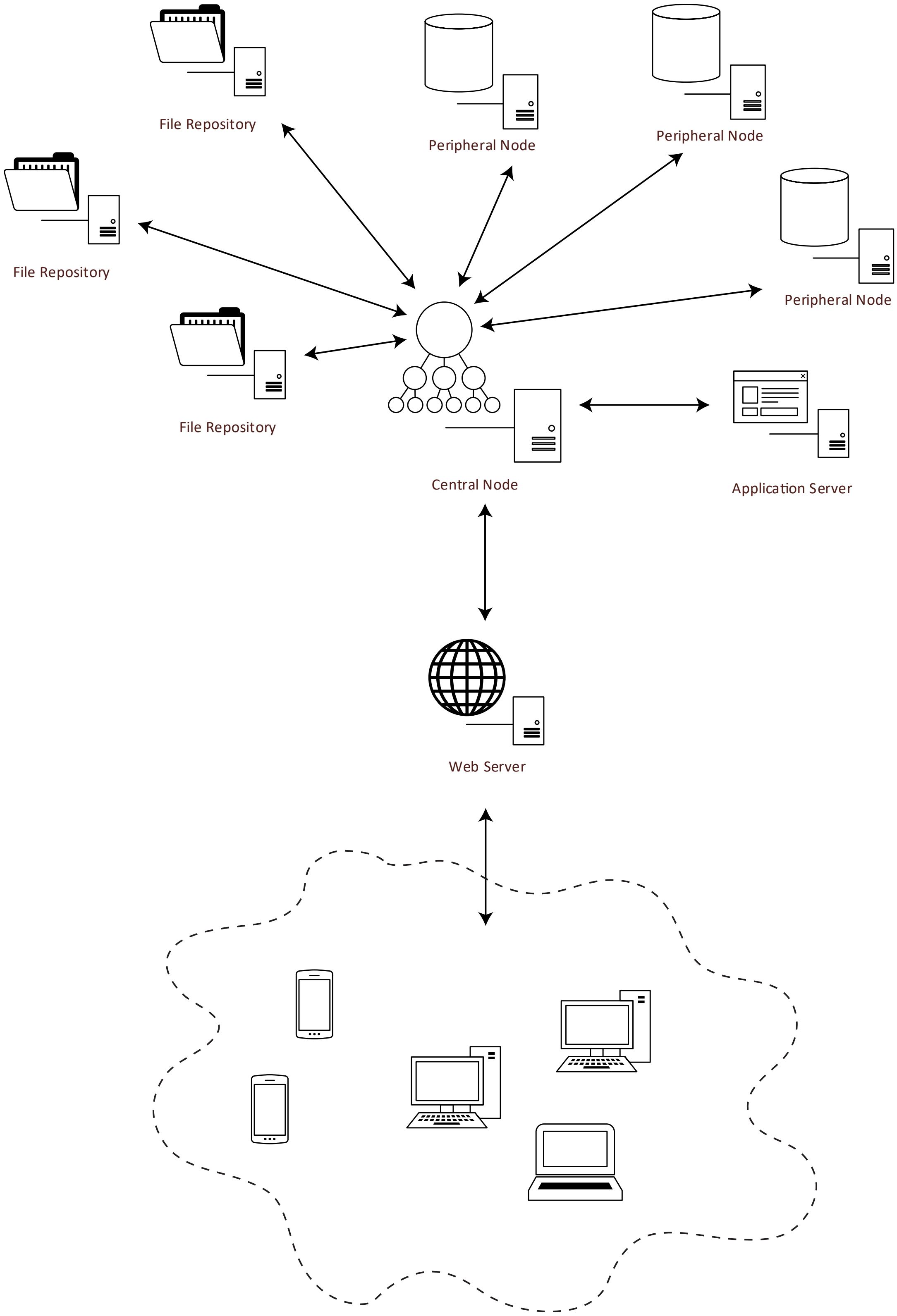

**Figure 3.1.:** *The proposed architecture for musical assets' stakeholders.*

Also the protection of intellectual property could be easily implemented, since each actor could decide a specific policy to grant access to its own assets; then, the *Central Node* would be in charge of filtering requests in order to meet such requirements, in accordance with the scenarios described in [207].

Is IEEE 1599 ready to support these ideas? Based on the state of the art, namely the 2008 version, the answer is only partially affirmative. For example, it is already possible to link remote files by providing a Uniform Resource Locator (URL), the General layer and multimedia-related layers allow to encode basic catalogue metadata, and a light support is offered to Digital Rights Management (DRM). Nevertheless, much remains to be done. For instance, some formats that have recently affirmed in the field of music interoperability (e.g., the revision of MusicXML in the context of the W3C Music Notation Community Group)[6] have not been integrated yet. Similarly, IEEE 1599 offers no native support to commonly in-use international standards that describe:

- core metadata (e.g., the Dublin Core Metadata Initiative – DCMI) [216];
- traditional items catalogued by libraries (e.g., MAchine Readable Cataloguing – MARC) [217];
- objects within a digital library (e.g., the Metadata Encoding and Transmission Standard – METS) [218];
- control policies for DRM (e.g., the Open Digital Rights Language – ODRL) [219].

[6]https://www.w3.org/community/music-notation/

# 4
# ASMD – Audio-Score Meta-Dataset

In this chapter, a new, automated framework for compiling multimodal datasets is presented. The framework, named Audio and Score Meta Dataset [4] (ASMD), works as main dataset source for the successive chapters. The principles of *extensibility* and *modularity* have inspired its design, while the main motivations behind its creation are *reproducibility* and *generalization* of music research.

The framework was first introduced in [4] and then updated with new datasets and features in [5]. Therefore, this chapter takes as reference the latest stable release of ASMD, namely version 0.5.

## 4.1. Introduction

Two issues that are more and more debated in several research fields are the ability to *reproduce* published results [220, 221] and to *generalize* the resulting models [222].

Reproducibility is associated with the differences occurring in various implementations of the same method. As an example, one issue is related to the different data formats used in music and in the available datasets, which might cause troubles in the translation between representation formats and, consequently, in the reproducibility of research.

The generalization problem instead is due, among other factors, to the need of large and well-annotated datasets for training effective models. In particular, the whole field of music information processing has only a limited number of large datasets, which could be much more useful if they could be merged together. Music itself, moreover, is particularly affected by the difficulty of creating accurate annotations to evaluate and train models, often hindering the collection of large datasets and causing a low generalization ability.

With these three keywords in mind (*multimodality*, *reproducibility* and *generalization*), we have built ASMD to help researchers in the standardization of music ground-truth annotations and in the distribution of datasets. ASMD is the acronym for Audio-Score Meta-Dataset and provides a framework for describing, converting, and accessing a single dataset that includes various datasets – hence the expression *Meta-Dataset*; it was born as a side-project of a research about audio-to-score alignment and, consequently, it contains audio recordings and music scores for all the data included in the official release – hence the *Audio-Score* part. However, we have endeavoured to make ASMD able to include any contribute from anyone. ASMD is available under free licenses.[1]

A similar effort is held by mirdata [223], a Python package for downloading and using common MIR datasets. However, our work is more focused on multimodality and tries to keep the entire framework easily extensible and modular. Section 4.5 compares these two tools in detail.

In Chapter 4, we describe a) the design principles, b) the implementation details, c) a few use cases and, d) possible future works.

## 4.2. Design Principles and Specifications

In this section we present the principles that guided the design of the framework. Throughout this paper, we are going to use the word *annotation* to refer to any music-related data complementing an audio recording. For instance, common types of annotations are music notes, f0 for each audio frame, beat position, separated audio tracks, etc.

### 4.2.1. Generalization

With *generalization*, we mean the ability of including different datasets that are distributed with various formats and annotation types in the model generation process. This is an important issue especially during the conversion procedure: since we aimed at distributing a conversion script to recreate annotations from scratch for the sake of *reproducibility*, we need to be able to handle all various storage systems – e.g. file name patterns, directory structures, etc. – and file formats – e.g. midi, csv, musicxml, ad-hoc formats, etc.

Also, our ground-truth format should be generic enough to represent all the information contained in the available ground-truths and, at the same time, it should permit to handle datasets with different ground-truth types – i.e. one dataset could provide *aligned notes* and *f0*, while another one could provide *aligned notes* and *beat-tracking*, and they should be completely accessible.

### 4.2.2. Modularity

*Modularity* refers to the re-use of parts of the framework in different contexts. Modularity is important during both addition of new datasets and usage of the API. To ease the conversion between ground-truth file formats, the user should be able to re-use existing utilities to include additional datasets. Moreover, the user should be allowed to use only some parts of the datasets and the corresponding annotations.

### 4.2.3. Extensibility

The purpose of the framework is to create a tool to help the standardization of music information processing research. Consequently, we aimed for a framework that is completely open to new additions: it should be easy for the user to add new datasets without editing sources from the framework itself. Also, it should be easy to convert from existing formats in order to take advantage of the API and to be able to merge existing datasets. Finally, the framework should provide a usable format to add new annotations so that new datasets can be natively created with the incorporated tools.

[1] Code is available at https://github.com/LIMUNIMI/ASMD/, documentation is available at https://asmd.readthedocs.io/

### 4.2.4. Set operability

Since the framework aims at merging multiple datasets, we wanted to add the ability to perform set operations over datasets. As an example, within the context of *automatic music transcription* research, several large datasets exist consisting of piano music [224, 225, 226], but only few and considerably smaller are available for other instruments [227, 228, 229, 230, 231]. Consequently, a useful feature of the framework would be the ability to select only some songs from multiple datasets based on particular attributes, such as the instrument involved, the number of instruments, the composer or the type of ground-truth available for that song.

### 4.2.5. Copyrights

A common issue with distributing music recordings and annotations are copyrights. Today, most of the datasets typically used for music information processing are released under Creative Commons Licenses, but there are many exceptions of datasets released under closed terms [228, 232] or not released at all because of copyright restrictions [2]. To overcome this problem, we wanted all datasets to be downloadable from their official sources, in order to avoid any form of redistribution. Nonetheless, all the annotations that we produced were redistributable under Creative Commons License.

### 4.2.6. Audio-score oriented

Besides the effort to produce a general framework for music processing experiments, this project was born as a utility during conducting research addressing the audio-to-score problem. The underlying idea is that we have various scores and large amounts of audio available to end-users, thus trained models could easily take advantage of such multimodality (i.e. the ability of the model to exploit both scores and audio). The main problem is the availability of data for training the models: there is abundance of aligned data, but without the corresponding scores; on the other hand, when scores are available, aligned performances are almost invariably missing. Thus, the choice of the datasets that are included at now has mainly been focused on datasets providing audio, symbolic scores and alignment annotations. However, since datasets fitting all these requirements are quite rare, we wanted to augment the data available to increase the alignment data usable in our research.

## 4.3. Implementation Details

This section details the implementation satisfying the design principles outlined in section 4.2. Figure 4.1 depicts the structure of the overall framework and the interactions between its modules.

**Listing 4.1:** *Example of* datasets.json *file*

```
{
    "author" : "Federico Simonetta",
    "install_dir" : "/datasets/AudioScoreDatasets-v0.4",
    "year" : 2020,
    "url" : "https://federicosimonetta.eu.org",
    "decompress_path" : "./"
}
```

### 4.3.1. The datasets.json file

The entire framework is based on a small-sized but fundamental JSON file loaded by the API and the installation script to get the path where files are installed. Moreover, the user can optionally set a custom directory where to decompress downloaded files if the hard-disk space is a critical issue. Once the installation path is found, the script looks for the existing directories in that path to discover which datasets are already installed and skips them. The API, instead, uses the information of the installation directory to decouple the definition of each single dataset from the directory structure of the user: a user can have the same dataset installed in multiple directories, or use the same dataset from different *datasets.json* without interfering with the API.

### 4.3.2. Definitions

In the context of this framework, a *dataset definition* is essentially a JSON[2] file that contains generic description of a dataset. *Definitions* are built by using a pre-defined schema allowing the definition of various information useful for the installation of the dataset and for the usage of the API – e.g. for filtering the dataset. If any of the information is not available for a dataset, the value unknown is offered as well.

Examples of information contained in *definitions* are:

(i) ensemble: if the dataset contains solo instrument music pieces or ensemble;
(ii) groups: the list of subset included in the dataset – e.g. *train*, *validation*, *test*.
(iii) instruments: a list of instruments that are used in the dataset;
(iv) sources: if source-separated tracks are available, their format can be added here;
(v) recording: the format of audio recordings;
(vi) install: field containing all information for installing the dataset: URL for downloading, shell commands for post-processing data, and so on;
(vii) ground-truth: field associated to each type of ground-truth supported by the framework indicating whether the specific annotation type is available or not – see Sec 4.3.3;
(viii) songs: a list of songs with subsets to which the song belongs to, the meta-data such as the composer name and instruments used in these songs, and the list of paths to the audio recordings and to the annotations.

[2]https://www.json.org/json-en.html

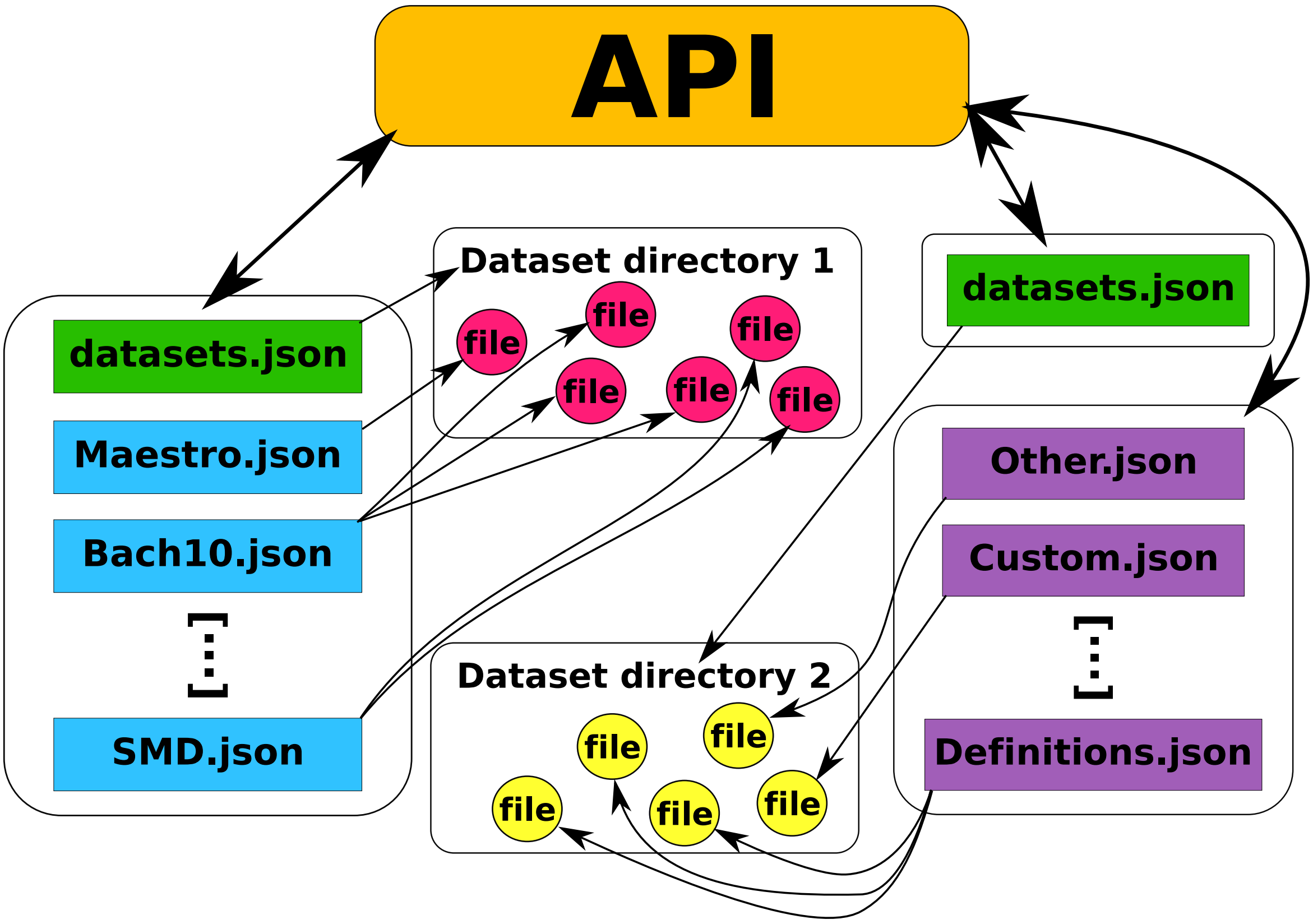

**Figure 4.1.:** *Block diagram of the proposed framework: API interacts with definitions and* datasets.json*; the former contain references to the actual sound recording files and annotations, while the latter contains references to the dataset root path.*

Once a dataset has been described in this schema, its definition can be used *out-of-the-box* by simply specifying to the API the path of its folder, possibly containing other dataset definitions. All the paths specified in a *definition* must be relative to the installation directory as described in Sec. 4.3.1.

For the sake of generalization, we had to deal with a wide heterogeneity in path management among datasets. For instance, Bach10 [230] provides one different annotation file per each instrument in a song; in such a case we list all the annotation files for each song and leave to the API the task of reassembling them. PHENICX [228], instead, only provides source-separated tracks and thus we list all of them to reference the mixed track; again, we leave to the API the task of mixing them. In general, we have kept the following principle: if a list of paths is provided where one would logically expect a single path – such as in mixed tracks or annotation files – it is intended that the files in the list should be "merged" whatever this means for that specific file-type. For instance, if multiple audio recordings are provided instead of only one, it is assumed that the *mixed* track is derivable by adding (and normalizing) all listed tracks; if multiple annotation files are provided, it is assumed that each annotation file refers to a different instrument.

**Listing 4.2:** *Example of usage for official* definitions

```
from asmd.asmd import Dataset
from asmd import dataset_utils as du

data = Dataset()
du.filter(data, instruments=['piano'], ensemble=False, composer='Mozart',
    → ground_truth=['precise_alignment'])

# get audio and all the annotations audio_array, sources_array,
ground_truth_array = data.get_item(1)

# get only the annotations you want audio_array = data.get_mix(2) source_array =
data.get_source(2)
ground_truth_list = data.get_gts(2)

# get a MIDI Toolbox-like numpy array
mat = du.get_score_mat(data, 2, score_type=['precise_alignment'])

# get a pianoroll numpy array
mat = data.get_pianoroll(2, score_type=['non_aligned'])

# or to process songs in parallel using joblib:
def processing(i, dataset, **args, *kwargs):
  mat = du.get_score_mat(data, 2, score_type=['precise_alignment'])
  # other stuffs here
  pass

du.filter(data, instruments=['violin']).parallel(processing, n_jobs= 1)
```

### 4.3.3. Annotations

Annotations are added in a custom JSON compressed format stored in the same directory of the audio track that they refer to. In fact, annotation files can be stored anywhere and their path must be provided in the dataset definition relatively to the installation path defined in *datasets.json*. Moreover, one annotation file must be provided for each instrument of the track; if multiple instruments should refer to the same annotations – e.g. first and second violins – the annotation file can be only one, but in the dataset definition file, its path should be repeated once for each instrument referring to it.

Multiple types of annotations are available, but not all of them are provided for all the datasets in the official collection. In the dataset definition, the type of annotations available should be explained. In our implementation, we used 3 different levels to describe ground truth availability and reliability:

**0:** annotation type not available

**1:** annotation type available and manually or mechanically annotated: this type of annotation has been added by a domain expert or some mechanical transducer – e.g. Disklavier.

**2:** annotation type available and algorithmically annotated: this type of annotation has been added by exploiting a state-of-art algorithm.

The types of annotations currently supported are:

(i) *precise alignment*: onsets and offsets times in seconds, pitches, velocities and note names for each note played in the recording, taking into account asynchronies inside chords;

(ii) *broad alignment*: same as *precise alignment* but the alignment does not consider asynchronies inside chords;

(iii) *misaligned*: same as *precise alignment* but artificially misaligned (see Section 4.3.4 for more information);

(iv) *score*: similar to *precise alignment* but it refers to a music score; times are in seconds and computed using 60 BPM where BPM is not available;

(v) *missing*: a boolean list that can be used to identify missing notes – i.e. notes that you can consider as being played but not in the score; these notes are chosen algorithmically.

(vi) *extra*: a boolean list that can be used to identify extra notes – i.e. notes that you can consider as not being played but in the score; these notes are chosen algorithmically.

(vii) *f0*: the f0 of this instrument for each audio frame in the corresponding track;

(viii) *sustain*, *soft* and *sostenuto*: the list of piano pedaling values used in the performance

(ix) *instrument*: General Midi program number associated with this instrument, starting from 0, while value 128 indicates a drums kit.

### 4.3.4. Alignment

As described in section 4.2.6, this project originated for music alignment research. One problem is the lack of large datasets containing audio recordings, aligned notes and symbolic non aligned scores.

The approach that we used to overcome this problem is to statistically analyze the available manual annotations and to augment the data by approximating them through the statistical model. To prevent biases, we also replaced the manual annotations with the approximated ones. In the latest release, a number of improvements were made to such statistical approximation.

The first improvement we made is the addition of the ASAP [233] dataset to enlarge the number of considered statistics. Second, we used A note matching method (EITA) [234] (EITA) to select matching notes against which we compute statistics as well. Third, instead of modeling the misalignment of onsets and offsets, we have now recorded statistics about the onsets and the duration ratio between score and performance. Fourth, statistics are computed with models trained on the "stretched" scores, so that the training data consists of scores at the same average BPM as the performance; as such, the misaligned data consists of times at that average BPM.

More precisely, we create statistical models as follows:

(i) we compute standardized onset misalignment and duration ratio for each note by subtracting the mean value for that piece and dividing by the standard deviation;

**Table 4.1.:** *L1 macro-average error between artificial misalignments and ground-truth scores.*

| | **Ons** | **Offs** |
|---|---|---|
| **GMM-HMM** | 18.6 ± 49.7 | 20.7 ± 50.6 |
| **Histograms** | 7.43 ± 15.5 | 8.95 ± 15.5 |

(ii) we collect two histograms, one for the standardized onset misalignments ($X_{ons}$) and one for the standardized duration ratios ($X_{dur}$);

(iii) we collect each piece-wise mean and standard deviation in four histograms: two for the onset misalignment means and standard deviations ($Y^{m}_{ons}$, $Y^{std}_{ons}$), and two for duration ratio means and standard deviations ($Y^{m}_{dur}$, $Y^{std}_{dur}$)

To infer a new misaligned onset or duration, we choose a standardized value for each note from histograms $X_{ons}$ and $X_{dur}$, and a mean and a standard deviation for each piece, using the corresponding histograms $Y^{m}_{ons}$, $Y^{std}_{ons}$, $Y^{m}_{dur}$, $Y^{std}_{dur}$; with these data, we compute a non-standardized onset misalignment and duration ratio for each note. These two latter values can be used in reference to the ground-truth performance to compute the misaligned timing values.

We actually tested two methods for choosing the standardized value: an HMM with Gaussian mixture emissions (GMM-HMM) and the above-described histogram-based sampling. We hand-tuned the HMMs finding an optimum in 20 states and 30 mixture components for onsets and 2 states and 3 components for duration models. We then compared GMM-HMM and histogram models on the notes matched by the EITA method. During this evaluation, we used the scores provided by ASMD for a total of 875 scores, namely "Vienna Corpus" [226], "Traditional Flute" [229], "MusicNet" [227], "Bach10" [230] and "asap" [233] group from "Maestro' [224] dataset. We divided the data into train and test sets with 70-30 proportion, resulting in 641 pieces for training and 234 for testing. As evaluation measure, we used the L1 macro-average error between matching note onsets and offsets in music score and performance. However, due to EITA's high computational cost, we removed the scores for which EITA terminates after 20 seconds. This resulted in a total of 347 songs for training and 143 songs for testing — ~54% and ~61% of the corresponding splits. Table 4.1 shows the results.

Misaligned data are finally created, using the histogram-based method for every dataset provided by ASMD by collecting the histograms corresponding to all 875 scores — 481 considering songs where EITA method took less than 20 sec. Thus, we set up a corpus of 1787 music recordings with misaligned and aligned MIDI data.

Artificially misaligned data is more similar to a different performance than to a symbolic score; however, for most of MIR applications, such misaligned data is enough to cover both training and evaluation needs.

In ASMD version 0.5, we provide randomly generated missing and extra notes as well. To this end, we chose the number $n$ of notes to be tagged as "missing" or "extra" as a random variable with uniform distribution in $(0.1 \times L, 0.5 \times L)$, where $L$ is the number of notes in the music piece. Then, we picked random contiguous sequences of notes until the total number $n$ was met and we labeled each of the chosen region as "missing" or "extra" according to two random variables $p_1$ and $p_2$ defined by a uniform distribution in $(0.25, 0.75)$ and $p_2 = 1 - p_1$.

An additional problem is due to the fact that the time units in the aligned data are seconds, while those in the scores are note lengths – e.g. breve, semibreve and so on. Usually, one translates a note length to seconds by using BPM; however, in some scores the BPM annotation is unavailable or is not reliable. To be consistent with the other representations of music performance and to bypass the lack of BPM information in some cases, in the annotated ground-truth, we always consider the tempo as 60 BPM. Note that one can still change the BPM timing of the music score data by using the beat timing provided in it. For the statistical analysis we have always used the average BPM of the approximated performance; the conversion to the average BPM is provided in the API.

### 4.3.5. API

The framework is complemented with a Python API. It allows in particular to load various dataset definitions aside of the official ones. The API provides methods to retrieve audio and annotations in various structures, such as a matrix list of notes similar to the one used by *Matlab MIDI Toolbox*[235] or pianorolls. Thanks to the API, one can also filter the loaded datasets' songs based on the original dataset, active instrument, ensemble or solo instrumentation, composer, available annotation types, etc. Moreover, in version 0.5, it is also possible to perform union, intersection and complement operations over `Dataset` objects, so that the user can easily create datasets from multiple filtering options.

Finally, since the API basically consists in a class representing a large dataset, it is very easy to extend it in order to use it in conjunction of PyTorch or TensorFlow frameworks for training neural network models. In Section 4.4 we provide an example of the specific functionality.

### 4.3.6. Conversion

To give the user the ability to write his/her own definitions without having to edit the framework code, we designed a conversion procedure as follows:

(i) the creator can use already developed conversion tools for the most common file formats (MIDI, sonic visualizer, etc.);

(ii) the creator can still write an ad-hoc function which converts a file from the original format to the ASMD one; in this case the creator has to decorate the conversion function with a special decorator provided by ASMD;

(iii) the creator adds the needed conversion function in the `install` section in the dataset definition;

(iv) the user can run the conversion script for only one specific dataset or for all other datasets.

All the technical details are available in the official documentation.[3]

[3]https://asmd.readthedocs.io/

**Listing 4.3:** *Example of usage for custom* definitions

```
from asmd import asmd

d = asmd.Dataset(['path/to/directory/containing/custom/definitions',
'path/to/the/official/definitions/']) d.filter(instruments=['piano'],
ensemble=False, composer='Mozart', ground_truth=['precise_alignment'])
```

**Listing 4.4:** *Example for using ASMD inside PyTorch*

```
import torch
from asmd import asmd
from asmd.dataset_utils import get_score_mat
torch.utils.data import Dataset as TorchDataset

class MyDataset(asmd.Dataset, TorchDataset):
  def __init__(self, *args, **kargs):
    super().__init__(['path/to/definitions']).filter(instruments=['piano'])

  def __getitem__(self, i):
    # for instance, return the MIDI Toolbox-like score
    return torch.tensor(get_score_mat(i))

  def another_awsome_method(self, *args, **kargs):
    print("Hello, world!")

for i, mat in enumerate(MyDataset()):
  # train your nn model here
```

## 4.4. Use Cases

This section demonstrates the efficacy of the ASMD framework through four different use cases.

### 4.4.1. Using API with the official dataset collection

To use the API, the user should carry out the following steps:

(i) import asmd.asmd module;

(ii) create an asmd.Dataset object, giving the path of the datasets.json file as an argument to the constructor;

(iii) use the filter method on the object to filter data according to his/her needs (conveniently, it is also possible to re-filter them at a later stage, without reloading the datasets.json file);

(iv) retrieve elements by calling the get_item method or similar ones.

After the execution of the filter method, the Dataset instance will contain a field paths repre-

**Listing 4.5:** *Example for writing a custom conversion function*

```
from asmd.convert_from_file import convert, prototype_gt
from copy import deepcopy

# use @convert
@convert(['.myext'])
def function_which_converts(filename, *args, **kargs):
   # prepare empty output
   out = deepcopy(prototype_gt)

   # open file
   data = csv.reader(open(filename), delimiter=',')

   # fill output dictionary for row in data:
   out[alignment]["onsets"].append(float(row[0]))
   out[alignment]["offsets"].append(float(row[0]) + float(row[2]))
   out[alignment]["pitches"].append(int(row[1])

   return out
```

senting the list of correct paths to the files requested by the user. Listing 4.2 shows an example of such an operation.

### 4.4.2. Using API with definitions for a customized dataset

Whenever the user wishes to apply customized definitions, he/she need simply to provide the list of directories to the Dataset constructor, as shown in listing 4.3.

### 4.4.3. Using ASMD with PyTorch

Integrate *ASMD* with *PyTorch* is straightforward. The user has to inherit from both PyTorch and ASMD Dataset classes and to implement the __getitem__ method. Listing 4.4 shows such an example.

### 4.4.4. Writing a conversion function and a custom dataset definition

Towards adding new definitions enabling users to download datasets, a user should also provide a conversion function. Listing 4.5 is an example of one can write its own conversion function. However, conversion functions for the most common file types – such as Midi and Sonic Visualizer – are already provided by the framework.

## 4.5. Comparison with similar tools

While ASMD was being developed, another Python module named miradata [236] was independently created and published, aiming at the management of various Music Information Processing (MIP) datasets.

The main distinction between ASMD and mirdata was that the former focuses on the multimodal representations. However, ASMD has received only a little engagement from the research community, while mirdata is slowly becoming a must-have tool for MIP projects. The large number of contributors participating to mirdata, led it to the ability of representing other kind of information about music, including scores; on the opposite, the little engagement of ASMD prevented the inclusion of modalities that mirdata is not considering now – e.g. video, images, lyrics. We think that the reason for such adoption disparity must be looked not in the features itself – mirdata was much different from the current version when it was released for the first time – but in the connections between the developers and the research community.

However, a few design choices trace differences between ASMD and mirdata giving the former some advantages over the latter.

The first asset of ASMD is its intrinsic distributed nature: if mirdata adopts a centralized management where developers decide which datasets should be included via github discussions, ASMD is designed from its grounds to allow researchers to distribute their datasets without interfacing with the developers of the API, leaving to the usual research channels – not to github community – the decision of which datasets should be disseminated.

A second main difference between mirdata and ASMD is the handling of annotation data. Both the two modules gives the users the ability to access the original dataset's data, but only ASMD implements its own file format for annotations. mirdata actually defines a data format, but it is aimed at representing annotations at runtime, even if they could be easily saved to file by using serialization functionalities of Python. The difference, however, is fundamental: designing the API to convert the annotations during the installation procedure allows to reduce the loading time, which may be a critical aspect in data-driven approaches.

The main weakness of ASMD is probably the annotation format, that lacked the contribution of other researchers until now, contrarily to mirdata that managed to receive more attention. For this reason, ASMD could take advantage from other experiences, including mirdata, to improve the data format and building a tool to make mirdata data format compatible with the distributed design of ASMD, easily cacheable, and extensible.

Other minor differences exist, such as the ability for ASMD to perform set-operations and to consider different groups of the same dataset. mirdata, instead, has a better dataset version management but an obscure management of the dataset files, making it harder to access the original data. In general, all the remaining features of mirdata and ASMD could easily be implemented in both the two modules. For this reason, if the development of newer features for ASMD is in the road-map, it is reasonable to think that such new features will also be added to mirdata.

Considering the much larger adoption of mirdata by the research community, we think that ASMD can be considered as an exploratory project that seeks to enhance existing data management tools with no ambition of replacing them.

## 4.6. Conclusions

Future works will focus on the enhancement of conversion and installation procedures, as well as on the definition of standards for music annotations. In addition, multimodal music processing often requires processing of annotation types not included in the current of the framework, but could instead be handled in a future release. Some annotation types could be stored in standalone formats and users should be able to distribute annotations focusing only on a specific ground truth kind, thus enhancing the distributed infrastructure of ASMD.

Moreover, the addition of newer modalities (video, images, and lyrics) is in the road map, as well as the implementation of music matching algorithms to match the same music piece across different datasets automatically; the objective is to potentially empowering users with the ability of using annotations available in a dataset A for the same music piece from dataset B.

Studying the user experience of the framework should also be a priority: for instance, users could be able to choose datasets also based on the estimation of the download time since for some datasets that is a relevant issue. Labels used in the annotation format are also relevant to ease the usage of the framework by new users, especially in a multidisciplinary field such as sound and music computing.

This Chapter presented a new framework for multimodal music processing. We hope that our efforts in easing the development of multimodal machine learning approaches for music information processing will be useful to the sound and music computing community. We are completely aware that for a truly general and usable framework, the participation of various and different perspectives is needed and we are therefore open to any contribution towards the creation of utilities that allow training and testing multimodal models, ensuring reasonable generalization ability and reliable reproducibility of scientific results.

# Part III.

# Improving multimodal music processing

# 5

# Multimodal music source separation: melody identification in symbolic scores

This Chapter introduces a first effort towards the enhancement in Music Information Processing (MIP) methods designed for "Music Interpretation Analysis" (MIA) and more in general for Music Performance Analysis (MPA). The underlying idea is that piano Automatic Music Transcription (AMT) can be seen as a source-separation task where each key is a different source. Towards enhancing the analysis of piano music performances, this Chapter proposes a source-separation method in the symbolic domain. Specifically, a method for melody identification in symbolic scores is presented. Even though the proposed method is still quite distant from the main topic of the Thesis, it can be used to assist multimodal melody separation in the audio domain, serving as guidance to any related application. Moreover, it delivers an example of a MIP application that, even though it may be perceived as a completely different task, can turn out to be useful in the MPA field as well thanks to multimodal approaches described in Chapter 1.

The Chapter is structured as follows: in Section 5.2, we discuss related work on voice separation and streaming. Section 5.3 briefly describes the baseline methods that we used for comparison against our model. Section 5.4 presents a description of the proposed method. Section 5.5 describes the three datasets used in this work. Section 5.6 describes the experimental evaluation of the proposed method. Section 5.7 discusses the results of the experimental evaluation. Finally, Section 5.8 concludes this paper and proposes some future research directions. A companion website was also created to show additional material for the sake of reproducibility.[1]

## 5.1. Introduction

Many musical traditions make use of melody-accompaniment structures. Generally, the melody line carries the most significant meaning, while the accompaniment provides harmonic and rhythmic support.

In Western art music – which, unlike music in some other traditions, is typically notated – special attention is paid to the construction of melodies during composition. Ideally, melodies in Western art music styles should involve an intervallic structure that is dependent on the specific tonal hierarchy defined by the piece [237, 238]. Musicians typically accentuate melody lines dur-ing performance as a way of clarifying the piece structure for listeners: for example, melody lines may be played louder and with more flexible timing than accompaniment [239, 240].

[1]https://limunimi.github.io/Symbolic-Melody-Identification/

The term melody can be defined in different ways, and is often thought of as the main theme of a piece. More broadly, a melody can be described as the most structurally salient part of a piece at any given moment. As an example, in Mozart's Sonata No. 16 in C major, K. 545, the melody comprises the main theme for the first four bars. A bridge passage follows, in which the melody comprises a series of scales, which cannot be considered a new theme, but are still of primary structural significance. Melodies may shift between voices as a piece unfolds. Fugues, for example, comprise multiple interwoven voices which develop one or more themes through imitation and repetition.

For the purpose of this research, we restrict our focus to the identification of *melody lines*. Our goal is therefore to develop a method that identifies the part or voice that is the most structurally salient overall , or carries the melody most of the time.

Most listeners readily distinguish melody lines from accompaniment. In contrast, identifying the melody line through visual inspection of a musical score – without hearing the piece – can be a difficult task, even for trained musicians [241]. In this paper, we propose a convolutional approach for identifying the melody line of a piece using a piano roll representation of the score. A solution for this task has potential implications for music information retrieval and musicology [215]. An effective algorithm could be applied to music retrieval tasks such as music transcription, query-by-humming, searching a database of MIDI files for melodies, developing performance models that account for melody in predicting musical expression, etc. Our focus is on music of the common practice period that uses melody-dominated homophonic textures (i.e., a single melody line plus accompaniment lines), rather than equal-voice polyphony (i.e., multiple independent melody lines) or monophony (i.e., unison melody shared by all voices). However, we provide extensive tests of the proposed method in styles other than common practice era, such as pop, baroque and contemporary art music.

## 5.2. Related Work

### 5.2.1. Voices and Streams

Music perception research has investigated listeners' abilities to distinguish between voices in homo- and polyphonic music, and has shown that the theoretical rules of voice leading are motivated by listeners' abilities to follow voices [242]. Cambouropoulos [243] proposed three ways of defining musical "voices":

(i) for multi-instrument music, each instrument can be said to constitute a separate voice; this would allow for the possibility of non-monophonic voices in instruments that produce chords;
(ii) voices can be assigned to melodic streams as they are perceived and segmented by listeners, following cognitive grouping principles;
(iii) in monophonic music, the harmonic content of the piece may imply a horizontal organization of polyphonic voices that unfold over time – e.g. multiple temporally-overlapping

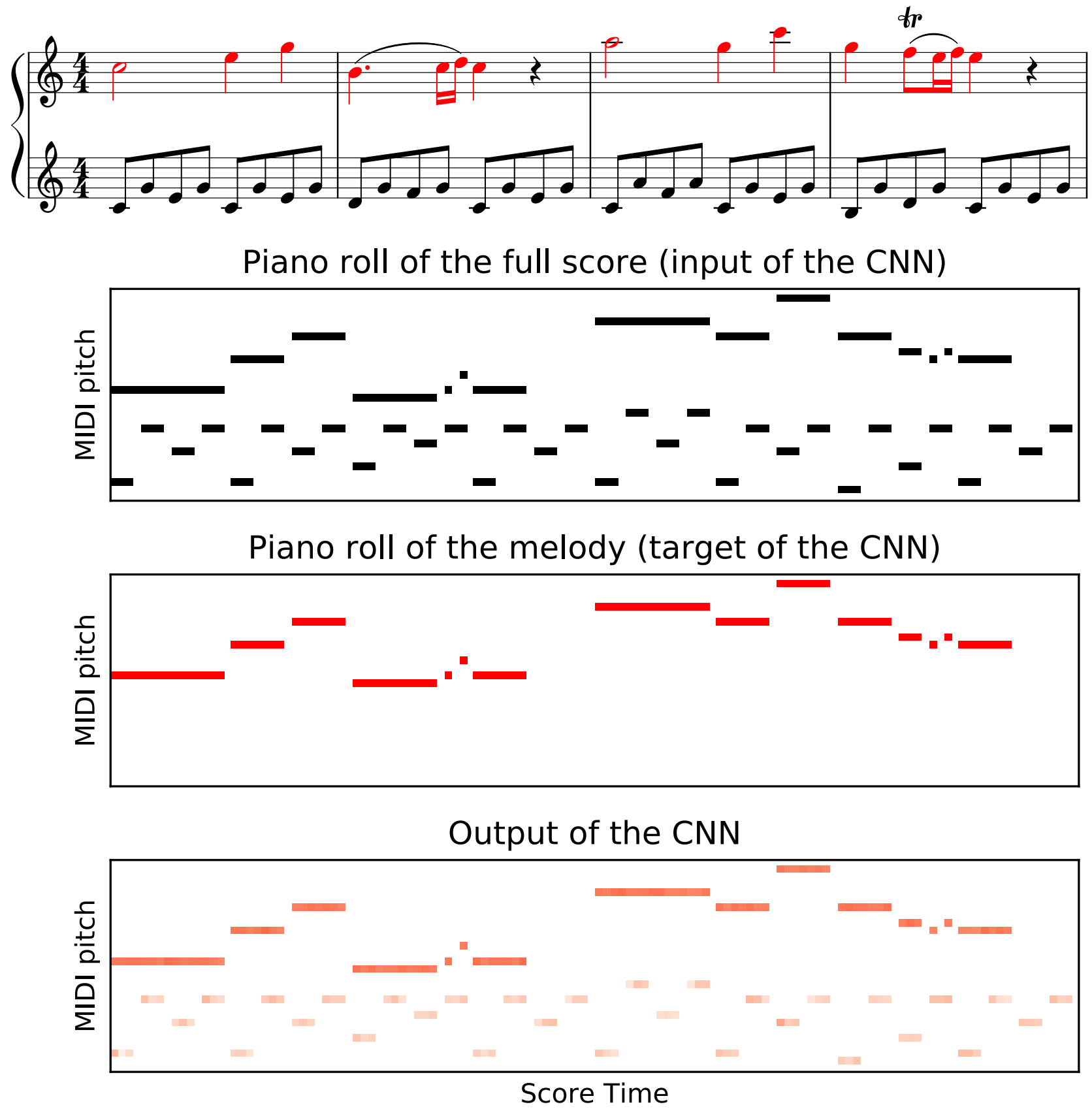

**Figure 5.1.:** *Top: Excerpt of Mozart's Sonata K. 545 (melody highlighted in red). Middle: Piano roll representation of the score (melody is highlighted in red). Bottom: Prediction of the CNN for this excerpt. In this piano roll, the intensity of the color of each cell represents its probability of belonging to the melody.*

voices could be assigned to passages of Bach's Cello Suites.

In this work, we use the second definition, and we define the melody line as the most salient voice.

In the music information retrieval literature, three corresponding tasks have been addressed:

(i) voice separation from symbolic scores [244, 245, 246, 247];

(ii) main track identification (from MIDI files with multiple tracks) [248, 249, 250, 251];

(iii) main melody identification from audio [252, 253, 254].

The latter is a different problem than that addressed here: it deals with the complex task of identifying notes from an audio file, but can use performance cues (e.g., contrasts in timbre and dynamics, which are not present in MIDI data) to facilitate melody identification.

Most relevant to the current study is the task of voice separation from symbolic scores. Some of the proposed methods are computational implementations that attempt to capture perceptual rules of segmentation [243, 244, 246, 247, 255] – in particular those rules codified by Huron [242].

For a more in-depth discussion on voice separation algorithms from symbolic scores, we refer the reader to [255, 256, 257, 258].

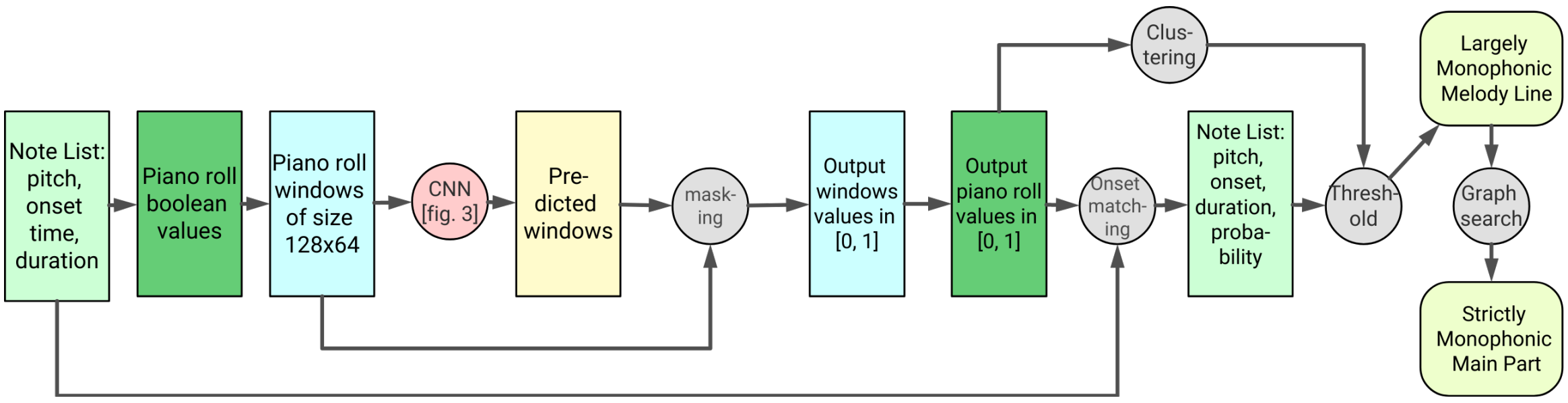

**Figure 5.2.:** *The pipeline of the proposed method (see Section 5.4). Starting from the note list, the CNN models the piano roll as boolean values. Its output is masked based on the available input and the melody-line probabilities are obtained. A clustering scheme follows and after a suitably-defined thresholding process, the largely monophonic main part is extracted. Finally, the strictly monophonic main part is retrieved by means of a graph-based search.*

## 5.3. Baseline Methods

### 5.3.1. Skyline Algorithm

The Skyline algorithm is a heuristic that takes the highest note at each point in time [259, 260]. In Western art music, pop and many folk traditions from around the world, melodies are often carried by the highest voice. After the submission of this paper, we discovered that a new method was being submitted for this same task [261], confirming the relevance of this topic.

### 5.3.2. VoSA

Proposed by Chew and Wu [247], VoSA is a successful voice separation method. In this approach, a piece is split into segments based on voice entry and exit points, so that the number of sounding notes is constant within each segment. The segment with the highest number of sounding notes defines the number of voices in the piece. Notes are then connected into voices using connection weights, equal to the absolute size of the interval between one note and the next. Like most voice separation methods, VoSA was designed to work with polyphonic rather than homophonic music. In spite of its apparent simplicity, VoSA has been favorably compared against more sophisticated computational models of voice separation [246, 255, 262].

## 5.4. Method

### 5.4.1. Music Score Modeling Using CNNs

A schematic representation of our method is given in Figure 5.2. The backbone of the method consists of a fully convolutional neural network (shown in Figure 5.3), which takes as input segments of a music score, represented as a piano roll, and estimates the probability that each note in the score (more precisely: each cell in the piano roll encoding) belongs to the melody line.

A piano roll can be described as a 2D representation of a musical score; the x-axis indicates score time and the y-axis indicates pitch. The piano rolls used in this study are constructed with a temporal resolution of 8 cells/beat (i.e., a cell represents a 32nd note in $\frac{4}{4}$). The piano roll of each piece is divided into overlapping fixed-length windows of 64 cells (i.e., 8 beats). The length of the window was determined using hyper-parameter optimization, (see Section 5.6.2). The overlap between windows is 50% (i.e., 2 beats), and windows shorter than this size are padded with zeros.

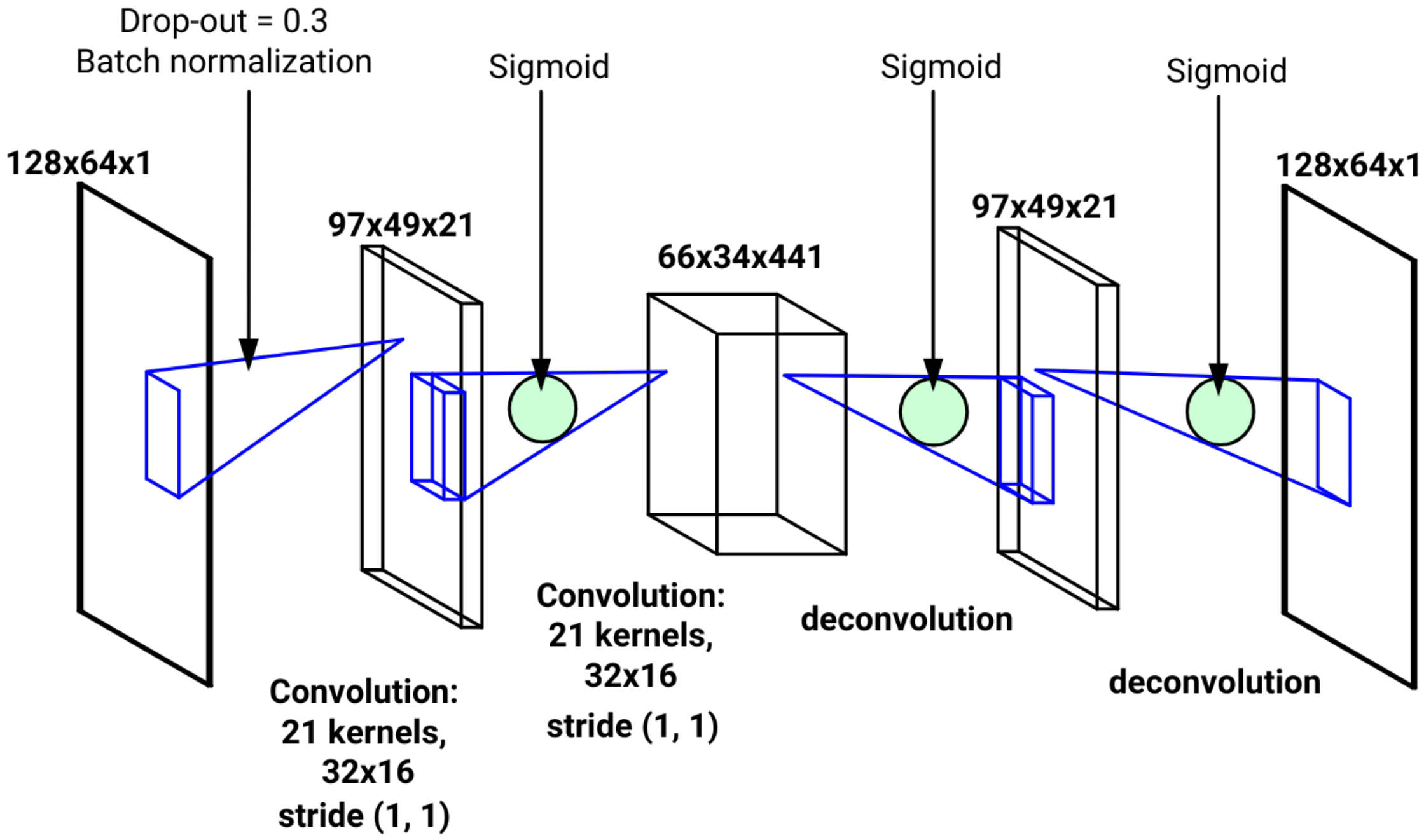

**Figure 5.3.:** *The architecture of the fully convolutional neural network used in the proposed method. The architecture of the network was determined using hyper-parameter optimization (see Section 5.6.2 for explanation). Input and output are two matrices of the same size (128 rows and 64 columns). The kernel size was fixed at* $32\times16$ *and an initial drop-out of* 0.3 *was used during the training. Moreover, all the parameters were regularized by adding the* $L_1$ *norm to each layer parameters.*

The task accomplished by the CNN can be seen as a filtering process, in which the input data are first compressed to select the relevant information – the encoder part of the network – and then uncompressed to restore the original size with the unwanted cells zeroed – the decoder part. However, it is hard for the network to filter out cells with high precision and non-zero probabilities arise in places where there are no notes, resulting in a slightly noisy output. To reduce noise, we apply a mask on the output piano roll by multiplying it by the (binary) input piano roll, so that areas with no notes take values of zero, and non-zero probabilities only remain where there are notes. The probability of each note belonging to the melody is then calculated as the median across the output values of its cells. In the following discussion, we will use *note probability* as a shorthand to refer to the probability of a note to belong to the melody.

In Figure 5.1 we show an excerpt of Mozart's Piano Sonata K. 545 and three vertically aligned piano rolls corresponding to the excerpt. The second row of this figure is the input piano roll, while the third row gives the ground truth melody line that we aim to identify in the input. The bottom gives the piano roll that we obtain as output. The output is color-coded with the notes that were identified as melody highlighted in red.

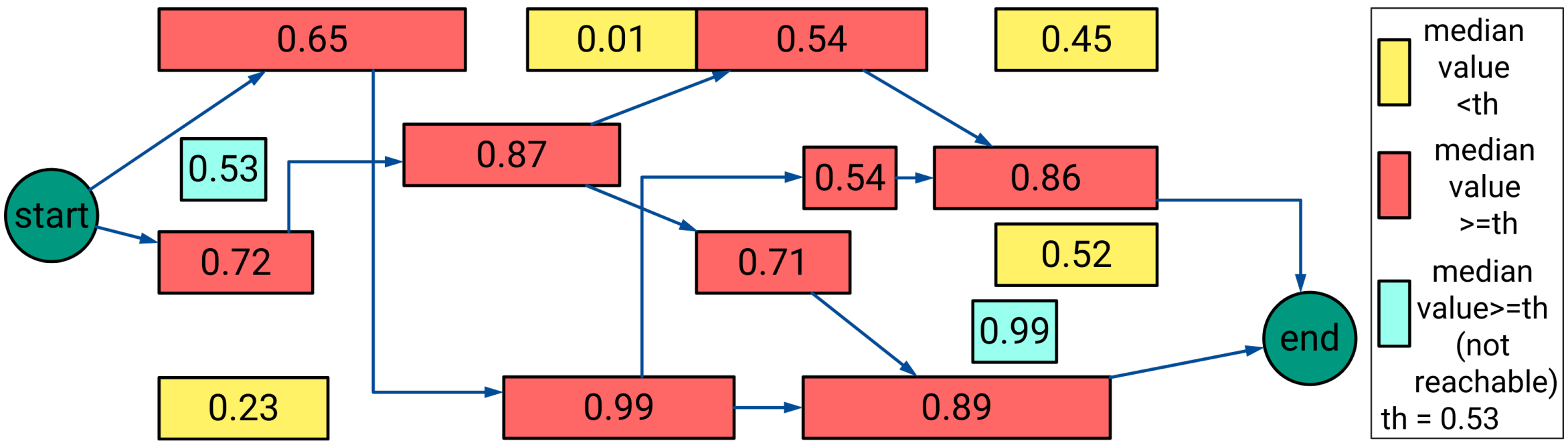

**Figure 5.4.:** *Example of graph built with Algorithm 1. Red notes are notes over threshold, yellow notes are under threshold, while blue notes are over threshold but are not reached by any path. The green circles are the starting and ending nodes. Numbers indicate note probabilities, which are computed as the median of their cells.*

A threshold is needed to determine which note probabilities should indicate melody notes. Distributions of probabilities differ between pieces, so a hard threshold (e.g. 0.5) would be inappropriate. Instead, we find a threshold for each piece using a statistical analysis of the values of the note probabilities. In the implementation of the proposed method we use hierarchical *single-linkage* clustering [263]: two clusters across the values of note probabilities are identified, and a piece-wise threshold is selected as the largest value of the lowest cluster. We then compare each note probability to this threshold and either retain the note as melody or filter it out as accompaniment. This produces largely (but not entirely) monophonic melody output – in some cases, multiple simultaneous notes pass the threshold. A graph-based method, explained next, was thus implemented to select a strictly monophonic melody line from this output. Such a method, however, should only be used when strictly monophonic melody lines are needed. For instance, in the case of voice or flute solo parts, we expect advancements by the usage of the graph-based processing. On the contrary, when the solo part is played by a string instrument such as a violin or a cello, we expect some polyphonic regions in the melody line; consequently, the graph-based method may decrease the accuracy of the processing.

### 5.4.2. Graph Search

Having identified notes that pass the threshold as defined above, we have to select a sequence of these notes that maximizes the probability of the sequence being monophonic. This is achieved using a graph-based approach. Algorithm 1 is used to build a *directed acyclic graph* (or digraph, see Figure 5.4). Such a graph consists of a set of nodes and a set of directed edges. Each of these edges specifies a connection from a node to another. In the graph defined by Algorithm 1, each note that passes the threshold is represented by a node, and the pitch, onset and duration information of this note are used to determine to which nodes is the note connected (in order to guarantee a strictly monophonic sequence). Note probabilities are used to determine the edge weights. Additionally, we set a start and end node at the beginning and end of the piece, respectively. The underlying idea is to connect a sequence of consecutive non-overlapping notes and use the note probabilities to weight the strength of the connection.

In Algorithm 1, an edge is placed between each note $n$ in the score and a set of notes $L'$. $L'$ could

**Algorithm 1** Melo-digraph building

$L \leftarrow$ list of notes
$\alpha \leftarrow$ starting node (end time = 0)
$\omega \leftarrow$ ending node (onset = $\infty$, probability = $-0.5$)
Push $\alpha$ to the beginning of $L$
Push $\omega$ to the end of $L$
**for** *note* in $L$ **do**
  $L' \leftarrow$ notes with onset $\geq$ end time of *note*
  $L' \leftarrow$ notes with onset = minimum onset in $L'$
  **for** *note'* in $L'$ **do**
    **if** probability of *note'* $\geq$ threshold **then**
      $p$ = probability of *note'*
      add an edge (*note*, *note'*) with weight $-p$
    **end if**
  **end for**
**end for**

be defined as the set of notes that have onset greater or equal than the offset of $n$, thus ensuring they are not overlapping with $n$. However, such a definition would create a large number of edges, increasing the problem complexity, especially for the first notes, which would be connected to almost all the notes in the score. For this reason, only the notes with the minimum onset in $L'$ are selected for being connected from $n$. Obviously, other strategies could be chosen.

We can then use a single-source shortest path algorithm to find the main melody line as the longest path from the start to the end nodes. Specifically, the connection weights must be the negative note probabilities and the Bellman-Ford algorithm[2] can be used to find the shortest path through the graph. Note that using complementary probabilities as connection weights would not allow to find the path that maximize the cumulative probability of the notes.

### 5.4.3. Training

The CNN is trained in a supervised fashion to filter out accompaniment parts. Inputs are provided in the form of piano roll segments and the targets are the corresponding piano rolls with only the melody notes. We also augmented the training dataset by 50% by creating copies of the original examples in the dataset with the melody transposed down for 2 octaves or up for 1 octave. Though the standard loss function for binary classification problems like this one is the binary cross entropy, during development of the model, we achieved more accurate models by minimizing mean squared error for the match between output and target piano rolls. The networks were trained using AdaDelta [264] with initial learning rate set to 1. In order to avoid overfitting, we use dropout with probability $p_{dropout} = 0.3$ and $L_1$-norm weight regularization. Additionally we use batch-normalization [265]. The training is stopped after 20 epochs without improvement in validation loss [266] .

[2]https://docs.scipy.org/doc/scipy-1.2.1/reference/generated/scipy.sparse.csgraph.bellman_ford.html

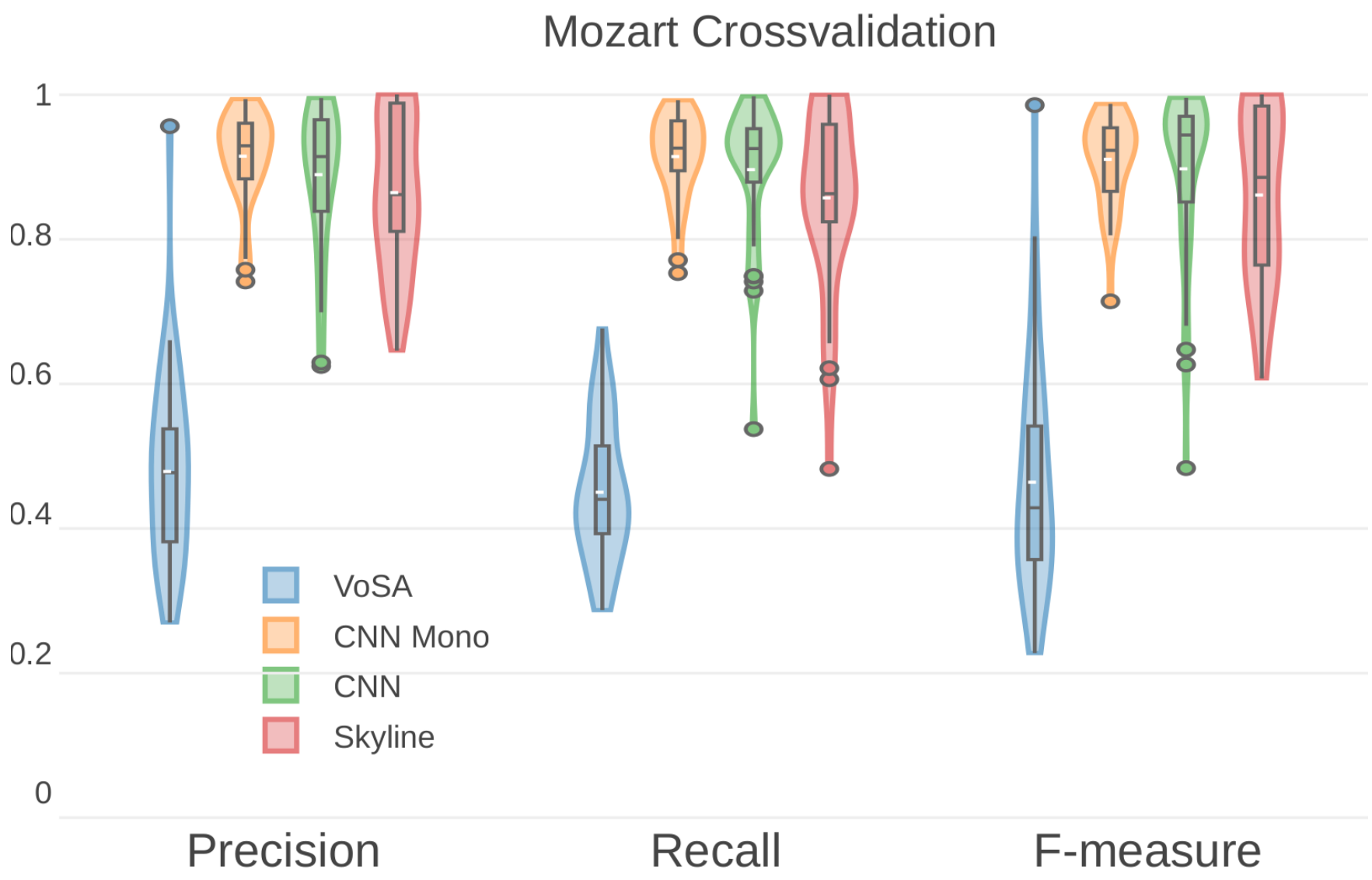

**Figure 5.5.:** *Cross-validation on the Mozart dataset. With the Wilcoxon test applied to F-measure, we found a significant difference between* CNN Mono *and* VoSA*,* CNN Mono *and* Skyline*, and* CNN Mono *and* CNN*. The mean is marked with a white dash.*

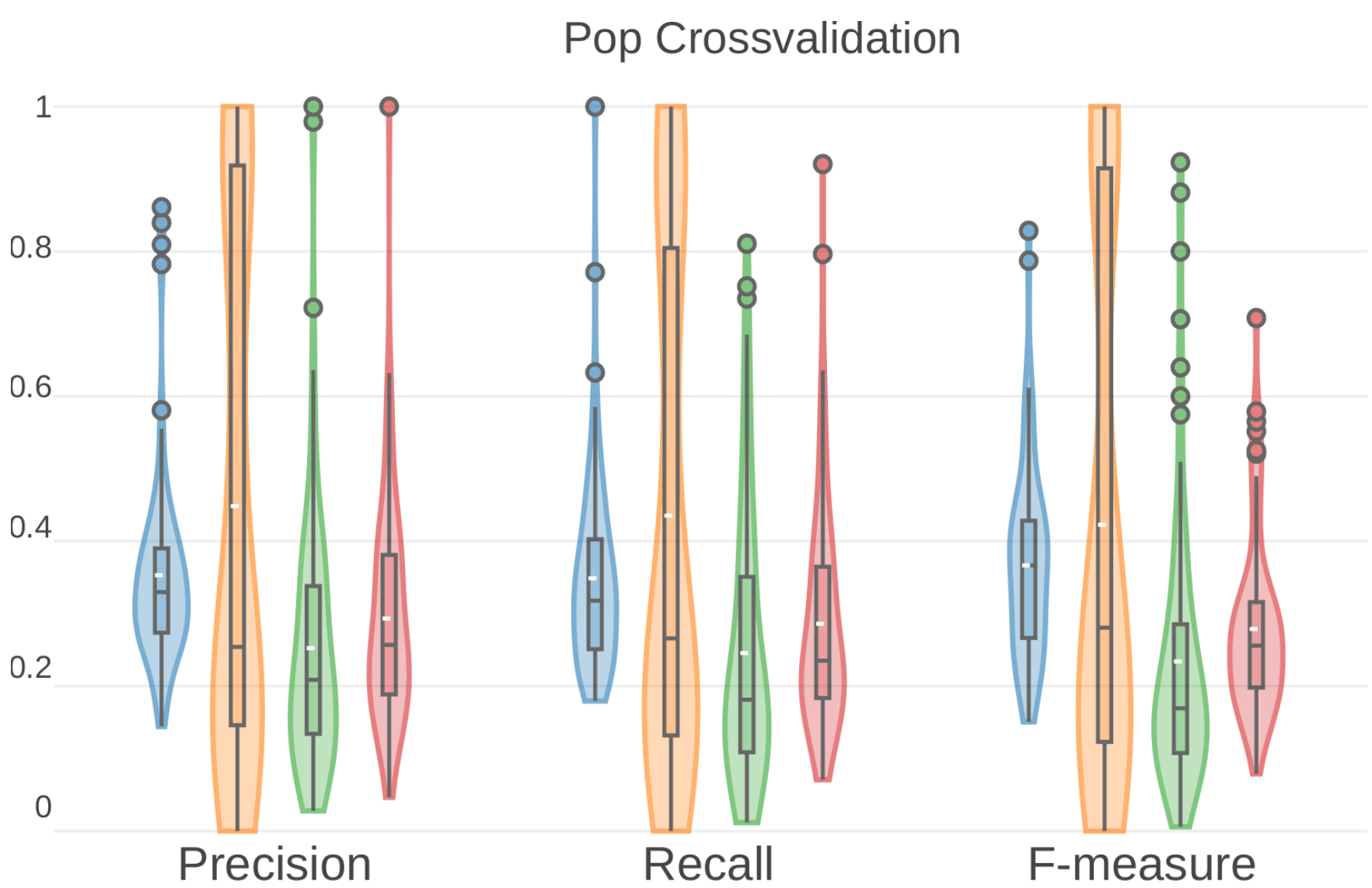

**Figure 5.6.:** *Cross-validation on Pop dataset. With the Wilcoxon test applied to F-measure, we found a significant difference between* CNN Mono *and* VoSA *and between* CNN Mono *and* CNN*, but no significant difference was found between* CNN Mono *and* Skyline*. The mean is marked with a white dash.*

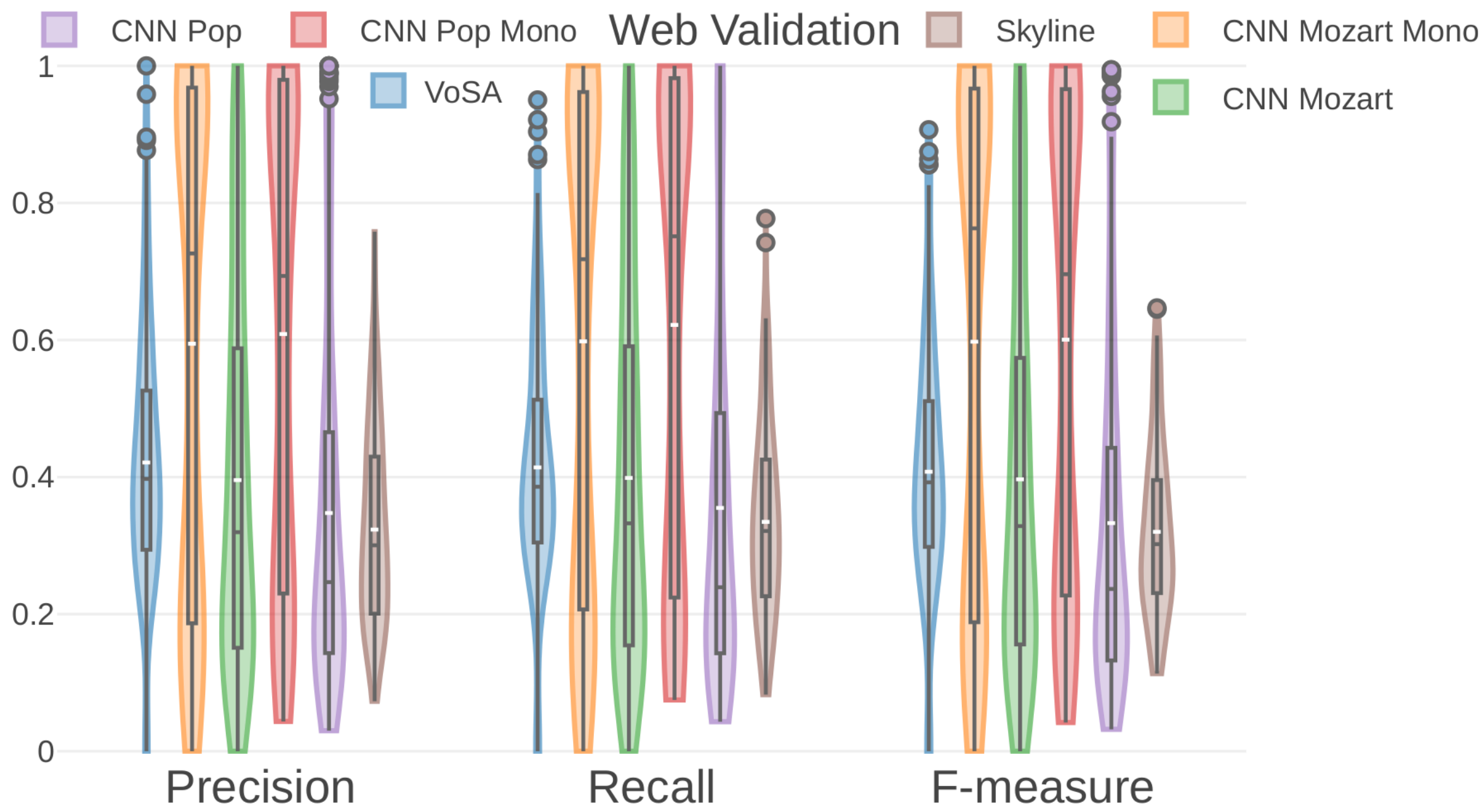

**Figure 5.7.:** *Validation on the Web music dataset. With the Wilcoxon test, we found a significant difference between* Mono *models and* Skyline/VoSA, *but there was not always a significant difference when comparing non-*Mono *models and* Skyline/VoSA. *The mean is marked with a white dash.*

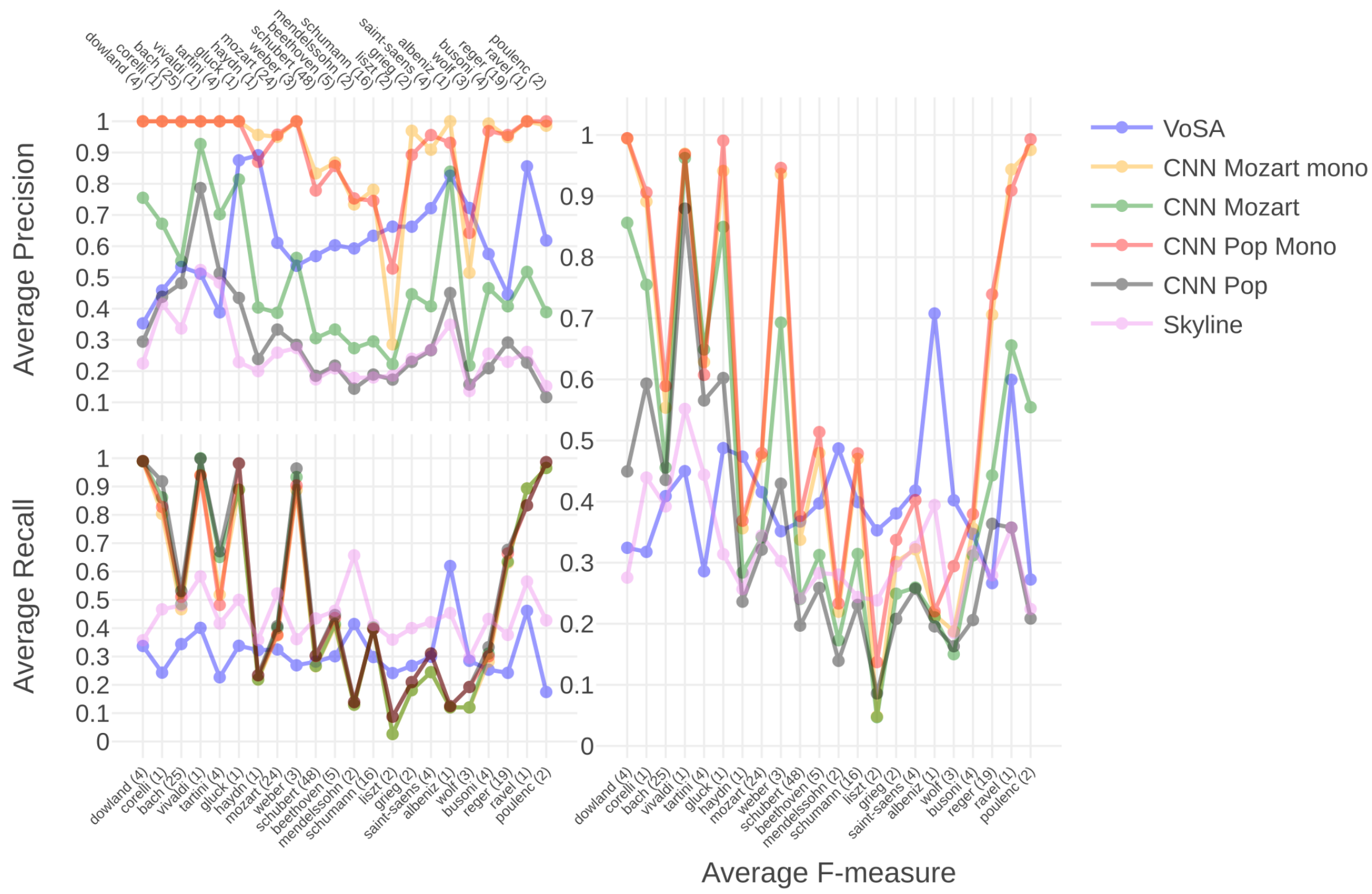

**Figure 5.8.:** *Visualization composer-by-composer of the validation on the Web music dataset. The number in round brackets is the number of music pieces included in the dataset. The order of the composers is by birth-date.*

## 5.5. Datasets

We used three different datasets to evaluate the performances of our method.

The first dataset ("Mozart") consists of 38 movements from Mozart piano Sonatas, while the majority of salient melodies was annotated manually by expert musicians.

The second dataset ("Pop") consists of 83 pop songs, where the singer part was considered to be the main one. It should be noted that this increases the difficulty in distinguishing the instrumental part from the symbolic data without having available knowledge of tracks.

Finally, the third dataset ("Web") consists of 169 pieces ranging from the English Renaissance of John Dowland to the 20th century music by Francis Poulenc – see Figure 5.8 for an overview of the composers. All of these pieces include a solo instrument and accompaniment; typical solo instruments are voice, flute, violin and clarinet, while typical accompaniment instruments are strings and piano. In this case, the proposed system was asked to predict the solo part.

The above-mentioned datasets are publicly available for research purposes in the companion site[3].

## 5.6. Experiments

### 5.6.1. Evaluation Metrics and Baseline Methods

In all experiments, we evaluated the quality of the predictions using the F-measure. We experimented on the largely monophonic (which we denote *CNN* in the following discussion) and strictly monophonic (denoted as *CNN Mono*) variants of the proposed model described in Sections 5.4.1 and 5.4.2, respectively. As a baseline comparison, we used the Skyline algorithm and VoSA (both described in Section 5.3). Since VoSA does not directly output the melody line, we first separate the piece into individual voices (as identified by VoSA), then select the voice with the highest F-measure as the melody. These modifications allowed us to consider the best case scenario of VoSA.

### 5.6.2. Network Architecture

To determine the architecture of the network, we used hyper-parameter optimization.[4] The number of convolutional layers, kernel size and number, and window lengths were optimized. This hyper-parameter optimization was done on 100 pieces randomly selected from across the three datasets plus 65 MIDI files collected online using the same criteria as the Web dataset. To compare models, we constructed training, validation, and test sets from the 100 pieces. A model configuration was selected that performed most successfully on the test set. The selected network architecture is shown in Figure 5.3: 2 convolutional layers, each with 21 kernels of size $32 \times 16$ (i.e., over two and a half octaves in the pitch dimension and 2 beats in the time dimension).

[3]As we do not have distribution rights w.r.t the pop song dataset, we provide the full list of pieces.
[4]Using the "hyperopt" library in Python (http://hyperopt.github.io/hyperopt/).

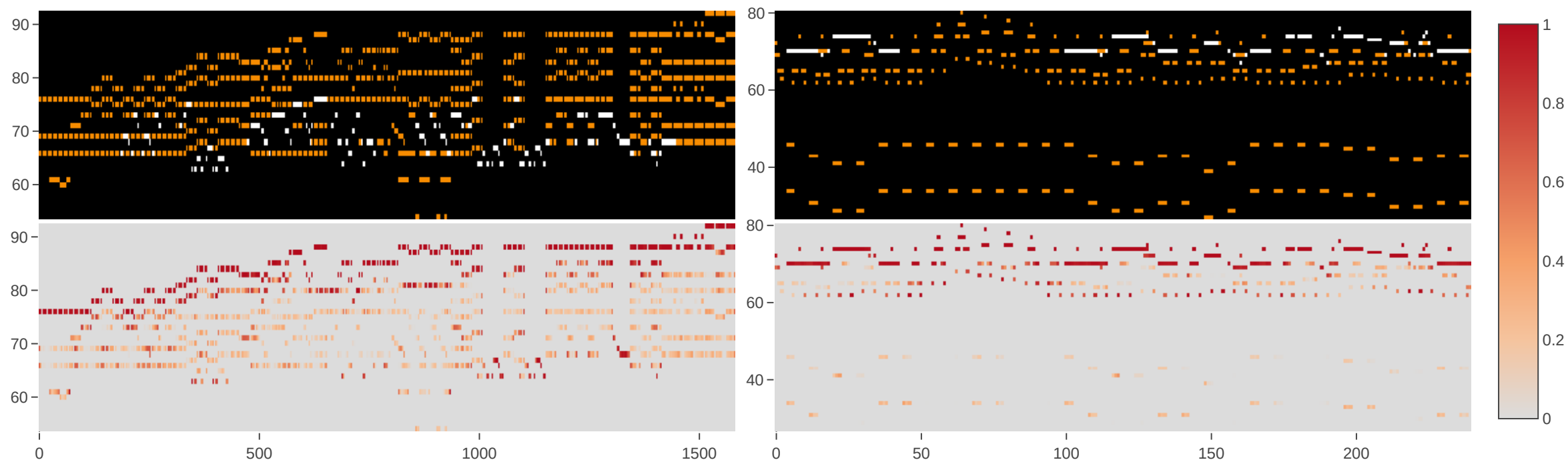

**Figure 5.9.:** *Liszt's* Ihr Glocken von Marling *(left) and an excerpt from Schubert's* Ave Maria *(right). Input piano roll (above), prediction of the CNN (middle). In Liszt, the model fails to identify the main part because the texture is rather different from the most common case and the melody is in the middle voices. In Schubert, instead, the texture changes but the model is not able to identify when the main part starts and stops because the accompaniment plays similar notes.*

### 5.6.3. Evaluation of the Proposed Method

To evaluate the quality of the predictions of the proposed method, we conducted two experiments. In the first experiment, we were interested in evaluating the predictive accuracy of the models trained on different datasets. In the second experiment we tested how well models generalize to different music styles. For the first experiment, we performed a 10-fold cross-validation on each of the Mozart and Pop datasets and reported precision, recall and f-measure for each piece. In each of these cross-validations, the dataset was split into 10 folds. For the second experiment, we trained two models, one on the Mozart dataset and one on the Pop dataset; then we evaluated the two models on the Web dataset.

In all these experiments, we compared the broadly monophonic version with the strictly monophonic one computed through the graph-based method described in – see Section 5.4.2. Moreover, we compared our method with the Skyline algorithm [260] and with a modified version of a state-of-art voice separation method (*chew*) [247]: we executed the voice separation and we considered the voice with the highest F1 score as the main part, substituting the track selection stage with an oracle.

## 5.7. Results and Discussion

### 5.7.1. Model Performance

The violin plots summarizing the results of these experiments are shown in Figures 5.5, 5.6, 5.7, and 5.8, while detailed results are available in the companion website (see footnote 1).

Our first experiment tested how well models predicted melody lines given training and testing on the same genre of music. Wilcoxon signed-rank tests were run on F-measures to assess potential differences between models. Test results are described in the caption of Figures 5.5 and 5.6. Overall, our proposed method that identified strictly monophonic melody lines (*CNN Mono*) performed better than the other models, but this difference was only significant for the Mozart dataset.

The Mozart pieces are highly structured and their melody lines tend to occur in the upper-most voice. The Pop dataset, in contrast, contains pieces with variable structure, with longer breaks in the melody (e.g., there is sometimes an interlude in the accompaniment part). Furthermore, the accompaniment part often overlaps in register with the melody line. It seems that without additional timbral information, our model could not sufficiently distinguish between melody and accompaniment lines when they shared a similar texture.

Our second experiment tested how well trained models generalize to new types of data (i.e., Web dataset). We hypothesized that models trained on the Mozart dataset would outperform models trained on the Pop dataset, as the Mozart and Web datasets are more similar in style (though the Web dataset is more heterogeneous). However, no significant difference between models was found – both models performed well on the Web dataset.

Regarding the less-successful performance of the two baseline methods, the Skyline method fails when the melody is not the highest voice; furthermore, this method cannot identify when pauses occur in the solo part. The VoSA method, which was developed for use with polyphonic music, tends to create too many voices and shows a bias towards connecting notes separated by small intervals – this is not surprising, as polyphonic music tends to assign voices to small pitch ranges. As a result, accompaniment notes are often wrongly included in the melody line that VoSA identifies.

Overall, the Mozart dataset turned out to be a much easier case than the Web and the Pop datasets. This is surely connected to the larger stylistic diversity of the Web and Pop datasets, but it is also linked to the fact that the melody-identification method proposed here does not take into consideration the possibility of large portions of music without melody, while instead it tries to identify a melody line in each window. This might explain why the proposed method should be improved for the identification of solo instrument parts as in the Pop and Web dataset. Nevertheless, the described experiments hold as preliminary evaluations, with an encouraging outcome.

### 5.7.2. Saliency Maps

To investigate what the CNNs are learning, we propose a method that evaluates the contribution of individual locations of the piano roll to predictions at other locations using saliency maps.[5] The approach is similar to the Permutation Feature Importance analysis used for Random Forests [267] and consists of investigating the change of inference values when the input changes. Specifically, the proposed method analyzes how the probability that a given note belongs to the melody shifts (i.e., increases or decreases) when certain other notes are removed (i.e., by converting the cells belonging to those notes to 0).

For example, take a rectangular input window $I$ and its prediction $P$. A new input window $I'$ with prediction $P'$ is created by converting the cells inside a given rectangle $R$ to 0. The difference between the original and new predictions is denoted as $d(P, P')$ and can be interpreted as the contribution given by the notes inside $R$ to the original prediction. By testing different input windows across the piano roll, we can see how different elements of the music contribute to the predictions that are obtained for individual notes.

[5] Kernels, saliency maps and additional material are available on the companion website – see footnote 1.

If we are interested in a particular note $n$, we can compute $d(P, P')$ specifically for the cells belonging to $n$ – what we will refer to with "query region". For our analysis, for certain notes of interest, we define 5 randomly-positioned rectangles $R$ and calculate $d(P, P')$. This difference is summed to the cells of the notes inside each rectangle $R$. This procedure is repeated $N$ times (where $N$ is a trade-off between computational complexity and resolution of the saliency map; in our case $N = 30000$), and we select only the iterations in which the cells in the query region are not converted to 0. Each cell is then normalized by the number of times it was converted to 0 and summed. The difference function is defined as follows:

$$d(P, P') = \frac{\sum_{i=n_{start}}^{i=n_{end}} P[i] - P'[i]}{Area(n_{start}, n_{end})} \tag{5.1}$$

where $n_{start}$ and $n_{end}$ identify the query region. In general, $n_{start}$ and $n_{end}$ indicate two opposite corners of any rectangle.

With this difference function, given a rectangle $R$, if $d(P, P') > 0$, then $P > P'$ in average across $n$ and, thus, removing the notes inside $R$ decreases the prediction values of $n$; conversely, if $d(P, P') < 0$, then $P < P'$ and removing the notes inside $R$ increases the prediction.

For example, in the bottom piano roll in Figure 5.10 (A), the blue high-pitched notes occurring around beats 20 and 35 have non-positive saliency values. Since they are higher pitched, these notes contribute negatively to the melody note highlighted with a green box, making it unlikely for this note to be identified as melody.

Similarly, in Figure 5.10 (B), the lower pitched note is correctly predicted as accompaniment because all the higher pitched notes have negative values, thus hinder the prediction of the query region as melody. The behavior in Figures 5.10 (A) and (B) is the same as the Skyline algorithm.

However, in Figure 5.10 (C), the query rectangle in green is correctly predicted as melody thanks to the lower pitched notes in the accompaniment; however, such prediction is not connected with lower pitched notes in the same register as the melody, but with regular patterns that constitute an arpeggio. The CNN is able to identify those patterns, that would instead be discarded by the Skyline method.

Finally, Figure 5.10 (D) shows that the system suffers from border-effects due to the inability at considering time-coherence in the music score. Indeed, when the query region is placed near to the border, a large portion of the window does not contribute to its prediction.

Overall, our model incorporates features of both the Skyline algorithm and VoSA. Like the Skyline algorithm it focuses on the highest notes of the piece; on the other hand, by allowing for different probabilities like VoSA, it is more successful at drawing coherent melody lines. Unlike VoSA, however, our model does not incorporate explicit perceptual constraints.

## 5.8. Conclusions

We implemented and analyzed a novel method to identify the melody line in a symbolic music score. Some of the functions of our model were found to be similar to functions of the Skyline algorithm and VoSA (in particular, focusing on the upper-most pitch, and defining a melody line as finding the sequence of notes that minimizes the connection cost). However, our method does not take into account the long-term sequential nature of music; it can compute windows in any order. While such a property might have some practical benefits, it also makes the network unable to generalize to diverse textures, leading to poor results when musical texture is varied (e.g., Figure 5.9).

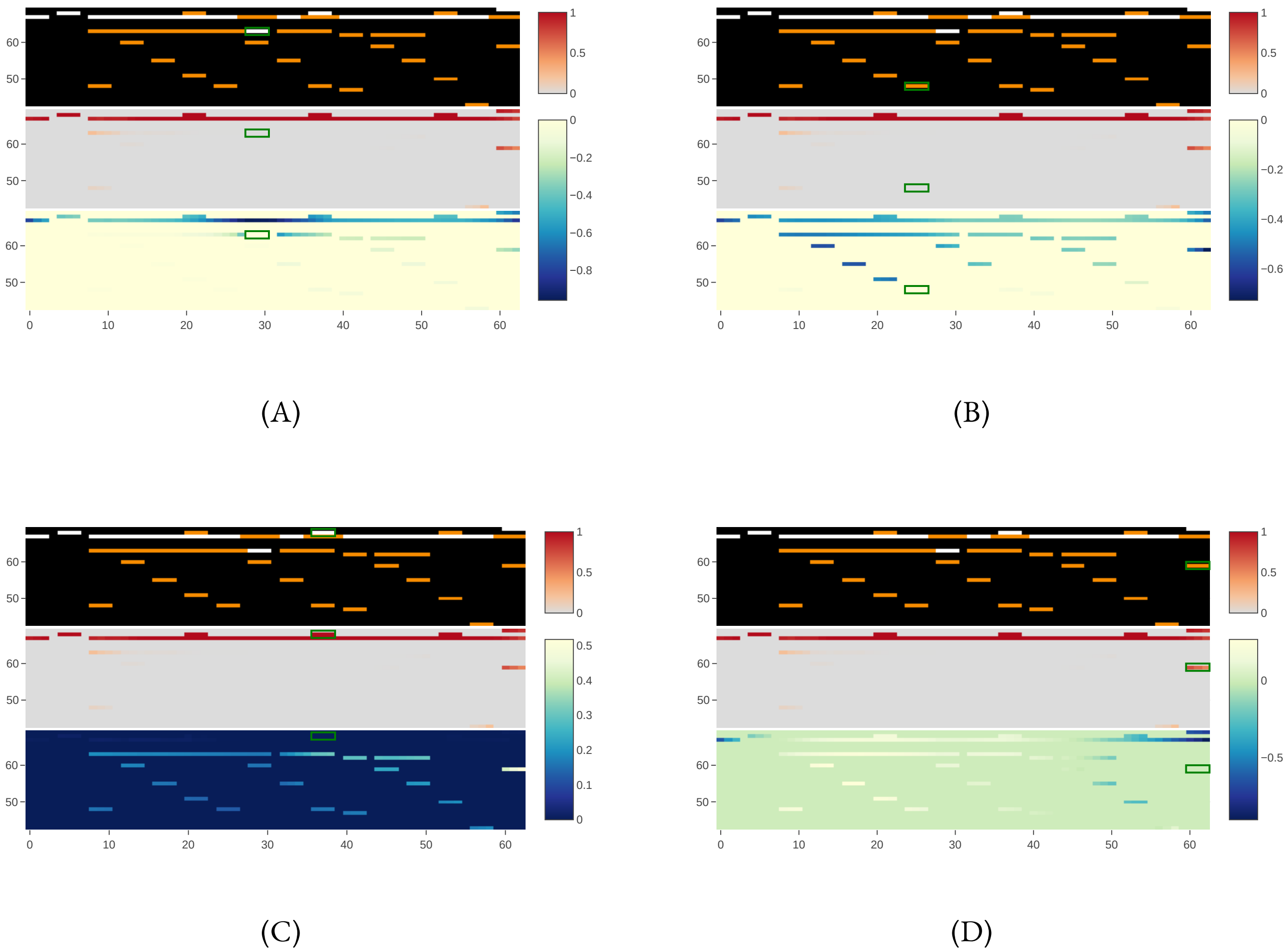

**Figure 5.10.:** *Various examples of saliency maps. Top: input piano roll with ground truth in white – Middle: prediction of the CNN – Bottom: and proposed saliency computed with respect to the query region (green rectangle). Positive values indicate notes that help the prediction of the query region as melody; negative values indicate regions that hinder the prediction of the query region as melody.*

The next step for this line of research would be to develop a model that can take into account a larger temporal context.

Convolutional recurrent networks with attention mechanisms could address this deficit. Another direction to pursue would be the creation of genre-based models, using classifiers to decide on suitable models. We implemented and analyzed a novel method to identify the main part in a symbolic music score. The method does not require knowledge of the entire piano roll, since its only input is one window of data. Moreover, it makes no assumptions regarding the evolution in time of the processed windows, while it is genre-agnostic. Even though such properties are desirable, at the same time, they render the network unable to generalize over diverse textures leading to failures when the texture changes during the same piece or some type of peculiarity is present, e.g. Figure 5.9.

In the future, we plan to improve the model with methodologies able to overcome these issues. Finally, we argue that the proposed inspection method can help in analyzing and understanding the human music cognition and we hope that the ability to understand the CNN way of working can help the development of even better and simpler algorithms.

# 6
# Audio-to-score alignment

An important task in Multimodal Music Information Processing (MIP) is Audio-to-Score Alignment (A2SA). Chapter 1 described how this task is useful for both end-user applications and feature extraction aimed at further score-informed processing.

Audio-to-score alignment is a Music Information Retrieval (MIR) task which aims at finding correspondences between time instants in a music recording and time instants in the associated music score. Such a technology facilitates various tasks, ranging from cultural heritage applications attempting to ease the fruition of music, to pre-processing stage for various multimodal MIR tasks – see Chapter 1 [1]. In the context of this Thesis, A2SA may be used to condition the performance analysis on the knowledge of the actually played notes. For instance, a musician could use a score-informed approach to reduce the error rates of the pitch inference of Automatic Music Transcription (AMT) models.

In the following, a novel method based on AMT is described. The proposed method reaches new state-of-art results for piano music alignment and seems promising for non-piano music[1].

## 6.1. Introduction

A major difference in A2SA methods is set between online and offline alignment. Online methods, often named "score-followers", try to predict the time instant in which a new note is played and track the change without future information about the performance. Offline methods, instead, try to match time instants by exploiting the knowledge of the full performance. In this work, we will concentrate on offline A2SA.

Similarly to other alignment problems, offline A2SA can be addressed using dynamic programming approaches and, as such, most of the literature focuses on Dynamic Time Warping [81] (DTW) based methods. DTW is an algorithm which is able to find the minimum cost path in a fully connected graph where nodes are the elements of two sequences and branches are weighted according to a given distance function. Even though DTW is effective and versatile, it has a strong requirement: the two input sequences must be sorted with the same element order. Formally, given any two pairs of corresponding elements $(a', b')$ and $(a'', b'')$, then $a' \geq a'' \implies b' \geq b''$.

[1] This Chapter is mainly based on a previous work [5] that contained an error in the evaluation code. Consequently, this Chapter also serves as reference for the updated experiments. The main novelty is that the proposed method is even more reliable than what looked in the previous publication, especially for non-piano music.

This requisite is met by music representations at sample or frame-level, but it hinders the alignment of polyphonic music at the note-level because the sequence of note onsets and offsets in a performance is not always the same as in the score. As consequence, most DTW methods use a sequence of frames as input. Moreover, since DTW is based on a distance function, such methods focus on discovering the optimal function and feature space.

There is a limited number of works that have faced the problem of note-level alignment, mainly with the objective of music performance analysis. Indeed, it is known that subtle asynchronies are generated during a human performance: notes in the same chord are written in musical scores as events having the same onset – and possibly the same offset –, but music players always introduce asynchronies of less than 0.05 seconds among the timings of such notes [268]. Other discrepancies between score and performance note order are related to the phrasing and articulation practices; for instance, the *legato* articulation consists of a slight overlap between two successive notes, even if in the musical score they are notated with no overlap. These almost imperceptible timing effects are considered to be responsible of the incredibly various expressiveness of music performances and are consequently of crucial importance in music performance analysis studies [269]. Methods used for note-level alignment so far include HMM, DTW, NMF and blob recognition in spectrograms [268, 270, 271, 272, 273].

The rise of Artificial Neural Networks in their Deep Learning (DL) paradigm has led several researchers to exploit DL models for feature learning tailored to DTW. Two methods are particularly noteworthy: one employs Siamese Networks for learning features that can be used for some distance function in DTW [274]; the second method relies on the improvements made in the field of AMT for converting the sequences to a common space — the space of the symbolic notation [275].

In this Chapter, we elaborate on the exploitation of AMT DL-based models for achieving note-level alignment. We propose a method which benefits from HMM-based score-to-score alignment and AMT, showing a remarkable advancement against the state-of-the-art. Finally, we perform a thorough comparison and extensive tests on multiple datasets. For reproducibility purposes, the implementation of the proposed method along with the present experiments is available online.[2]

## 6.2. Baseline method

DTW requires as input a distance matrix representing every possible matching between sorted sequence elements. If $N$ and $M$ are the number of elements in the input sequences, the distance matrix will have size $N \times M$. DTW finds the shortest path from element $(1, 1)$ to $(N, M)$ according to so-called *local* and *global* constraints. *Local* constraints list all possible moves among which the algorithm can chose during the computation of the path, while *global* constraints limit the computational complexity of the procedure (which in the no-constrained form is dominated by $O(M \times N)$ in both time and memory). As a consequence, DTW is highly expensive for long sequences: for instance, to align sample-to-sample two audio recordings lasting 10 minutes with sample rate 22050 Hz, DTW needs a distance matrix with $1.75 \times 10^{14}$ elements, meaning 318

[2]See Appendix A.

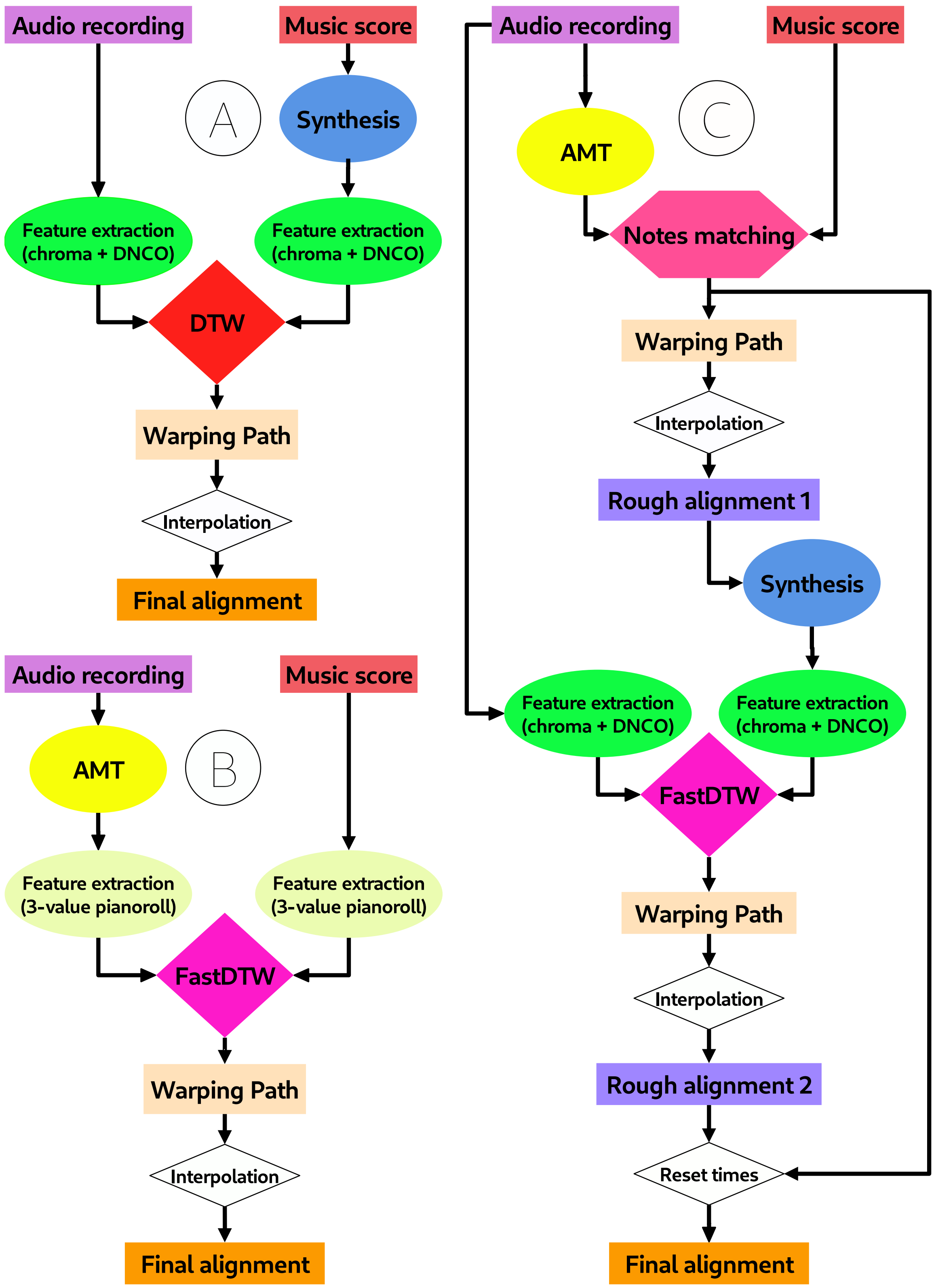

**Figure 6.1.:** *Flow chart of the three methods used here: Ⓐ SEBA method; Ⓑ TAFE method; Ⓒ EIFE method*

Terabytes using 16-bit floating point numbers.

Apart from the *global* constraints, various approximated alternatives for common local constraints have been proposed to relax the high complexity in time and memory – for an extensive overview see [276] – , with *FastDTW* [277] being one of the most widely adopted solutions.

Interestingly, one of the most widespread methods for A2SA consists in converting all the data to the audio level (usually by synthesizing the music score) and subsequently extracting some audio-related features (see Figure 6.1Ⓐ). Notably, one method [278] uses the sum of two distance matrices computed with two different combinations of audio features and distance functions; the main objective is to consider both percussive and harmonic features of musical instrument acoustics. In this paper, we will refer to such method with the name SEBA[3].

## 6.3. The proposed alignment methods

### 6.3.1. AMT-based frame-level alignment

AMT consists in the analysis of music audio recordings to discover semantically meaningful events, such as notes, instruments and chords. Two main methodologies for note-level AMT exist, e.g. Non-negative Matrix Factorization (NMF) and Deep Learning (DL) (for a thorough review see [279]). During the last 4 years, DL has tremendously advanced the state-of-art of AMT, especially for piano music recordings [194, 280, 281]. Due to the high variability of timbres, instrumental acoustics, playing practice, and difficulties in collecting data, multi-instrument AMT remains a hard challenge.

To our knowledge, the state-of-art of A2SA for piano music [275] is based on (1) AMT of recorded audio, and, (2) alignment of piano-roll representations of music. This approach can be seen as the opposite of classical DTW methods since it converts data to the symbolic domain instead of the audio domain.

Piano-rolls are 2D boolean matrices with $K$ rows and $N$ columns in which the entry $(k, n)$ is 1 if pitch $k$ at time $n$ is playing, and 0 otherwise. Usually, $K$ is set to 128 so that it is directly related to the MIDI specifications.

In [275], an AMT system is used to infer a MIDI performance; from there, a piano-roll is constructed. Piano-rolls coming from the transcribed audio and from the score can then be aligned using FastDTW to create a mapping between columns (frames) in the score domain and columns in the audio domain. The mapping, so-called "warping path", can be used to recompute the correct duration of the notes found in the score without relying on the AMT output, which is prone to errors in pitch identification [279].

Following the same line of thought, we used the new state-of-art piano AMT model [282] for the alignment, which by itself has a greater precision than the method previously used. The second improvement we made is the use of 3-valued piano-rolls, that is, we introduced a new value to represent the onset of a note bringing two advantages: (1) in the boolean piano-roll, repeated

[3]Here and in the following sections, SEBA, EITA, TAFE and, EIFE refer to the first syllables of the researcher first name who worked at the corresponding method — i.e. SEBAstian [278], EITA [234], TAegyun [275] and FEderico (this Thesis author)

notes are not distinguishable if the onset is immediately after the offset of the previous note, and, (2) the introduction of a new value works as "anchor" for the DTW algorithm, which tries to find correspondences between the alternations of three values instead of only two. We also attempted to use the same approach for multi-instrument A2SA by using a state-of-art multi-instrument AMT model [283].

Here, we will refer to the specific method with the name TAFE. A schematic representation of the method is depicted in Figure 6.1Ⓑ.

### 6.3.2. AMT-based note-level alignment

The merit of the above method was to highlight that DL models for piano AMT are tremendously effective in identifying onsets; however, the accuracy with pitch identification is low, due to issues such as false octaves and fifths. Moreover, as we will show in the results, offset time identification is still an unresolved obstacle.

On the other side, the TAFE method relies on DTW algorithm for aligning score and audio at the frame-level. This not only is a high computationally demanding task, but also suffers from the DTW requirements; in other words, it cannot align transcribed notes to score notes because of the discrepancies between score note order and audio note order. As such, it fails in two important tasks:

- it cannot handle correctly the subtle asynchronies that a human performer introduces among onsets of notes in the same chord and that are fundamental for performance analysis [269];
- it cannot correctly align scores that differ from the recorded performance or from the transcribed one — e.g. has some missing/extra note(s).

In the music alignment domain, a few studies have faced the problem of aligning scores and music performances that refer to the same music piece but differ in terms of presence/absence of a few notes — e.g. wrong performances, different score editions, etc. Such methods usually try to classify if a certain note in a score/performance is a *missing* note or an *extra* note compared to another score/performance. Commonly used approaches are DTW and HMM, with the latter being so far the best-performing approach [234].

To overcome the TAFE issues, we propose to use what we here call EITA method [234]. EITA uses HMM to create a mapping between two sequences of notes in order to identify missing/extra notes and notes that have different pitches — e.g. wrong pitch inferred by the AMT.

However, after having identified extra notes in the performance/transcription, missing notes remain in the score. Here, we assume that such notes were actually played but not identified by the AMT. In this perspective, we build a warping path from the mapping between notes matched by EITA between score and AMT output; then, we use the warping path to linearly interpolate the onsets and offsets times of the non-matched notes — i.e. notes in the score that are not present in the transcription according to EITA. To align such notes with even higher precision, we apply the standard SEBA method based on the synthesis of the score. Moreover, to reduce the computational cost, we use FastDTW instead of the classic version. Since we expect that the AMT output contains precise onsets, after SEBA processing, we set the EITA matched notes onsets using AMT output and keep SEBA alignment only for non-matched notes. The flow-chart of this method is shown in Figure 6.1Ⓒ. In the next, this method is referred to as EIFE.

## 6.4. The Employed Datasets

Unfortunately, there is not a great variety of datasets providing exact matches between score and midi performances. Thus, we used a systematic approach to generate misaligned sequences of notes as similar as possible to a musical score. The drawback of our method is that the resulting evaluation will not produce reliable values for real-world applications. However, it ensures that data does not contain manual annotation errors regarding matching notes. Moreover, here we are interested in the comparison of the considered approaches and leave the perceptual assessment of a performance on real-world score for future work.

In our previous work – see Chapter 4 [4] – we proposed a simple way for statistically modeling misalignments between scores and performances, and used such models to recreate similar misalignments for datasets not including scores, collecting them in the "ASMD" framework.

ASMD provides artificially misaligned notes that are more similar to a different performance than to a symbolic score; however, for most of MIR applications, such misaligned data is enough to cover both training and evaluation needs. To achieve an even more accurate evaluation, in this work we also applied a single-linkage clustering to the onsets of each misaligned score. We stopped the agglomerative procedure when a certain minimum distance $t$ among clusters was reached. We have randomly chosen such threshold in $[0.03, 0.07]$ seconds, representing broad interval around 0.05 seconds that is assumed as upper-bound of usual chord asynchronies [284]. Subsequently, we set the onsets of the notes in each cluster equal to the average onset time of that cluster so that the final misaligned note sequence contains chords made by notes having the same onset. This is a crucial difference between scores and performance data, in which chords are played with light asynchronies between same-onset notes.

ASMD also provides sets of artificially generated missing and extra notes that can be used for simulating performances and music scores that differ in some regions of the data.

## 6.5. Experimental Set-Up

We conducted four experiments to cover every aspect of the problem space as generated by the combination of two different conditions. Namely, we observed how alignment methods change in case (1) missing/extra notes between the score and the performance are introduced, and (2) instruments other than piano are present. To ensure a fair comparison of AMT-based A2SA methods, we used two state-of-art models, namely one trained on piano solo music [194] (BYTEDANCE) and one trained on ensemble music [285] (OMNIZART).

We used the Audio and Score Meta Dataset [4] (ASMD) Python API to retrieve missing and extra notes computed as explained in section 4.3.4. To simulate notes unavailable in the score, we removed the "extra" notes from the artificially misaligned score, while to simulate notes not played in the recording – "missing" –, we generated ad-hoc notes using the same procedure and removed them from the transcribed performance. However, since the SEBA method does not rely on AMT, it is tested without extra notes. Note that even though we remove notes in the input data, we still have them in the ground-truth, allowing to correctly assess all inferred timing.

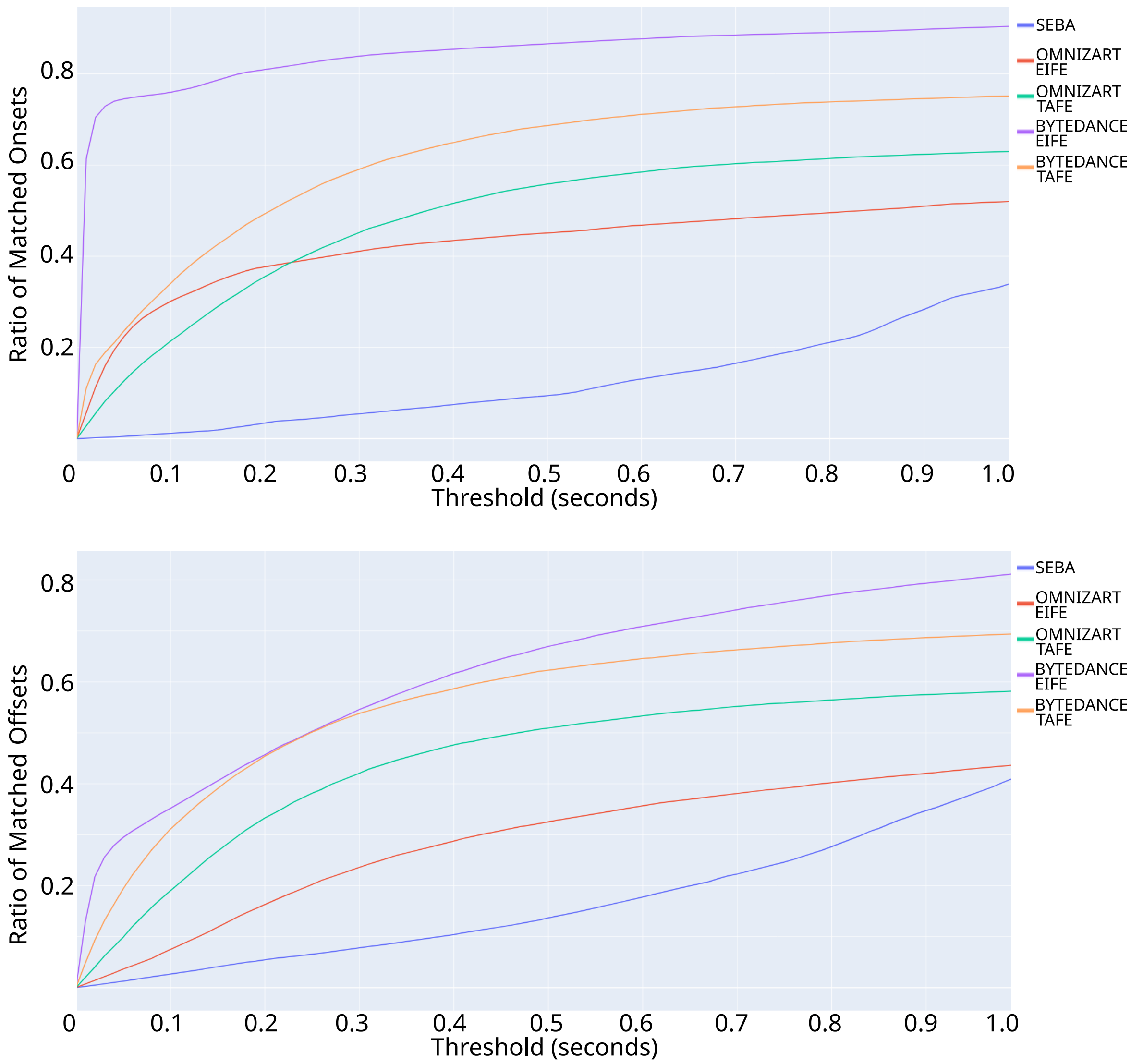

**Figure 6.2.:** *Evaluation on piano-solo music (SMD dataset) without missing/extra note. Curves are the ratio macro-averaged curves of ratios between the number of matched notes at a given threshold and the total number of notes. Top-performing curve is BYTEDANCE EIFE.*

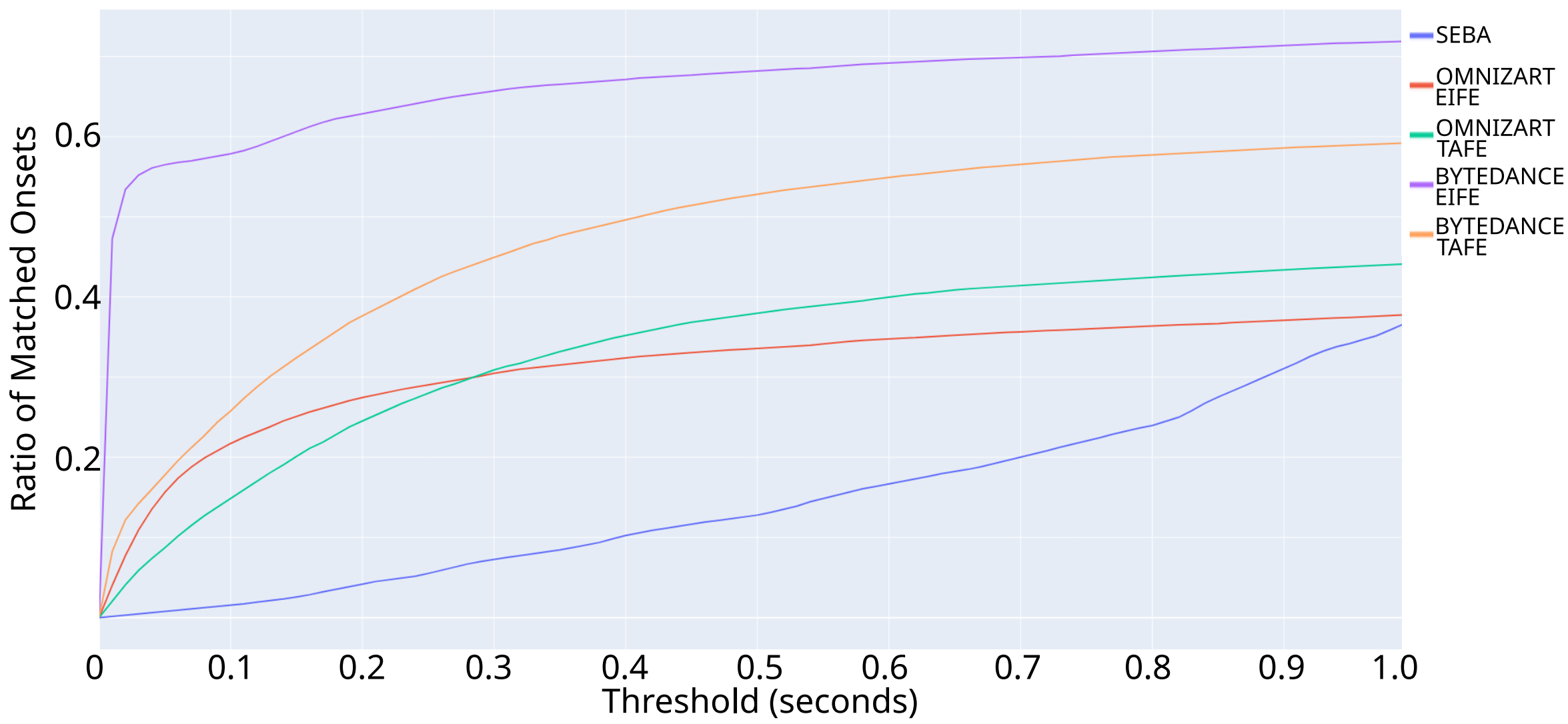

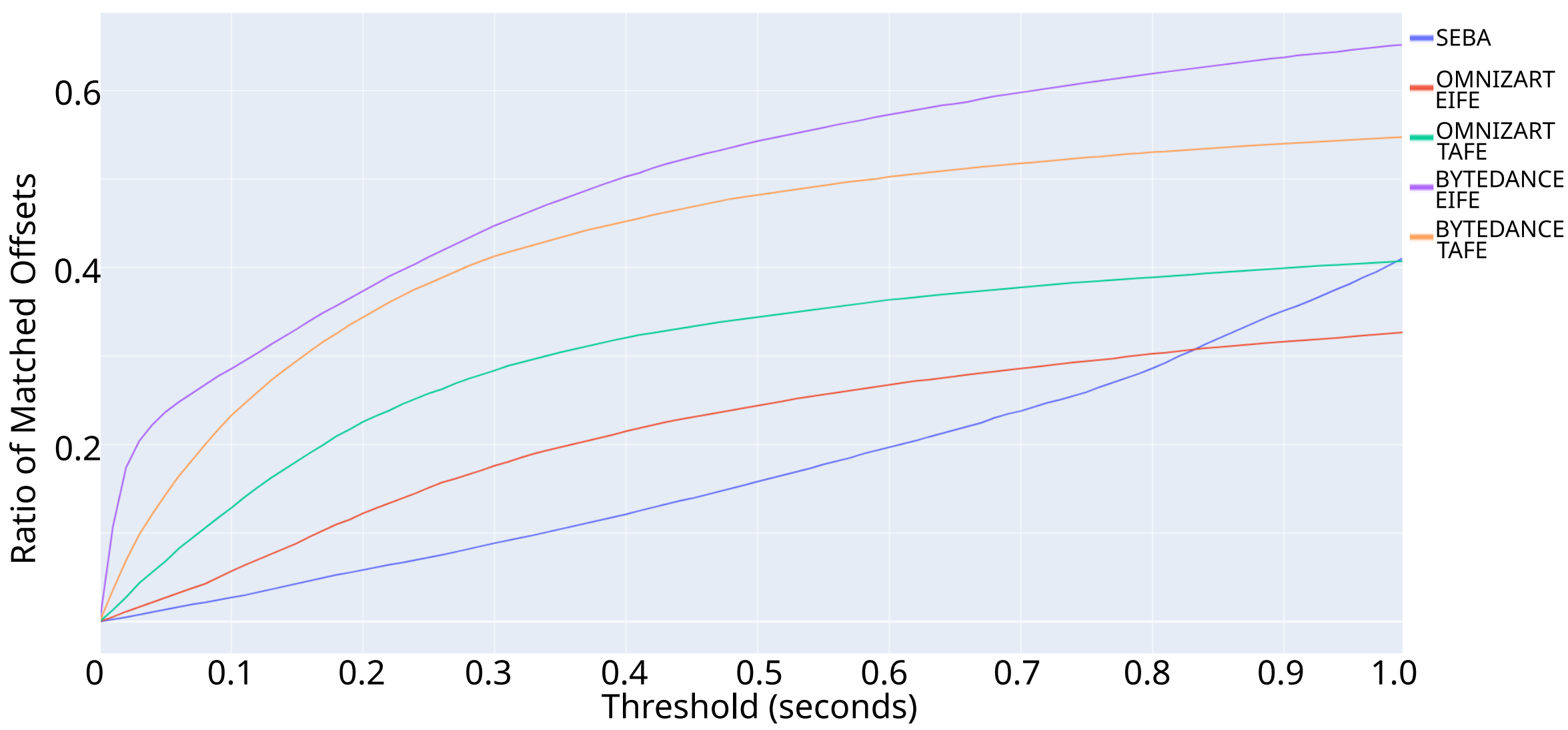

**Figure 6.3.:** *Evaluation on piano-solo music (SMD dataset) with missing/extra note. Curves are the ratio macro-averaged curves of ratios between the number of matched notes at a given threshold and the total number of notes. Top-performing curve is BYTEDANCE-EIFE.*

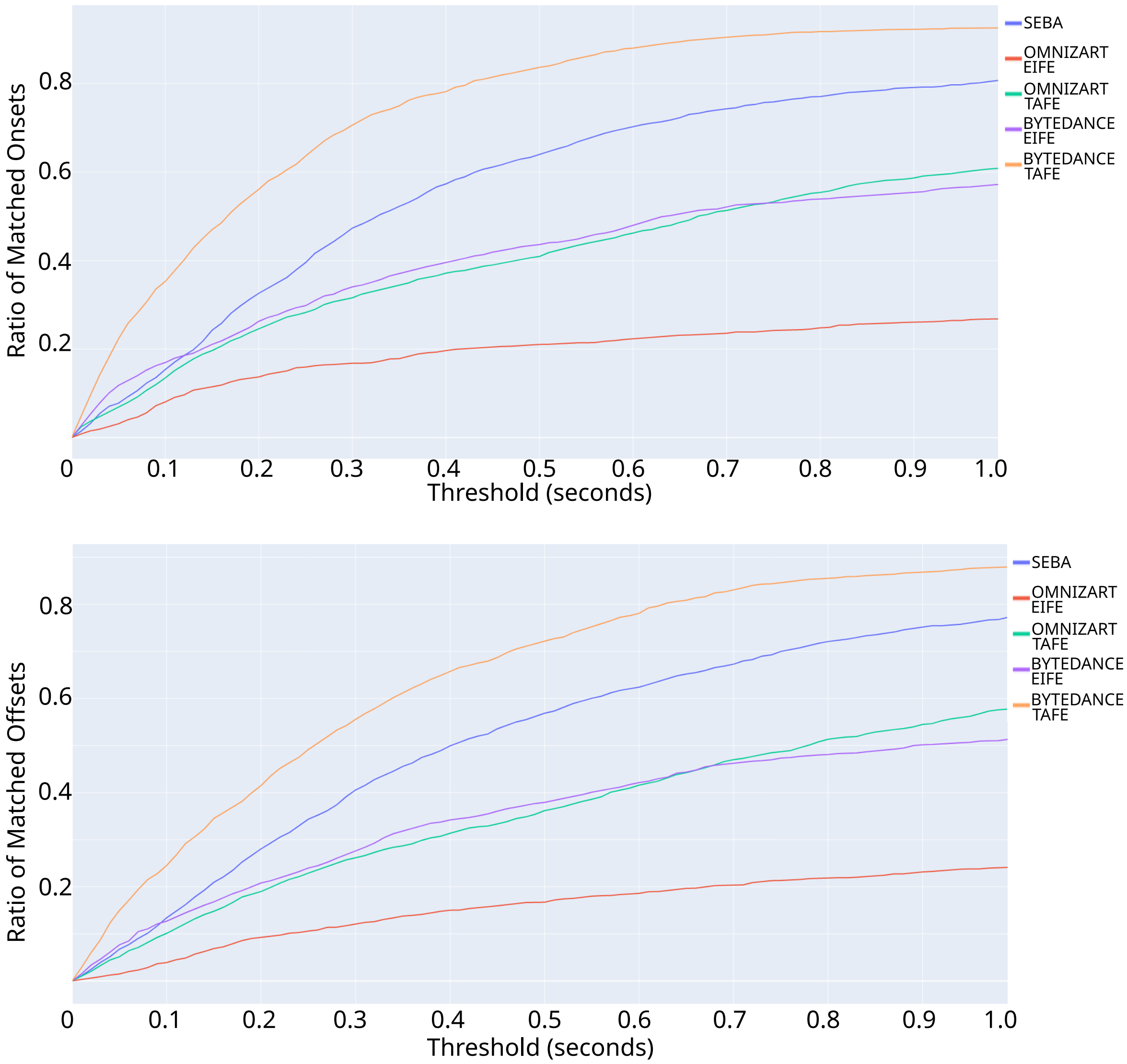

**Figure 6.4.:** *Evaluation on multi-instrument music (Bach10 dataset) without missing/extra note. Curves are the ratio macro-averaged curves of ratios between the number of matched notes at a given threshold and the total number of notes. BYTEDANCE EIFE is slightly better than SEBA fot threshold* < 0.1.

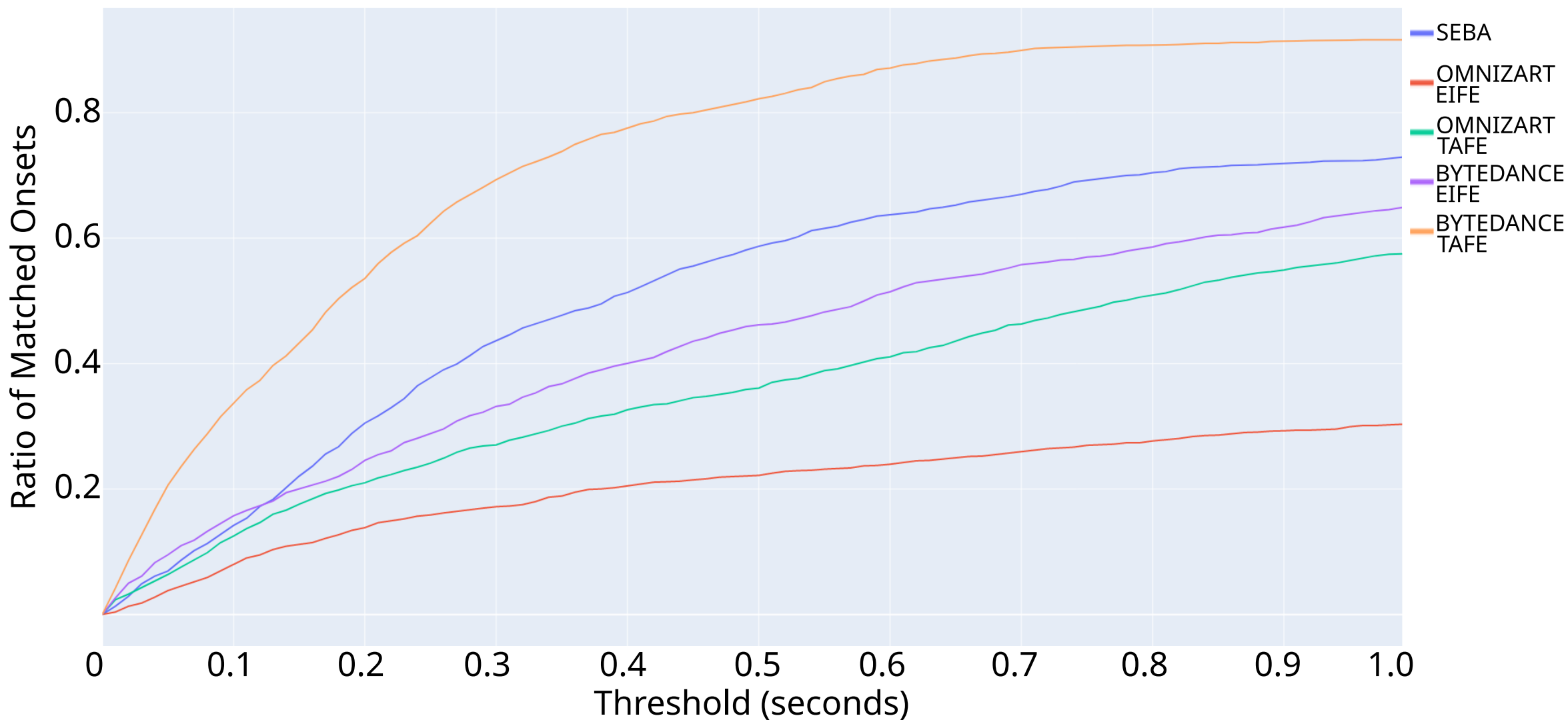

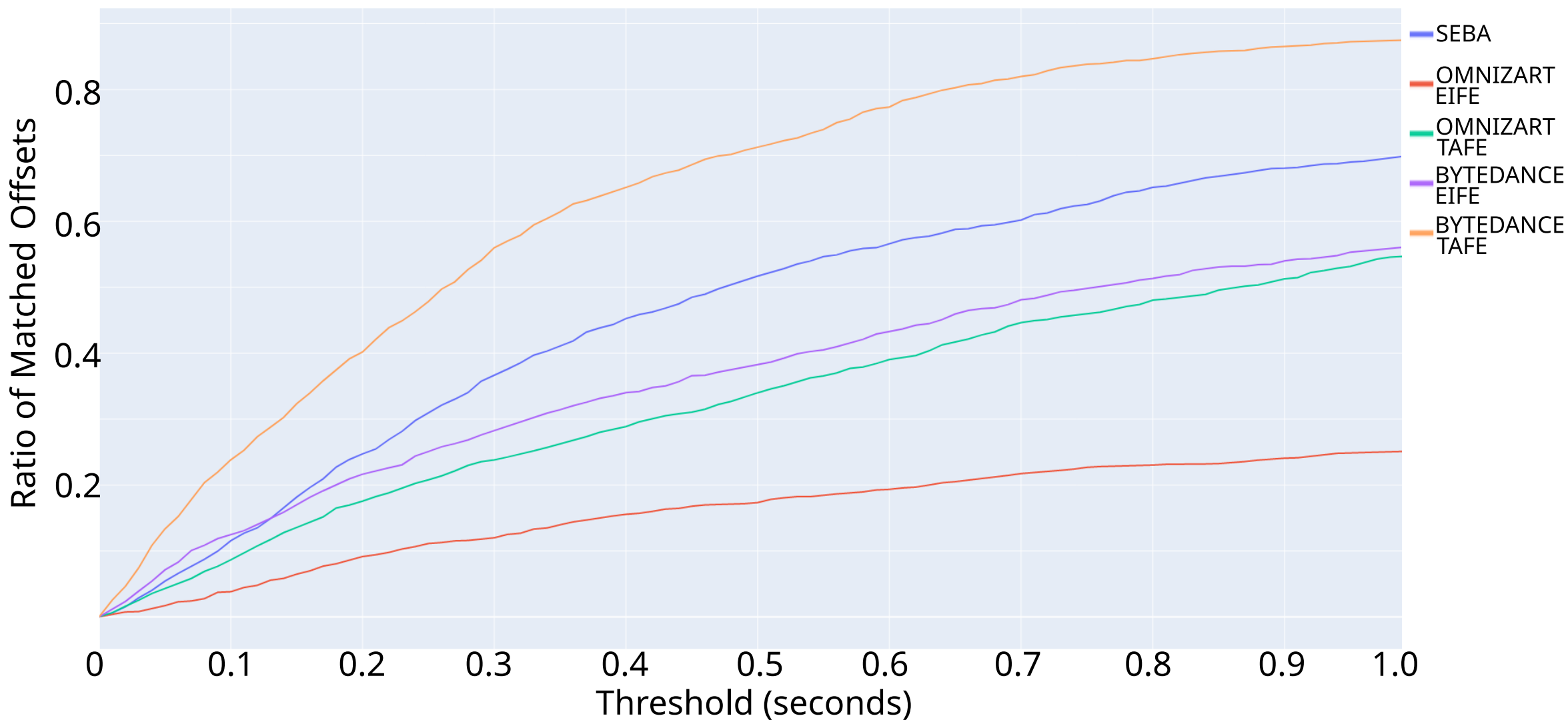

**Figure 6.5.:** *Evaluation on multi-instrument music (Bach10 dataset) with missing/extra notes. Curves are the ratio macro-averaged curves of ratios between the number of matched notes at a given threshold and the total number of notes. BYTEDANCE EIFE is slightly better than SEBA fot threshold* $< 0.1$.

We also used the ASMD Python API to select the proper datasets for our experiments. To avoid over-fitting during the evaluation stage, we did not use the Maestro [191] and MusicNet [286] datasets because the AMT models were trained on them. Instead, we used the "Saarland Music Dataset" [287] for evaluating piano A2SA. It consists of 50 piano audio recordings along with the associated MIDI performances, recorded with high-quality piano equipped with MIDI transducers. As regards to multi-instrument music, we relied on another well known dataset: the "Bach10" [288] dataset, which includes 10 different Bach chorales synthesized with virtual chamber instruments. Even though Bach10 dataset provides non-aligned scores, we used our artificially misaligned data to obtain results comparable with the other datasets.

To reduce the computational cost, we constrained each method inside 32 GB of RAM and 600s. Whenever a method failed for an out-of-ram/out-of-time error, the specific piece was removed from the evaluation. In doing so, we also get a rough reliability estimation of the various approaches here tested. Hence, due to the high resources required by EITA, the SMD dataset size is reduced to 26 music pieces when testing without missing/extra notes and 31 music pieces when considering them.

To ease alignment, we preprocessed both score and audio by stretching the note timings so that the score duration was the same as the trimmed audio. This operation corresponds to enforcing in the music score the performance average BPM.

We tuned the TAFE method by using the 5% of the available pieces sampled with a uniform distribution from the entire ASMD. We ran the TAFE method to find the best parameters for aligning the misaligned data to the ground-truth performance, after having removed missing and extra notes. We adopted a Bayesian Optimization approach with an Extra Trees surrogate model, Expected Improvement acquisition function, and exploitation-exploration factor set to 0.01. We used 180 calls and let the space of parameters being defined by 7 different distance functions and the *FastDTW* radius in [1, 200]. We found *cosine* distance and radius 178 as optimum parameters. We then used the same radius size for *FastDTW* in the EIFE method, while using the distance defined by SEBA.

For synthesizing MIDI files in the SEBA and EIFE methods, we employed the freely available MuseScore SoundFont[4]. As evaluation measure, we observed the ratio of matched onsets and offsets under several different thresholds in each music piece; we then averaged the obtained curves to get a macro-curve representing the overall performance of each method.

## 6.6. Experimental Results

With regard to piano-solo music, methods based on BYTEDANCE model are outperforming the rest – see Fig. 6.2 and 6.3. EIFE manages to exploit the good onset prediction of BYTEDANCE better than TAFE. However, the performance decreases when considering offsets due to the poor inference of onset positions of generic AMT models. OMNIZART does not perform well, as shown in Figure 6.6, which is expected as it is trained on multi-instrument music. Finally, SEBA method seems more robust in offset prediction and maintains a similar score for both offsets and onsets.

[4]MuseScore 2.2 SF2 version: https://musescore.org/en/handbook/3/soundfonts-and-sfz-files

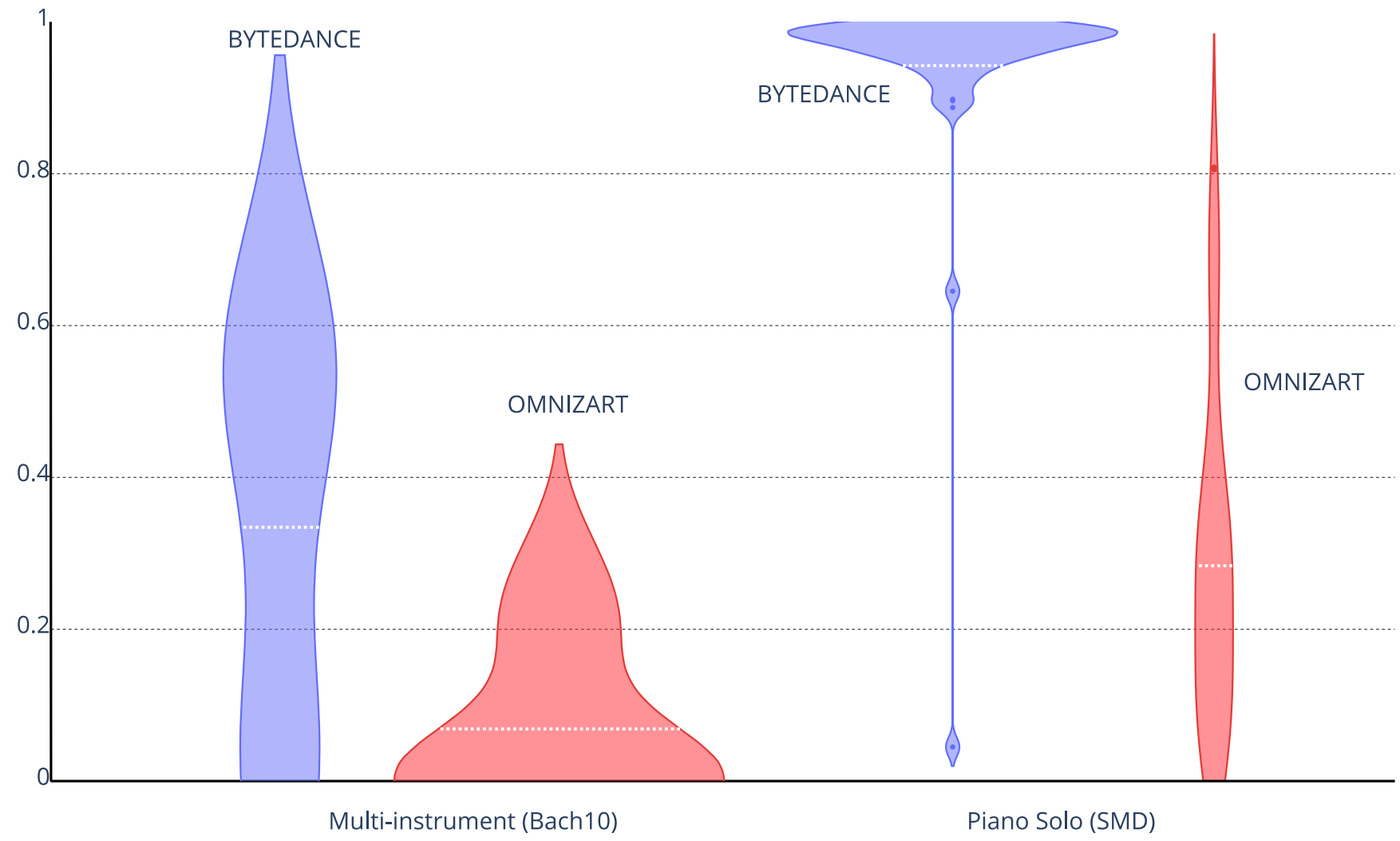

**Figure 6.6.:** *Onsets F1-measure for the two AMT models*

Furthermore, we observe that in non-piano music TAFE method is the best-performing one – see Fig. 6.4 and 6.5. Indeed, the good performance of AMT models, makes EIFE and TAFE approaches still reliable, especially for little thresholds – i.e. $< 0.1$ seconds. Moreover, even though we were expecting a useful input from OMNIZART multi-instrument model, we observed better performance with BYTEDANCE; this could be due to the low generalization ability of OMNIZART — see Fig. 6.6.

Every considered model suffers in case of missing notes, while retaining a similar curve shape and proportions. As such, we think that the most promising option for increasing the performance of A2SA with missing and extra notes is to increase the overall alignment accuracy.

## 6.7. Conclusion

We designed a methodology to compare various alignment systems and proposed two methods for frame and note-level alignment. After preliminary experiments, it was shown that the proposed method for note-level alignment brings notable advancement to the state-of-art thanks to the AMT models. Moreover, even if AMT is still not reliable for non-piano solo music, the top-performing approach among those tested is still based on the AMT model trained on piano-solo music. Our intuition is that the size of the training dataset is extremely relevant for the good performance of the model.

Our future studies will focus on the quality of the alignment with perceptual measures to confirm the results obtained through the present assessment.

# Part IV.

# Disentangling Performance and Interpretation

# 7 Perception of Performance Resynthesis

This Chapter focuses on the perception of music performances when acoustic contextual factors, such as room acoustics and instrument, change. We propose to distinguish the concept of "performance" from the one of "interpretation", which expresses the "artistic intention". Towards assessing this distinction, we carried out an experimental evaluation where 91 subjects were invited to listen to various audio recordings created by resynthesizing MIDI data obtained through Automatic Music Transcription (AMT) systems and a sensorized acoustic piano. During the resynthesis, we simulated different contexts and asked listeners to evaluate how much the interpretation changes when the context changes. Results show that: (1) MIDI format alone is not able to completely grasp the artistic intention of a music performance, (2) the usual AMT output does not completely grasp the interpretation information of the performance, and (3) usual objective evaluation measures based on MIDI data present low correlations with the average subjective evaluation. To bridge this gap, we propose a novel measure for assessing AMT systems' effectiveness from a perceptual perspective; the proposed measure is meaningfully correlated with the outcome of the tests. In addition, we investigate multimodal machine learning by providing a new score-informed AMT method and propose an approximation algorithm for the $p$-dispersion problem.

## 7.1. Introduction

Automatic Music Transcription (AMT) can be broadly defined as the process elaborating on digital audio recordings in order to infer a specific set of relevant musical parameters of the sounds, and to convert them in some form of notation [289]. Nowadays, AMT is a broad signal processing field encompassing a wide gamut of tasks and approaches. As an example, the output of an AMT system can be a traditional score, a Standard MIDI File (SMF), or a set of ad-hoc features [171]. A traditional score is a sequence of symbols that describes music according to the western notation and focuses on expressing music in a human-readable way so that it can be easily reproduced. SMFs instead describe the performance itself, possibly sacrificing precision at the semantic level (which is useful to the musician for performing the piece) while gaining precision in the description of the physical events that happened during the execution – i.e. velocity and duration with which the keyboard keys were pressed, the pedaling timing, etc.

SMFs originate from the description of keyboard music and it hardly adapts to other instruments; moreover, the constantly increasing importance of Music Information Retrieval (MIR) created the need for a different symbolic representation of music sounds: hence, several AMT systems extract MIR features – e.g. $f0$ estimation, intensity levels, and timbral descriptors [290, 291]. The input to AMT systems is a variable itself: most authors focus on mono-modal methods which take as input only the audio recordings, while other methods tackle the problem with multimodal approaches [1] such as audio and scores (score-informed) [292, 293, 294, 295].

Regarding the resynthesis of MIDI transcribed recordings, many studies have shown that performers change their way of playing according to acoustic contextual conditions, such as physical properties of the instrument, reverberation and room acoustics, often even unconsciously – see Section 2.4.1. Recently, the authors of [186] proposed a method to automatically transfer a piano performance across different contexts (instruments and environments) in order to make the reproduced sound as similar as possible to the original one, by adapting MIDI velocities and duration to the new context. However, they assume knowledge of the original piano parameters to carry out the adaptation; moreover, they use the same microphones and post-processing pipelines in every different acoustic context.

There are very few attempts addressing the subjective evaluation of AMT systems, with the notable exception of [296]. The authors prepared more than 150 questions asking subjects to chose the best transcription of a reference audio clip lasting 5-10 seconds and managed to collect 4 answers per question. We use the results of the above work as a main reference for the present study.

We propose a methodology for evaluating the resynthesis of MIDI recordings extracted through AMT systems, taking into account contextual conditions in the resynthesis. Specifically, we wish to adapt the performance to the new resynthesized context while having knowledge only of the target acoustic context and of the original recording. In doing so, we propose a conceptual framework which distinguishes between the actual performance and the underlying artistic intention (i.e., the interpretation), and we design a methodology assessing to which extent such interpretation is perceivable in the resynthesized recording. Possible applications that may be impacted by the proposed study are manifolds. The long-term objective of this study is the resynthesis of music for production and restoration purposes – see Chapter 2.

The contributions of the present work are:

(i) an indication of MIDI format's inability to completely capture the artistic intention of a music performance and

(ii) a perceptually-motivated measure for the evaluation of AMTs

In addition, we investigate multimodal machine learning technologies applied to AMT by providing a new score-informed method and propose an approximation algorithm for the $p$-dispersion problem to optimally-chose the excerpts for the test.

For the purpose of comparability and reproducibility of the results, the code is made available online and the full set of the computed statistics are available in the Supplementary Materials – see Appendix A.

## 7.2. Restoration, Performance and Interpretation

One of the long-term motivations behind the present work is the automatic restoration of old and contemporary music recordings by reproducing the performances as accurately as possible.

The audio restoration literature is dominated by two general approaches: the first aims at reconstructing the sound as it was originally "reproduced and heard by the people of the era", while the second and most ambitious one aims at reconstructing "the original sound source exactly as received by the transducing equipment (microphone, acoustic horn, etc.)" [297, 298] – see Section 2.2 for a complete discussion. However, an exact restoration is impossible in both cases. Particularly regarding the second approach, aiming at recovering the so-called "true sound of the artist" [299] exposes the restoration to subjective interpretations regarding the performer's artistic intention. Indeed, the artist's original intention is never completely captured by the recording because of the recording equipment limitations, such as microphones compression, noises, and degradation [186]. To get over this issue, several studies tried to exploit the timbral features of the audio recording to compute original sound characteristics such as note intensities [300, 301, 302], but this is a particularly challenging problem hindered by the variability of MIDI velocity mappings [303].

Thus, we propose not to recover the original artistic intention but the intention *survived* until today and *perceivable* by the listener, as this is the best case scenario which is not influenced by subjective factors. The proposed idea consists in:

(i) analyzing a recording via an AMT system so as to estimate the parameters of the performance;

(ii) resynthesizing it using modern technologies.

In other words, we wish to retain the perceivable effort of the performer in a resynthesized version of the automatically-extracted music transcription. Note that this is different from wishing to reconstruct the true performer's interpretation or the "performer's idea of music" [304] – Storm's Type II approach, see Section 2.2; instead, we assume that such information, in their entirety, are not recoverable, and we limit our expectations to the restoration of the *survived* interpretation .

Towards defining the specific problem, two concepts need to be distinguished, i.e. *interpretation* and *performance*. The *performance* is the set of physical events that result in the activity of playing a music piece. It is bijectively associated with a certain time, place and performer, so that it is a unique unrepeatable act. *Interpretation*, instead, refers to the *ideal* performance that the performer has in mind and tries to realize. Thus, an interpretation could be repeated in different performances and differs from the performance because it lacks the adaptation to the acoustic context. It comprises the ultimate goal of the performer and thus, we identify it with the performer's artistic intention. During the restoration process, we seek to generate a new performance based on the interpretation extracted from the audio recording.

This operative definition is completely unrelated to the musicological debate about what an interpretation is – e.g. when the notion "interpretation" was introduced with reference to music [305] or whether the interpretation is the "performer's idea of the music" [304]. We rather focus on tracing the difference between desired and realized performance, which differ due to external causes. Such distinction is in line with the state-of-art research in the field of Music Performance Analysis that focuses on the acoustics of concert stages and rooms – see Section 2.4.1.

## 7.3. Designing the Test

This section analyzes the conducted experiment from the technical point of view as well as the reasoning behind the presented choices.

### 7.3.1. Research questions

Given the definitions of Sec. 7.2, we assume the following:

(i) MIDI – and consequently SMF – is able to record every aspect characterizing a piano performance;

(ii) using the same interpretation, a musician is going to create different performances given that the context (instrument, room, audience) changes;

(iii) during the audio recording process, there is information loss up to a certain extent, while a different type of information, related to the context (including microphones), may be introduced. Contextual alterations render practically impossible to extract the exact MIDI performance from the audio.

The first research question that we seek to answer with the listening test is to which extent the interpretation is still identifiable when changing the context and retaining the performance, i.e. the MIDI recording. According to our second assumption and to psychological studies described in Section 2.4.1, if the context changes and the performance stays constant, we assume that the two performances are generated from two different interpretations. Consequently, while retaining the same interpretation in two different contexts, we expect that the listener will perceive two different interpretations. With this research question, we aim at assessing whether MIDI is effective in representing aspects related to interpretation. In this case, we do not mind about the true original performer's interpretation. Instead, we are interested in comparing the perception of the interpretations.

The second research question is whether state-of-art AMT systems, typically trained to extract performance parameters, are effective in the extraction of the perceptually-relevant interpretation data as well. Since it is practically impossible to recreate the exact performance, it is interesting to investigate whether a slightly modified MIDI coming from AMT systems is able to encompass interpretation aspects. In case this is true, we could resynthesize a given interpretation with a different context and obtain a different performance of the same interpretation.

Finally, following the line of thought presented in [296], we want to provide a perceptual evaluation of AMT systems.

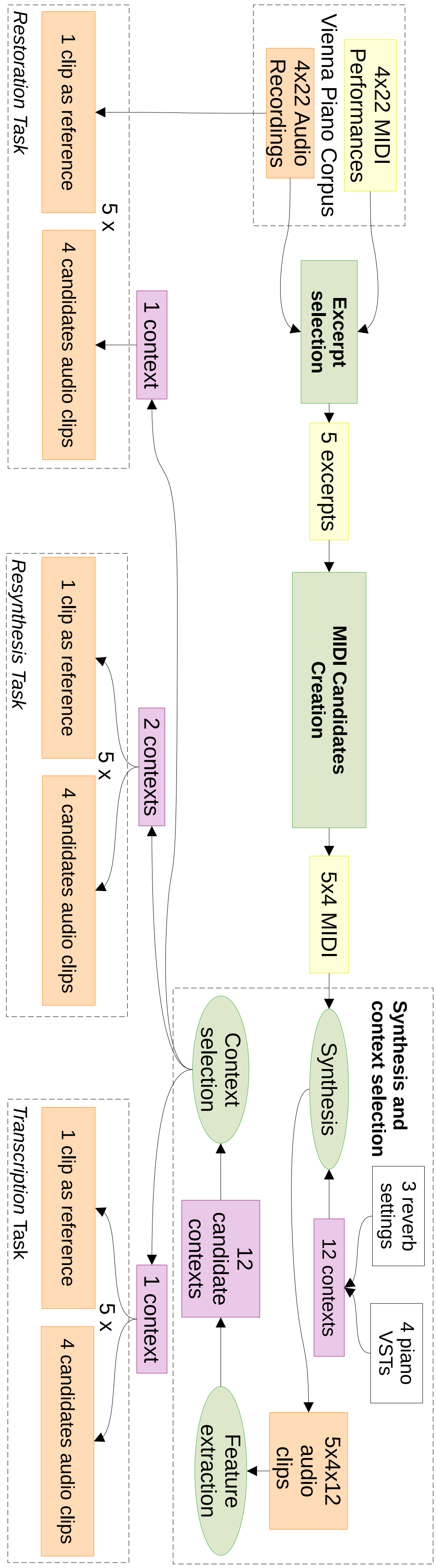

**Figure 7.1.:** *The workflow used for creating the restoration, resynthesis, and transcription tasks. Legend: a) Yellow: MIDI data; b) Orange: audio data; c) Purple: contexts; d) Green: Operations.*

### 7.3.2. Tasks

Three different tasks were designed addressing the above-mentioned questions. They consist in assessing the similarity in the interpretation between a reference audio excerpt and several candidates. The tasks differed in the way the audio clips were generated, as follows:

(i) in the first task, named "transcription", all audio excerpts including the reference were synthesized from MIDI using the same context (i.e., same virtual instrument and reverberation); in [296];

(ii) in the second task, named "resynthesis", all audio excerpts were still synthesized from MIDI but we used two different acoustic contexts – i.e. two different virtual instruments – for the reference and the candidates;

(iii) in the third task, named "restoration", the reference was a real-world recording, while the candidates were synthesized from MIDI with a virtual instrument. Since the original recording contains substantially more noise than the synthesized candidates, this specific task is representative of a restoration process.

All tasks used the same 5 excerpts extracted following the process explained in section 7.4, where we also describe the reasoning behind the choice of the virtual instruments.

### 7.3.3. Protocol and interface

Due to COVID-19 restrictions, we designed an online test using the "Web Audio Evaluation Tool" (WAET) [306].

First, subjects were prompted with some introductory slides explaining the difference between interpretation and performance. Specifically, after the formal definitions, they were suggested to adopt the following way of thinking: the interpretation is associated with the pianist, while the performance is related to a particular concert. Then, they listened to the first 30 seconds of two different performances by Maurizio Pollini of the Sonata No.30 op.109 by L. van Beethoven, and to a performance of the same piece by Emil Gilels; they were told that the first two were the same interpretation but different performances while the latter was both a different performance and interpretation. Such a distinction could introduce a bias in the test, since the question could be interpreted as a performer identification task. Nevertheless, it is unlikely that two different performers play the same interpretation, and we checked that the reference and the negative reference (NR) contained not only two different performances but also two different interpretations – see Section 7.4. In this way, the two tasks fundamentally coincide. Moreover, for a proper test about performer identification, additional candidate should have been proposed to the users, consisting of the same performer playing another interpretation. In this work, we instead focused only on the interpretation perception. Overall, the identification of the interpretation and the identification of the performer are two tasks highly related one each other.

After the introductory slides, they were asked to:

- use headphones or headset;
- stay in a quiet place;

- consent to the use of their answers in an anonymous form.

Next, subjects were asked preliminary questions, namely:

(i) what level of expertise they have with music (options: none/hobbyist vs. undergraduate/-graduate/professional); note that the related literature suggests agreement between raters with expertise in different instruments, especially for piano assessment [307, 308]

(ii) how often they listen to classical music (options: less vs. more than 1 hour per week)

(iii) how often they listen to music other than classical (options: less vs. more than 1 hour per week)

(iv) what is the cost of the headphones they were using (options: less vs. more than 20 euros)

At this point, subjects were exposed to the experimental interface through an example question, with guitar instead of piano recordings. Similarly to MUSHRA test [309], subjects could play back a reference audio file and four additional candidates containing the same musical excerpt resynthesized with various contexts and performances; they were asked to rate each candidate with a horizontal slider on a continuous variable evaluating the extent to which the candidate clip contained the same interpretation as the reference. Three labels were put along the slider: "different interpretations", "don't know", and "same interpretation" – see Fig. 7.2 for a screenshot. Users were instructed to experiment with the example question until they felt comfortable. As shown in Figure 7.2, users could adjust the audio level at any time, even during the example question. Audio clips were normalized to -23 LUFS level in both the example and the actual test. However, authors are not aware of studies about loudness influence in Music Performance Assessment studies.

Finally, subjects started the actual test, with piano recordings. The order of questions and candidates were randomized, as well as the initial positions of the sliders to prevent biases. Since the test lasted approximately 30 minutes on average, if a subject decided to leave the session her/his answers were recorded. For this, we used a feature provided by WAET to first prompt new subjects with questions for which fewer answers were collected, so that the number of answers per question was uniform.

### 7.3.4. Number vs. duration of excerpts

In any listening test which deals with the artistic expression of the performer, an issue arises concerning the length of the excerpts. In general, it can be expected that longer duration will lead to more accurate subjective judgments. However, one second competing factor is the total duration of the test: the longer the test is, the more difficult is to find volunteers willing to take the full test and be able to keep their concentration for the entire duration [310].

One study observed that the duration of excerpts did not influence the emotional response of the listeners [311]. On the other hand, it was shown that in the context of piano performance assessment, graduate music students and faculty professors rated 60 s excerpts higher than 20 s excerpts, while no significant difference was observed for non-graduate students [312]. Another study conducted on wind bands took into account the level of the performers and showed that music majors rated 25 s and 50 s excerpts higher than 12 s excerpts in the case of university or professional level performances, while the opposite happened for performances at the high-school level [313]. A study on children chorale music revealed that 60 s excerpts were rated higher then 20 s excerpts by music majors [314]. Finally, another study on band performances was conducted with excerpts 12 s, 15 s, and 30 s long, and found that ratings from music majors were higher for long excerpts of bad performances [315].

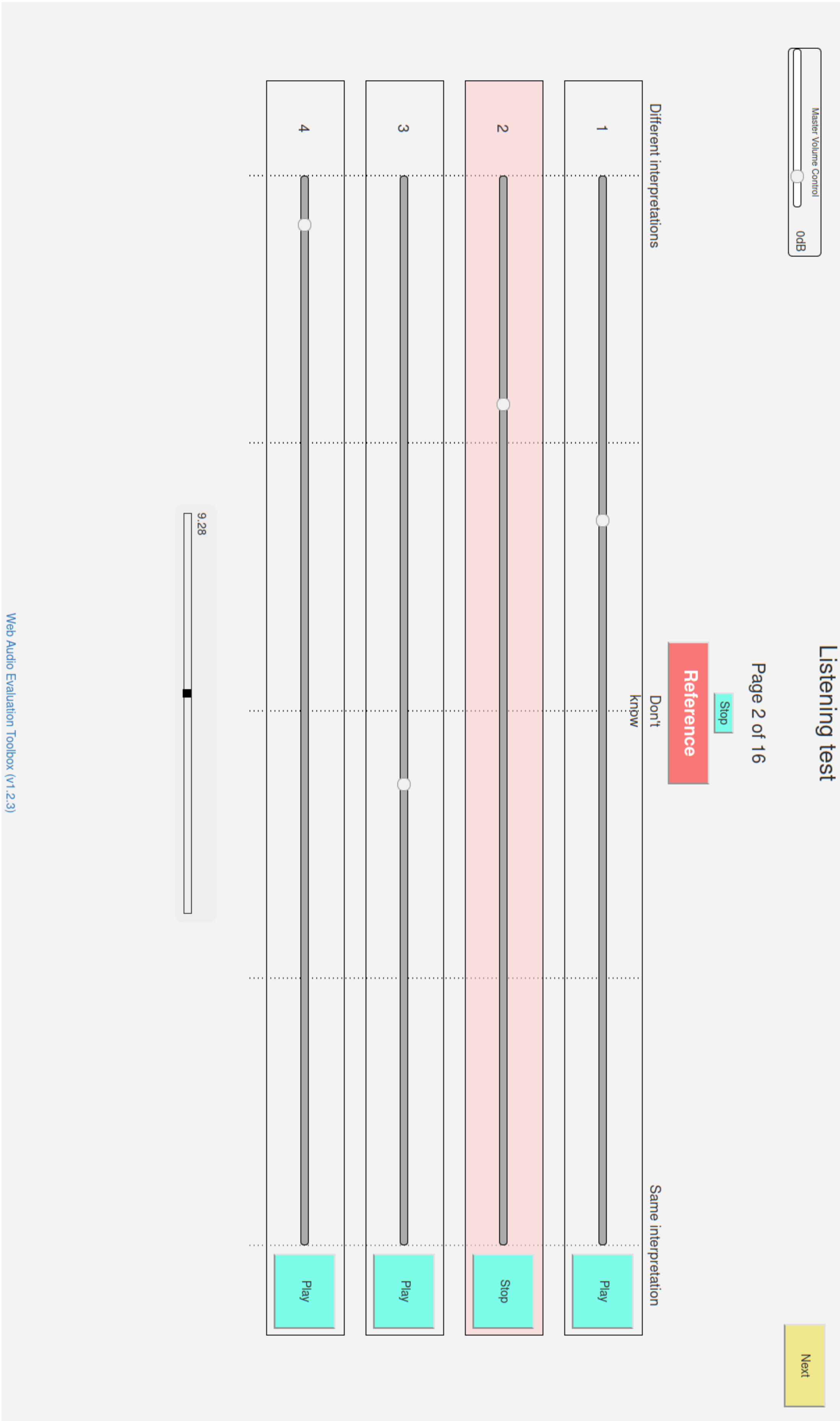

**Figure 7.2.:** *Screenshot of the interface created using the "Web Audio Evaluation Tool"* [306]

It comes from the cited literature, that a minimum sufficient duration to observe the difference between ratings given by expert and non-expert listeners is in the range $15 - 25$ s. Thus, we used excerpts lasting 20 s. Since each subject performed all of the three tasks – see Sec. 7.3.2 – and we aimed at keeping the test less than 30 minutes long, we opted for 5 excerpts, resulting in 15 questions (5 questions per task) each with 5 audio clips (1 reference and 4 candidates) lasting 20 s, for a total minimum duration of 25 minutes.

The studies discussed above also suggest that expert listeners tend to base their judgements on longer time features with respect to non expert raters. We expect a similar behavior for the task of comparing two interpretations. As a consequence, if no significant difference is found between expert and non-expert ratings, a difference may still be observed when using excerpts longer than 20 s. The only way to rule out such hypothesis would be by using full song excerpts.

## 7.4. Generating excerpts and contexts

Since we used a limited number of excerpts, we wanted to chose them in a way that minimizes any type of subjective bias. The same also applies for the choice of the acoustic contexts. In Fig. 7.1, we show the overall workflow used for solve this problem.

### 7.4.1. The $p$-dispersion problem and uniform selection

For choosing the excerpts, we built a dataset where each sample was represented by features extracted from audio clips and MIDI symbols. Since we expected that the perception of music changes as these features vary, we aim at maximizing the distance between feature vectors of the chosen excerpts so that perceptual variations are revealed.

This means that we are looking for the $p$ samples in the feature space which are distributed uniformly, while the distance between them is maximal. This problem can be seen as a variation of the *p-dispersion problem* [317] or *max-min facility dispersion problem* [318]. However, the $p$-dispersion problem does not impose any restriction with respect to data distribution, thus we derived four *ad hoc* algorithms solving the present problem based on Ward-linkage clustering [319]. Fig. 7.3 compares them on our excerpt dataset.

### 7.4.2. $p$-dispersion problem

Here we briefly describe the algorithm and compare it with a state-of-art method for solving the $p$-dispersion problem but we leave to future works the mathematical study of the method – see Appendix A.

Our approach consists in finding $p$ subsets using hierarchical clustering. We used Ward method because it tries to minimize the variance inside each cluster and, consequently, each cluster is well represented by its centroid. In contrast to $k$-means clustering, instead, it is well suitable even for little sized datasets and does not depends on initialization heuristic. After having partitioned the data in $p$ clusters with the Ward method, we chose one point per cluster as follows. We chose the point in each cluster which maximized the distance from the centroid of all the other points in the dataset (*Method A*). Successively, we have also considered other strategies to chose the point in each cluster, namely: the point which maximize the minimum distance from the centroids of the other clusters (*Method B*), the point which maximize the minimum distance from the points in the other clusters (*Method C*), the point which maximize minimum distance from all the other points (*Method D*). In table 7.1 we compare these methods to the state-of-art method for the $p$-dispersion problem by Contardo, for $p = [4, 5, 10]$, using the code provided us by the author; we used the datasets of the Contardo's work containing less than 10.000 samples with the addition of our dataset created with PCA output of 2 features per sample [316]. For method [316], we used the Julia implementation provided us by the author. We have always used the euclidean distance – or sometime the sum of the absolute differences for improving the computational time, without affecting the results.

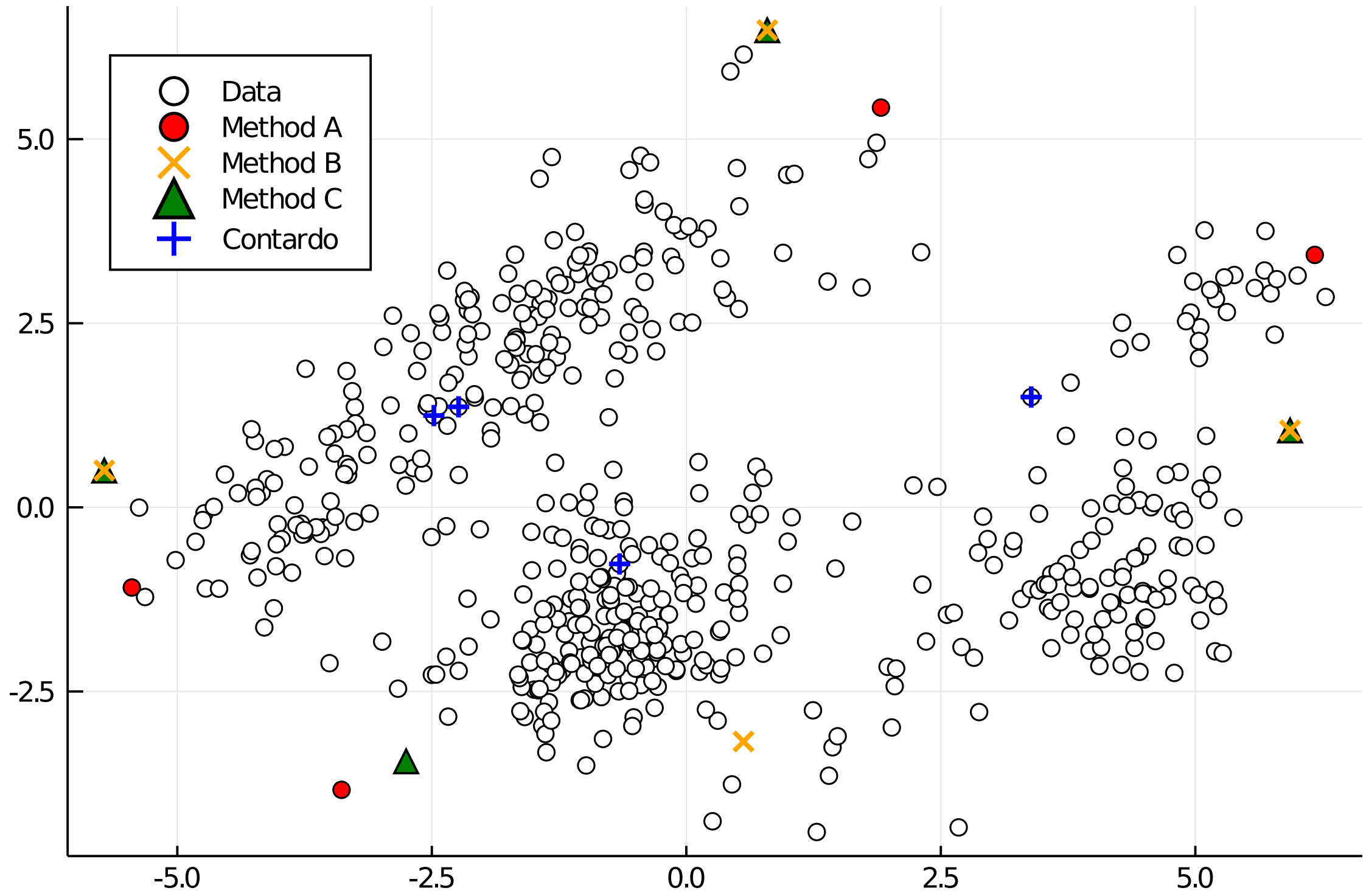

**Figure 7.3.:** *Comparison of methods for the p-dispersion problem with $p = 4$. Data are the windows extracted from the* Vienna 4x22 Piano Corpus*. PCA was used in this plot to reduce the dimensionality from 15 down to 2 for demonstration purposes. For the listening test, we used* Method A*.* Contardo *is the state-of-art mathematically-proven method for the p-dispersion problem [316]. For the comparison, we used the original Julia code provided by the author.*

**Table 7.1.:** *Comparison of methods for solving the p-dispersion problem. Columns are: average minimum distance in the output set, average time in seconds needed, percentage of instances in which each method had* Min Dist *grater or equal than any other (all) or than the Contardo's method [316]*

| | Min Dist | Time (s) | % wins vs all | % wins vs [316] |
|---|---|---|---|---|
| Method A | 1.58E+04 | 2.00 | 9.52% | 60.32% |
| Method B | 1.33E+04 | **1.66** | 25.40% | 61.90% |
| Method C | **1.90E+04** | 9.34 | **41.27%** | **63.49%** |
| Method D | 8.43E+03 | 10.44 | 1.59% | 50.79% |
| Contardo | 9.21E+03 | 153.30 | 34.92% | 100.00% |

### 7.4.3. Excerpt selection

The selection of the audio excerpts started from the *Vienna 4x22 Piano Corpus* [320], which consists of 88 audio and corresponding MIDIs recordings of 4 famous pieces highly representative of the classical-romantic music period, played by 22 professional and advanced student pianists. This corpus was useful in order to have a negative reference (NR) available for any chosen excerpt – i.e. a different interpretation. We used the *ASMD* framework [4] to handle the loading of files and dispatching parallel processing routines. Every audio clip was converted to monophonic, downsampled to 22050 Hz and normalized using ReplayGain [1]. MIDI files were loaded using pianorolls where each pixel contained the velocity value of the ongoing note and each column had a resolution of 5 ms. In order to compensate for temporal misalignments, for each pair (*audio, MIDI*), we identified the audio frame of first onset and last offset using an energy threshold of -60 *dB* under which the sound was identified as silence, and trimmed the files accordingly.

We split the audio recordings in windows and considered each window as a possible excerpt

[1] https://web.archive.org/web/20211007074859/https://wiki.hydrogenaud.io/index.php?title=ReplayGain_1.0_specification

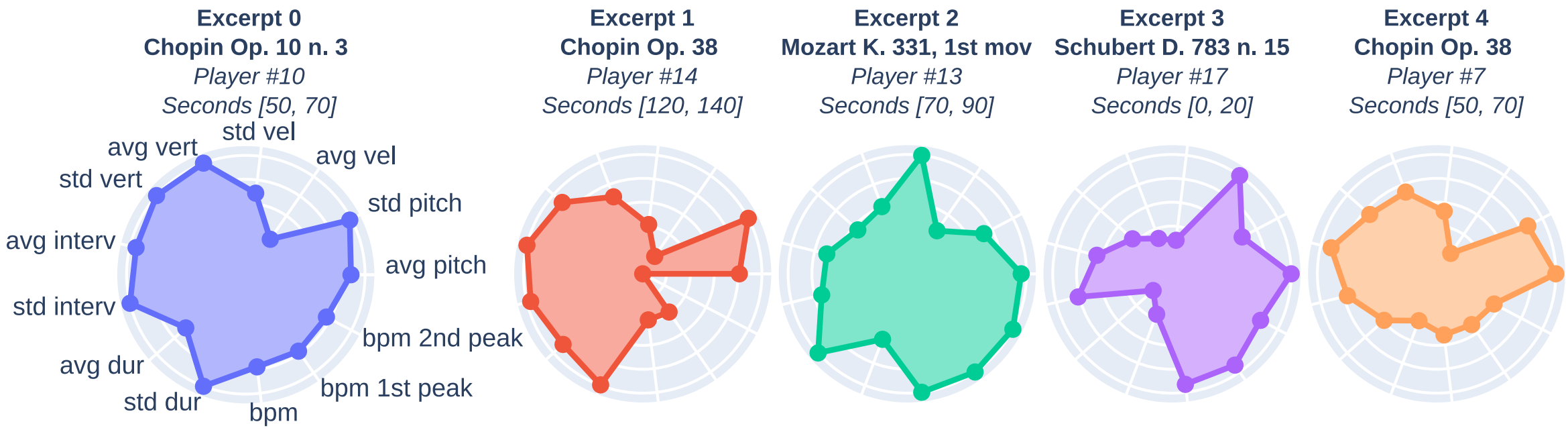

**Figure 7.4.:** *Some of the features extracted from the chosen excerpts that allow to conceptualize the difference between the excerpts. To visualization purposes, we excluded from this plot the features that are hardly understandable in terms of musical concepts – i.e. MFCCs, rhythmic descriptors, and pitch difference in each column. They are normalized in [0,1]. Seconds are relative to the audio recordings.*

for the listening test. Each window lasted exactly 20 s with a hop-size of 10 s, resulting in 564 total windows. For each window we extracted a set of audio and symbolic features. The first were extracted from the audio recordings and consisted of high-level features among the most used in the MIR field, extracted using the state-of-the-art library Essentia [321].To take into account the timbral characteristics, we extracted 13 MFCCs [322]; we used a state-of-the-art onset detection method to extract 7 rhythmic descriptors [323]. Furthermore, we used the Essentia library to estimate BPM, along with the first and second peak values, spreads, and weights of the corresponding histogram; such features are related to the timing characteristics of the performance as rendered in the audio. Regarding the symbolic features, we used the non-zero pixels in the window pianorolls to extract information about the performance as recorded by the sensorized piano. Specifically, we extracted the average and standard deviation of pitches, velocities, duration, number of contemporaneous notes in each column and pitch difference in each column relatively to the lowest pitch in the column – which relates to the type of harmony. The resulting features were concatenated in an array of 30 features. The variability of the features of the chosen excerpts are shown in Fig. 7.4

The 564 windows were then standardized (mean removal and variance scaling) and passed through PCA to obtain linearly separable features. Considering that all the 30 features used had different musical meaning, to select a number of features that would have not prevented the subsequent steps, we tried to half the number of features. After PCA, we obtained 15 relevant features explaining 92% of variance.

We applied the methods described earlier to look for 4 dispersed windows. To ensure that the 4 selected points well represented the whole dataset, we chose the only method that managed to select one window for each of the 4 pieces in the *Vienna* corpus (*Method A*). Then, we added one excerpt computed as the medoid of the dataset, obtaining in total 5 excerpts.

It should be noted all the excerpts last exactly 20 s; this may be criticized as it implies that NR and realignments could lead to slightly different contents, since a given excerpt could be played and transcribed in a different time lapse. On the other hand, having the NR lasting a different amount of time would have produced a potential bias at the listener side, who would have been able to identify the NR based on duration rather than audio content.

### 7.4.4. MIDI Candidates Creation

For each question in the test, we used four candidates in addition to the reference excerpt: a negative reference (NR) containing a performance by another pianist of the same excerpt but conveying a different interpretation from the reference, a hidden reference (HR) containing the same performance as the reference, and two performances extracted using two different AMT systems described next. For the NR, we realigned its MIDI recording to the chosen reference MIDI using FastDTW [324], before trimming. For the *restoration* task, where the reference was a real-world recording, the HR was not the same audio recording (which would be immediately recognizable from the remaining candidates), but rather the associated MIDI available in the Vienna corpus.

#### 7.4.4.1. Score-Informed AMT

Various AMT models were published in recent years; unfortunately, only the ones trained for solo piano music achieve satisfactory performances. In particular one model, here called *onsets and frames* (O&F), has been extensively evaluated on various datasets and has been shown to overcome the rest in almost every piece [280, 296]. Recently, it has been shown that AMT models could be enhanced by pre-stacking a U-Net [325]. U-Nets were first used for image segmentation and then for audio source separation. By pre-stacking a U-net, the network tends to learn knowledge regarding the sparse structure of the spectrogram-like input representation. However, we are more interested in understanding how audio-based AMT differs from score-informed AMT; intuitively, since score-informed AMT models exploit more information, the output should be more accurate. Thus, we compared O&F with a score-informed model (SI) which we developed based on a previous work [292].

In SI, inputs are a non-aligned score and an audio recording, while the output is a list of MIDI notes, each associated with onset, offset and velocity. SI performs audio-to-score alignment using a method based on Dynamic Time Warping that improves a previous system for piano music [326], and subsequently executes a Non-negative Matrix Factorization as source-separation method for each piano note. Then, it employs a neural network for estimating the velocities of each aligned note using as input the spectrogram of the note, computed thanks to the source-separation. Since the SI method requires the score, one was obtained from the World Wide Web for each of the 5 excerpts. For further details see the Supplementary Materials – see Appendix A.

Both the AMT systems predict pitches, onsets, offsets and velocities, while no other MIDI parameter (e.g., pedaling, etc.) is considered.

### 7.4.5. Synthesis and Context selection

After producing the MIDI files, we synthesized them using 4 high quality different virtual pianos:

(i) the free *Salamander Grand Piano*[2];

(ii) two Pianoteq instruments freely available for research purposes (*Grand Steinway B* and *Grand Grotrian*);

[2]free as in speech: https://musical-artifacts.com/artifacts/533

**Table 7.2.:** *Number of questions with* $p > 0.05$ *for each pair of methods for both Sudents' t and Wilcoxon tests*

| | | | |
|---|---|---|---|
| SI | 4 | | |
| O&F | 6 | 0 | |
| HR | 7 | 1 | 9 |
| | NR | SI | O&F |

(iii) the *Steinway* piano from Garritan Personal Orchestra 4SI.

We post-processed every synthesized MIDI using SoX[3] in order to add reverberation using two different settings (values 50 and 100 of the SoX's reverb option). Thus, 12 different contexts, i.e. 4 without and 8 with reverberation, were formed.

We synthesized all the MIDI files with each context obtaining 12 different sets of audio clips. We extracted 13 MFCC and 7 rhythmic descriptors from each audio clip and computed the mean features to represent each context. We chose the medoid context for the *transcription* task and the most distant context from the average features of the original audio recordings for the *restoration* task. For the *resynthesis* task, two contexts were needed: in this case PCA explaining 99% of variance was applied to obtain a 10 dimensional representation from the original 20, and then searched for the two farthest points in the feature space based on the euclidean distance.

As a result of this process, the selected contexts were:

a) the Pianoteq *Grand Steinway* with SoX reverb set to 100 for the transcription task (44.1KHz / 16bit stereo audio for both reference and candidates);
b) the Pianoteq *Grand Steinway* without reverb and the *Salamander Grand Piano* with SoX reverb set to 50 for the resynthesis task, candidates and reference respectively (44.1KHz / 16bit stereo for the reference and 44.1KHz / 24bit stereo for candidates);
c) the *Salamander Grand Piano* without reverb for the restoration task (44.1KHz / 16bit mono for the reference and 44.1KHz / 24bit stereo for candidates).

Note that no perceptual difference is known between 24 and 16 bit depth at the same sample rate [327, 328, 329].

## 7.5. Results

The listening test was communicated through mailing lists, chats, university courses, etc.; in total, 91 subjects responded to the entire test. Thanks to JavaScript-based WAET, we observed the subjects' behavior during the test, so that we were able to discard the answers where subjects listened to the excerpt for less than 5 seconds or where they did not move the cursor. After such filtering, we obtained more than 40 answers per question. Since these did not result in enough answers for each class of the initial questionnaires described in 7.3.3, we focused on two groupings only:

[3]sox.sourceforge.net

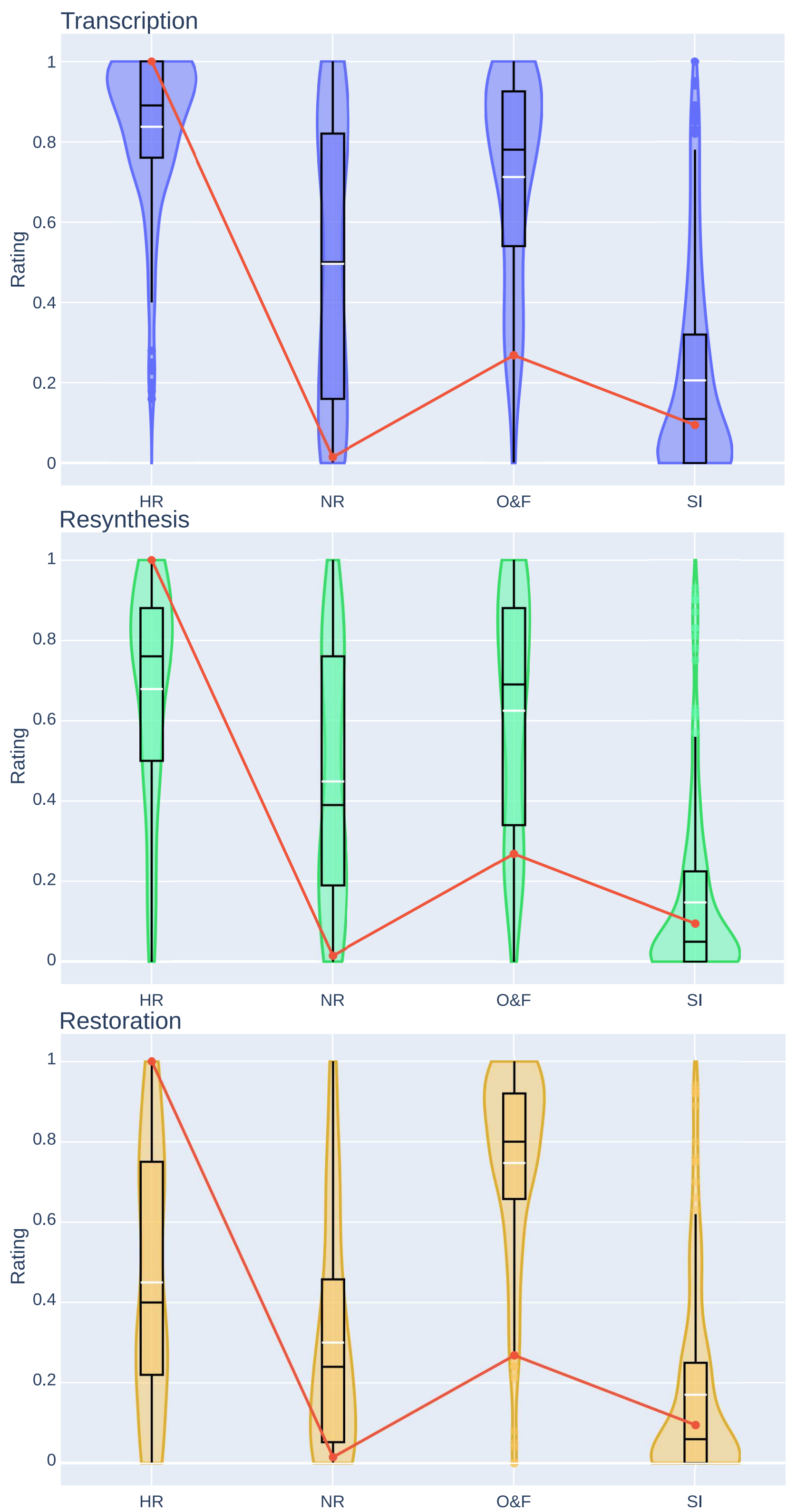

**Figure 7.5.:** *Ratings per task averaged over all the excerpts. The red line identical in all tasks is the objective F-measure. White horizontal line is the mean, the black horizontal line is the median. All plotted distributions pass the pairwise significance tests against the other distributions in the same task, except for O&F and HR in the resynthesis task.*

- subjects listening to classical music less than 1 hour per week (50) vs. subjects listening to classical music more often (41);
- subjects who have never studied music/hobbyists (57) vs. subjects who studied music professionally or having a degree or working as musicians (34);

We observed a general trend of non-experts providing higher ratings, meaning that, with respect to more experienced listeners or musicians, they rated candidates to be more like the same interpretation as the reference. However, this difference was not always statistically significant, thus not useful for the sake of our research questions. According to the literature discussed in Sec. 7.3, we could expect that by using longer excerpts the difference in the ratings would become significant and that expert listeners could give more accurate ratings.

We collected an imbalanced number of answers per type of headphones, namely 22 for headphones costing less than 20 euros and 69 for headphones costing more than 20 euros. Since we have found contrasting studies in literature about possible correlations between headphone retailing cost and sound quality, we have decided to disregard this factor during the successive analysis [330, 331].

During control group based analysis, we had more than 20 answers available per question and control group. We first applied Shapiro-Wilk normality tests with Bonferroni-Holm correction and $\alpha = 0.05$ to the collected responses for each control group, question, and method. We observed that the null-hypothesis – i.e. the collected answers are normally distributed – was rejected depending on the method and on the question. Consequently, we have performed the whole statistical analysis with both parametric and non-parametric methods and leave to the reader the ability to decide which test should be taken into account based on the single case. The following discussion and conclusions, however, hold in both the two cases – i.e. parametric and non-parametric analysis.

We computed error margins at 95% of confidence, that is a quantification of the accuracy for the estimated mean $\hat{\mu}$: we can say that by resampling the distribution, 95% of the collected populations will have the mean in $\hat{\mu} \pm e$, where $e$ is the error margin. Error margins were computed with both normal distribution assumption [332] – i.e. parametric estimation – and bootstrapping methods [333] – i.e. non-parametric estimation. The full set of error margins is available in the Supplementary Materials – see Appendix A. For our discussion, we can say that parametric and non-parametric error margins were rarely different when rounded at the 2nd decimal digit and that they ranged between 2% and 17%. Without using control groups, the error margins were between 3% and 10%, and between 2% and 5% when we average the ratings over the questions of the same task.

We analyzed results using ANOVA and Kruskal-Wallis tests with $\alpha = 0.01$ and we rejected the null hypothesis in all considered questions. We further analyzed the data using the Wilcoxon and the Student's t-test for related variables with the Bonferroni-Holm correction and $\alpha = 0.05$. In case the test condition is not satisfied, we cannot reject the hypothesis according to which the perception of two candidates is explainable by the same model. This is usually observable for non-expert subjects. Table 7.2 shows a general overview of the $p$-values computed for each pair of methods.

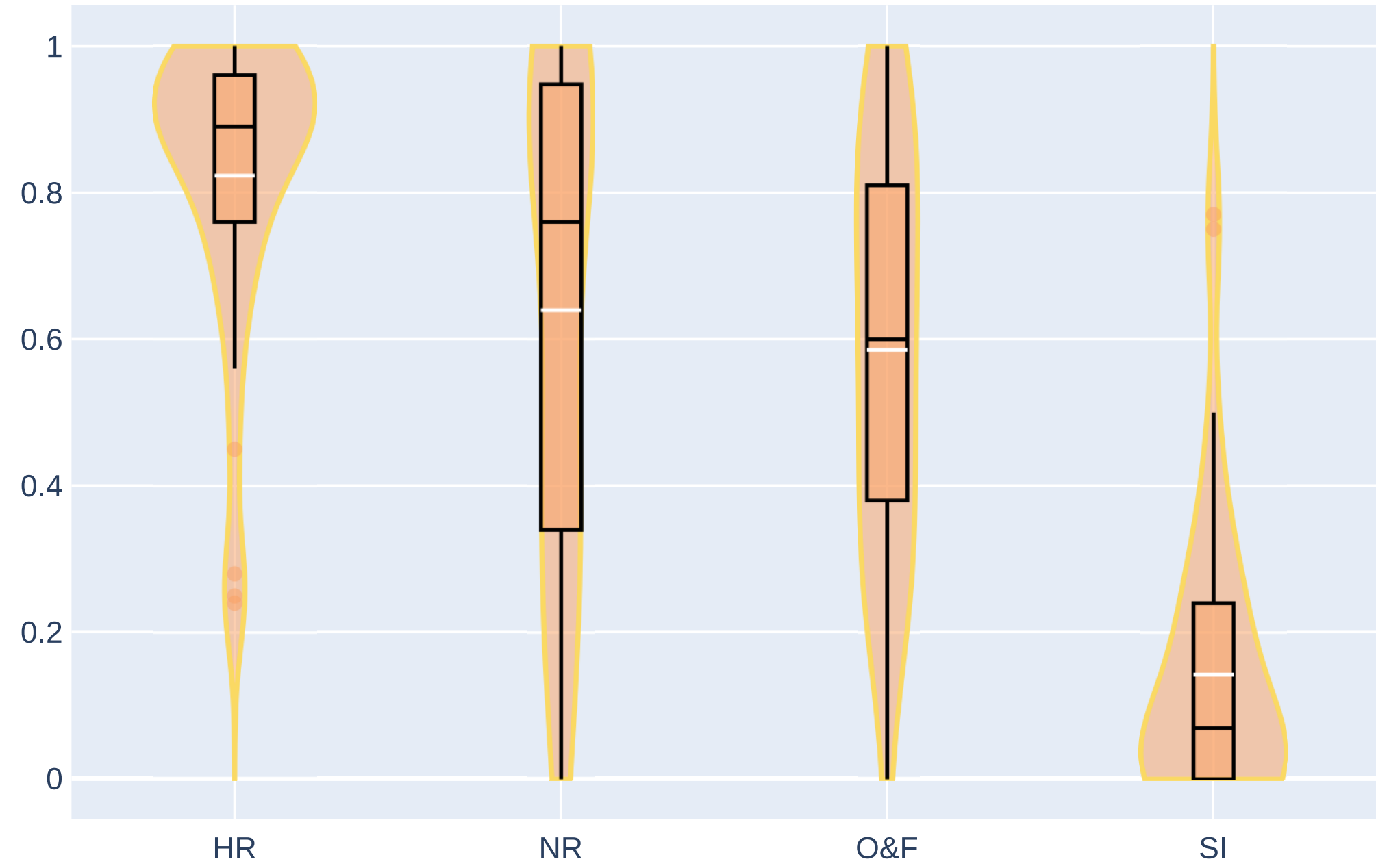

**Figure 7.6.:** *Ratings for Excerpt 3 in the Transcription task. For this excerpt, all distributions pass the pairwise significance test except O&F and NR.*

Fig. 7.5 illustrates the average ratings for every task. The HR and the NR were recognized in all tasks, and O&F performed always better than NR and SI. In a post-experimental interview, several subjects reported that SI was hardly comparable to the reference due to bad alignment. Indeed, in some excerpts, notes were distributed by SI in a very short time, producing a correspondingly long-lasting silence; this was often caused by missing/extra notes appearing in the music score. Thus, we conclude that the main reason for which SI was always rated worse than O&F is related to mis-alignments; we can consequently answer the third research question by stating that score-informed approaches are generally limited by the alignment stage and that, as of now, monomodal AMT approaches provide improved performance assessed from a perceptual point of view. However, recent works carried out in this Thesis should still be evaluated – see Chapter 6.

In the *restoration* task, O&F was rated higher than HR. Since this behavior is not observable in the *transcription* and *resynthesis* tasks, and since the MIDI files were identical throughout all tasks, we attribute this outcome to the specificity of the *restoration* task, where the HRs (MIDI) were different from the references (audio), unlike the other tasks. In particular, at the time of writing, performance annotations in the *ASMD* framework do not include information about the pedaling used by the players, and recorded in the audio. Thus the HRs in the *restoration* task were synthesized from MIDI with no sustain control changes, whereas the audio references contained them. On the other hand, O&F is not able to transcribe the pedaling, but its authors enlarged the duration of sustained notes in the training ground-truth, so that the prediction of note duration is temporally tied to the duration of the resonance of the note rather than the onset/offset of the key. Such a durational enlargement allows O&F (and SI as well, as it uses O&F for the alignment) to predict duration perceptually more accurate than the HR in the *restoration* task. analysis in the *restoration* task which reveals that excerpts 0, 3, and 4 show no statistically significant difference ($p > 0.23$) between HR and NR.

The *resynthesis* task is also worth some further discussion. In this task, HR and O&F are perceived similarly, especially by non-expert listeners, and there are no statistically significant differences between the distributions of their ratings – see Fig. 7.5. Even though HR is rated slightly higher than O&F, it can be noted to score lower than in the *transcription* task. We have thus analyzed how the HR was perceived across various tasks, finding that, for all the 4 excerpts, in the *resynthesis* task, HR candidates were rated lower than in the *transcription* task, meaning that listeners have found a substantial difference between the interpretative content of the original and resynthesized audio ($p < 0.06$ when each single excerpt is analyzed, $p = 4.7\text{e-}12$ when all answers from all excerpts are analyzed at once). This suggests that the whole reference interpretation was hardly recognised in the HR, and that part of the interpretation was perceptually lost. Based on this outcome, we can try to address negatively the first research question, that is: when the context changes, MIDI representation seems not adequate to reproduce the same interpretation. However, other experiments are needed to confirm this hypothesis.

Analysing results question by question, we discovered an interesting behavior in the *transcription* task for excerpt 3 – see Fig. 7.4 and 7.6. There, O&F is associated with lower ratings than NR with $p = 0.38$ for Wilcoxon test and $p = 0.47$ for Student's t-test, meaning that the transcription is so inadequate that another interpretation resembles better the original one. Similar results are derived when investigating excerpt 0 in *resynthesis* and *transcription* task, where NR and O&F present almost identical ratings ($p > 0.23$). This behavior is more evident when looking at less expert listeners. Such results can also answer negatively to the second research question: the state of the art for piano AMT may not be able to extract parameters usable for reproducing the original interpretation, regardless changes in context (resynthesis and transcription tasks).

## 7.6. A new measure

Having answered our research questions, we looked for correlations between ratings and typical measures used for evaluating AMT systems. In particular, we adopted the widely-used measures available in a Python package [334]. In Fig. 7.5, the red line represents the F1-score computed considering as matches notes whose parameters lie within a certain range around the true value; in this case, we considered as parameters:

(i) the onset and offset times with a range of ± 50ms;

(ii) the linearly re-scaled velocity so that the L2 error is minimized with tolerance of 10%;

(iii) the pitch with tolerance of 1 quarter-tone [280].

This measure, hereafter denoted as *OBJ*, demonstrated low correlation with subjective ratings, mainly due to the following factors:

- excerpt 0 has a low OBJ rating for O&F (almost 0); however, subjects rated it much higher than O&F;
- in the restoration task, O&F received higher ratings than HR, which always has OBJ equal to 1 – see Sec. 7.5;
- occasionally, O&F was rated lower than NR, which always has OBJ equal to 0.

**Table 7.3.:** *Correlations of various measures with the average rating of the subjects. Values are percentages.*

| | | | OBJ | PEAMT | Ours |
|---|---|---|---|---|---|
| **Transcription** | **Pearson** | **min** | 49 | 62 | **95** |
| | | **max** | 78 | **99** | 97 |
| | | **avg** | 75 | 89 | **97** |
| | **Spearman** | **min** | 40 | 40 | **80** |
| | | **max** | **100** | 80 | 80 |
| | | **avg** | **80** | **80** | **80** |
| **Resynthesis** | **Pearson** | **min** | 41 | 64 | **95** |
| | | **max** | 73 | **100** | 99 |
| | | **avg** | 66 | 85 | **98** |
| | **Spearman** | **min** | 40 | **60** | **60** |
| | | **max** | **100** | **100** | 80 |
| | | **avg** | **80** | **80** | **80** |
| **Restoration** | **Pearson** | **min** | -4 | 55 | **78** |
| | | **max** | 57 | 94 | **100** |
| | | **avg** | 29 | 78 | **89** |
| | **Spearman** | **min** | 0 | 0 | **80** |
| | | **max** | 60 | 60 | **100** |
| | | **avg** | 60 | 60 | **100** |
| **Average (leave-one-out)** | **Pearson** | | 45 | 71 | **85** |
| | **Spearman** | | 44 | 54 | **74** |

Another interesting measure, named *PEAMT*, reflecting subjective ratings was proposed in [296]. We computed Pearson and Spearman correlation coefficients between OBJ and PEAMT measures and the median and average values of the collected ratings. It was discovered that the average generally presents slightly higher correlation than the median, therefore the former was used for subsequent analyses. Table 7.3 shows that PEAMT correlates more strongly than OBJ to subjective ratings, especially for the Pearson coefficient. especially in the *restoration* task, correlation remains poor, motivating the search for an alternative measure.

We considered the features already used for the excerpt selection phase – see Sec. 7.4 – except for audio-based features to be consistent with the existing evaluation methodologies. However, in doing so, the searched measure will be context-unaware and will represent the amount of interpretation information kept by a MIDI sequence when it is resynthesized in a different context without modifications. We computed BPMs using the MIDI representation by counting how many onsets were present in sliding windows of size 0.1, 1.0, and 10 s with a hop-size of 50%. More precisely, we counted the mean and standard deviation for each window size. Importantly, to improve the portability of our measure, the features were first standardized using parameters computed on a large set of piano MIDI created by extracting piano solo performances from various datasets using *ASMD* [4]. After standardization, features of the predicted performances were subtracted from the target and the OBJ measure was appended. We performed linear regression on the dataset that we collected using various methods: Bayesian Ridge, Automatic Relevance Determination, Lasso Lars, Lasso, ElasticNet, Ridge and basic Linear regression. ElasticNet provided the best perfor-mance in terms of average L1 error. To further improve the generalization ability, we trained a model using ElasticNet while removing features with low weights, i.e. < 0.1.average L1 error.

We finally measured the average L1 error in a leave-one-out experiment, which provided 0.12 for our measure, 0.19 for PEAMT and 0.34 for OBJ. Table 7.3 (bottom rows) shows correlation coefficients for each task and measure.

When comparing PEAMT and our measure, one needs to consider differences in the design of the related tests. PEAMT is based on a test using audio clips lasting $5 - 10$ s, while, following the discussion summarized in 7.4, this work employs clips lasting 20 s. PEAMT authors' created 150 questions and collected 4 answers for each one; we instead preferred to collect more than 20 answers per question for plurality, while considering a control group which led us to reduce to 15 the total number of questions. At the same time, the space of possible note combinations is covered optimally (see Sec. 7.4). Furthermore, PEAMT is based on categorical questions – subjects could chose between two systems – with no HR and NR, while we measured a linear variable and included hidden and negative references. Finally, we focused on changing the context of the recordings and synthesis, but we included in our Transcription task the scenario used by PEAMT.

In general, we can state that PEAMT results agree with ours in finding low correlations between subjective ratings and OBJ, and that the two evaluation measures that we have built are rather similar in our preliminary tests. However, our test highlights new aspects that we think fundamental for audio restoration and that only our measure is able to tackle.

## 7.7. Conclusion

After conducting a thoroughly designed perceptual test, this work proposed a new approach for audio restoration: in the light of recent developments in audio signal processing, it becomes imaginable to recreate performances in the real world or through virtual instruments. We have therefore designed a perceptual test to assess to which extent existing technologies allow for such a methodology. It was discovered that the main limit lies in the usage of the MIDI format itself. Nonetheless, we proposed a new evaluation measure that seems consistent with the perception of context changes.

In case Standard MIDI Format is used as basis for the resynthesis, knowledge regarding contextual factors is required. Consequently, we argue that the future challenge for resynthesis-based audio restoration is in the conversion of the existing audio and music score in a new restored audio without the use of mid-level representations such as MIDI.

In this work, we have also identified limits for score-informed AMT, that, despite exploiting more information, lacks an effective feature fusion stage. Audio-to-score alignment should therefore become a main challenge for score-informed AMT; overcoming this problem could lead to improvements related to the exact knowledge of pitches and timings, leaving space for focusing on other parameters. For this sake, the method described in Chapter 6 may be an important advancement. Finally, we proposed a generic method to meaningfully choose excerpts when conducting music listening experiments.

# 8 A Mathematical Formalization

In this Chapter, a mathematical formalization describing context-aware transcription and resynthesis is proposed. Through such a formalization, empirical experiments can be designed to assess the validity of the proposed theory regarding "Music Interpretation Analysis" (MIA). We outline an experiment which does not require collecting new data but only the resynthesis of existing MIDI recordings. Using such a pipeline, the hypothesis according to which acoustics characteristics of the room can be used to reduce performance analysis error is empirically verified regarding the velocity estimation of piano notes. To this end, a novel score-informed Automatic Music Transcription (AMT) method able to take into account the synthesis context is proposed. Further experiments are required to investigate other performance parameters such as pedaling and other instruments, while new data collection is needed to consolidate the MIA framework. Overall, the proposed mathematical formalization is shown to be a promising approach to tackle realistic automated music resynthesis.

## 8.1. Introduction

In recent years, various studies about Music Performance Analysis (MPA) [170] have faced the problem of Automatic Music Transcription (AMT) [171], especially in piano music. Almost every year, a new state-of-the-art model is published in major conferences and journals [194, 280, 281, 325, 335, 336], and, importantly, current models have achieved an increase of almost 10% in F1-measure since 2018.

Despite the recent trend of the Music Information Processing (MIP) research in AMT, to our knowledge, no one has approached the problem of assessing how the environment acoustics influence the performance. Outside the AMT field, various works showed that musicians adapt their performance to the acoustic context in which they play – see 2.4.1 for a review.

AMT models able to deal with the acoustic context, the performance, and the interpretation may significantly impact a variety of applications. The long-term objective of this study is the resynthesis of music for production and restoration purposes [7]. Other possible use-case scenarios include musicological studies and music teaching applications, such as the analysis and comparison of the interpretation that in turns can pave completely new paths for these research fields; moreover, the ability of transcribing both performance and interpretation would also allow the comparison of the manifold ways in which different performers adapt their interpretation. Not least, architectural studies could be impacted from robust context-aware AMT models.

For all the above mentioned applications, tools offering high precision are needed. For this reason, we will focus on multimodal mip approaches, which currently constitute the most effective tools for precise analysis in the long term [1]. Among the multimodal approaches for AMT, most of the research deals with leveraging music score information in the so-called score-informed AMT [171], and they will be the center of our attention.

This work represents a first attempt to approach Automatic Music Transcription complemented with the knowledge deriving from the previous discussion. To this end, we first propose a suitable mathematical formalization of the adaptation and transcription of music performances. Subsequently, we propose an empirical experiment demonstrating the effectiveness of exploiting context acoustics for AMT models. During this phase, special attention was placed towards automatic music resynthesis – Chapter 2 [7] – since this is the main long-term objective of this work. Moreover, even though the proposed formalization makes no assumption about the playing instrument, we only performed empirical experiments only on piano music.

For the sake of reproducibility, the whole code used for the experiments is available online – see Appendix A.

## 8.2. A mathematical formalization

### 8.2.1. Notation

Let $\mathcal{I}$ be the set of all possible interpretations for a given musical score, $\mathcal{C}$ the set of acoustic contexts, and $\mathcal{P}$ the set of performances[1]. Then, we can define the set of all the possible adaptation functions as:

$$\mathcal{A} := \{\alpha_{c_0}(i) : \mathcal{I} \rightarrow \mathcal{P}\}$$

In other words, an adaptation function is defined as a function that takes as input an interpretation and an acoustic context and returns a performance. With $c_0$, we refer to the context with index 0. In the following, when an element or a function is referred to a single context $n$, we will simply use the subscript $n$ instead of $c_n$, such as $\alpha_0(i)$ in the previous definition of $\mathcal{A}$.

Once the performance has been generated, it should be synthesized. The synthesis function can be defined as:

$$\sigma_0(\cdot) := \mathcal{P} \rightarrow \mathcal{W},\ w_0 = \sigma_0(p_0) \quad (8.1)$$

where $\mathcal{W}$ is the set of all the possible audio waves and 0 is the synthesis' acoustic context. As said, in the proposed conceptual model, $p_0$ is generated by an adaptation function starting from an interpretation:

$$p_0 = \alpha_0(i)$$

To better explain the proposed notation, suppose having one interpretation $i$ and two different contexts $c_0$ and $c_1$. Since $c_0 \neq c_1$ by hypothesis, we can associate to each context a synthesis function: $\sigma_0(\cdot) \not\approx \sigma_1(\cdot)$. The adaptation of $i$ to contexts $c_0$ and $c_1$ will generate two different performances $p_0 = \alpha_0(i)$ and $p_1 = \alpha_1(i)$. Consequently, two different audio waves will be generated by the synthesis in the two corresponding contexts, namely $w_0 = \sigma_0(\alpha_0(i))$ and $w_1 = \sigma_1(\alpha_1(i))$. Now, suppose that $w_0$ and $w_1$ are identical; in such case, if $c_0 \neq c_1$, then $\sigma_0(\cdot) \not\approx \sigma_1(\cdot)$, and so $p_0 \neq p_1$. Indeed, if $p_0 = p_1$, then $w_0 = \sigma_0(p_0) = \sigma_0(p_1) \neq \sigma_1(p_1) = w_1 \Rightarrow w_0 \neq w_1$.

[1] In our formalization we will use upper case letters such as $\mathcal{A}$, $\mathcal{B}$ to indicate sets, while the corresponding lower case letters $a$, $b$ will refer to elements of that set. If the element is a function, then it will be indicated with the corresponding lower case Greek letters $\alpha$, $\beta$.

### 8.2.2. Automatic Music Transcription

We now have the elements to formalize the usual Automatic Music Transcription process. In the related literature, the function that takes as input an audio wave and returns the underlying performance parameters is usually named "transcription". As such, we will name transcription a function $\tau := \mathcal{W} \to \mathcal{P}$; comparing the latter definition to Equation (8.1), one could note that the domain and co-domain of $\tau$ and $\sigma$ are inverted. However, $\tau$ is not exactly $\sigma_0^{-1}$, because the transcription functions in literature are not context-aware. Indeed, usual transcription functions are models learning to infer the same performance for two identical audio $w_0$ and $w_1$ even if they are generated from synthesis from two different contexts. According to our previous discussion, instead, $w_0$ and $w_1$ have a different underlying performance. As such, we provide a more inclusive definition as follows:

$$\tau(w_0) = \frac{1}{K} \sum_{k=0}^{K} \sigma_k^{-1}(w_0), \tag{8.2}$$

where $k = 1, 2, \ldots, K$ are all the possible contexts seen during training and belonging to $\mathcal{C}$, with $K = |\mathcal{C}|$. In other words, we are assuming that the usual context-unaware transcription function extracts an *average* performance across all the contexts. As such, there will be an error connected with the original synthesis context $c$ between the true underlying performance and the transcribed one:

$$\hat{p}_c = \tau(w_c) = \sigma_c^{-1}(w_c) + \delta_c = p_c + \delta_c \tag{8.3}$$

### 8.2.3. Automatic Music Resynthesis

The goals of this work are to:

(i) extract the interpretation content from an audio wave $w_0$ obtained in the context $c_0$;

(ii) adapt it to the target context $c_1$ generating a performance;

(iii) synthesize the performance data in the target context.

Following the proposed formalization, we describe an audio wave as an interpretation adapted to the recording context $c_0$:

$$w_0 = \sigma_0(\alpha_0(i)).$$

To achieve goal (i), we estimate $i$ with an extraction function $\hat{\imath}(w) := \mathcal{W} \to \mathcal{I}$. Goal (ii) corresponds to the application on the estimated $\hat{\imath}$ of an adaptation function $\hat{\alpha}_1$ estimated for the target context $c_1$, producing an estimated performance $\hat{p}_1$. Finally, goal (iii) is the synthesis in the context $c_1$ via the synthesis function $\sigma_1$ of the estimated performance $\hat{p}_1$, obtaining a new audio wave $w_1 = \sigma_1(\hat{\alpha}_1(\hat{\imath}(w_0)))$. Considering that both $\alpha_1$ and $\iota$ are estimated functions, they will produce a certain degree of error, and as such $w_1$ should actually be written as follows:

$$w_1 = \sigma_1(\alpha_1(\iota(w_0) + \delta') + \delta'') \tag{8.4}$$

Suppose that we define $\tau(\cdot) = \alpha_1(\iota(\cdot))$. This definition corresponds to the hypothesis that the output of context-unaware transcriptions corresponds to the adapted performance, that is $\tau$ already takes care of adapting the output to the target context, or that the context has no influence on the final performance. Of course, our thesis is opposite to such a definition and indeed, considering Equation (8.3) and adding a further error $\delta'''$ connected with the imperfect optimization of the transcription model, we would obtain the following resynthesized audio wave:

$$w_1' = \sigma_1(\tau(w_0)) = \sigma_1(p_0 + \delta_0 + \delta''') = \sigma_1(\alpha_0(i) + \delta_0 + \delta''') \tag{8.5}$$

Equation (8.5) is not what we were looking for in Equation (8.4); even when discarding the errors, Equation (8.5) is still using the $\alpha_0$ function instead of $\alpha_1$. The proposed function $\iota(\cdot)$, instead, should be context-aware and should return:

$$\begin{cases} \hat{i} = \iota(\sigma_0(\alpha_0(i))) + \delta' = i + \delta'' \\ \iota(w_h) = \iota(w_k)\ \forall c_h, c_k \in \mathcal{C} \end{cases} \tag{8.6}$$

Note that $\iota(w_1'') = \iota(w_0)$, so that $\iota$'s context-awareness takes the form of context-independence. It could be observed that, according to our definition, $\tau$ is context-independent and that, consequently, we can define $\iota := \tau$ and $i := \bar{p}$ where $\bar{p}$ is the average performance across all the contexts. However, there is a fundamental distinction between the context independence of $\tau$ and the one of $\iota$: while $\tau$ is trained without considering the acoustic context conditions, $\iota$ should be trained to completely disregard them, so that $\iota$ *is* context-aware because it recognize context factors and eliminates them. Moreover, defining $\iota := \tau$ leads to consider $\mathcal{I} = \mathcal{P}$, meaning that performance and interpretation have the same representation format. In the following we will consider such situation a degenerate case of our framework.

### 8.2.4. Reward

To understand if the usage of Equation (8.4) will lead to better results than Equation (8.5), we need to define a distance function $\varepsilon(\cdot,\cdot)$ between performances and to require the following condition:

$$e' = \varepsilon\big(\alpha_1(i + \delta') + \delta'', \alpha_1(i)\big) \ll \varepsilon\big(\alpha_0(i) + \delta_0 + \delta''', \alpha_1(i)\big) = e'' \tag{8.7}$$

If $\varepsilon(\cdot,\cdot)$ satisfies the triangle inequality property – e.g. $\varepsilon(\cdot,\cdot)$ is a metric or even a simple subtraction – we can rewrite the right side of (8.7) to derive the following condition:

$$e' \ll e'' \leq \varepsilon\big(\alpha_0(i) + \delta_0 + \delta''', \alpha_1(i + \delta') + \delta''\big) + e'$$

$$\Downarrow$$

| Sample | $\iota(\cdot)$ | $\tau(\cdot)$ |
|---|---|---|
| $w_0^0$ | $i_0$ | $p_0^0$ |
| $w_0^1$ | $i_0$ | $p_0^1$ |
| $w_1^0$ | $i_1$ | $p_1^0$ |
| $w_1^1$ | $i_1$ | $p_1^1$ |

**Table 8.1.:** *Expected output from various audio waves from a collected dataset. The superscripts represent the index of the underlying interpretation, while the subscripts refer to the acoustic context.*

$$\varepsilon\big(\alpha_0(i) + \delta_0 + \delta''', \alpha_1(i + \delta') + \delta''\big) \gg 0$$

We say that the left side of the previous inequality is the "reward" for having estimated the $\iota$ and $\alpha_c$ functions

$$R := \varepsilon\big(\alpha_0(i) + \delta_0 + \delta''', \alpha_1(i + \delta') + \delta''\big) = \varepsilon\big(\hat{\tau}(w_0), \hat{\alpha}_1(\hat{\iota}(w_0))\big) \tag{8.8}$$

If the Reward is large, the condition (8.7) will be met. If the optimization errors $\delta'$, $\delta''$, and $\delta'''$ become near to 0 and $\varepsilon(\cdot, \cdot)$ is a simple subtraction, then the reward is:

$$R = \alpha_0(i) - \alpha_1(i) + \delta_0$$

Recalling Equations (8.2) and (8.3), we can also quantify $\delta_0$:

$$\delta_0 = \tau(w_0) - p_0 = \frac{1}{K}\sum_{k=0}^{K} \sigma_k^{-1}(w_0) - \alpha_0(i) = \overline{\sigma^{-1}}(w_0) - \sigma_0^{-1}(w_0),$$

where $\overline{\sigma^{-1}}(w_0)$ is the average inverted synthesis function. Overall, the reward for having estimated the $\alpha_1$ and $\iota$ functions instead of the unified $\tau$ is proportional to two components: 1) the difference between the adaptation functions in the original and target contexts and 2) the difference between the target and the average synthesis functions, or equivalently the error generated by the transcription function that is not connected with the approximation error $\delta'''$. In general, both terms depend on the difference between contexts $c_0$ and $c_1$: if the two contexts are enough different and the estimation succeed, the reward will be high. However, when the reward is 0, $\alpha_1(i) = \alpha_0(i) + \delta_0$ and Equations (8.5) and (8.4) coincide. Showing that the reward is high, thus, proves that (8.4) enables better results than (8.5), which is the ultimate goal of the proposed methodology.

### 8.2.5. Computing the reward

Unfortunately, the computation of $\delta_0$ in a realistic scenario is hard, because it is difficult to show that the approximation errors $\delta$ are near to 0 while the error due to $\delta_0$ is large. One option is to compare all the estimated functions as in Equation (8.8): if the reward is high, we can assume that the approximation errors cancel each other and that the remaining difference is due to $\delta_0$. However, such an approach will not guarantee that $\delta_0 \gg \delta'''$.

| Sample | $\iota(\cdot)$ | $\tau(\cdot)$ |
|---|---|---|
| $w_0^0$ | $i_0^{\cdot 0}$ | $p_0$ |
| $w_0^1$ | $i_0^{\cdot 1}$ | $p_0$ |
| $w_1^0$ | $i_1^{\cdot 0}$ | $p_1$ |
| $w_1^1$ | $i_1^{\cdot 1}$ | $p_1$ |

**Table 8.2.:** *Expected output from various audio waves from a resynthesized dataset. The superscripts represent the index of the underlying interpretation, while the subscripts refer to the acoustic context.*

We assume that proving such a statement in a logical and deterministic manner is almost impossible. Instead, we seek for a methodology to empirically prove that the previous framework provides significant improvements in realistic scenarios. In other words, we want to show that Equation (8.7) holds and that, consequently, $R \gg 0$ in the average case. To this end, we also need to build $\iota$ and $\alpha$ according to the explained definitions. We now show how such statements could be proved in practice. In the following, elements of sets will be indexed using superscripts such as $a^0, a^1$, except for indices representing contexts that are referenced with subscripts like $a_0, a_1$, as in the previous sections.

The empirical method proposed in this work consists in the estimation of $\tau$, $\iota$, and $\alpha_c$ in an experimental setting using Neural Network models that are universal approximators with sufficient accuracy [337, 338]. However, for training accurate Neural Network models, proper datasets are required. While for the estimation of $\tau$ several datasets exist, the most direct way to approximate $\iota$ and $\alpha_c$ would require the collection of data containing the same interpretation $i$ adapted to different contexts $c_0, c_1, \dots c_K$; such a dataset could be assembled by asking different performers to play the same music pieces in various contexts while keeping constant their interpretation.

To understand how the dataset should be created, let $i^{\cdot 0}$ and $i^{\cdot 1}$ be two different interpretations of the same musical piece and let $c_0$ and $c_1$ be two different acoustic contexts. We can therefore obtain 4 different performances by adapting each interpretation to each context and obtaining $p_0^0 = \alpha_0(i^{\cdot 0}), p_0^1 = \alpha_0(i^{\cdot 1}), p_1^0 = \alpha_1(i^{\cdot 0}), p_1^1 = \alpha_1(i^{\cdot 1})$. Then, each performance will be synthesized – or played – in the relative context, obtaining 4 audio waves $w_0^0 = \sigma_0(p_0^0), w_0^1 = \sigma_0(p_0^1), w_1^0 = \sigma_1(p_1^0)$, $w_1^1 = \sigma_1(p_1^1)$. We can then explore functions that accomplish the given definitions. Table 8.1 shows the expected output of the approximated $\iota$ and $\tau$ functions. By enforcing such outputs while training $\iota$ and $\alpha_c$ and optimizing $\tau$ with the usual context-unaware approach, we are enforcing the framework's hypothesis. Consequently, we should expect to find two functions that fulfill constraint (8.7), proving the validity of the proposed framework.

However, the collection of such a dataset is expensive and, since the proposed approach is still in its infancy, it would be preferable an easier method to perform empirical experiments. As such, we propose an alternative system that is able to address the same questions without requiring the collection of new data. The idea is to leverage the existing datasets for AMT by resynthesizing the various performances in multiple contexts. In such a case, starting from two different performances in two different unknown contexts $p_{c'}^0$ and $p_{c''}^1$, we obtain 4 different audio waves again, namely $w_0^0 = \sigma_0(p_{c'}^0)$, $w_0^1 = \sigma_0(p_{c''}^1)$, $w_1^0 = \sigma_1(p_{c'}^0)$, $w_1^1 = \sigma_1(p_{c''}^1)$. When synthesizing in a certain context $c$, we should think the argument of the $\sigma_c$ function as generated by the $\alpha_c$ function. For instance, the interpretation underlying $w_0^0$ and $w_1^0$ will be so that $\alpha_0(i_0^0) = \alpha_1(i_1^0) = p_{c'}^0$. Assuming that contexts $c_0$ and $c_1$ are different enough, $\alpha_0 \not\approx \alpha_1$ and, consequently, $i_0^0 \neq i_1^0$. To put it simply, all obtained audio waves have a different underlying interpretation. Table 8.2 shows the expected output of the approximated $\iota$ and $\tau$ functions on the resynthesized data. As such, the training strategy for $\iota$ and $\alpha$ must reflect these new expectations to accomplish the framework definitions. Showing that $\hat{\iota}$ and $\hat{\alpha}$ improve the model performance is equivalent to show that $R > 0$, which in turns proves Equation (8.7).

## 8.3. Experiments

### 8.3.1. Computing the Reward, in practice

We designed an ad-hoc piano AMT model for estimating the performance parameters. Since our aim is to understand whether the performance transcription error can be reduced by considering the acoustic context, we focus on performance parameter estimation in a controlled setting were note timings are known. In this Chapter, we limit our analysis to velocity estimation, but the proposed method can be extended to the estimation of new parameters – see Section 8.5.

To realize an extensible AMT method, we use the perfectly aligned MIDI files recorded by the Disklavier and available in the Maestro dataset to inform the transcription process. In a real-case scenario, a precise alignment of a score can be obtained using the Audio-to-score method presented in Chapter 6. We applied Non-negative Matrix Factorization (NMF) [339] to perform a source-separation of each single key of the piano and then analyzed the spectral representation of each source-separated note. Namely, we compute the MFCC features characterizing timbral characteristics that are connected to the note velocity due to the piano acoustics [197] and that are independent of the non-linear amplitude distortions of the microphones [292, 322]. Then, we employ a Convolutional Neural Network (CNN) model to infer the velocity of each note.

The CNN model is split in two parts as follows:

(i) the first part, the "encoder", serves to approximate the $\iota$ function and takes as input the note-separated spectrogram with size $13 \times 30$ and one channel;

(ii) the second part, the "performer", approximates the $\alpha_c$ function by taking the latent output of $\iota$ and computing the estimated velocity in the source acoustic context.

Since we want to show that $R > 0$ regardless the approximation error, we repeat the experiments with various types of function. More precisely, we define a set of hyper-parameters that determine the shape of the neural network model and then perform a grid-search to explore how the reward changes when different model structures are used for the estimation.

We also considered various methods to enforce the rules defined in Table 8.2 in the $\iota$ and $\alpha_c$ functions. Specifically, we considered the following strategies:

(i) the performer can be the same for all contexts, i.e. context-unaware, or it can be context-specific, meaning that in the model there is one performer for each context;

(ii) the latent space found by $\iota$ should be context-aware, meaning that it should still include the information required for identifying the input context; as such, it is interesting to understand if enforcing this property helps the learning process.

The implementation of the first variable consists in building one performer per *each* context or one performer for *all* contexts. As regards to the second variable, we add an additional branch to the model to classify the input context based on the $\iota$ output; this branch works as a learnable loss function that separates the latent space according to the input context.

We compared the 4 different strategies resulting by the combinations of the above 2 boolean variables. We will name them with two boolean values, namely:

- **False-False**: one single performer and no context classification – this case corresponds to the estimation of $\tau$;
- **True-False**: one performer per each context without using a context classifier;
- **False-True**: one single performer with context classifier ;
- **True-True**: one performer per each context and context classification.

It should be noted that setting both the two variables to False, makes the model to be fully context-unaware. Such a strategy represents the estimation of $\tau$. The four strategies are depicted in Figure 8.4.

All four strategies were tested for each point in the hyper-parameter space, resulting in a highly computationally demanding experiment. We thoroughly tested 36 different model shapes, each with 4 different training strategies, summing up to 144 trained models.

## 8.3.2. Dataset

In this section, we describe the data creation process facilitating the proposed experiment.

### Resynthesis

Following the method described in Section 8.2.5, we designed an experiment based on the resynthesis of existing datasets. To this end, we developed pycarla[2], a Python module that leverages the excellent Carla[3] plugin host to synthesize MIDI messages both in real-time and offline using the major audio plugin formats – such as VST, AU, LV2, LADSPA, DSSI, SF2, SFZ. Even if pycarla has been designed to support long-running processes, such as the resynthesis of large music datasets, it is based on a complex pipeline, starting with the Jack[4] audio server, passing through Carla and ending with selected audio plugins. In this pipeline, various bugs, crashes, and errors may occur negatively affecting the synthesized audio. For this reason, various checks on the resynthesized audio have been carried out to identify potential synthesis errors.

We used 6 different presets for the physically modeled virtual piano by Pianoteq, kindly provided for research purposes by Modartt[5]. Table 8.3 includes a summary of the main characteristics

[2]https://web.archive.org/web/20211213132835/https://pypi.org/project/pycarla/
[3]https://web.archive.org/web/20211205195725/https://kx.studio/Applications:Carla
[4]https://web.archive.org/web/20211211061956/https://jackaudio.org/
[5]https://web.archive.org/web/20211112075858/https://www.modartt.com/

| ID | Velocity Map | Reverb | Instrument |
|---|---|---|---|
| 0 | Linear | Jazz Studio | Steinway B Prelude |
| 1 | Logarithmic | Jazz Studio | Steinway B Prelude |
| 2 | Logarithmic | Cathedral | Steinway B Prelude |
| 3 | Linear | Jazz Studio | Grotrian Cabaret |
| 4 | Logarithmic | Jazz Studio | Grotrian Cabaret |
| 5 | Logarithmic | Cathedral | Grotrian Cabaret |

**Table 8.3.:** *Summary of the main characteristics of the 6 presets used for resynthesizing the Maestro [191] dataset.*

of each preset.

### Clustering

The source dataset was Maestro [191] and was used as provided by ASMD library – Chapter 4 [4]. The Maestro dataset was selected using the ASMD Python API; then, *train*, *validation*, and *test* splits were partitioned in 6 different subsets, each associated to one of the presets in Table 8.3, for a total of $6 \times 3 = 18$ subsets. From 18 subsets, 6 sets were generated by unifying subsets associated to the same preset, so that each set was still split in *train*, *validation*, and *test* sets. Each generated set was resynthesized and saved to a new ASMD definition file.

The 6 new subsets in each split were chosen as follows. First, one split at a time among the already defined *train*, *validation*, and *test* was selected. Supposing that the chosen split has cardinality $K$, $C$ clusters were created with $C = \lfloor K/6 + 1 \rfloor$ and a target cardinality $t = 6$ was set. Then, a redistribution policy is applied to the points of the clusters: for each cluster with cardinality $< t$ – a "poor" cluster – we look for the point nearest to that cluster's centroid and belonging to a cluster with cardinality $> t$ – a rich cluster – and move that point to the poorer cluster. The redistribution stops when all clusters have cardinality $\geq t$. Since the redistribution algorithm moves points from rich clusters to the poor ones, we named it "Robin Hood" redistribution policy. Having obtained $C$ clusters each with 6 samples, we partitioned the chosen split in 6 subsets as follows:

(i) we randomized the order of subsets and clusters
(ii) we selected one point from each cluster using a random uniform distribution and assigned it to the one of the 6 subsets
(iii) we did the same for the other 5 subsets
(iv) we restarted from point (ii) until every point is assigned to a cluster.

Clustering was performed as follows: first, a set of features was extracted from the recorded MIDI of each music piece available in Maestro in order to describe the available ground-truth performance parameters. Namely, the velocity and pedaling values were extracted for a total of 4 raw features – i.e. 1 velocity + 3 pedaling controls. For each raw feature, 3 high-level features were computed by fitting a generalized normal distribution ($\mathcal{N}$) and obtaining its parameters ($\alpha$, $\beta$, and $\mu$). Another high-level feature was computed as the entropy of each raw features ($H$). Finally, for each pedaling MIDI control, the ratio of 0-values and 127-values was added ($Q$); this feature is meaningful because most of the research about piano pedaling transcription considers the pedaling as an on-off switch, even if in real pianos it can be used as a continuous controller. Overall, each Maestro music piece was represented by concatenating the high-level features, including $3 \times 4$ features $N$, 4 features $H$, and $2 \times 3$ features $Q$, totaling 22 features. Then, PCA was performed to identify 11 coefficients while retaining 89%, 93%, and 91% of variance for the train, validation and test splits. Isolation Forest algorithm with 200 estimators and bootstrap were also employed to detect potential outliers, resulting in 81, 39, and 36 removed points from the train, validation, and test splits. Finally $k$-means clustering with kmeans++ initialization routine was applied to find $C$ centroids in the data with the outliers removed; from the found centroids, $C$ clusters were reconstructed from the full dataset (outliers included).

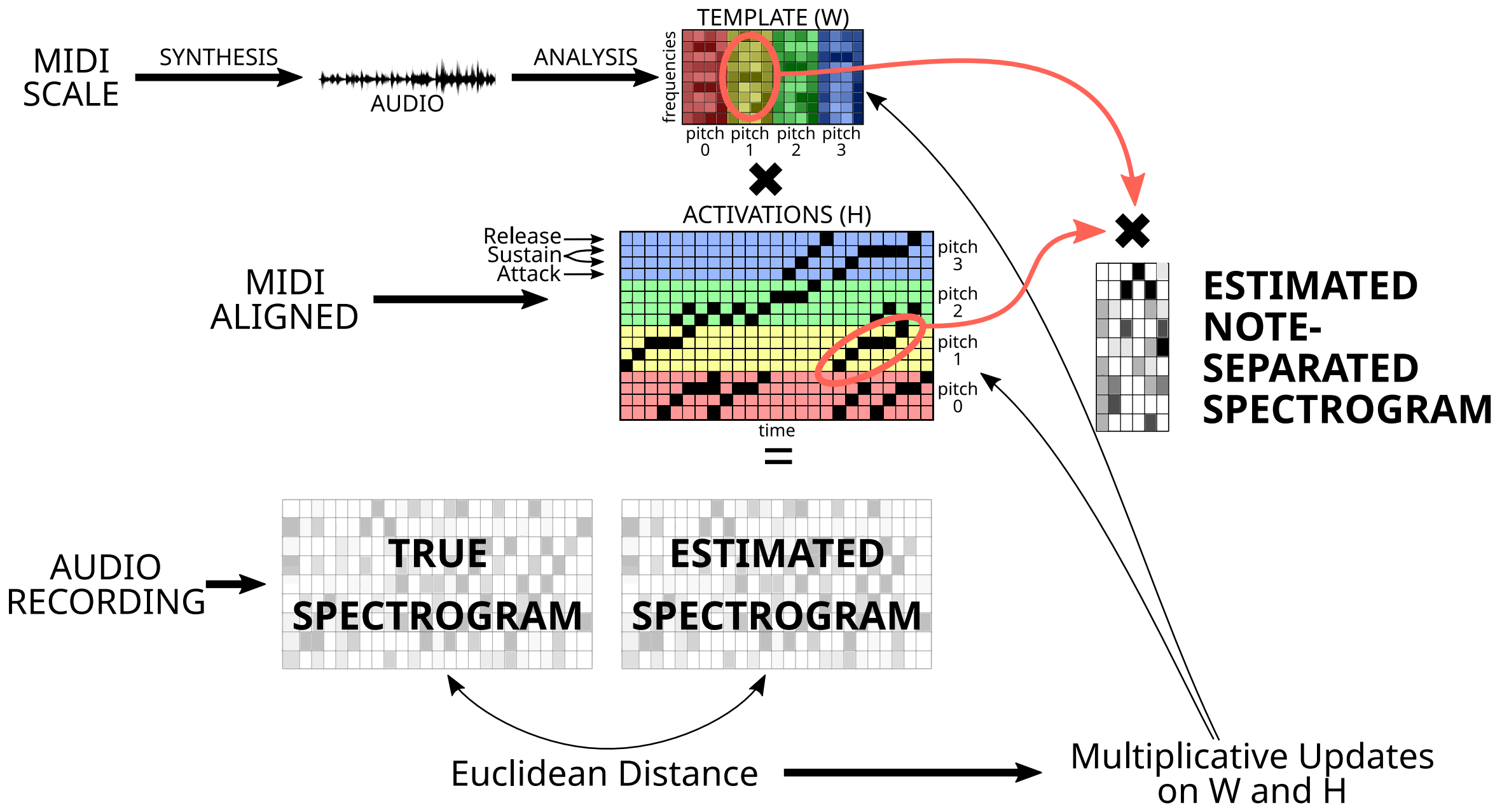

**Figure 8.1.:** *The full NMF workflow. First, the initial template and activation matrices are computed. Then, the Euclidean distance between the estimated and the true spectrograms is used for multiplicatively update both the template and the activation matrix. Finally, only the relevant part of the template and activations are used for estimating the note-separated spectrogram. In this image, 4 columns are used to represent each note. In the real work, we used 30 columns.*

### 8.3.3. Automatic Music Transcription

#### Note-separation

NMF has largely been used for score-informed AMT [6, 292, 295, 340] and our application is mainly based on the existing literature. Using NMF, a target non-negative matrix $S$ can be approximated with the multiplication between a non-negative template matrix $W$ and a non-negative activation matrix $H$. When applied to audio, $S$ is usually a time-frequency representation of the audio recording, $W$ is the template matrix representing each audio source, and $H$ represents the instants in which each source is active. As such, the rows of $W$ represent frequency bins, the columns of $W$ and the rows of $H$ refer to sound sources, and the columns of $H$ are time frames.

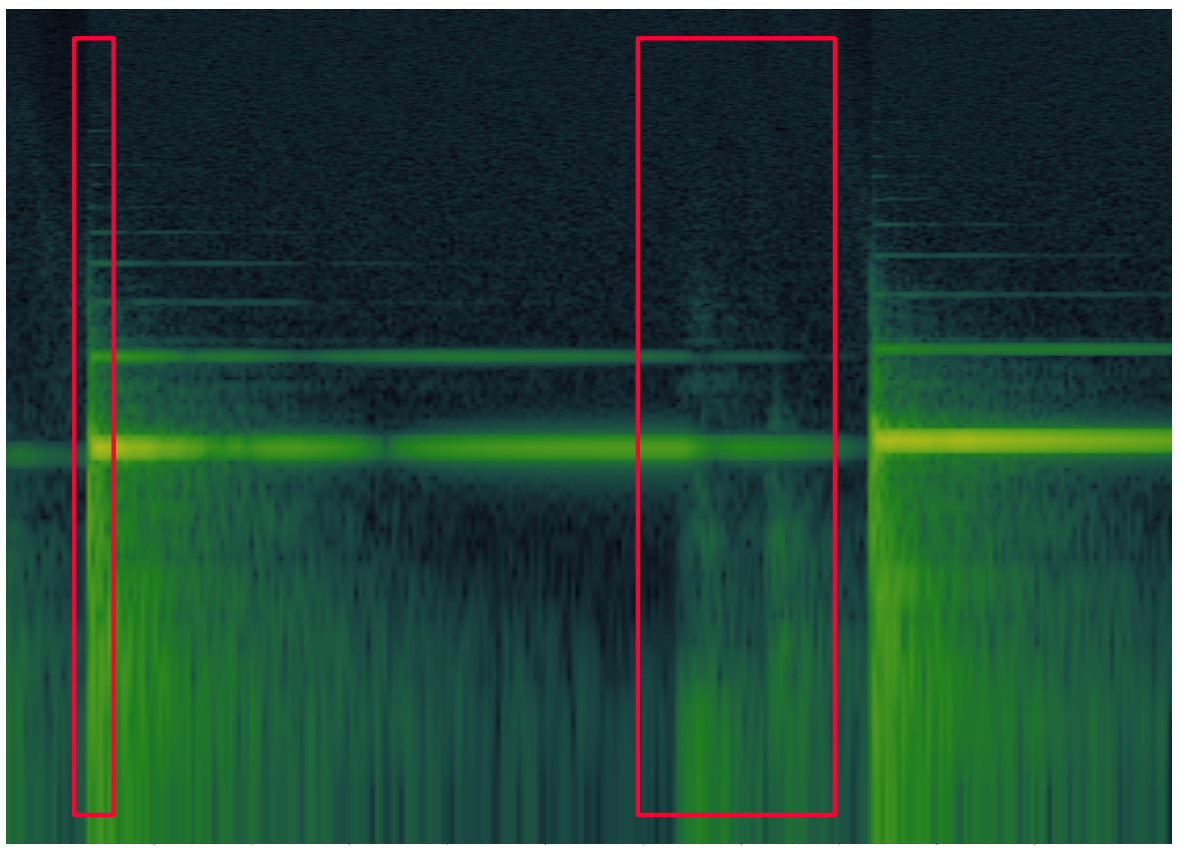

**Figure 8.2.:** *Log-Spectrogram of a piano note. The two red rectangles highlight the impulse connected with the attack of the note, and the impulses connected with the hammer release of the note. The image was obtained using Sonic Visualizer and the audio scales synthesized with Pianoteq "Steinway B Prelude" used for the computation of the initial NMF template. The Sonic Visualizer project used for extracting this image is available in the online repository. The spectrogram has been computed using windows of 1024 frames and 50% of overlap; the intensity scale is in dB, while the frequency scale is logarithmic; no normalization was applied. The pitch is 73, velocity is 22, and duration is 1.5 seconds.*

The $W$ and $H$ matrices are first initialized with some initial values and then updated until some loss function comparing $S$ and $W \times H$ is minimized.

The proposed method for NMF is shown in figure 8.1. Similarly to previous works [6, 292] we split the temporal evolution of each piano key in multiple sectors, namely the attack, the release, and the sustain. Each sector can be represented by one or more columns in $W$ (and correspondingly rows in $H$). We used the following subdivision:

- 1 column for the attack part, representing the first frame of the note envelope ($\sim$ 23 ms)
- 14 columns for the sustain section, representing 2 frames each; if the note sustain part lasts more than the 28 frames ($\sim$ 644 ms), the remaining part is modeled with the last column;
- 15 columns for the release part($\sim$ 345 ms), starting from the recorded MIDI offset and representing one audio frame each; the reason is that, as shown in Figure 8.2, after the MIDI offset, there are still about 350 ms before the hammer comes back, producing a characteristic impulsive sound and definitely stopping the sound.

The initial $W$ matrix is constructed by averaging the values obtained from a piano scale synthesized with pycarla and Pianoteq's "Steinway B Prelude" default preset. The scale contained all 88 pitches played with 20 velocity layers, 2 different note duration – 0.1 and 1.5 seconds – and 2 different inter-note silence duration – 0.5 and 1.5 seconds. First, the amplitude spectrogram is computed using the library Essentia [321]. We used 22050Hz sample-rate, frame size of 2048 samples (93 ms), hop-size of 512 samples (23 ms), and Hann window types.

The initial $H$ matrix, instead, is generated from the perfectly aligned MIDI data, by splitting each pitch among the various subdivisions as explained above.

Finally, the NMF optimization is performed using Euclidean distance and multiplicative updates [339]. Our NMF optimization routine is adjusted as follows: a first step (A) is performed separately on 5 windows with no overlap and with $H$ fixed; splitting the input activation matrix and the target spectrogram in windows is beneficial because the multiplicative contribute for $W$ depends on $SH^T$ [339], which can be repeatedly adjusted on each window, increasing both accuracy and generalization of the $W$ estimation. Having slightly optimized $W$, we applied the second step (B), consisting in 5 iteration of the usual NMF. Before both step A and B, $W$ and $H$ were normalized to the respective maximum value so that their values lay in $[0, 1]$.

Once the NMF algorithm is finished, we use the original perfectly aligned activation matrix to select the region of a note in $H$ and $W$ to obtain its approximated spectrogram separated from the rest of the recording. We consider the first 30 frames (690 ms) of each note, padding with 0 if the note is shorter. We finally compute the first 13 MFCC features in each column of the spectrogram using Essentia.

### Neural Network Models

For every function estimation, we use Convolutional Neural Networks (CNN) with skip connections similarly to ResNet [341]. A schematic representation of the proposed model architecture is shown in Figure 8.3.

In ResNet, a building block is defined so that the output can have the same size as the input ("not reducer" block) or can be reduced ("reducer" block); in both cases, the output of each block is summed to the input to prevent the vanishing gradients phenomenon and other degradation problems connected with the increase of the network complexity [341]. Since they can maintain the output size equal to the input, a virtually infinite number of blocks can be put one after the other, and multiple stacks of blocks can be concatenated to create arbitrarily large and complex networks without depending on the input size.

In the proposed model, each block consists of the following elements:

- a grouped convolutional layer with kernel size $K$; if the block is a not-reducer, a padding is used;
- a batch-normalization layer;
- a ReLU non-linear activation;
- a non-grouped convolutional layer with kernel size 1 – corresponding to a linear combination of each data entry across channels;
- another batch-normalization layer;
- a final ReLU activation.

Furthermore, each block sums its output to the input processed with a grouped convolutional layer having kernel size 1 if the block is a not reducer and $K$ otherwise. Figure 8.3 better depicts the building of a single block.

Multiple blocks can be put one after the other forming a stack. In each stack, the first block changes the number of channels, while the rest keep it constant. Moreover, all blocks in a stack are not reducers except the last one. As such, each stack can increase or decrease the number of channels in the data representation and at the same time it decreases the size of the data with only one convolution. Figure 8.3 represents a stack.

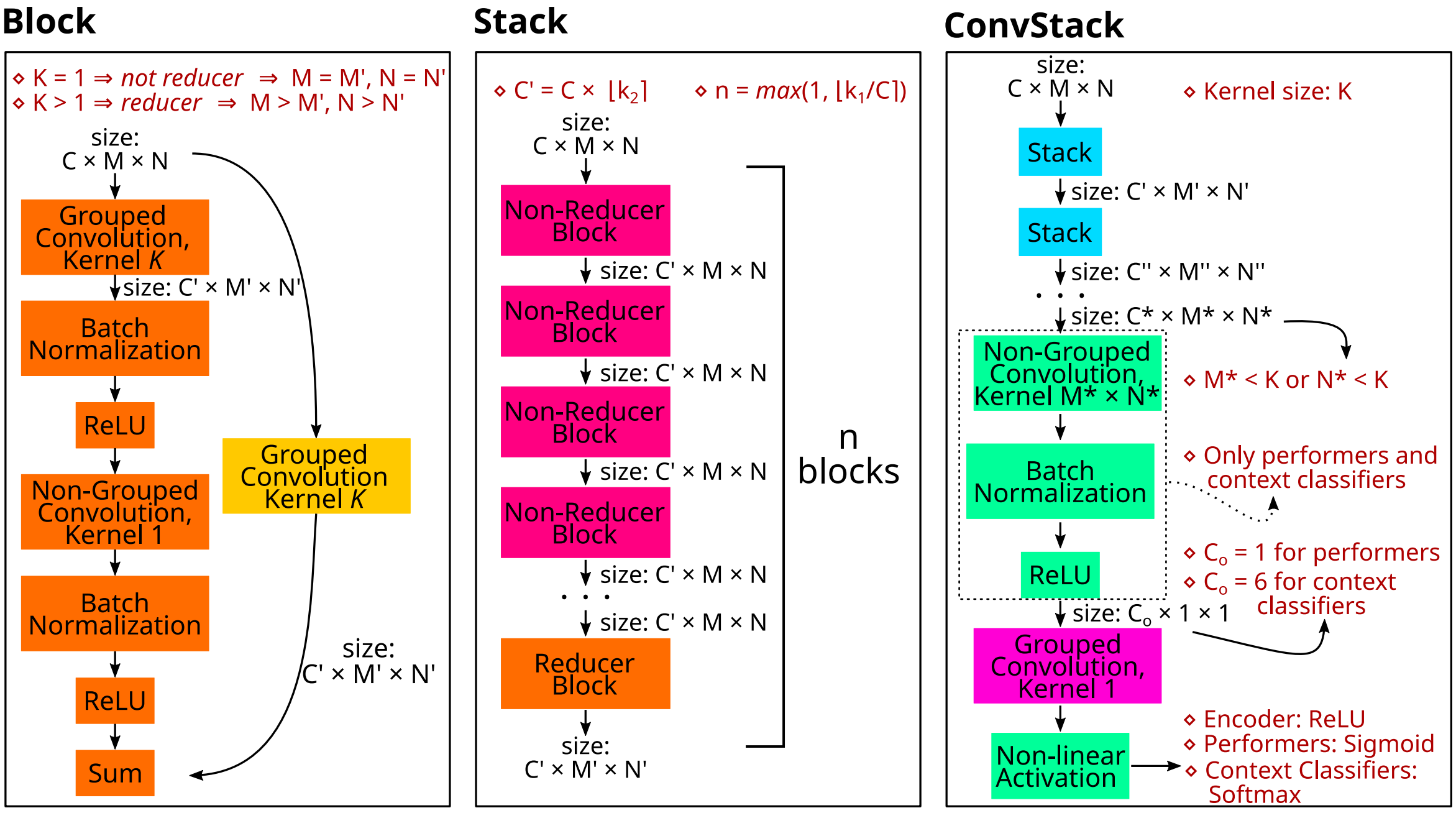

**Figure 8.3.:** *Architecture of the Residual blocks, stacks and convolutional networks used in this work. The final model architectures are represented in Figure 8.4.*

In order to control the complexity of the network, we designed a family of CNNs that vary the ratio between the number of blocks and the number of channels in each stack based on two parameters $k_1$ and $k_2$. By setting $k_1$ and $k_2$, one may find Residual CNNs that approximate various types of functions. Specifically, the algorithm used for shaping one CNN operates as follows. In each stack, the number of blocks is defined as $max(1, \lfloor \frac{2^{k_1}}{l} \rceil)$, where the $\lfloor \cdot \rceil$ represents the rounding operation, $l$ is the number of input channels and $k_1$ is an hyper-parameter. Similarly, the number of output channels in each stack was computed as $l \times \lfloor k_2 \rceil$. For instance, the first stack will always have $\lfloor 2^{k_1} \rceil$ blocks and $\lfloor k_2 \rceil$ output channels, because the input channel size is 1; the $i$-th stack, instead, will have $\lfloor \frac{2^{k_1}}{\lfloor k_2 \rceil^{(i-1)}} \rceil$ blocks and $\lfloor k_2 \rceil^i$ channels. In our experiments, we manually found that $k_1 = 4$ comprises an effective parameter and observed the way the models perform when $k_2$ changes. Following this algorithm, multiple stacks were concatenated until the output size has at least one dimension $< k_0$, where $k_0$ is the kernel size, which is fixed across the stacks.

We use multiple of such CNNs in each model to estimate the $\iota$ and $\alpha_c$ functions and an additional one for the context classifier. After the stacks, a further convolutional layer followed by batch normalization and ReLU is added aiming at reducing the data size to 1 and at compressing all existing information into the channel dimension; in the performer and context classifier, this last block also takes care of reducing the number of channels to the expected output dimension, i.e. 1 for the velocity and 6 for the context classifier. Finally, we apply a linear transformation using a grouped convolution and activation block with kernel size 1; the last activation is a ReLU in the encoder, a Sigmoid in the performers, and a SoftMax in the context classifier.

The considered hyper-parameters were 4:

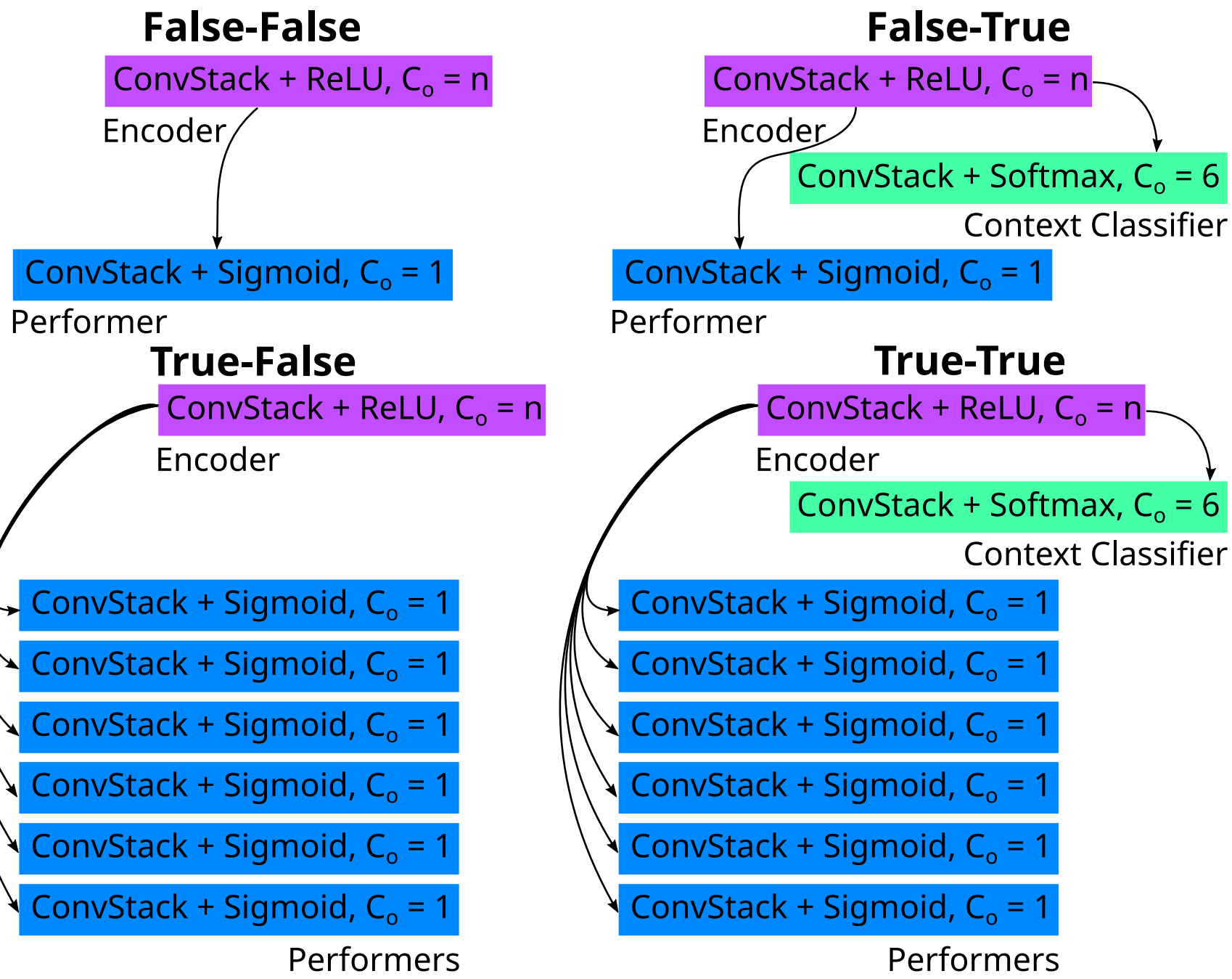

**Figure 8.4.:** *The 4 strategies tested in this work. Each strategy corresponds to a different model architecture during training. See Figure 8.3 for the representation of a* ConvStack*. $C_o$ is the number of outputs of each* ConvStack*.*

(i) the kernel size in the encoder CNN (considered values: 3, 5)
(ii) the kernel size in the performer CNNs (considered values: 3, 5)
(iii) the $k_2$ parameter in the encoder CNN (considered values: 1, 2, 3)
(iv) the $k_2$ parameter in the performer CNNs (considered values: 1, 2, 4)

The context classifier branch is built with the same performer kernel; however, due to the higher computational complexity needed for classifying 6 labels, $\{k_1, \ k_2\}$ were multiplied by 1.25 – i.e. $k_1 = 5$ and $\lfloor k_2 \rceil \in \{1, 3, 5\}$.

## Training

The training datasets include millions of music notes. To make the problem computationally accessible, we use only 0.1% of the available data with a batch size of 10, resulting in 703 batches (7030 notes). Subsampling was performed with a uniform distribution and was repeated on all the 6 contexts and splits (train, validation, and test sets). Overall, the training set is made of 566 batches, the validation counted 63 batches, and the test set is composed of 74 batches.

Training is performed using Adadelta [342] optimizer with initial learning rate set according to an existing algorithm designed to find its optimal value based on repeated small experiments with increasing learning rates [343]. When the algorithm fails, the initial learning rate is automatically set to 1e-5. The loss function for the performers is the L1 error, while for the context classifier we use the Cross-Entropy loss. When $\iota$ is trained with the context classifier, we treat the problem from a multi-task perspective. For this reason, we sum the two losses and use the recently proposed RotoGrad algorithm [344] to stabilize the gradients. Moreover, when using multiple performers, to speed up the training process, we load data so that each batch contains notes related to only one context at a time and we cycle across context so that all of them are equally represented. However, this strategy leads to unstable losses both in training and in validation, making it hard to understand when the model is actually overfitting. For this reason, we also imposed an early-stop procedure with a patience of 20 by observing the Exponential Moving Average of the validation loss on a window of 15 epochs.

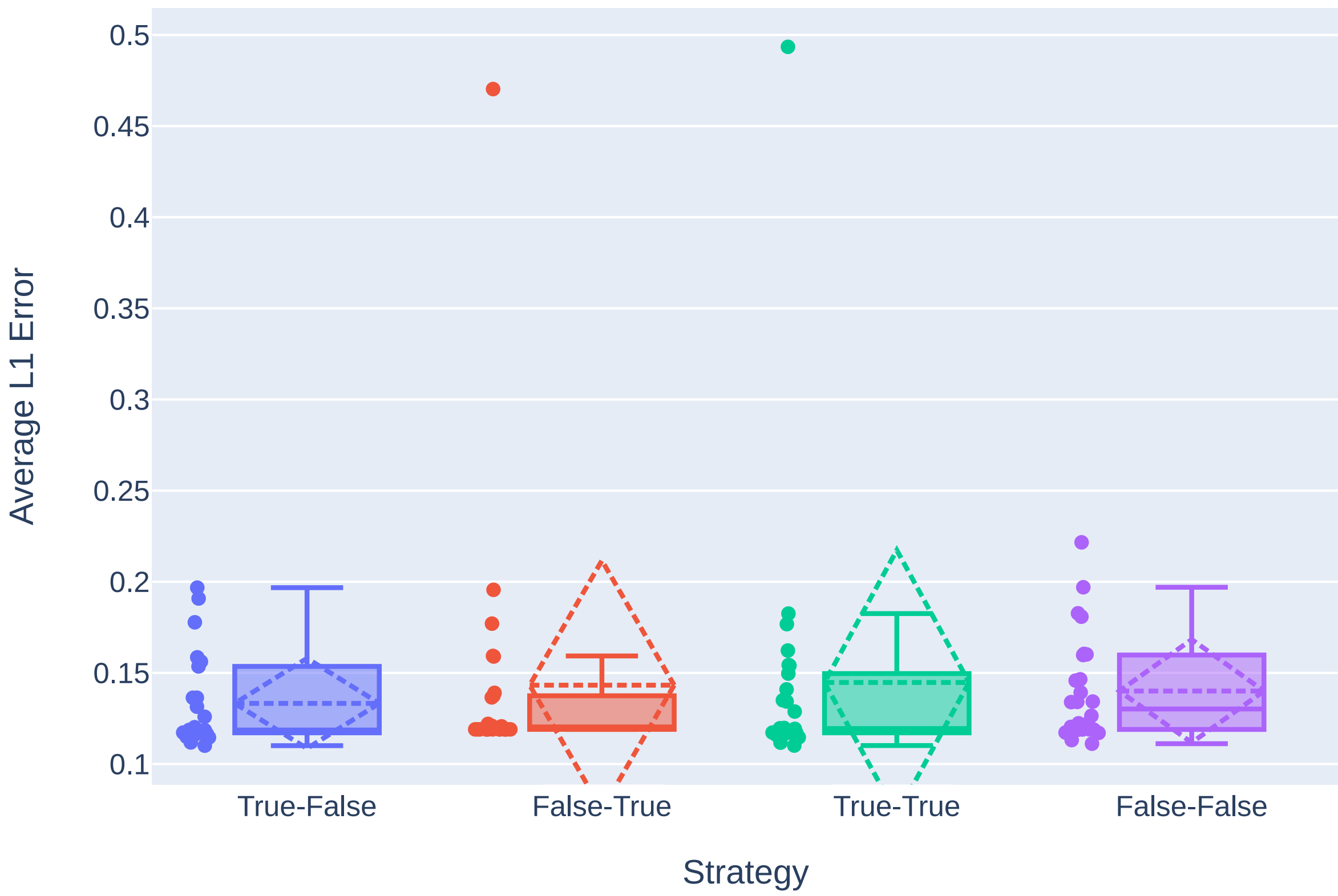

**Figure 8.5.:** *Comparison between the 4 proposed strategies.* ***False-False*** *strategy corresponds to the traditional context-unaware transcription function. The dotted rhombus represents average and standard deviation. The continuous line represents the median and box-plot.*

## 8.4. Results

The results we obtained are shown in Figure 8.5 and 8.6. We computed the average L1 error for velocity estimation in each tested hyper-parameter set, discarding those configurations that generated models exceeding GPU or even CPU RAM or that returned invalid losses. Overall, we considered 26 hyper-parameter sets corresponding to 104 runs. Moreover, to reduce the computational burden, we stopped each training at the 40th epoch, if the training procedure was not terminated by the early-stopping criterion.

Note that since L1 error is a metric, it satisfies the triangle inequality property, and as such it can be used as the $\varepsilon$ function in Equations (8.7) and (8.8). Consequently, showing that False-False strategy has a larger L1 error than the other strategies, confirms that $R > 0$.

For evaluating the statistical significance of the results in Figure 8.5, we applied the Shapiro-Wilk normality test to each strategy distribution and then Kruskal-Wallis and Wilcoxon signed-rank test for post-hoc analysis. We found that all analyzed distributions rejected the null hypothesis of normality tests with $p < 4\text{e-}3$, meaning that the distributions are not normal. We found no significant difference according to the omnibus Kruskal-Wallis test ($p = 1.74\text{e-}1$). However, we also computed the Wilcoxon p-values using the Bonferroni-Holm correction and found a statistically significant difference with confidence of 95% only between True-False and False-False, True-True and False-False. Note that the $p > 0.05$ found with the Kruskal-Wallis test is coherent with the pairwise significance found using the corrected Wilcoxon test [345, 346]. Given the statistical analysis, we argue that $R > 0$ for the True-True and True-False strategies. To further assess such conclusions, we also computed the number of hyper-parameter sets won by each strategy. Table 8.4 shows this analysis and highlights how in only 1 configuration the best training strategy was False-False.

| | False-False | True-False | False-True | True-True | All |
|---|---|---|---|---|---|
| **False-False** | - | 2 | 12 | 3 | 1 |
| **True-False** | 24 | - | 19 | 11 | 11 |
| **False-True** | 14 | 7 | - | 6 | 5 |
| **True-True** | 23 | 12 | 20 | - | 12 |

**Table 8.4.:** *Win analysis. The table must be ridden as: "strategy at row* x *is better than strategy at column* y *in* n *hyper-parameter configurations"*

However, no optimal strategy was found. Indeed, considering True-False, False-True, and True-True, there is no agreement about which one is the most effective method. To obtain a deeper understanding of the problem, we tried to check what would happen in case an oracle could indicate the optimal strategy depending on the model shape. Results were highly statistically significant ($p = 6\text{e-}5$) and showed far improved results when one of the proposed context-aware strategy was used – see Figure 8.6. This result highlights how, in theory, $R$ can be definitely larger than 0.

## 8.5. Conclusions

This chapter reports an attempt to deepen the understanding of acoustic factors on music performance analysis. The problem was formalized from a mathematical perspective and analyzed with empirical tools. However, further experiments are required to assess the mathematical formalization and to understand how the acoustic context can be successfully exploited to improve AMT models. For this reason, the proposed framework can be easily extended to include other parameters and to consider the performer-specific models of the context-adaptation functions [181], as well as other instruments. According to the discussion in Sections 8.1 and 2.4.1, future works could estimate non-MIDI parameters that are relevant for the timbre realization of pianists [197]. Another attractive addition would be the note offset precise inference based on the hammer second and third impulsive sound – see Figure 8.2; given the low accuracy of the note offset inference in state-of-the-art AMT models, such an addition could be useful for precisely defining the performer interpretation. A third addition could be performer-specific adaptation functions, as suggested in previous experiments [181]. Finally, an important parameter that we plan to focus

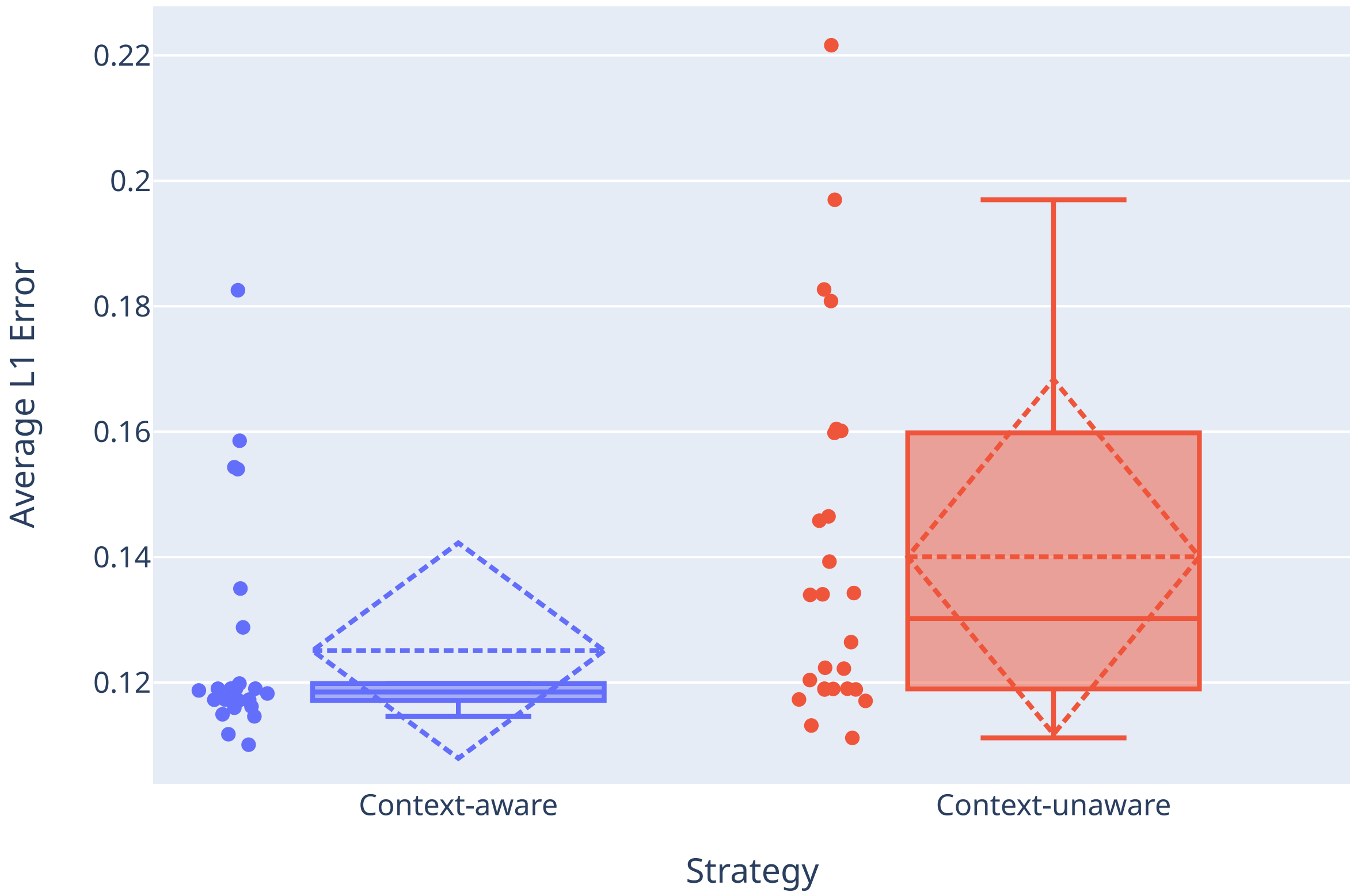

**Figure 8.6.:** *Comparison between the best context-aware strategy and the traditional context-unaware transcription in each hyper-parameter point. The dotted rhombus represents average and standard deviation. The continuous line represents the median and box-plot.*

in next works is the pedaling level estimation.

In this work, we extensively evaluated 4 different strategies for velocity estimation of single notes. We demonstrated that considering context-aware strategies consistently improves model performance. However, no context-aware strategy was found to outperform the rest; it was shown that they complement each other.

Finally, the present work is an initial attempt to tackle with a rigorous approach the interpretation analysis within the AMT context. Various applications may benefit from such studies, beginning with the main application under study in this Thesis – i.e. Automatic Music Resynthesis.

# Conclusions

In this Doctoral Thesis, a deep analysis about Multimodal Music Information Processing (MIP), with particular reference to music performance studies, has been presented. Once defined the main problem, consisting in the resynthesis of music performances by using Automatic Music Transcription (AMT) methods, the main gaps in the literature have been identified and filled – see Part I. The analysis of the existing scientific literature on the topic has revealed that the multimodal approach may be a promising method for facing various currently unsolved issues and that Multimodal MIP is worth of attention. For the problem under discussion, two main issues have been identified and addressed, namely data availability and feature fusion.

The second step has been to focus on the existing literature and approaches for the faithful resynthesis of music and to design hypothetical solutions to the problem. The audio restoration literature is dominated by two general approaches: the first aims at reconstructing the sound as it was originally "reproduced and heard by the people of the era", while the second and most ambitious one aims at reconstructing "the original sound source exactly as received by the transducing equipment (microphone, acoustic horn, etc.)" [297, 298]. A novel resynthesis approach has been proposed in this Thesis, so that the aim becomes the recovering of the artistic intention survived until today and perceivable by the listener. These topics were approached in Chapter 2, where the distinction between "performance" and "interpretation" is first introduced. In this Thesis, in line with various existing studies, the performance is the set of physical events that take place during a music execution, while the interpretation is the performers' interior and ideal representation of the performance. The two differ because of the adaptations that performer apply to their interior representation based on the environment acoustics. Pros and cons of automated music resynthesis for music restoration and for music production are discussed and possible solutions are proposed.

Part II faces the first issue identified in the MIP review, namely the data availability.

First, the IEEE 1599 standard has been analyzed and proposed as a representation format for archiving multimodal documents – Chapter 3. The main strength of the IEEE 1599 is the capability of referring to geographically distributed documents and to describe the multimodal information consisting in the synchronization – temporal and/or spatial – between remote documents; this feature is of primary importance in the contemporary society, where cloud infrastructures could theoretically allow the worldwide distribution of digital information but intellectual copyrights hinder the efforts of sharing knowledge, even for scientific purpose. The IEEE 1599 standard could be an answer to such needing, especially in the proposed configuration, in which a *Central Node* coordinates the requests from user clients by redirecting them to *Peripheral Nodes*, that provide the multimodal information, and *File Repositories*, containing the digital objects that are subject to copyrights.

The second work dealing with multimodal data distribution is the proposal of *ASMD*, a Python framework for compiling, distributing and creating multimodal music datasets – Chapter 4. The framework provides predefined conversion functions from the most common representation format, while still allowing the user to write its own custom functions. Moreover, the framework defines a JSON-based representation for whole datasets, including information for downloading and converting remote contents. *ASMD* defines an extensible JSON annotation format usable for machine-learning tasks; for now, the included annotations are limited to audio and MIDI-like data, with the main focus of score-informed music processing, but it will be easy to extend the number of modalities in future. An important concept in *ASMD* is the attempt of easing the reproducibility of research by substituting manual and possibly erroneous annotations with machine-generated ground-truths where possible. *ASMD* comes with powerful API for retrieving, querying and manipulating data; using the API, the user is able to define its own datasets from the intersection, union and filtering of the whole amount of data. Moreover, a user can use the framework on custom data distributions, different from the official ones; this is a fundamental features in research contexts, where new datasets are continuously created.

The literature analysis conducted in Part I has also highlighted the feature fusion as a main problem in multimodal information processing. For this reason, Part III tries to improve the current state-of-art methods for multimodal music processing.

Considering the general focus on Automatic Music Transcription (AMT) of this Thesis, the first work shows a source-separation method for the music symbolic level. The proposed model, described in Chapter 5, is able to separate the melody and accompaniment notes in music from the common practice era, by using an encoder-decoder architecture, the piano-rolls as guides for the inference and a graph-based search for tuning the predictions. Moreover, a newer method for inspecting "same-size" networks – i.e. networks whose output has the same size as the input – has been proposed. Source-separation is a fundamental task in audio analysis, especially in piano music, where each piano key can be considered a different sound source. Indeed, the model may be exploited for multimodal AMT to improve the accuracy of the most salient notes and for further music performance studies.

For a successful feature fusion, however, a good Audio-to-Score alignment (A2SA) should be available. In our perceptual tests – Chapter 7 – the state-of-art A2SA for piano music has been found not enough effective. Moreover, in the perspective of the main problem of the Thesis, the endemic uncertainty about the correctness of the AMT output should be faced with a musicologically curated music score, that should work as main guide for the audio analysis. Consequently, the notes in the music score must be completely matched with the greatest accuracy possible. For this reason, a novel state-of-art method for A2SA has been proposed in Chapter 6. The proposed method exploits AMT for extracting features from audio; it then matches the notes from AMT with the notes in the score and successively performs a new alignment for aligning the not matched notes. The evaluation shows how the proposed method overcomes the state-of-art method for piano music, while for non-piano music, the low generalization ability of AMT in non-piano music makes the proposed approach unusable.

As already described, the main problem faced by the Thesis is to resynthesize a music performance. From the analysis of Chapter 2, the most promising approach has been found to be the resynthesis of the output of an AMT model. Specifically, the aim is to transfer a music performance in a context different from the one where the recording was taken, while maintaining the original artistic intention. In the last section of Thesis, Part IV, this problem has been faced more directly.

Chapter 7 attempts to perceptually evaluate the resynthesis of AMT output in different contexts and to answer if such "artistic intention" is kept in the resynthesized version of the audio. Interpretation and performance concepts were perceptually evaluated with a listening test in the first attempt to understand the influence of the acoustic context on the music performance with the perspective of AMT field. The main outcomes of the perceptual tests are that MIDI format alone is not able to completely grasp the artistic content of a music performance and that such performance adaptations are meaningful for conveying the "interpretation".

The concepts of interpretation and performance have been formalized in Chapter 8 with a mathematical framework able to describe the signal transformation during transcription and synthesis. The formalization has been used to design an empirical experiment that brings the acoustic context influence on the music performance into the AMT field for the first time. With this experiment, it is demonstrated that the consideration of the acoustic context influence in AMT models is beneficial.

In conclusion, this Thesis has proposed a novel approach for the resynthesis of music recordings and has carried out an exploratory study about its feasibility. While analyzing the problem, a number of contributions were raised, as summarized in Preface. The most important achievement is the theorization, formalization, and analysis of the concept of "interpretation", hence the title of the Thesis.

# A
# Supplementary Materials

Here are listed additional materials that are relevant to the single works included in this Thesis.

- **Chapter 1**
  - online spreadsheet categorizing all the works dealing with Multimodal MIP was created when the related publication [1] was submitted; therefore, it is not completely up-to-date: https://tinyurl.com/multimodalmir
- **Chapter 4**
  - the whole code is freely available online: https://github.com/LIMUNIMI/ASMD/
  - documentation is available at https://asmd.readthedocs.org/
- **Chapter 5**
  - a companion website was created containing many additional information, such as plots and detailed textual results; from the website it is also possible to view and download the source code: https://limunimi.github.io/Symbolic-Melody-Identification/
- **Chapter 6**
  - the full source code is available in git repository: https://github.com/LIMUNIMI/MMSP2021-Audio2ScoreAlignment
- **Chapter 7**
  - the full source code, including the introductory slides and the source files of the web interface, are available at https://github.com/LIMUNIMI/PerceptualEvaluation
  - an external mega folder is used to store datasets and audio: https://tinyurl.com/perceptualeval
  - supplementary information about the score-informed method and the statistical analysis of the results is available at the publisher page – in open access – and on ArXiv:
    (i) https://doi.org/10.1007/s11042-022-12476-0
    (ii) https://arxiv.org/e-print/2202.12257

- code related to the $p$-dispersion problem is available in a separate git repository: https://framagit.org/sapo/selection_test

- **Chapter 8**

  - the code for reproducing the experiments, including the dataset resynthesis, is available at https://github.com/LIMUNIMI/ContextAwareAMT
  - pyCarla, a python module developed for resynthesis using audio plugins, is available separately:

    * git repository: https://github.com/00sapo/pycarla
    * documentation: https://pycarla.readthedocs.io

- A generic website whose aim is to collect my works in the "Music Interpretation Analysis" (MIA) field is available at https://limunimi.github.io/MIA/

# B
# List of Figures

# C
# List of Tables

# D
# List of Acronyms

**A2SA** Audio-to-Score Alignment. 85

**AMR** Automatic Music Resynthesis. 23, 43

**AMT** Automatic Music Transcription. 2, 3, 34, 69, 85, 99, 119

**ASMD** Audio and Score Meta Dataset [4]. 2, 53, 90

**BYTEDANCE** Automatic Piano Music Transcription method [194]. 90

**DDSP** Differentiable Digital Signal Processing [164]. 32

**DSP** Digital Signal Processing. 23

**DTW** Dynamic Time Warping [81]. 19, 85

**EIFE** An audio-to-score alignment method based on AMT and note matching – see Chapter 6. 90

**EITA** A note matching method (EITA) [234]. 59, 89

**MIA** “Music Interpretation Analysis”. 1, 2, 3, 69, 119, 142

**MIP** Music Information Processing. 1, 2, 7, 23, 64, 69, 85, 119, 141

**MIR** Music Information Retrieval. 2, 85, 99

**MPA** Music Performance Analysis. 69, 119

**NMF** Non-negative Matrix Factorization. 19, 86

**O&F** Onsets and Frame, an AMT model [280]. 111

**OMNIZART** Automatic Multi-instrument Music Transcription method [285]. 90

**PEAMT** A perceptual-based evaluation measure for AMT systems. 117

**SEBA** An audio-to-score alignment method based on synthesis and DTW [347]. 88

**SI** A Score-Informed AMT method. 111, 112

**TAFE** An audio-to-score alignment method based on AMT and DTW, mainly based on a previous work [326] – see Chapter 6. 89

**WAET** Web Audio Evaluation Tool [306]. 104

# Bibliography

[1] Federico Simonetta, Stavros Ntalampiras, and Federico Avanzini. "Multimodal Music Information Processing and Retrieval: Survey and Future Challenges". In: *Proceedings of 2019 International Workshop on Multilayer Music Representation and Processing*. Int. Work. on Multilayer Music Representation and Processing. Milan, Italy: IEEE Conference Publishing Services, 2019, pp. 10–18 (cit. on pp. i, 23, 85, 100, 120, 141).

[2] Federico Simonetta, Carlos Cancino-Chacón, Stavros Ntalampiras, and Gerhard Widmer. "A Convolutional Approach to Melody Line Identification in Symbolic Scores". In: *20th Int. Conf. on Music Information Retrieval Conference (ISMIR)*. 2019 (cit. on pp. i, 55).

[3] Luca Andrea Ludovico, Adriano Baratè, Federico Simonetta, and Davide Andrea Mauro. "On the Adoption of Standard Encoding Formats to Ensure Interoperability of Music Digital Archives: The IEEE 1599 Format." In: *6th International Conference on Digital Libraries for Musicology*. ACM, Nov. 2019, pp. 20–24 (cit. on p. i).

[4] Federico Simonetta, Stavros Ntalampiras, and Federico Avanzini. "ASMD: An Automatic Framework for Compiling Multimodal Datasets with Audio and Scores". In: *Proceedings of the 17th Sound and Music Computing Conference*. Torino, 2020 (cit. on pp. i, 2, 39, 53, 90, 109, 117, 127, 151).

[5] Federico Simonetta, Stavros Ntalampiras, and Federico Avanzini. "Audio-to-Score Alignment Using Deep Automatic Music Transcription". In: *Proceeddings of the IEEE MMSP 2021*. 2021 (cit. on pp. i, 33, 53, 85).

[6] Federico Simonetta, Federico Avanzini, and Stavros Ntalampiras. "A Perceptual Measure for Evaluating the Resynthesis of Automatic Music Transcriptions". In: *Multimed. Tools Appl.* (2022) (cit. on pp. i, 34, 35, 128, 129).

[7] Federico Simonetta. "Towards Faithful Automatic Music Resynthesis". In: *Comput. Music J.* (2022) (cit. on pp. i, 119, 120).

[8] Ian D. Bent et al. "Notation". In: *Grove Music Online*. Oxford University Press, 2014 (cit. on p. 7).

[9] Howard Mayer Brown, Ellen Rosand, Reinhard Strohm, Michel Noiray, Roger Parker, Arnold Whittall, Roger Savage, and Barry Millington. "Opera (i)". In: *Grove Music Online*. Oxford University Press, Nov. 2001 (cit. on p. 7).

[10] Gordon Mumma, Howard Rye, Barry Kernfeld, and Chris Sheridan. "Recording". In: *Grove Music Online*. Oxford University Press, Nov. 2003 (cit. on p. 8).

[11] David A. Cook and Robert Sklar. "History of the Motion Picture". In: *Britannica Academic*. Encyclopaedia Britannica, 2018 (cit. on p. 8).

[12] Tetsuro Kitahara. “Mid-Level Representations of Musical Audio Signals for Music Information Retrieval”. In: *Advances in Music Information Retrieval*. Springer Berlin Heidelberg, 2010, pp. 65–91 (cit. on p. 8).

[13] Hugues Vinet. “The Representation Levels of Music Information”. In: *Computer Music Modeling and Retrieval*. Ed. by Uffe Kock Wiil. Berlin, Heidelberg: Springer Berlin Heidelberg, 2004, pp. 193–209 (cit. on p. 8).

[14] François Pachet. “Musical Metadata and Knowledge Management”. In: *Encyclopedia of Knowledge Management, Second Edition*. Ed. by David G. Schwartz and Dov Te'eni. IGI Global, 2005, pp. 1192–1199 (cit. on p. 8).

[15] Diana Deutsch. *The Psychology of Music (Third Edition)*. Ed. by Diana Deutsch. Third Edition. Academic Press, 2013 (cit. on p. 8).

[16] Pradeep K. Atrey, M. Anwar Hossain, Abdulmotaleb El Saddik, and Mohan S. Kankanhalli. “Multimodal Fusion for Multimedia Analysis: A Survey”. In: *J Multimed. Syst.* 16.6 (2010), pp. 345–379 (cit. on pp. 8, 18, 20).

[17] Tadas Baltrusaitis, Chaitanya Ahuja, and Louis-Philippe Morency. “Multimodal Machine Learning: A Survey and Taxonomy”. In: *IEEE Trans. Pattern Anal. Mach. Intell.* 41.2 (2018), pp. 1–1 (cit. on pp. 8, 18, 20).

[18] Marvin Minsky. “Logical Versus Analogical or Symbolic Versus Connectionist or Neat Versus Scruffy”. In: *AI Mag.* 12.2 (1991), pp. 34–51 (cit. on p. 8).

[19] Markus Schedl, Emilia Gómez, and Julián Urbano. “Music Information Retrieval: Recent Developments and Applications”. In: *Found. Trends® Inf. Retr.* 8.2-3 (2014), pp. 127–261 (cit. on pp. 9, 11).

[20] G. Tzanetakis and P. Cook. “Musical Genre Classification of Audio Signals”. In: *IEEE Trans. Speech Audio Process.* 10.5 (July 2002), pp. 293–302 (cit. on p. 9).

[21] Loris Nanni, Yandre M. G. Costa, Alessandra Lumini, Moo Young Kim, and SeungRyul Baek. “Combining Visual and Acoustic Features for Music Genre Classification”. In: *Expert Syst. Appl.* 45 (2016), pp. 108–117 (cit. on p. 9).

[22] Slim Essid and Gaël Richard. “Fusion of Multimodal Information in Music Content Analysis”. In: *Multimodal Music Processing*. Schloss Dagstuhl - Leibniz-Zentrum fuer Informatik GmbH, Wadern/Saarbruecken, Germany, 2012 (cit. on p. 9).

[23] Li Zhonghua. “Multimodal Music Information Retrieval: From Content Analysis to Multimodal Fusion”. PhD thesis. 2013 (cit. on pp. 10, 13).

[24] Meinard Müller. *Fundamentals of Music Processing: Audio, Analysis, Algorithms, Applications*. 1st ed. Springer Publishing Company, Incorporated, 2015 (cit. on pp. 11, 14, 17, 19).

[25] Hiromasa Fujihara and Masataka Goto. “Lyrics-to-Audio Alignment and Its Application”. In: *Multimodal Music Processing*. Ed. by Meinard Müller, Masataka Goto, and Markus Schedl. Vol. 3. Dagstuhl Follow-Ups. Dagstuhl, Germany: Schloss Dagstuhl–leibniz-zentrum Fuer Informatik, 2012, pp. 23–36 (cit. on pp. 11, 19).

[26] Matthias Dorfer, Andreas Arzt, and Gerhard Widmer. "Learning Audio-Sheet Music Correspondences for Score Identification and Offline Alignment". In: *Proceedings of the 18\textsuperscriptth International Society for Music Information Retrieval Conference, ISMIR 2017, Suzhou, China*. Ed. by Sally Jo Cunningham, Zhiyao Duan, Xiao Hu, and Douglas Turnbull. 2017, pp. 115–122 (cit. on pp. 11, 13).

[27] Alexios Kotsifakos, Panagiotis Papapetrou, Jaakko Hollmén, Dimitrios Gunopulos, and Vassilis Athitsos. "A Survey of Query-by-Humming Similarity Methods". In: *Proceedings of the 5th International Conference on PErvasive Technologies Related to Assistive Environments*. PETRA '12. Heraklion, Crete, Greece: ACM, 2012 (cit. on p. 13).

[28] MIREX. *2016:Query by Singing/Humming*. 2016 (cit. on p. 13).

[29] Geoffray Bonnin and Dietmar Jannach. "Automated Generation of Music Playlists: Survey and Experiments". In: *ACM Comput. Surv.* 47.2 (Nov. 2014), 26:1–26:35 (cit. on p. 13).

[30] Puja Deshmukh and Geetanjali Kale. "A Survey of Music Recommendation System". In: *Int. J. Sci. Res. Comput. Sci. Eng. Inf. Technol. IJSRCSEIT* 3.3 (Mar. 2018), pp. 1721–1729 (cit. on p. 13).

[31] M. Müller, A. Arzt, S. Balke, M. Dorfer, and G. Widmer. "Cross-Modal Music Retrieval and Applications: An Overview of Key Methodologies". In: *IEEE Signal Process. Mag.* 36.1 (Jan. 2019), pp. 52–62 (cit. on p. 13).

[32] Albin Correya, Romain Hennequin, and Mickaël Arcos. "Large-Scale Cover Song Detection in Digital Music Libraries Using Metadata, Lyrics and Audio Features". In: *Arxiv E-prints* (Aug. 2018) (cit. on p. 13).

[33] Iman S. H. Suyoto, Alexandra L. Uitdenbogerd, and Falk Scholer. "Searching Musical Audio Using Symbolic Queries". In: *IEEE Trans. Audio Speech Lang. Process.* 16.2 (Feb. 2008), pp. 372–381 (cit. on pp. 13, 14).

[34] S. Balke, V. Arifi-Müller, L. Lamprecht, and M. Mü ller. "Retrieving Audio Recordings Using Musical Themes". In: *2016 IEEE International Conference on Acoustics, Speech and Signal Processing (ICASSP)*. Mar. 2016, pp. 281–285 (cit. on pp. 13, 14).

[35] O. Gillet, S. Essid, and G. Richard. "On the Correlation of Automatic Audio and Visual Segmentations of Music Videos". In: *Ieee Trans. Circuits Syst. Video Technol.* 17.3 (Mar. 2007), pp. 347–355 (cit. on pp. 13, 14).

[36] Youngmoo E. Kim, Erik M. Schmidt, Raymond Migneco, Brandon G. Morton, Patrick Richardson, Jeffrey J. Scott, Jacquelin A. Speck, and Douglas Turnbull. "State of the Art Report: Music Emotion Recognition: A State of the Art Review". In: *Proceedings of the 11\textsuperscriptth International Society for Music Information Retrieval Conference, ISMIR 2010, Utrecht, Netherlands, August 9-13, 2010*. Ed. by J. Stephen Downie and Remco C. Veltkamp. International Society for Music Information Retrieval, 2010, pp. 255–266 (cit. on p. 13).

[37] Tao Li and Mitsunori Ogihara. "Music Artist Style Identification by Semi-Supervised Learning from Both Lyrics and Content". In: *Proceedings of the 12\textsuperscriptth Annual ACM International Conference on Multimedia*. MULTIMEDIA '04. New York, NY, USA: ACM, 2004, pp. 364–367 (cit. on p. 13).

[38] Rudolf Mayer, Robert Neumayer, and Andreas Rauber. "Combination of Audio and Lyrics Features for Genre Classification in Digital Audio Collections". In: *Proceedings of the 16\textsuperscriptth ACM International Conference on Multimedia*. MM '08. New York, NY, USA: ACM, 2008, pp. 159–168 (cit. on p. 13).

[39] Rudolf Mayer, Robert Neumayer, and Andreas Rauber. "Rhyme and Style Features for Musical Genre Classification by Song Lyrics". In: *ISMIR 2008, 9th International Conference on Music Information Retrieval, Drexel University, Philadelphia, PA, USA, September 14-18, 2008*. 2008, pp. 337–342 (cit. on p. 13).

[40] Rudolf Mayer and Andreas Rauber. "Multimodal Aspects of Music Retrieval: Audio, Song Lyrics – and Beyond?" In: *Advances in Music Information Retrieval*. Ed. by Zbigniew W. Raś and Alicja A. Wieczorkowska. Berlin, Heidelberg: Springer Berlin Heidelberg, 2010, pp. 333–363 (cit. on p. 13).

[41] Chao Zhen and Jieping Xu. "Multi-Modal Music Genre Classification Approach". In: *Proc. 3rd Int. Conf. Computer Science and Information Technology*. Vol. 8. July 2010, pp. 398–402 (cit. on p. 13).

[42] Rudolf Mayer and Andreas Rauber. "Musical Genre Classification by Ensembles of Audio and Lyrics Features". In: *Proceedings of the 12\textsuperscriptth International Society for Music Information Retrieval Conference (ISMIR 2012)*. University of Miami, 2011, pp. 675–680 (cit. on p. 13).

[43] Alexander Schindler and Andreas Rauber. "Harnessing Music-Related Visual Stereotypes for Music Information Retrieval". In: *ACM Trans. Intell. Syst. Technol.* 8.2 (Oct. 2016), pp. 1–21 (cit. on p. 13).

[44] Sergio Oramas, Francesco Barbieri, Oriol Nieto, and Xavier Serra. "Multimodal Deep Learning for Music Genre Classification". In: *Trans. Int. Soc. Music Inf. Retr.* 1.1 (2018), pp. 4–21 (cit. on pp. 13, 17, 20).

[45] Kamelia Aryafar and Ali Shokoufandeh. "Multimodal Music and Lyrics Fusion Classifier for Artist Identification". In: *2014 13\textsuperscriptth International Conference on Machine Learning and Applications*. IEEE, Dec. 2014 (cit. on pp. 13, 15, 19).

[46] Jordan B. L. Smith, M. Hamasaki, and M. Goto. "Classifying Derivative Works with Search, Text, Audio and Video Features". In: *Proc. IEEE Int. Conf. Multimedia and Expo (ICME)*. July 2017, pp. 1422–1427 (cit. on p. 13).

[47] Olga Slizovskaia, Emilia Gómez, and Gloria Haro. "Musical Instrument Recognition in User-Generated Videos Using a Multimodal Convolutional Neural Network Architecture". In: *Proceedings of the 2017 ACM on International Conference on Multimedia Retrieval - ICMR '17*. ACM Press, 2017 (cit. on p. 13).

[48] Angelica Lim, Keisuke Nakamura, Kazuhiro Nakadai, Tetsuya Ogata, and Hiroshi G. Okuno. *Audio-Visual Musical Instrument Recognition*. 2011 (cit. on pp. 13, 20).

[49] Sertan Sentürk, Sankalp Gulati, and Xavier Serra. "Score Informed Tonic Identification for Makam Music of Turkey". In: *Proceedings of the 14\textsuperscriptth International Society for Music Information Retrieval Conference, ISMIR 2013, Curitiba , Brazil, November 4-8, 2013*. Ed. by Alceu de Souza Britto Jr., Fabien Gouyon, and Simon Dixon. 2013, pp. 175–180 (cit. on pp. 13, 20).

[50] Pei-Ching Li, Li Su, Yi-Hsuan Yang, and Alvin W. Y. Su. "Analysis of Expressive Musical Terms in Violin Using Score-Informed and Expression-Based Audio Features". In: *Proceedings of the 16\textsuperscriptth International Society for Music Information Retrieval Conference, ISMIR 2015, Má Laga, Spain, October 26-30, 2015*. Ed. by Meinard Müller and Frans Wiering. 2015, pp. 809–815 (cit. on pp. 13, 14, 19).

[51] Gil Weinberg, Aparna Raman, and Trishul Mallikarjuna. "Interactive Jamming with Shimon: A Social Robotic Musician". In: *Proceedings of the 4\textsuperscriptth ACM/IEEE International Conference on Human Robot Interaction*. HRI '09. New York, NY, USA: ACM, 2009, pp. 233–234 (cit. on p. 14).

[52] A. Lim, T. Mizumoto, L. Cahier, T. Otsuka, T. Takahashi, K. Komatani, T. Ogata, and H. G. Okuno. "Robot Musical Accompaniment: Integrating Audio and Visual Cues for Real-Time Synchronization with a Human Flutist". In: *Proc. IEEE/RSJ Int. Conf. Intelligent Robots and Systems*. Oct. 2010, pp. 1964–1969 (cit. on p. 14).

[53] T. Itohara, T. Otsuka, T. Mizumoto, T. Ogata, and H. G. Okuno. "Particle-Filter Based Audio-Visual Beat-Tracking for Music Robot Ensemble with Human Guitarist". In: *Proc. IEEE/RSJ Int. Conf. Intelligent Robots and Systems*. Sept. 2011, pp. 118–124 (cit. on p. 14).

[54] David Ross Berman. "AVISARME: Audio Visual Synchronization Algorithm for a Robotic Musician Ensemble". PhD thesis. University of Maryland, 2012 (cit. on p. 14).

[55] M. Ohkita, Y. Bando, Y. Ikemiya, K. Itoyama, and K. Yoshii. "Audio-Visual Beat Tracking Based on a State-Space Model for a Music Robot Dancing with Humans". In: *Proc. IEEE/RSJ Int. Conf. Intelligent Robots and Systems (IROS)*. 2015, pp. 5555–5560 (cit. on pp. 14, 16).

[56] Siying Wang, Sebastian Ewert, and Simon Dixon. "Identifying Missing and Extra Notes in Piano Recordings Using Score-Informed Dictionary Learning". In: *IEEEACM Trans. Audio Speech Lang. Process.* 25.10 (Oct. 2017), pp. 1877–1889 (cit. on p. 14).

[57] Tsubasa Fukuda, Yukara Ikemiya, Katsutoshi Itoyama, and Kazuyoshi Yoshii. "A Score-Informed Piano Tutoring System with Mistake Detection and Score Simplification". In: *Proc of the Sound and Music Computing Conference (SMC)*. Zenodo, 2015 (cit. on p. 14).

[58] Emmanouil Benetos, Anssi Klapuri, and Simon Dixon. "Score-Informed Transcription for Automatic Piano Tutoring". In: *European Signal Processing Conference*. European Signal Processing Conference. IEEE, 2012, pp. 2153–2157 (cit. on p. 14).

[59] O. Mayor, J. Bonada, and A. Loscos. “Performance Analysis and Scoring of the Singing Voice”. In: *AES 35\textsuperscriptth International Conference: Audio for Games*. 2009 (cit. on p. 14).

[60] W. Tsai and H. Lee. “Automatic Evaluation of Karaoke Singing Based on Pitch, Volume, and Rhythm Features”. In: *Lang. Process. IEEE Trans. Audio Speech* 20.4 (May 2012), pp. 1233–1243 (cit. on p. 14).

[61] Jakob Abeßer, Johannes Hasselhorn, Christian Dittmar, Andreas Lehmann, and Sascha Grollmisch. “Automatic Quality Assessment of Vocal and Instrumental Performances of Ninth-Grade and Tenth-Grade Pupils”. In: (2013) (cit. on p. 14).

[62] Johanna Devaney, Michael I. Mandel, and Ichiro Fujinaga. “A Study of Intonation in Three-Part Singing Using the Automatic Music Performance Analysis and Comparison Toolkit (AMPACT)”. In: *Proceedings of the 13th International Society for Music Information Retrieval Conference, ISMIR 2012, Mosteiro S.Bento Da Vitória, Porto, Portugal, October 8-12, 2012*. Ed. by Fabien Gouyon, Perfecto Herrera, Luis Gustavo Martins, and Meinard M üller. FEUP Edições, 2012, pp. 511–516 (cit. on p. 14).

[63] Yongwei Zhu, Kai Chen, and Qibin Sun. “Multimodal Content-Based Structure Analysis of Karaoke Music”. In: *Proceedings of the 13\textsuperscriptth Annual ACM International Conference on Multimedia*. MULTIMEDIA ’05. New York, NY, USA: ACM, 2005, pp. 638–647 (cit. on p. 14).

[64] Heng-Tze Cheng, Yi-Hsuan Yang, Yu-Ching Lin, and Homer H. Chen. “Multimodal Structure Segmentation and Analysis of Music Using Audio and Textual Information”. In: *2009 IEEE International Symposium on Circuits and Systems*. IEEE, May 2009 (cit. on pp. 14, 20).

[65] Jeff Gregorio and Youngmoo Kim. “Phrase-Level Audio Segmentation of Jazz Improvisations Informed by Symbolic Data.” In: *Proceedings of the 17\textsuperscriptth International Society for Music Information Retrieval Conference, ISMIR 2016, New York City, United States*. 2016, pp. 482–487 (cit. on pp. 14, 17, 19, 20).

[66] M. Paleari, B. Huet, A. Schutz, and D. Slock. “A Multimodal Approach to Music Transcription”. In: *Proc. 15th IEEE Int. Conf. Image Processing*. Oct. 2008, pp. 93–96 (cit. on pp. 14, 15).

[67] Alex Hrybyk. “Combined Audio and Video Analysis for Guitar Chord Identification”. In: *11\textsuperscriptth International Society for Music Information Retrieval Conference (ISMIR 2010)*. 2010 (cit. on p. 14).

[68] Bernardo Marenco, Magdalena Fuentes, Florencia Lanzaro, Martín Rocamora, and Alvaro Gómez. “A Multimodal Approach for Percussion Music Transcription from Audio and Video”. In: *Algebr. Coalgebraic Methods Math. Program Constr.* (Jan. 2015) (cit. on p. 14).

[69] Verena Konz and Meinard Müller. "A Cross-Version Approach for Harmonic Analysis of Music Recordings". In: *Multimodal Music Processing*. Ed. by Meinard Müller, Masataka Goto, and Markus Schedl. Vol. 3. Dagstuhl Follow-Ups. Schloss Dagstuhl - Leibniz-Zentrum fuer Informatik, Germany, 2012, pp. 53–72 (cit. on pp. 14, 15, 20).

[70] Bochen Li, Chenliang Xu, and Zhiyao Duan. "Audiovisual Source Association for String Ensembles through Multi-Modal Vibrato Analysis". In: *Proc Sound Music Comput. Smc* (2017) (cit. on pp. 14, 15).

[71] Bochen Li, Karthik Dinesh, Zhiyao Duan, and Gaurav Sharma. "See and Listen: Score-Informed Association of Sound Tracks to Players in Chamber Music Performance Videos". In: *Proc. Speech and Signal Processing (ICASSP) 2017 IEEE Int. Conf. Acoustics*. Mar. 2017, pp. 2906–2910 (cit. on pp. 14, 15, 20).

[72] K. Dinesh, B. Li, X. Liu, Z. Duan, and G. Sharma. "Visually Informed Multi-Pitch Analysis of String Ensembles". In: *Proc. Speech and Signal Processing (ICASSP) 2017 IEEE Int. Conf. Acoustics*. Mar. 2017, pp. 3021–3025 (cit. on pp. 14, 15, 20).

[73] Markus Schedl, Emilia Gómez, and Julián Urbano. "Music Information Retrieval: Recent Developments and Applications". In: *Found. Trends® Inf. Retr.* 8.2-3 (2014), pp. 127–261 (cit. on p. 15).

[74] Francesc Alías, Joan Socoró, and Xavier Sevillano. "A Review of Physical and Perceptual Feature Extraction Techniques for Speech, Music and Environmental Sounds". In: *Appl. Sci.* 6.5 (May 2016), p. 143 (cit. on p. 15).

[75] Stavros Ntalampiras, Ilyas Potamitis, and Nikos Fakotakis. "Exploiting Temporal Feature Integration for Generalized Sound Recognition". In: *EURASIP J. Adv. Signal Process.* 2009.1 (Dec. 28, 2009), p. 807162 (cit. on p. 16).

[76] D. Brezeale and D. J. Cook. "Automatic Video Classification: A Survey of the Literature". In: *IEEE Trans. Syst. Man Cybern. Part C Appl. Rev.* 38.3 (May 2008), pp. 416–430 (cit. on p. 16).

[77] S. Wang and Q. Ji. "Video Affective Content Analysis: A Survey of State-of-the-Art Methods". In: *IEEE Trans. Affect. Comput.* 6.4 (Oct. 2015), pp. 410–430 (cit. on p. 16).

[78] W. Bruce Croft, Donald Metzler, and Trevor Strohman. *Search Engines: Information Retrieval in Practice*. Vol. 283. Pearson Education, Inc., 2015 (cit. on p. 17).

[79] Evgeniy Gabrilovich and Shaul Markovitch. "Computing Semantic Relatedness Using Wikipedia-Based Explicit Semantic Analysis." In: *Proceedings of the 20th International Joint Conference on Artifical Intelligence*. Vol. 7. San Francisco, CA, USA, 2007, pp. 1606–1611 (cit. on p. 17).

[80] Federico Simonetta, Filippo Carnovalini, Nicola Orio, and Antonio Rodà. "Symbolic Music Similarity through a Graph-Based Representation". In: *Proceedings of the Audio Mostly 2018 on Sound in Immersion and Emotion - AM'18*. ACM Press, 2018 (cit. on pp. 17, 36).

[81] Meinard Müller. "Dynamic Time Warping". In: *Information Retrieval for Music and Motion*. Springer Berlin Heidelberg, 2007, pp. 69–84 (cit. on pp. 19, 85, 151).

[82] Christopher M. Bishop. *Pattern Recognition and Machine Learning, 5\textsuperscriptth Edition*. Information Science and Statistics. Springer, 2007 (cit. on p. 19).

[83] N. Degara, A. Pena, M. E. P. Davies, and M. D. Plumbley. "Note Onset Detection Using Rhythmic Structure". In: *2010 IEEE International Conference on Acoustics, Speech and Signal Processing*. Mar. 2010, pp. 5526–5529 (cit. on p. 20).

[84] Gabriel Meseguer-Brocal, Alice Cohen-Hadria, and Peeters Geoffroy. "Dali: A Large Dataset of Synchronized Audio, Lyrics and Notes, Automatically Created Using Teacher-Student Machine Learning Paradigm." In: *Proceedings of the International Society for MusicInformation Retrieval Conference*. Ed. by ISMIR. Vol. abs/1906.10606. Sept. 2018 (cit. on pp. 20, 48).

[85] Esteban Maestre, Panagiotis Papiotis, Marco Marchini, Quim Llimona, Oscar Mayor, Alfonso Pérez, and Marcelo M. Wanderley. "Enriched Multimodal Representations of Music Performances: Online Access and Visualization". In: *Ieee Multimed.* 24.1 (Jan. 2017), pp. 24–34 (cit. on p. 20).

[86] Bochen Li, Xinzhao Liu, Karthik Dinesh, Zhiyao Duan, and Gaurav Sharma. "Creating a Multitrack Classical Music Performance Dataset for Multimodal Music Analysis: Challenges, Insights, and Applications." In: *IEEE Trans. Multimed.* 21.2 (Feb. 2019), pp. 522–535 (cit. on p. 20).

[87] Sinno Jialin Pan and Qiang Yang. "A Survey on Transfer Learning". In: *IEEE Trans. Knowl. Data Eng.* 22.10 (Oct. 2010), pp. 1345–1359 (cit. on p. 21).

[88] Stavros Ntalampiras. "A Transfer Learning Framework for Predicting the Emotional Content of Generalized Sound Events". In: *J. Acoust. Soc. Am.* 141.3 (Mar. 2017), pp. 1694–1701 (cit. on p. 21).

[89] S. Schreibman, R. Siemens, and J. Unsworth. *A New Companion to Digital Humanities*. Blackwell Companions to Literature and Culture. Wiley-Blackwell, 2015 (cit. on p. 23).

[90] William Storm. "The Establishment of International Re-Recording Standards". In: *Phonogr. Bull.* 27 (1980), pp. 5–12 (cit. on p. 24).

[91] Robert Philip. *Performing Music in the Age of Recording*. New Haven, USA: Yale University Press, 2004 (cit. on p. 25).

[92] Dietrich Schüller. "The Ethics of Preservation, Restoration, and Re-Issues of Historical Sound Recordings". In: *J. Audio Eng. Soc.* 39.1212 (Dec. 1991), pp. 1014–1017 (cit. on p. 25).

[93] Angelo Orcalli. "Recorded Music: From the Ethics of Preservation to the Critical Editing". In: *Sounds, Voices and Codes from the Twentieth Century. The Critical Editing of Music at Mirage*. Ed. by Luca Cossettini and Angelo Orcalli. Department of Languages, Literatures, Communication, Education, and Society. University of Udine, 2017 (cit. on p. 26).

[94] Simon J. Godsill and Peter J. W. Rayner. *Digital Audio Restoration*. Springer London, 1998 (cit. on pp. 27–29).

[95] S.V. Vaseghi and P.J.W. Rayner. "A New Application of Adaptive Filters for Restoration of Archived Gramophone Recordings". In: *ICASSP-88., International Conference on Acoustics, Speech, and Signal Processing*. Vol. 5. 1988, 2548–2551 vol.5 (cit. on p. 27).

[96] Chingshun Lin, Zongchao Cheng, and Dongliang Shih. "Music Enhancement Using Nonnegative Matrix Factorization with Penalty Masking". In: *2013 IEEE 16th International Conference on Computational Science and Engineering*. 2013, pp. 125–129 (cit. on p. 27).

[97] Paulo AA Esquef and Guilherme S Welter. "Audio De-Thumping Using Huang's Empirical Mode Decomposition". In: *Proceedings of the International Conference on Digital Audio Effects*. 2011, pp. 19–23 (cit. on p. 27).

[98] Hugo T. de Carvalho, Flavio R. Avila, and Luiz W. P. Biscainho. "Bayesian Restoration of Audio Degraded by Low-Frequency Pulses Modeled via Gaussian Process". In: *IEEE J. Sel. Top. Signal Process.* 15.11 (2021), pp. 90–103 (cit. on p. 27).

[99] Christoph F. Stallmann and Andries P. Engelbrecht. "Gramophone Noise Detection and Reconstruction Using Time Delay Artificial Neural Networks". In: *IEEE Trans. Syst. Man Cybern. Syst.* 47.66 (June 2017), pp. 893–905 (cit. on p. 27).

[100] Sahar Sadrizadeh, Nematollah Zarmehi, Ehsan Asadi Kangarshahi, Hamidreza Abin, and Farokh Marvasti. "A Fast Iterative Method for Removing Impulsive Noise From Sparse Signals". In: *IEEE Trans. Circuits Syst. Video Technol.* 31.11 (2021), pp. 38–48 (cit. on p. 27).

[101] Philipos C. Loizou. *Speech Enhancement*. CRC Press, June 2007 (cit. on p. 28).

[102] Qing Zhou and Zuren Feng. "Robust Sound Event Detection Through Noise Estimation and Source Separation Using NMF". In: *Proceedings of the Detection and Classification of Acoustic Scenes and Events 2017 Workshop (DCASE2017)*. 2017, pp. 138–142 (cit. on p. 28).

[103] Emad M. Grais and Mark D. Plumbley. "Single Channel Audio Source Separation Using Convolutional Denoising Autoencoders". In: *2017 IEEE Global Conference on Signal and Information Processing (GlobalSIP)*. 2017, pp. 1265–1269 (cit. on pp. 28, 29).

[104] Hiroki Tanji, Takahiro Murakami, and Hiroyuki Kamata. "Laplace Nonnegative Matrix Factorization with Application to Semi-Supervised Audio Denoising". In: *2019 27th European Signal Processing Conference (EUSIPCO)*. 2019, pp. 1–5 (cit. on p. 28).

[105] Samuel Sonning, Christian Schüldt, Hakan Erdogan, and Scott Wisdom. "Performance Study of a Convolutional Time-Domain Audio Separation Network for Real-Time Speech Denoising". In: *ICASSP 2020 - 2020 IEEE International Conference on Acoustics, Speech and Signal Processing (ICASSP)*. IEEE, 2020, pp. 831–835 (cit. on pp. 28, 29).

[106] Ruilin Xu, Rundi Wu, Yuko Ishiwaka, Carl Vondrick, and Changxi Zheng. "Listening to Sounds of Silence for Speech Denoising". In: *Advances in Neural Information Processing Systems*. Ed. by H. Larochelle, M. Ranzato, R. Hadsell, M. F. Balcan, and H. Lin. Vol. 33. Curran Associates, Inc., 2020, pp. 9633–9648 (cit. on p. 29).

[107] Dario Rethage, Jordi Pons, and Xavier Serra. "A Wavenet for Speech Denoising". In: *2018 IEEE International Conference on Acoustics, Speech and Signal Processing (ICASSP)*. IEEE, 2018, pp. 5069–5073 (cit. on p. 29).

[108] Andrew Maas, Quoc V. Le, Tyler M. O'Neil, Oriol Vinyals, Patrick Nguyen, and Andrew Y. Ng. "Recurrent Neural Networks for Noise Reduction in Robust ASR". In: *INTERSPEECH*. ISCA, 2012, pp. 22–25 (cit. on p. 29).

[109] Ying-Hui Lai et al. "Deep Learning–Based Noise Reduction Approach to Improve Speech Intelligibility for Cochlear Implant Recipients". In: *Ear Hear.* 39.44 (July 2018), pp. 795–809 (cit. on p. 29).

[110] Ying-Hui Lai, Fei Chen, Syu-Siang Wang, Xugang Lu, Yu Tsao, and Chin-Hui Lee. "A Deep Denoising Autoencoder Approach to Improving the Intelligibility of Vocoded Speech in Cochlear Implant Simulation". In: *IEEE Trans. Biomed. Eng.* 64.77 (2017), pp. 1568–1578 (cit. on p. 29).

[111] Serkan Kiranyaz, Turker Ince, Osama Abdeljaber, Onur Avci, and Moncef Gabbouj. "1-D Convolutional Neural Networks for Signal Processing Applications". In: *ICASSP 2019 - 2019 IEEE International Conference on Acoustics, Speech and Signal Processing (ICASSP)*. IEEE, 2019, pp. 8360–8364 (cit. on p. 29).

[112] Saurabh Kataria, Jesús Villalba, and Najim Dehak. "Perceptual Loss Based Speech Denoising with an Ensemble of Audio Pattern Recognition and Self-Supervised Models". In: *ICASSP 2021 - 2021 IEEE International Conference on Acoustics, Speech and Signal Processing (ICASSP)*. June 2021, pp. 7118–7122 (cit. on p. 29).

[113] François G. Germain, Qifeng Chen, and Vladlen Koltun. "Speech Denoising with Deep Feature Losses". In: *Proc. Interspeech 2019*. ISCA, 2019, pp. 2723–2727 (cit. on p. 29).

[114] Pavel Záviška, Pavel Rajmic, Alexey Ozerov, and Lucas Rencker. "A Survey and an Extensive Evaluation of Popular Audio Declipping Methods". In: *IEEE J. Sel. Top. Signal Process.* 15.11 (2021), pp. 5–24 (cit. on p. 29).

[115] Ioannis Pitas and Anastasios N Venetsanopoulos. "Median Filters". In: *Nonlinear Digital Filters: Principles and Applications*. Springer Science+Business Media, LLC, 1999 (cit. on p. 29).

[116] A. Janssen, R. Veldhuis, and L. Vries. "Adaptive Interpolation of Discrete-Time Signals That Can Be Modeled as Autoregressive Processes". In: *IEEE Trans. Acoust. Speech Signal Process.* 34.22 (1986), pp. 317–330 (cit. on p. 29).

[117] Marco Ruhland, Jörg Bitzer, Matthias Brandt, and Stefan Goetze. "Reduction of Gaussian, Supergaussian, and Impulsive Noise by Interpolation of the Binary Mask Residual". In: *IEEE ACM Trans. Audio Speech Lang. Process.* 23.1010 (2015), pp. 1680–1691 (cit. on p. 29).

[118] Renjie Tong, Yingyue Zhou, Long Zhang, Guangzhao Bao, and Zhongfu Ye. "A Robust Time-Frequency Decomposition Model for Suppression of Mixed Gaussian-Impulse Noise in Audio Signals". In: *IEEE ACM Trans. Audio Speech Lang. Process.* 23.11 (2015), pp. 69–79 (cit. on p. 29).

[119] Marcelo Bertalmio, Guillermo Sapiro, Vincent Caselles, and Coloma Ballester. "Image Inpainting". In: *Proceedings of the 27th Annual Conference on Computer Graphics and Interactive Techniques*. ACM Press, 2000 (cit. on p. 29).

[120] Amir Adler, Valentin Emiya, Maria G. Jafari, Michael Elad, Rémi Gribonval, and Mark D. Plumbley. "Audio Inpainting." In: *IEEE Trans. Audio Speech Lang. Process.* 20.33 (2012), pp. 922–932 (cit. on p. 29).

[121] André Marafioti, Nicki Holighaus, Piotr Majdak, and Nathanaë Perraudin. "Audio Inpainting of Music by Means of Neural Networks". In: *J. Audio Eng. Soc.* (Mar. 2019) (cit. on p. 29).

[122] Andrés Marafioti, Nathanaël Perraudin, Nicki Holighaus, and Piotr Majdak. "A Context Encoder For Audio Inpainting". In: *IEEE ACM Trans. Audio Speech Lang. Process.* 27.1212 (2019), pp. 2362–2372 (cit. on p. 29).

[123] Andrés Marafioti, Piotr Majdak, Nicki Holighaus, and Nathanaël Perraudin. "GACELA: A Generative Adversarial Context Encoder for Long Audio Inpainting of Music". In: *IEEE J. Sel. Top. Signal Process.* 15.11 (Jan. 2021), pp. 120–131 (cit. on p. 29).

[124] Hang Zhou, Ziwei Liu, Xudong Xu, Ping Luo, and Xiaogang Wang. "Vision-Infused Deep Audio Inpainting." In: *Proceedings of the IEEE/CVF International Conference on Computer Vision (ICCV)*. 2019 (cit. on p. 29).

[125] Giovanni Morrone, Daniel Michelsanti, Zheng-Hua Tan, and Jesper Jensen. "Audio-Visual Speech Inpainting with Deep Learning". In: *ICASSP 2021 - 2021 IEEE International Conference on Acoustics, Speech and Signal Processing (ICASSP)*. 2021, pp. 6653–6657 (cit. on p. 29).

[126] Luca A. Ludovico, Giorgio Presti, and Alessandro Rizzi. "Audio Dynamics Automatic Equalization Inspired by Visual Perception". In: *Multimed. Tools Appl.* 80.88 (Mar. 1, 2021), pp. 11903–11915 (cit. on p. 29).

[127] Andrzej Czyzewski and Przemyslaw Maziewski. "Some Techniques for Wow Effect Reduction". In: *2007 IEEE International Conference on Image Processing*. Vol. 4. IEEE, 2007, pp. 29–32 (cit. on p. 29).

[128] Andrzej Czyzewski, Przemyslaw Maziewski, and Adam Kupryjanow. "Reduction of Parasitic Pitch Variations in Archival Musical Recordings". In: *Signal Process.* 90.44 (2010), pp. 981–990 (cit. on p. 29).

[129] Giovanni De Poli and Federico Avanzini. *Sound Modeling: Signal-Based Approaches*. 2006, p. 62 (cit. on p. 30).

[130] D. Schwarz. “Corpus-Based Concatenative Synthesis”. In: *IEEE Signal Process. Mag.* 24.22 (Mar. 2007), pp. 92–104 (cit. on p. 30).

[131] Nick Collins and Bob L Sturm. “Sound Cross-Synthesis and Morphing Using Dictionary-Based Methods”. In: *Proceedings of the International Computer Music Conference*. Michigan Publishing, 2011 (cit. on p. 30).

[132] Juan José Burred. “Cross-Synthesis Based on Spectrogram Factorization”. In: *International Computer Music Conference*. Michigan Publishing, 2013 (cit. on p. 30).

[133] Chris Donahue, Tom Erbe, and Miller S Puckette. “Extended Convolution Techniques for Cross-Synthesis”. In: *International Computer Music Conference*. Michigan Publishing, 2016 (cit. on p. 30).

[134] Thierry Rochebois and Gérard Charbonneau. “Cross-Synthesis Using Interverted Principal Harmonic Sub-Spaces”. In: *Music, Gestalt, and Computing*. Ed. by Marc Leman. Berlin, Heidelberg: Springer Berlin Heidelberg, 1997, pp. 375–385 (cit. on p. 30).

[135] Francisco Daniel Andrade Fonseca. “Musical Cross Synthesis Using Matrix Factorisation”. In: (2019). In collab. with Faculdade De Engenharia (cit. on p. 30).

[136] Homer Dudley. “Remaking Speech”. In: *J. Acoust. Soc. Am.* 11.22 (Oct. 1939), pp. 169–177 (cit. on p. 30).

[137] Xavier Serra. “Musical Sound Modeling with Sinusoids plus Noise”. In: *Music Signal Processing*. Ed. by Curtis Roads, Stephen Travis Pope, Aldo Piccialli, and Giovanni De Poli. Routledge, 1997 (cit. on p. 30).

[138] Xavier Rodet and Diemo Schwarz. “Spectral Envelopes and Additive + Residual Analysis/Synthesis”. In: *Analysis, Synthesis, and Perception of Musical Sounds: The Sound of Music*. Ed. by James W. Beauchamp. New York, NY: Springer New York, 2007, pp. 175–227 (cit. on p. 30).

[139] Qiong Hu, Korin Richmond, Junichi Yamagishi, and Javier Latorre. “An Experimental Comparison of Multiple Vocoder Types”. In: *Eighth ISCA Workshop on Speech Synthesis*. 2013 (cit. on p. 30).

[140] Marco Liuni and Axel Röbel. “Phase Vocoder and Beyond”. In: *Musica/Tecnologia* (2013), pp. 73–89 (cit. on p. 30).

[141] John Makhoul, R Viswanathan, Richard Schwartz, and A. W. F. Huggins. “A Mixed-Source Model for Speech Compression and Synthesis”. In: *J. Acoust. Soc. Am.* 64.66 (1978), pp. 1577–1581 (cit. on p. 30).

[142] M.R. Schroeder. “Vocoders: Analysis and Synthesis of Speech”. In: *Proc. IEEE* 54.55 (1966), pp. 720–734 (cit. on p. 30).

[143] J. L. Flanagan and R. M. Golden. “Phase Vocoder”. In: *Bell Syst. Tech. J.* 45.99 (Nov. 1966), pp. 1493–1509 (cit. on p. 30).

[144] Zhen-Hua Ling, Shi-Yin Kang, Heiga Zen, Andrew Senior, Mike Schuster, Xiao-Jun Qian, Helen M. Meng, and Li Deng. "Deep Learning for Acoustic Modeling in Parametric Speech Generation: A Systematic Review of Existing Techniques and Future Trends". In: *IEEE Signal Process. Mag.* 32.33 (May 2015), pp. 35–52 (cit. on p. 30).

[145] X. Xiao and R. M. Nickel. "Speech Enhancement With Inventory Style Speech Resynthesis." In: *IEEE Trans. Audio Speech Lang. Process.* 18.66 (2010), pp. 1243–1257 (cit. on p. 30).

[146] Soumi Maiti and Michael I Mandel. "Speech Denoising by Parametric Resynthesis". In: *ICASSP 2019 - 2019 IEEE International Conference on Acoustics, Speech and Signal Processing (ICASSP)*. 2019, pp. 6995–6999 (cit. on p. 30).

[147] Aäron van den Oord, Sander Dieleman, Heiga Zen, Karen Simonyan, Oriol Vinyals, Alex Graves, Nal Kalchbrenner, Andrew W. Senior, and Koray Kavukcuoglu. *WaveNet: A Generative Model for Raw Audio*. 2016 (cit. on p. 31).

[148] Shulei Ji, Jing Luo, and Xinyu Yang. *A Comprehensive Survey on Deep Music Generation: Multi-Level Representations, Algorithms, Evaluations, and Future Directions*. 2020 (cit. on p. 31).

[149] Jesse Engel, Cinjon Resnick, Adam Roberts, Sander Dieleman, Mohammad Norouzi, Douglas Eck, and Karen Simonyan. "Neural Audio Synthesis of Musical Notes with WaveNet Autoencoders." In: *Proceedings of the 34th International Conference on Machine Learning*. Ed. by Doina Precup and Yee Whye Teh. Vol. 70. Proceedings of Machine Learning Research. PMLR, Aug. 6, 2017, pp. 1068–1077 (cit. on p. 31).

[150] Aäron van den Oord et al. "Parallel WaveNet: Fast High-Fidelity Speech Synthesis". In: *ICML*. Ed. by Jennifer G. Dy and Andreas Krause. Vol. 80. Proceedings of Machine Learning Research. PMLR, 2018, pp. 3915–3923 (cit. on p. 31).

[151] Florin Schimbinschi, Christian J. Walder, Sarah Monazam Erfani, and James Bailey. "SynthNet: Learning to Synthesize Music End-to-End". In: *IJCAI*. ijcai.org, 2019, pp. 3367–3374 (cit. on p. 31).

[152] Alexandre Défossez, Neil Zeghidour, Nicolas Usunier, Léon Bottou, and Francis R. Bach. "SING: Symbol-to-Instrument Neural Generator". In: *NeurIPS*. 2018 (cit. on p. 31).

[153] Aäron van den Oord, Oriol Vinyals, and Koray Kavukcuoglu. "Neural Discrete Representation Learning". In: *NIPS*. 2017 (cit. on p. 31).

[154] Sander Dieleman, Aaron van den Oord, and Karen Simonyan. "The Challenge of Realistic Music Generation: Modelling Raw Audio at Scale". In: *Advances in Neural Information Processing Systems*. Ed. by S. Bengio, H. Wallach, H. Larochelle, K. Grauman, N. Cesa-Bianchi, and R. Garnett. Vol. 31. Curran Associates, Inc., 2018 (cit. on p. 31).

[155] Stefan Lattner and Maarten Grachten. "High-Level Control of Drum Track Generation Using Learned Patterns of Rhythmic Interaction". In: *2019 IEEE Workshop Appl. Signal Process. Audio Acoust. WASPAA* (2019), pp. 35–39 (cit. on p. 31).

[156] Soroush Mehri, Kundan Kumar, Ishaan Gulrajani, Rithesh Kumar, Shubham Jain, Jose M. R. Sotelo, Aaron C. Courville, and Yoshua Bengio. "SampleRNN: An Unconditional End-to-End Neural Audio Generation Model". In: *ArXiv* abs/1612.07837 (2017) (cit. on p. 31).

[157] Nal Kalchbrenner et al. "Efficient Neural Audio Synthesis". In: *Proceedings of the 35th International Conference on Machine Learning*. Ed. by Jennifer Dy and Andreas Krause. Vol. 80. Proceedings of Machine Learning Research. PMLR, July 10, 2018, pp. 2410–2419 (cit. on p. 31).

[158] Chris Donahue, Julian McAuley, and Miller Puckette. "Adversarial Audio Synthesis". In: *ICLR*. 2019 (cit. on p. 31).

[159] Jesse Engel, Kumar Krishna Agrawal, Shuo Chen, Ishaan Gulrajani, Chris Donahue, and Adam Roberts. "GANSynth: Adversarial Neural Audio Synthesis". In: *Proceedings of the International Conference on Learning Representations*. 2019 (cit. on p. 31).

[160] Sean Vasquez and Mike Lewis. "MelNet: A Generative Model for Audio in the Frequency Domain". In: *ArXiv* abs/1906.01083 (2019) (cit. on p. 31).

[161] Javier Nistal, Stefan Lattner, and Gaël Richard. "DrumGAN: Synthesis of Drum Sounds With Timbral Feature Conditioning Using Generative Adversarial Networks". In: *ISMIR*. 2020 (cit. on p. 31).

[162] Merlijn Blaauw and Jordi Bonada. "A Neural Parametric Singing Synthesizer Modeling Timbre and Expression from Natural Songs". In: *Appl. Sci.* 7.12 (12 Dec. 2017), p. 1313 (cit. on p. 32).

[163] Krishna Subramani, Preeti Rao, and Alexandre D'Hooge. "Vapar Synth - A Variational Parametric Model for Audio Synthesis". In: *ICASSP 2020 - 2020 IEEE International Conference on Acoustics, Speech and Signal Processing (ICASSP)*. May 2020, pp. 796–800 (cit. on p. 32).

[164] Jesse Engel, Lamtharn Hantrakul, Chenjie Gu, and Adam Roberts. "DDSP: Differentiable Digital Signal Processing". In: *ArXiv* abs/2001.04643 (2020) (cit. on pp. 32, 151).

[165] Edmond T. Johnson. "Player Piano". In: *Oxford Music Online*. Oxford University Press, Oct. 2013 (cit. on pp. 32, 36).

[166] David Moffat and Joshua D. Reiss. "Perceptual Evaluation of Synthesized Sound Effects". In: *ACM Trans. Appl. Percept.* 15.2 (Apr. 2018), pp. 1–19 (cit. on p. 32).

[167] Paulo Esquef, Vesa Välimäki, and Matti Karjalainen. "Restoration and Enhancement of Solo Guitar Recordings Based on Sound Source Modeling". In: *J. Audio Eng. Soc.* 50.4 (2002), pp. 227–236 (cit. on p. 32).

[168] Balazs Bank and Juliette Chabassier. "Model-Based Digital Pianos: From Physics to Sound Synthesis". In: *IEEE Signal Process. Mag.* 36.1 (Jan. 2019), pp. 103–114 (cit. on p. 32).

[169] Leonardo Gabrielli, Stefano Squartini, and Vesa Välimäki. "A Subjective Validation Method for Musical Instrument Emulation". In: *J. Audio Eng. Soc.* (2011) (cit. on p. 32).

[170] Alexander Lerch, Claire Arthur, Ashis Pati, and Siddharth Gururani. "Music Performance Analysis: A Survey". In: *Prooceedings of the International Society for Music Information Retrieval Conference*. Delft, The Netherlands, Nov. 2019 (cit. on pp. 33, 119).

[171] Emmanouil Benetos, Simon Dixon, Zhiyao Duan, and Sebastian Ewert. "Automatic Music Transcription: An Overview". In: *IEEE Signal Process. Mag.* 36.1 (Jan. 2019), pp. 20–30 (cit. on pp. 33, 99, 119, 120).

[172] Georg Von Békésy. *Feedback Phenomena between the Stringed Instrument and the Musician*. Rockefeller University Press, 1968 (cit. on p. 34).

[173] Sten Ternström. "Long-Time Average Spectrum Characteristics of Different Choirs in Different Rooms". In: *Voice UK* 2 (1989), pp. 55–77 (cit. on p. 34).

[174] G. M. Naylor. "A Laboratory Study of Interactions Between Reverberation, Tempo and Musical Synchronization". In: *Acta Acust. United Acust.* (1992) (cit. on p. 34).

[175] S. Bolzinger, O. Warusfel, and E. Kahle. "A Study of the Influence of Room Acoustics on Piano Performance". In: *J. Phys. Iv* 04 (1994) (cit. on p. 34).

[176] Alf Gabrielsson. "The Performance of Music". In: *The Psychology of Music (Second Edition)*. Ed. by Diana Deutsch. Second Edition. Cognition and Perception. San Diego: Academic Press, 1999, pp. 501–602 (cit. on pp. 34, 35).

[177] Kanako Ueno, Takako Kanamori, and Hideki Tachibana. "Experimental Study on Stage Acoustics for Ensemble Performance in Chamber Music". In: *Acoust. Sci. Technol.* 26.4 (2005), pp. 345–352 (cit. on p. 34).

[178] Kanako Ueno and Hideki Tachibana. "Cognitive Modeling of Musician's Perception in Concert Halls". In: *Acoust. Sci. Technol.* 26.2 (2005), pp. 156–161 (cit. on p. 34).

[179] Kanako Ueno, Kosuke Kato, and Keiji Kawai. "Effect of Room Acoustics on Musicians' Performance. Part I: Experimental Investigation with a Conceptual Model". In: *Acta Acust. United Acust.* 96.33 (2010), pp. 505–515 (cit. on pp. 34, 35).

[180] Kosuke Kato, Kanako Ueno, and Keiji Kawai. "Effect of Room Acoustics on Musicians' Performance. Part Ii: Audio Analysis of the Variations in Performed Sound Signals". In: *Acta Acust. United Acust.* 101.44 (2015), pp. 743–759 (cit. on p. 34).

[181] Zora S. Kalkandjiev and Stefan Weinzierl. "The Influence of Room Acoustics on Solo Music Performance: An Experimental Study". In: *Psychomusicology Music Mind Brain* 25.3 (2015), pp. 195–207 (cit. on pp. 34, 134).

[182] Paul Luizard, Erik Brauer, Stefan Weinzierl, and Nathalie Henrich Bernardoni. "How Singers Adapt to Room Acoustical Conditions". In: *Proceedings of the Institute of Acoustics*. 2018 (cit. on p. 34).

[183] Zoran Potočan. "Aesthetic Perception of the Singing Voice in Relation to the Acoustic Conditions". PhD thesis. University of Ljubljana, 2020 (cit. on p. 34).

[184] S. V. A. Garí, Malte Kob, and Tapio Lokki. "Analysis of Trumpet Performance Adjustments Due to Room Acoustics". In: *Proceedings of the International Symposium on Room Acoustics*. 2019 (cit. on p. 34).

[185] Malte Kob, Sebastià V. Amengual Garí, and Zora Schärer Kalkandjiev. "Room Effect on Musicians' Performance". In: *The Technology of Binaural Understanding*. Ed. by Jens Blauert and Jonas Braasch. Cham: Springer International Publishing, 2020, pp. 223–249 (cit. on p. 34).

[186] M. Xu, Z. Wang, and G. G. Xia. "Transferring Piano Performance Control across Environments". In: *ICASSP 2019 - 2019 IEEE International Conference on Acoustics, Speech and Signal Processing (ICASSP)*. Proceedings of the IEEE International Conference on Acoustics Speech and Signal Processing. 2019, pp. 221–225 (cit. on pp. 35, 100, 101).

[187] Zora Schärer Kalkandjiev. "The Influence of Room Acoustics on Solo Music Performances : An Empirical Investigation". PhD thesis. TU Berlin, 2015 (cit. on p. 35).

[188] Kanako Ueno, Kosuke Kato, and Keiji Kawai. "Effect of Room Acoustics on Musicians' Performance. Part I: Experimental Investigation with a Conceptual Model". In: *Acta Acust. United Acust.* 96.33 (2010), pp. 505–515 (cit. on p. 35).

[189] John Butt. "Authenticity". In: *Oxford Music Online*. Oxford University Press, 2001 (cit. on p. 36).

[190] Harry Haskell. "Early Music". In: *Oxford Music Online*. Oxford University Press, 2001 (cit. on p. 36).

[191] Curtis Hawthorne, Andriy Stasyuk, Adam Roberts, Ian Simon, Cheng-Zhi Anna Huang, Sander Dieleman, Erich Elsen, Jesse H. Engel, and Douglas Eck. "Enabling Factorized Piano Music Modeling and Generation with the MAESTRO Dataset." In: *International Conference on Learning Representations*. 2019 (cit. on pp. 37, 40, 95, 127).

[192] Eric Brochu, Vlad M. Cora, and Nando de Freitas. *A Tutorial on Bayesian Optimization of Expensive Cost Functions, with Application to Active User Modeling and Hierarchical Reinforcement Learning*. 2010 (cit. on p. 37).

[193] Reza Alizadeh, Janet K. Allen, and Farrokh Mistree. "Managing Computational Complexity Using Surrogate Models: A Critical Review". In: *Res. Eng. Des.* 31.3 (July 1, 2020), pp. 275–298 (cit. on p. 37).

[194] Qiuqiang Kong, Bochen Li, Xuchen Song, Yuan Wan, and Yuxuan Wang. "High-Resolution Piano Transcription with Pedals by Regressing Onset and Offset Times". In: *IEEEACM Trans. Audio Speech Lang. Process.* (2021), pp. 1–1 (cit. on pp. 39, 40, 88, 90, 119, 151).

[195] Werner Goebl, Roberto Bresin, and Alexander Galembo. "Once Again: The Perception of Piano Touch and Tone. Can Touch Audibly Change Piano Sound Independently of Intensity?" In: *Proceedings of the International Symposium on Musical Acoustics*. 2004 (cit. on p. 39).

[196] Werner Goebl, Roberto Bresin, and Ichiro Fujinaga. "Perception of Touch Quality in Piano Tones". In: *The Journal of the Acoustical Society of America* 136.5 (Nov. 2014), pp. 2839–2850 (cit. on p. 39).

[197] Michel Bernays and Caroline Traube. "Investigating Pianists Individuality in the Performance of Five Timbral Nuances through Patterns of Articulation, Touch, Dynamics, and Pedaling". In: *Front. Psychol.* 5 (Mar. 2014) (cit. on pp. 39, 125, 134).

[198] Michel Bernays and Caroline Traube. "Expressive Production of Piano Timbre: Touch and Playing Techniques for Timbre Control in Piano Performance". In: *Sound Music Comput. Conf. SMC* (2013) (cit. on p. 39).

[199] Camille Adkison, Eric Rokni, Lauren Neldner, and Thomas Moore. "Controlling Piano Tone by Varying the "Weight" Applied on the Key". In: *Proceedings of the 2017 International Symposium on Musical Acoustics*. Ed. by Gary Scavone, Esteban Maestre, Connor Kemp, and Song Wang. Montreal, Canada, 2017 (cit. on p. 39).

[200] Goffredo Haus and Maurizio Longari. "A Multi-Layered, Time-Based Music Description Approach Based on XML". In: *Comput. Music J.* 29.1 (2005), pp. 70–85 (cit. on p. 44).

[201] Adam Lindsay and Werner Kriechbaum. "There's More than One Way to Hear It: Multiple Representations of Music in MPEG-7". In: *J. New Music Res.* 28.4 (1999), pp. 364–372 (cit. on p. 44).

[202] Jacques Steyn. "Framework for a Music Markup Language". In: *Proceeding of the First International IEEE Conference on Musical Application Using XML (MAX2002)*. 2002, pp. 22–29 (cit. on p. 44).

[203] Denis L. Baggi and Goffredo Haus. *Music Navigation with Symbols and Layers: Toward Content Browsing with IEEE 1599 XML Encoding*. John Wiley & Sons, 2013 (cit. on p. 44).

[204] Adriano Baratè, Goffredo Haus, and Luca A. Ludovico. "A Critical Review of the IEEE 1599 Standard". In: *Comput. Stand. Interfaces* 46 (May 2016), pp. 46–51 (cit. on pp. 44, 46).

[205] *Introducing MNX | Music Notation Community Group*. July 21, 2020 (cit. on p. 45).

[206] Denis L. Baggi and Goffredo Haus. "IEEE 1599: Music Encoding and Interaction." In: *IEEE Comput.* 42.3 (2009), pp. 84–87 (cit. on p. 45).

[207] Adriano Baratè, Goffredo Haus, and Luca A. Ludovico. "A Critical Review of the IEEE 1599 Standard". In: *Comput. Stand. Interfaces* 46 (May 2016), pp. 46–51 (cit. on pp. 46, 51).

[208] Adriano Baratè and Luca Andrea Ludovico. "IEEE 1599 Applications for Entertainment and Education". In: *Music Navigation with Symbols and Layers: Toward Content Browsing with IEEE 1599 XML Encoding*. Ed. by Denis Baggi and Goffredo Haus. Hoboken: John Wiley and Sons, 2013, pp. 115–132 (cit. on p. 46).

[209] Adriano Baratè, Goffredo Haus, Luca Andrea Ludovico, and Davide Andrea Mauro. "IEEE 1599 for Live Musical and Theatrical Performances". In: *J. Multimed.* 7.2 (2012), pp. 170–178 (cit. on p. 46).

[210] Adriano Baratè, Luca Andrea Ludovico, Stavros Ntalampiras, and Giorgio Presti. *2019 International Workshop on Multilayer Music Representation and Processing (MMRP).* IEEE Conference Publishing Services (CPS), Jan. 2019 (cit. on p. 46).

[211] Bochen Li, Xinzhao Liu, Karthik Dinesh, Zhiyao Duan, and Gaurav Sharma. "Creating a Multitrack Classical Music Performance Dataset for Multimodal Music Analysis: Challenges, Insights, and Applications." In: *IEEE Trans. Multimed.* 21.2 (Feb. 2019), pp. 522–535 (cit. on p. 48).

[212] Matthias Dorfer, Jan Hajič jr., Andreas Arzt, Harald Frostel, and Gerhard Widmer. "Learning Audio–Sheet Music Correspondences for Cross-Modal Retrieval and Piece Identification". In: *Trans. Int. Soc. Music Inf. Retr.* 1.1 (Sept. 2018), p. 22 (cit. on p. 48).

[213] Katerina Kosta, Oscar F. Bandtlow, and Elaine Chew. "MazurkaBL: Score-Aligned Loudness, Beat, and Expressive Markings Data for 2000 Chopin Mazurka Recordings". In: *Proceedings of the International Conference on Technologies for Music Notation and Representation*. Montréal, Canada: Concordia University, May 2018, pp. 85–94 (cit. on p. 48).

[214] Adriano Baratè and Luca Andrea Ludovico. "Local and Global Semantic Networks for the Representation of Music Information". In: *J. E-Learn. Knowl. Soc.* 12.4 (2016), pp. 109–123 (cit. on p. 49).

[215] Federico Simonetta, Carlos Cancino-Chacón, Stavros Ntalampiras, and Gerhard Widmer. "A Convolutional Approach to Melody Line Identification in Symbolic Scores". In: *20th Int. Conf. on Music Information Retrieval Conference (ISMIR)*. 2019 (cit. on pp. 49, 70).

[216] Stuart L. Weibel and Traugott Koch. "The Dublin Core Metadata Initiative". In: *-Lib Mag.* 6.12 (2000), pp. 1082–9873 (cit. on p. 51).

[217] Henriette D. Avram. "Machine-Readable Cataloging (MARC) Program". In: *Encycl. Libr. Inf. Sci.* 3 (2003), p. 1712 (cit. on p. 51).

[218] Linda Cantara. "METS: The Metadata Encoding and Transmission Standard". In: *Cat. Classif. Q.* 40.3-4 (2005), pp. 237–253 (cit. on p. 51).

[219] Renato Ianella. "Open Digital Rights Language (ODRL)". In: *Open Content Licens. Cultiv. Creat. Commons* (2007) (cit. on p. 51).

[220] Monya Baker. "1,500 Scientists Lift the Lid on Reproducibility". In: *Nat. News* 533.7604 (2016), p. 452 (cit. on p. 53).

[221] Matthew Hutson. "Artificial Intelligence Faces Reproducibility Crisis". In: *Science* 359.6377 (2018), pp. 725–726 (cit. on p. 53).

[222] Yaser S. Abu-Mostafa, Malik Magdon-Ismail, and Hsuan-Tien Lin. *Learning From Data*. Vol. 4. AMLBook, 2012 (cit. on p. 53).

[223] Rachel M Bittner, Magdalena Fuentes, David Rubinstein, Andreas Jansson, Keunwoo Choi, and Thor Kell. “MIRDATA: SOFTWARE FOR REPRODUCIBLE USAGE OF DATASETS”. In: *International Conference on Music Information Retrieval (ISMIR)*. 2019, p. 8 (cit. on p. 54).

[224] Curtis Hawthorne, Andriy Stasyuk, Adam Roberts, Ian Simon, Cheng-Zhi Anna Huang, Sander Dieleman, Erich Elsen, Jesse H. Engel, and Douglas Eck. “Enabling Factorized Piano Music Modeling and Generation with the MAESTRO Dataset.” In: (2019) (cit. on pp. 55, 60).

[225] Meinard Müller, Verena Konz, Wolfgang Bogler, and Vlora Arifi-Müller. “Saarland Music Data (SMD)”. In: *ISMIR* (2011) (cit. on p. 55).

[226] Werner Goebl. *The Vienna 4x22 Piano Corpus*. 1999 (cit. on pp. 55, 60).

[227] John Thickstun, Zaïd Harchaoui, Dean P. Foster, and Sham M. Kakade. “Invariances and Data Augmentation for Supervised Music Transcription.” In: (2018), pp. 2241–2245 (cit. on pp. 55, 60).

[228] Marius Miron, Julio J. Carabias-Orti, Juan J. Bosch, Emilia G ómez, and Jordi Janer. “Score-Informed Source Separation for Multichannel Orchestral Recordings.” In: *J Electr. Comput. Eng.* 2016 (2016), 8363507:1–8363507:19 (cit. on pp. 55, 57).

[229] Juan P. Braga Brum. *Traditional Flute Dataset for Score Alignment”*. 2018 (cit. on pp. 55, 60).

[230] Zhiyao Duan and Bryan Pardo. “Soundprism: An Online System for Score-Informed Source Separation of Music Audio.” In: *J Sel Top. Signal Process.* 5.6 (2011), pp. 1205–1215 (cit. on pp. 55, 57, 60).

[231] Joachim Fritsch. *The TRIOS Score-Aligned Multitrack Recordings Dataset*. 2012 (cit. on p. 55).

[232] Federico Simonetta. “Graph Based Representation of the Music Symbolic Level. A Music Information Retrieval Application”. Thesis. Università di Padova, Apr. 2018 (cit. on p. 55).

[233] Francesco Foscarin, Andrew Mcleod, Philippe Rigaux, Florent Jacquemard, and Masahiko Sakai. “ASAP: A Dataset of Aligned Scores and Performances for Piano Transcription”. In: (2020) (cit. on pp. 59, 60).

[234] Eita Nakamura, Kazuyoshi Yoshii, and Haruhiro Katayose. “Performance Error Detection and Post-Processing for Fast and Accurate Symbolic Music Alignment.” In: (2017). Ed. by Sally Jo Cunningham, Zhiyao Duan, Xiao Hu, and Douglas Turnbull, pp. 347–353 (cit. on pp. 59, 88, 89, 151).

[235] Tuomas Eerola and Petri Toiviainen. “MIR In Matlab: The MIDI Toolbox.” In: *5th Int. Conf. on Music Information Retrieval (ISMIR)*. 2004 (cit. on p. 61).

[236] Rachel M Bittner, Magdalena Fuentes, David Rubinstein, Andreas Jansson, Keunwoo Choi, and Thor Kell. "MIRDATA: SOFTWARE FOR REPRODUCIBLE USAGE OF DATASETS". In: *International Conference on Music Information Retrieval (ISMIR)*. 2019, p. 8 (cit. on p. 64).

[237] Felix Salzer and Carl Schachter. *Counterpoint in Composition*. New York, NY, USA: Columbia University Press, 1989 (cit. on p. 69).

[238] Walter Piston. *Counterpoint*. W. W. Norton & Company, 1947 (cit. on p. 69).

[239] Werner Goebl. "Melody Lead in Piano Performance: Expressive Device or Artifact?" In: *J. Acoust. Soc. Am.* 110.1 (2001), pp. 563–572 (cit. on p. 70).

[240] Werner Goebl. "Geformte Zeit in Der Musik". In: *Zeit in Den Wissenschaften*. Ed. by W. Kautek, R. Neck, and H. Schmidinger. Vol. 19. Wien, Köln, Weimar: Böhlau Verlag, 2016, pp. 179–199 (cit. on p. 70).

[241] Warren Brodsky, Avishai Henik, Bat-Sheva Rubinstein, and Moshe Zorman. "Auditory Imagery from Musical Notation in Expert Musicians". In: *Percept. Psychophys.* 65.4 (2003), pp. 602–6012 (cit. on p. 70).

[242] David Huron. "Tone and Voice: A Derivation of the Rules of Voice-Leading from Perceptual Principles". In: *Music Percept.* 19.1 (2001), pp. 1–64 (cit. on pp. 70, 71).

[243] Emilios Cambouropoulos. "Voice and Stream: A Perceptual and Computational Modeling of Voice Separation". In: *Music Percept.* 26.1 (2008), pp. 75–94 (cit. on pp. 70, 71).

[244] Patrick Gray and Razvan C. Bunescu. "A Neural Greedy Model for Voice Separation in Symbolic Music". In: *Proceedings of the 17th International Society for Music Information Retrieval Conference, ISMIR 2016, New York City, United States, August 7-11, 2016*. Ed. by Michael I. Mandel, Johanna Devaney, Douglas Turnbull, and George Tzanetakis. 2016, pp. 782–788 (cit. on p. 71).

[245] John M. Geringer, K. Madsen Clifford, and Rebecca McLeod. "Effects of Articulation Styles on Perception of Modulated Tempos in Violin Excerpts". In: *Int. J. Music Educ.* 25.2 (2006), pp. 165–175 (cit. on p. 71).

[246] Nicolas Guiomard-Kagan, Mathieu Giraud, Richard Groult, and Florence Levé. "Improving Voice Separation by Better Connecting Contigs". In: *Proceedings of the 17\textsuperscriptth International Society for Music Information Retrieval Conference, ISMIR 2016, New York City, United States, August 7-11, 2016*. Ed. by Michael I. Mandel, Johanna Devaney, Douglas Turnbull, and George Tzanetakis. 2016, pp. 164–170 (cit. on pp. 71, 72).

[247] Elaine Chew and Xiaodan Wu. "Separating Voices in Polyphonic Music: A Contig Mapping Approach". In: *Computer Music Modeling and Retrieval: Second International Symposium, CMMR 2004, Esbjerg, Denmark, May 26-29, 2004, Revised Papers*. Ed. by Uffe Kock Wiil. Vol. 3310. Lecture Notes in Computer Science. Berlin, Heidelberg: Springer Berlin Heidelberg, 2005, pp. 1–20 (cit. on pp. 71, 72, 79).

[248] Zhi-gang Huang, Chang-le Zhou, and Min-juan Jiang. "Melodic Track Extraction for MIDI". In: *J. Xiamen Univ. Nat. Sci.* 1 (2010) (cit. on p. 71).

[249] Raúl Martín, Ramón A. Mollineda, and Vicente Garc\'i a. "Melodic Track Identification in MIDI Files Considering the Imbalanced Context". In: *Pattern Recognition and Image Analysis, 4\textsuperscriptth Iberian Conference, IbPRIA 2009, Póvoa de Varzim, Portugal, June 10-12, 2009, Proceedings*. Ed. by Helder Araújo, Ana Maria Mendonça, Armando J. Pinho, and M. Inés Torres. Vol. 5524 LNCS. Lecture Notes in Computer Science. Springer, 2009, pp. 489–496 (cit. on p. 71).

[250] Jiangtao Li, Xiaohong Yang, and Qingcai Chen. "MIDI Melody Extraction Based on Improved Neural Network". In: *Proceedings of the 2009 International Conference on Machine Learning and Cybernetics*. Vol. 2. Machine Learning and Cybernetics, 2009 International Conference On. IEEE, July 2009, pp. 1133–1138 (cit. on p. 71).

[251] Anders Friberg and Sven Ahlbäck. "Recognition of the Main Melody in a Polyphonic Symbolic Score Using Perceptual Knowledge". In: *J. New Music Res.* 38.2 (2009), pp. 155–169 (cit. on p. 71).

[252] Rachel M. Bittner, Justin Salamon, Slim Essid, and Juan Pablo Bello. "Melody Extraction by Contour Classification". In: *Proceedings of the 16th International Society for Music Information Retrieval Conference (ISMIR 2015)*. Malaga, Spain, 2015 (cit. on p. 71).

[253] Juan J. Bosch, Rachel M. Bittner, Justin Salamon, and Emilia G\'o mez. "A Comparison of Melody Extraction Methods Based on Source-Filter Modelling". In: *Proceedings of the 17th International Society for Music Information Retrieval Conference (ISMIR 2016)*. New York, New York, USA, 2016 (cit. on p. 71).

[254] Justin Salamon, Rachel M. Bittner, Jordi Bonada, Juan J. Bosch, Emilia Gómez, and Juan Pablo Bello. "An Analysis/Synthesis Framework for Automatic F0 Annotation of Multitrack Datasets." In: *Proceedings of the 18th International Society for Music Information Retrieval Conference (ISMIR 2017)*. Ed. by Sally Jo Cunningham, Zhiyao Duan, Xiao Hu, and Douglas Turnbull. Suzhou, China, 2017, pp. 71–78 (cit. on p. 71).

[255] Nicolas Guiomard-Kagan, Mathieu Giraud, Richard Groult, and Florence Levé. "Comparing Voice and Stream Segmentation Algorithms". In: *Proceedings of the 16th International Society for Music Information Retrieval Conference (ISMIR 2015)*. Malaga, Spain, 2015 (cit. on pp. 71, 72).

[256] Cihan Isikhan and Giyasettin Ozcan. "A Survey of Melody Extraction Techniques for Music Information Retrieval". In: *Proceedings of the Fourth Conference on Interdisciplinary Musicology (CIM08)*. Thessaloniki, Greece, 2008 (cit. on p. 71).

[257] Frans Wiering, Justin de Nooijer, Anja Volk, and Hermi J. M. Tabachneck-Schijf. "Cognition-Based Segmentation for Music Information Retrieval Systems". In: *J. New Music Res.* 38.2 (2009), pp. 139–154 (cit. on p. 71).

[258] Reinier de Valk and Tillman Weyde. "Deep Neural Networks with Voice Entry Estimation Heuristics for Voice Separation in Symbolic Music Representations". In: *Proceedings of the 19th International Society for Music Information Retrieval Conference (ISMIR 2018)*. Paris, France, 2018, pp. 281–288 (cit. on p. 71).

[259] Wei Chai and Barry Vercoe. “Melody Retrieval on the Web”. In: *Multimedia Computing and Networking 2002.* Ed. by Martin G. Kienzle and Prashant J. Shenoy. SPIE, Dec. 2001 (cit. on p. 72).

[260] Alexandra L. Uitdenbogerd and Justin Zobel. “Manipulation of Music for Melody Matching”. In: *Proceedings of the 6th ACM International Conference on Multimedia ’98, Bristol, England, September 12-16, 1998.* Ed. by Wolfgang Effelsberg and Brian C. Smith. ACM, 1998, pp. 235–240 (cit. on pp. 72, 79).

[261] Zheng Jiang and Roger B. Dannenberg. “Melody Identification in Standard MIDI Files”. In: *Proceedings of the 16\textsuperscriptth Sound & Music Computing Conference*. Màlaga, Spain, May 28, 2019, pp. 65–71 (cit. on p. 72).

[262] Andrew McLeod and Mark Steedman. “HMM-Based Voice Separation of MIDI Performance”. In: *J. New Music Res.* 45.1 (Jan. 2016), pp. 17–26 (cit. on p. 72).

[263] Daniel Müllner. “Fastcluster: Fast Hierarchical, Agglomerative Clustering Routines for R and Python”. In: *J. Stat. Softw.* 53.9 (2013), pp. 1–18 (cit. on p. 74).

[264] Matthew D. Zeiler. “AdaDelta: An Adaptive Learning Rate Method”. In: (2012) (cit. on p. 75).

[265] Sergey Ioffe and Christian Szegedy. *Batch Normalization: Accelerating Deep Network Training by Reducing Internal Covariate Shift*. 2015 (cit. on p. 75).

[266] N. Morgan and H. Bourlard. “Generalization and Parameter Estimation in Feedforward Nets: Some Experiments”. In: *Advances in Neural Information Processing Systems 2.* Ed. by D. S. Touretzky. Morgan-Kaufmann, 1990, pp. 630–637 (cit. on p. 75).

[267] Leo Breiman. “Random Forests”. In: *Mach. Learn.* 45 (2004), pp. 5–32 (cit. on p. 80).

[268] Johanna Devaney. “Estimating Onset and Offset Asynchronies in Polyphonic Score-Audio Alignment”. In: *J. New Music Res.* 43.3 (July 2014), pp. 266–275 (cit. on p. 86).

[269] Alexander Lerch, Claire Arthur, Ashis Pati, and Siddharth Gururani. “Music Performance Analysis: A Survey”. In: (Nov. 2019) (cit. on pp. 86, 89).

[270] Julio José Carabias-Orti, Francisco J. Rodríguez-Serrano, Pedro Vera-Candeas, Nicolás Ruiz-Reyes, and Francisco J. Cañadas-Quesada. “An Audio to Score Alignment Framework Using Spectral Factorization and Dynamic Time Warping.” In: *Proceedings of the 16\textsuperscriptth International Society for Music Information Retrieval Conference*. Ed. by Meinard Müller and Frans Wiering. Málaga, Spain: ISMIR, Oct. 2015, pp. 742–748 (cit. on p. 86).

[271] Marius Miron, Julio José Carabias-Orti, and Jordi Janer. “Audio-to-Score Alignment at the Note Level for Orchestral Recordings”. In: (2014). Ed. by Hsin-Min Wang, Yi-Hsuan Yang, and Jin Ha Lee, pp. 125–130 (cit. on p. 86).

[272] Siying Wang, Sebastian Ewert, and Simon Dixon. “Compensating for Asynchronies between Musical Voices in Score-Performance Alignment”. In: *2015 IEEE International Conference on Acoustics, Speech and Signal Processing (ICASSP).* IEEE, Apr. 2015 (cit. on p. 86).

[273] Xueyang Wang, Ryan Stables, Bochen Li, and Zhiyao Duan. "Score-Aligned Polyphonic Microtiming Estimation". In: (2018) (cit. on p. 86).

[274] Ruchit Agrawal and Simon Dixon. "Learning Frame Similarity Using Siamese Networks for Audio-to-Score Alignment". In: (Jan. 2021), pp. 141–145 (cit. on p. 86).

[275] Taegyun Kwon, Dasaem Jeong, and Juhan Nam. "Audio-to-Score Alignment of Piano Music Using RNN-Based Automatic Music Transcription". In: (2017) (cit. on pp. 86, 88).

[276] Abdullah Mueen and Eamonn Keogh. "Extracting Optimal Performance from Dynamic Time Warping". In: *Proceedings of the 22nd ACM SIGKDD International Conference on Knowledge Discovery and Data Mining*. KDD '16. New York, NY, USA: Association for Computing Machinery, Aug. 13, 2016, pp. 2129–2130 (cit. on p. 88).

[277] Stan Salvador and Philip Chan. "Toward Accurate Dynamic Time Warping in Linear Time and Space". In: *Intell. Data Anal.* 11.5 (2007), pp. 561–580 (cit. on p. 88).

[278] Sebastian Ewert, Meinard Muller, and Peter Grosche. "High Resolution Audio Synchronization Using Chroma Onset Features". In: (Apr. 2009) (cit. on p. 88).

[279] Emmanouil Benetos, Simon Dixon, Zhiyao Duan, and Sebastian Ewert. "Automatic Music Transcription: An Overview". In: *IEEE Signal Process. Mag.* 36.1 (Jan. 2019), pp. 20–30 (cit. on p. 88).

[280] Curtis Hawthorne, Erich Elsen, Jialin Song, Adam Roberts, Ian Simon, Colin Raffel, Jesse Engel, Sageev Oore, and Douglas Eck. "Onsets and Frames: Dual-Objective Piano Transcription". In: *Proc. 19th Int. Soc. Music Inf. Retr. Conf. ISMIR* (2018). Ed. by Emilia Gómez, Xiao Hu, Eric Humphrey, and Emmanouil Benetos, pp. 50–57 (cit. on pp. 88, 111, 116, 119, 151).

[281] Yujia Yan, Frank Cwitkowitz, and Zhiyao Duan. "Skipping the Frame-Level: Event-Based Piano Transcription With Neural Semi-CRFs". In: *NeurIPS*. 2021 (cit. on pp. 88, 119).

[282] Qiuqiang Kong, Bochen Li, Xuchen Song, Yuan Wan, and Yuxuan Wang. "High-Resolution Piano Transcription with Pedals by Regressing Onset and Offset Times". In: *IEEEACM Trans. Audio Speech Lang. Process.* (2021), pp. 1–1 (cit. on p. 88).

[283] Y. T. Wu, B. Chen, and L. Su. "Multi-Instrument Automatic Music Transcription With Self-Attention-Based Instance Segmentation". In: *IEEEACM Trans. Audio Speech Lang. Process.* 28 (2020), pp. 2796–2809 (cit. on p. 89).

[284] Johanna Devaney. "Estimating Onset and Offset Asynchronies in Polyphonic Score-Audio Alignment". In: *J. New Music Res.* 43.3 (July 2014), pp. 266–275 (cit. on p. 90).

[285] Y. T. Wu, B. Chen, and L. Su. "Multi-Instrument Automatic Music Transcription With Self-Attention-Based Instance Segmentation". In: *IEEEACM Trans. Audio Speech Lang. Process.* 28 (2020), pp. 2796–2809 (cit. on pp. 90, 151).

[286] John Thickstun, Zaïd Harchaoui, Dean P. Foster, and Sham M. Kakade. "Invariances and Data Augmentation for Supervised Music Transcription." In: *International Conference on Acoustics, Speech, and Signal Processing (ICASSP)*. IEEE, 2018, pp. 2241–2245 (cit. on p. 95).

[287] Meinard Müller, Verena Konz, Wolfgang Bogler, and Vlora Arifi-Müller. "Saarland Music Data (SMD)". In: *ISMIR* (2011) (cit. on p. 95).

[288] Zhiyao Duan and Bryan Pardo. "Soundprism: An Online System for Score-Informed Source Separation of Music Audio." In: *J Sel Top. Signal Process.* 5.6 (2011), pp. 1205–1215 (cit. on p. 95).

[289] Anssi P. Klapuri. "Automatic Music Transcription as We Know It Today". In: *J. New Music Res.* 33.3 (2004) (cit. on p. 99).

[290] A. Rizzi, M. Antonelli, and M. Luzi. "Instrument Learning and Sparse NMD for Automatic Polyphonic Music Transcription". In: *IEEE Trans Multimed.* 19.7 (2017), pp. 1405–1415 (cit. on p. 100).

[291] Z. Fu, G. Lu, K. M. Ting, and D. Zhang. "A Survey of Audio-Based Music Classification and Annotation". In: *IEEE Trans Multimed.* 13.2 (2011), pp. 303–319 (cit. on p. 100).

[292] Dasaem Jeong, Taegyun Kwon, and Juhan Nam. "Note-Intensity Estimation of Piano Recordings Using Coarsely Aligned MIDI Score". In: *J. Audio Eng. Soc.* 68.1/2 (Feb. 2020), pp. 34–47 (cit. on pp. 100, 111, 125, 128, 129).

[293] Johanna Devaney and Michael I. Mandel. "An Evaluation of Score-Informed Methods for Estimating Fundamental Frequency and Power from Polyphonic Audio". In: *2017 IEEE International Conference on Acoustics, Speech and Signal Processing, ICASSP 2017, New Orleans, LA, USA, March 5-9, 2017*. IEEE, 2017, pp. 181–185 (cit. on p. 100).

[294] M. Akbari and H. Cheng. "Real-Time Piano Music Transcription Based on Computer Vision". In: *IEEE Trans Multimed.* 17.12 (2015), pp. 2113–2121 (cit. on p. 100).

[295] Siying Wang, Sebastian Ewert, and Simon Dixon. "Identifying Missing and Extra Notes in Piano Recordings Using Score-Informed Dictionary Learning". In: *IEEEACM Trans. Audio Speech Lang. Process.* 25.10 (Oct. 2017), pp. 1877–1889 (cit. on pp. 100, 128).

[296] Adrien Ycart, Lele Liu, Emmanouil Benetos, and Marcus T Pearce. "Investigating the Perceptual Validity of Evaluation Metrics for Automatic Piano Music Transcription". In: *Trans. Int. Soc. Music Inf. Retr. TISMIR Accept.* (2020) (cit. on pp. 100, 102, 104, 111, 117).

[297] William Storm. "The Establishment of International Re-Recording Standards". In: *Phonogr. Bull.* 27 (1980), pp. 5–12 (cit. on pp. 101, 137).

[298] Simon J. Godsill and Peter J. W. Rayner. *Digital Audio Restoration*. Springer London, 1998 (cit. on pp. 101, 137).

[299] Angelo Orcalli. "On the Methodologies of Audio Restoration". In: *J. New Music Res.* 30.4 (2001) (cit. on p. 101).

[300] Luca Marinelli, Athanasios Lykartsis, Stefan Weinzierl, and Charalampos Saitis. "Musical Dynamics Classification with CNN and Modulation Spectra". In: *Proceedings of the 17th Sound and Music Computing Conference*. Proceedings of the Sound and Music Computing Conference. Torino, 2020 (cit. on p. 101).

[301] Stefan Weinzierl, Steffen Lepa, Frank Schultz, Erik Detzner, Henrik von Coler, and Gottfried Behler. "Sound Power and Timbre as Cues for the Dynamic Strength of Orchestral Instruments". In: *J. Acoust. Soc. Am.* 144.3 (2018) (cit. on p. 101).

[302] Dasaem Jeong and Juhan Nam. "Note Intensity Estimation of Piano Recordings by Score-Informed Nmf." In: *Int Conf Semantic Audio* (2017). Ed. by Christian Dittmar, Jakob Abeßer, and Meinard Müller (cit. on p. 101).

[303] Roger B. Dannenberg. "The Interpretation of MIDI Velocity." In: *Proceedings of the 2006 International Computer Music Conference*. Michigan Publishing, Nov. 2006 (cit. on p. 101).

[304] Stephen Davies and Stanley Sadie. "Interpretation". In: *Grove Music Online*. Oxford University Press, 2001 (cit. on p. 101).

[305] Laurence Dreyfus. "Beyond the Interpretation of Music". In: *J. Musicol. Res.* 0.0 (2020), pp. 1–26 (cit. on p. 101).

[306] Nicholas Jillings, David Moffat, Brecht De Man, and Joshua D. Reiss. "Web Audio Evaluation Tool: A Browser-Based Listening Test Environment". In: *12th Sound and Music Computing Conference*. July 2015 (cit. on pp. 104, 106, 152).

[307] Martin J. Bergee. "Faculty Interjudge Reliability of Music Performance Evaluation". In: *Journal of Research in Music Education* 51.2 (July 1, 2003), pp. 137–150 (cit. on p. 105).

[308] Michael P. Hewitt and Bret P. Smith. "The Influence of Teaching-Career Level and Primary Performance Instrument on the Assessment of Music Performance". In: *Journal of Research in Music Education* 52.4 (Dec. 1, 2004), pp. 314–327 (cit. on p. 105).

[309] B. Feiten, I. Wolf, Eunmi Oh, Jeongil Seo, and Hae-Kwang Kim. "Audio Adaptation According to Usage Environment and Perceptual Quality Metrics". In: *IEEE Trans Multimed.* 7.3 (2005), pp. 446–453 (cit. on p. 105).

[310] Diemo Schwarz, Guillaume Lemaitre, Mitsuko Aramaki, and Richard Kronland-Martinet. "Effects of Test Duration in Subjective Listening Tests." In: *ICMC*. Michigan Publishing, 2016 (cit. on p. 105).

[311] E. Bigand, S. Vieillard, F. Madurell, J. Marozeau, and A. Dacquet. "Multidimensional Scaling of Emotional Responses to Music: The Effect of Musical Expertise and of the Duration of the Excerpts". In: *Cogn. Emot.* 19.8 (2005), pp. 1113–1139 (cit. on p. 105).

[312] Joel Wapnick, Charlene Ryan, Louise Campbell, Patricia Deek, Renata Lemire, and Alice-Ann Darrow. "Effects of Excerpt Tempo and Duration on Musicians' Ratings of High-Level Piano Performances". In: *J. Res. Music Educ.* 53.2 (July 2005), pp. 162–176 (cit. on p. 105).

[313] John M. Geringer and Christopher M. Johnson. "Effects of Excerpt Duration, Tempo, and Performance Level on Musicians Ratings of Wind Band Performances". In: *J. Res. Music Educ.* 55.4 (Dec. 2007), pp. 289–301 (cit. on p. 107).

[314] Jessica Napoles. "The Effect of Excerpt Duration and Music Education Emphasis on Ratings of High Quality Children's Choral Performances". In: *Bull. Counc. Res. Music Educ.* 179 (2009), pp. 21–32 (cit. on p. 107).

[315] Matthew Williams. "Effect of Excerpt Duration on Adjudicator Ratings of Middle School Band Performances". In: *Res. Perspect. Music Educ.* 18.1 (2016), pp. 16–25 (cit. on p. 107).

[316] Claudio Contardo. "Decremental Clustering for the Solution of P-Dispersion Problems to Proven Optimality". In: *Inf. J. Optim.* 2.2 (Jan. 2020), pp. 134–144 (cit. on pp. 108, 109).

[317] Erhan Erkut. "The Discrete P-Dispersion Problem". In: *Eur. J. Oper. Res.* 46.1 (1990), pp. 48–60 (cit. on p. 107).

[318] S. S. Ravi, Daniel J. Rosenkrantz, and Giri Kumar Tayi. "Heuristic and Special Case Algorithms for Dispersion Problems." In: *Oper. Res.* 42.2 (Apr. 1994), pp. 299–310 (cit. on p. 107).

[319] Brian S Everitt, Sabine Landau, Morven Leese, and Daniel Stahl. "Hierarchical Clustering". In: *Cluster Analysis*. 2011 (cit. on p. 107).

[320] Werner Goebl. *The Vienna 4x22 Piano Corpus*. 1999 (cit. on p. 109).

[321] Dmitry Bogdanov et al. "ESSENTIA: An Open-Source Library for Sound and Music Analysis". In: *Proceedings of the 21\textsuperscriptst ACM International Conference on Multimedia - MM '13*. International Society for Music Information Retrieval (ISMIR), 2013, pp. 855–858 (cit. on pp. 110, 129).

[322] Francesc Alías, Joan Socoró, and Xavier Sevillano. "A Review of Physical and Perceptual Feature Extraction Techniques for Speech, Music and Environmental Sounds". In: *Appl. Sci.* 6.5 (May 2016), p. 143 (cit. on pp. 110, 125).

[323] J. R. Zapata, M. E. P. Davies, and E. Gómez. "Multi-Feature Beat Tracking". In: *IEEEACM Trans. Audio Speech Lang. Process.* 22.4 (Apr. 2014), pp. 816–825 (cit. on p. 110).

[324] Stan Salvador and Philip Chan. "Toward Accurate Dynamic Time Warping in Linear Time and Space". In: *Intell. Data Anal.* 11.5 (2007), pp. 561–580 (cit. on p. 111).

[325] Y. Wu, B. Chen, and L. Su. "Polyphonic Music Transcription with Semantic Segmentation". In: *ICASSP 2019 - 2019 IEEE International Conference on Acoustics, Speech and Signal Processing (ICASSP)*. ICASSP 2019 - 2019 IEEE International Conference on Acoustics, Speech and Signal Processing (ICASSP). May 2019, pp. 166–170 (cit. on pp. 111, 119).

[326] Taegyun Kwon, Dasaem Jeong, and Juhan Nam. "Audio-to-Score Alignment of Piano Music Using RNN-Based Automatic Music Transcription". In: *14th Sound and Music Computing Conference*. 2017 (cit. on pp. 111, 152).

[327] Richard Repp. “Recording Quality Ratings by Music Professionals”. In: *ICMC*. Michigan Publishing, 2006 (cit. on p. 112).

[328] Jan-Erik Mörtberg. *Is Dithered Truncation Preferred over Pure Truncation at a Bit Depth of 16-Bits When a Digital Re-Quantization Has Been Performed on a 24-Bit Sound File?* 2007 (cit. on p. 112).

[329] Mitsunori Mizumachi, Ryuta Yamamoto, and Katsuyuki Niyada. “Discussion on Subjective Characteristics of High Resolution Audio”. In: *J. Audio Eng. Soc.* (2017) (cit. on p. 112).

[330] Jeroen Breebaart. “No Correlation between Headphone Frequency Response and Retail Price”. In: *The Journal of the Acoustical Society of America* 141.6 (June 2017), EL526–EL530 (cit. on p. 114).

[331] Pablo Gutierrez-Parera and Jose J. Lopez. “Perception of Nonlinear Distortion on Emulation of Frequency Responses of Headphones”. In: *The Journal of the Acoustical Society of America* 143.4 (Apr. 2018), pp. 2085–2088 (cit. on p. 114).

[332] Judith M. Tanur. “Margin of Error”. In: *International Encyclopedia of Statistical Science*. Ed. by Miodrag Lovric. Berlin, Heidelberg: Springer Berlin Heidelberg, 2011, pp. 765–765 (cit. on p. 114).

[333] Michael R. Chernick, Wenceslao González-Manteiga, Rosa M. Crujeiras, and Erniel B. Barrios. “Bootstrap Methods”. In: *International Encyclopedia of Statistical Science*. Ed. by Miodrag Lovric. Berlin, Heidelberg: Springer Berlin Heidelberg, 2011, pp. 169–174 (cit. on p. 114).

[334] Colin Raffel, Brian McFee, Eric J. Humphrey, Justin Salamon, Oriol Nieto, Dawen Liang, and Daniel P. W. Ellis. “MIR_EVAL: A Transparent Implementation of Common MIR Metrics.” In: *Proceedings of the 15th International Society for Music Information Retrieval Conference*. Ed. by Hsin-Min Wang, Yi-Hsuan Yang, and Jin Ha Lee. 2014, pp. 367–372 (cit. on p. 116).

[335] Jong Wook Kim and Juan Pablo Bello. *Adversarial Learning for Improved Onsets and Frames Music Transcription*. 2019 (cit. on p. 119).

[336] Kin Wai Cheuk, Yin-Jyun Luo, Emmanouil Benetos, and Dorien Herremans. *The Effect of Spectrogram Reconstruction on Automatic Music Transcription: An Alternative Approach to Improve Transcription Accuracy.* 2020 (cit. on p. 119).

[337] Hongzhou Lin and Stefanie Jegelka. “ResNet with One-Neuron Hidden Layers Is a Universal Approximator”. In: *NeurIPS*. 2018 (cit. on p. 124).

[338] Ding-Xuan Zhou. “Universality of Deep Convolutional Neural Networks”. In: *Appl. Comput. Harmon. Anal.* 48.22 (2020), pp. 787–794 (cit. on p. 124).

[339] Daniel D. Lee and H. Sebastian Seung. “Algorithms for Non-Negative Matrix Factorization”. In: *Advances in Neural Information Processing Systems 13*. Ed. by T. K. Leen, T. G. Dietterich, and V. Tresp. MIT Press, 2001, pp. 556–562 (cit. on pp. 125, 130).

[340] Emmanouil Benetos, Anssi Klapuri, and Simon Dixon. "Score-Informed Transcription for Automatic Piano Tutoring". In: *European Signal Processing Conference*. European Signal Processing Conference. IEEE, 2012, pp. 2153–2157 (cit. on p. 128).

[341] Kaiming He, X. Zhang, Shaoqing Ren, and Jian Sun. "Deep Residual Learning for Image Recognition". In: *2016 IEEE Conf. Comput. Vis. Pattern Recognit. CVPR* (2016), pp. 770–778 (cit. on p. 130).

[342] Matthew D. Zeiler. "AdaDelta: An Adaptive Learning Rate Method". 2012 (cit. on p. 132).

[343] Leslie N. Smith. "Cyclical Learning Rates for Training Neural Networks". In: *2017 IEEE Winter Conference on Applications of Computer Vision (WACV)*. Mar. 2017, pp. 464–472 (cit. on p. 132).

[344] Adrián Javaloy and Isabel Valera. "Rotograd: Dynamic Gradient Homogenization for Multi-Task Learning". In: *ArXiv* abs/2103.02631 (2021) (cit. on p. 132).

[345] Jason Hsu. *Multiple Comparisons*. Chapman and Hall/CRC, Feb. 1996 (cit. on p. 134).

[346] Scott E. Maxwell, Harold D. Delaney, and Ken Kelley. *Designing Experiments and Analyzing Data*. Routledge, Sept. 2017 (cit. on p. 134).

[347] Sebastian Ewert, Meinard Muller, and Peter Grosche. "High Resolution Audio Synchronization Using Chroma Onset Features". In: *2009 IEEE International Conference on Acoustics, Speech and Signal Processing*. IEEE, Apr. 2009 (cit. on p. 152).